\pdfoutput=1
\documentclass[twoside,letterpaper,titlepage,nofootinbib,prd,11pt,floatfix,unsortedaddress,superscriptaddress]{revtex4-1}

\usepackage{ifpdf}
\usepackage[english]{babel}
\usepackage{subfigure}
\usepackage{graphicx}
\usepackage{epsfig}
\usepackage{amsmath}
\usepackage{amssymb}
\usepackage{amsfonts}
\usepackage{slashed}
\usepackage{bbm}
\usepackage{ctable,longtable}
\usepackage{cancel}
\usepackage{fancyhdr,graphicx}
\usepackage{color}
\usepackage{lineno}
\usepackage{xcolor}
\def\mum{\,\mu\text{m}}

\def\SU2U1{{\rm SU}(2)\times{\rm U}(1)}

\mathchardef\qsm=63
\mathchardef\pls=43
\mathchardef\mns=512
\mathchardef\plm=518
\mathchardef\eql=61
\mathchardef\smallleft=300
\mathchardef\smallright=301
\mathchardef\perslsh=47
\mathchardef\les=316
\mathchardef\gre=318
\mathchardef\leq=532
\mathchardef\grq=533
\chardef\usc=95
\chardef\til=126

\def\sqr#1#2#3{{\vcenter{\hrule height.#3ex\hbox{\vrule width.#2ex height#1ex
    \kern#1ex\vrule width.#3ex}\hrule height.#2ex}}}

\def\angleto{\vrule width.035em height2.1ex depth-.56ex\unskip\kern-.6ex\to}
\def\perchc#1{{\raise.4ex\hbox{$\mkern4mu#1{\it\perslsh}_
             {\mkern-5mu\scriptscriptstyle{{\rm o}\!{\rm o}}}^
             {\mkern-12.8mu\scriptscriptstyle{\rm o}}$}}}

\catcode`\@=11 
\def\parenbar{\mathpalette\p@renb@r}
\def\p@renb@r#1#2{\vbox{%
  \ifx#1\scriptscriptstyle \dimen@.7em\dimen@ii.2em\else
  \ifx#1\scriptstyle \dimen@.8em\dimen@ii.25em\else
  \dimen@1em\dimen@ii.4em\fi\fi \offinterlineskip
  \ialign{\hfill##\hfill\cr
    \vbox{\hrule width\dimen@ii}\cr
    \noalign{\vskip-.3ex}%
    \hbox to\dimen@{$\mathchar300\hfil\mathchar301$}\cr
    \noalign{\vskip-.3ex}%
    $#1#2$\cr}}}
\catcode`\@=12 

\newbox\struttbox
\setbox\struttbox=\hbox{\vrule height1.65ex depth.485ex width0pt}
\def\strutt{\relax\ifmmode\copy\struttbox\else\unhcopy\struttbox\fi}
\def\stru#1#2{\relax\ifmmode\hbox{\vrule height#1 depth#2 width0pt}
\else\vrule height#1 depth#2 width0pt\fi}

\def\ronum#1{\uppercase\expandafter{\romannumeral#1}}
\def\ronuml#1{\expandafter{\romannumeral#1}}

\DeclareMathAlphabet{\mathbf}{OT1}{cmr}{bx}{sl}

 {\end{list}}
 {\end{list}}
 {\end{list}}

\catcode`\@=11 
\newlength{\@fninsert}
\newlength{\@fnwidth}
\renewcommand{\@makefntext}[1]%
  {\noindent\makebox[\@fninsert][r]{\@makefnmark}\hfil%
  \parbox[t]{\@fnwidth}{#1}}
\catcode`\@=12 
\catcode`\@=11 
\renewcommand\section{\@startsection{section}{1}{\z@}%
                                   {-3.5ex \@plus -1ex \@minus -.2ex}%
                                   {2.3ex \@plus.2ex}%
                                   {\normalfont\Large\bfseries}}
\renewcommand\subsection{\@startsection{subsection}{2}{\z@}%
                                   {-3.25ex\@plus -1ex \@minus -.2ex}%
                                   {1.5ex \@plus .2ex}%
                                   {\normalfont\large\bfseries}}
\renewcommand\subsubsection{\@startsection{subsubsection}{3}{\z@}%
                                   {-3.25ex\@plus -1ex \@minus -.2ex}%
                                   {1.5ex \@plus .2ex}%
                                   {\normalfont\large\bfseries}}
\renewcommand\paragraph{\@startsection{paragraph}{4}{\z@}%
                                   {3.25ex \@plus1ex \@minus.2ex}%
                                   {1.2ex \@plus .2ex}%
                                   {\normalfont\normalsize\bfseries}}
\catcode`\@=12 

\newcommand{\bi}{\begin{itemize}}
\newcommand{\ei}{\end{itemize}}
\newcommand{\be}{\begin{equation}}
\newcommand{\ee}{\end{equation}}
\newcommand{\bea}{\begin{eqnarray}}
\newcommand{\eea}{\end{eqnarray}}

\begin{document}
\hyphenation{nuSTORM}
\selectlanguage{english}
\makeatletter
\begin{center}
{\bf{ \Large \vspace{-3mm} Neutrinos from STORed Muons\\} {\large Proposal to the Fermilab PAC\\}}
\end{center}
%
%
%
\author{D.~Adey}
\affiliation{Fermi National Accelerator Laboratory, Box 500, Batavia, IL 60510-5011, USA}
\author{S.K.~Agarwalla}
\affiliation{Institute of Physics, Sachivalaya Marg, Sainik School Post, Bhubaneswar 751005, Orissa, India}
\author{C.M.~Ankenbrandt}\thanks{Also at Fermilab, P.O. Box 500, Batavia, IL 60510-5011, USA}
\affiliation{Muons Inc., 552 N. Batavia Avenue, Batavia, IL 60510, USA} 
\author{R.~Asfandiyarov}
\affiliation{University de Geneve, 24, Quai Ernest-Ansermet, 1211 Geneva 4, Switzerland}
\author{J.J.~Back}
\affiliation{Department of Physics, University of Warwick, Coventry, CV4 7AL, UK}
\author{G.~Barker}
\affiliation{Department of Physics, University of Warwick, Coventry, CV4 7AL, UK}
\author{E. Baussan}
\affiliation{IPHC, Universit\'e de Strasbourg, CNRS/IN2P3, F-67037 Strasbourg, France}
\author{R.~Bayes}
\affiliation{School of Physics and Astronomy, Kelvin Building, University of Glasgow, Glasgow G12 8QQ, Scotland, UK}
\author{S.~Bhadra}
\affiliation{Department of Physics and Astronomy, York University, 4700 Keele Street, Toronto, Ontario, M3J 1P3, Canada}
\author{V.~Blackmore}
\affiliation{Oxford University, Subdepartment of Particle Physics, Oxford, UK}
% ...
\collaboration{The nuSTORM Collaboration}
\date{\today}
\thispagestyle{plain}
\maketitle
%
%
\parindent 10pt
\pagenumbering{roman}                   
\setcounter{page}{1}
\thispagestyle{plain}
\pagestyle{plain}
%
\tableofcontents
\pagenumbering{arabic}                   
\setcounter{page}{1}
\cleardoublepage
\pagestyle{plain}
\section{Executive Summary}
\label{sec:ES}

\begin{figure}
  \begin{center}
    \includegraphics[width=0.9\textwidth]{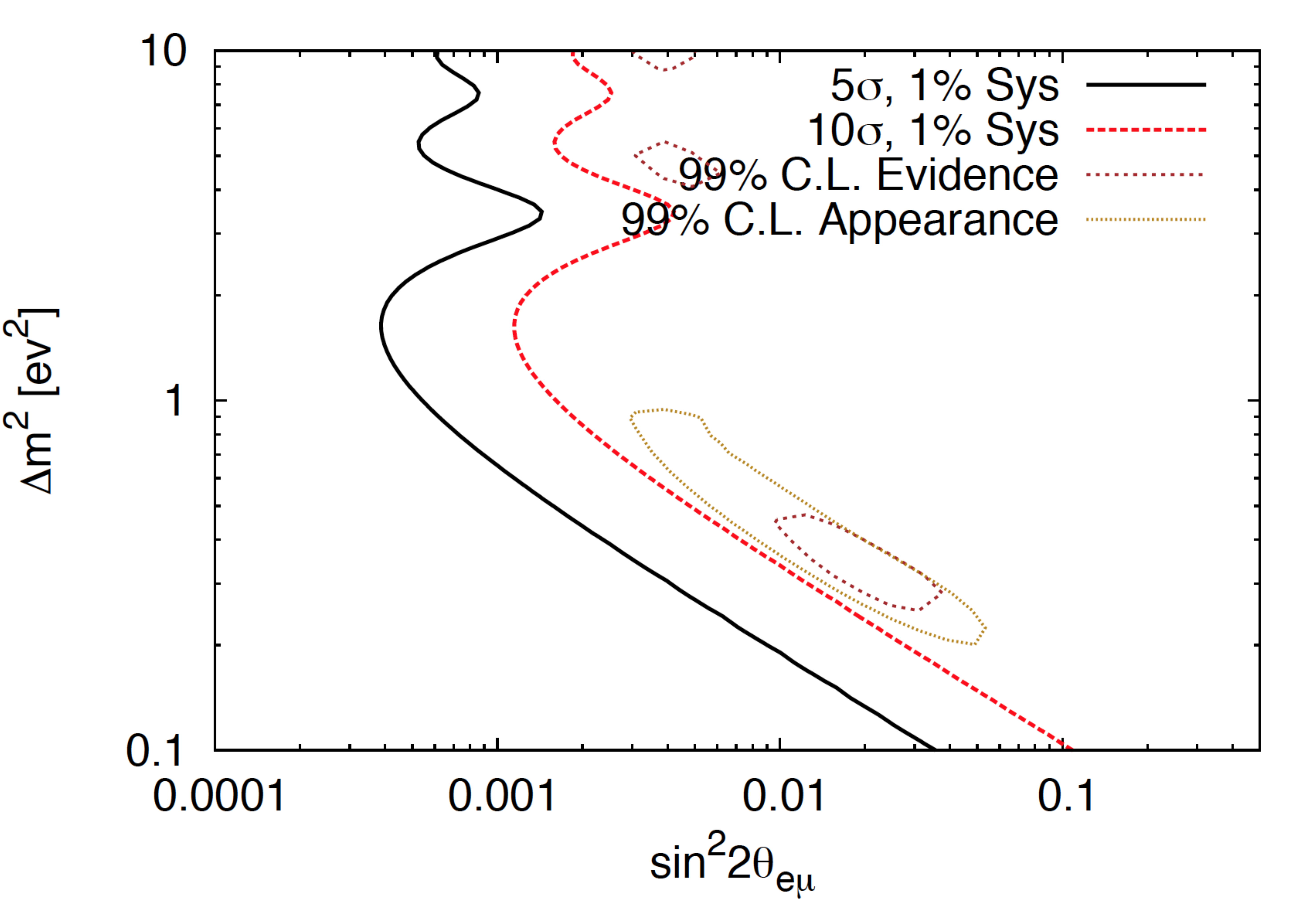}
  \end{center}
  \caption{ Contours of the $\chi^{2}$ deviation from the no-sterile
    neutrino hypothesis corresponding to 5$\sigma$
    and 10$\sigma$ variations with 1\% systematic uncertainties. The 99\% confidence
    level contours from a global fit to all experiments showing evidence for unknown
    signals (appear + reactor + Gallium) and the contours derived from the accumulated data from all
    applicable neutrino appearance experiments \cite{Kopp:2013vaa} are also shown. }
  \label{fig:TenSig}
\end{figure}

The nuSTORM facility has been designed to deliver beams of
$\parenbar{\nu}_e$ and $\parenbar{\nu}_\mu$ from the decay of a
stored $\mu^\pm$ beam with a central momentum of 3.8\,GeV/c and
a momentum acceptance of 10\% \cite{Kyberd:2012iz}.
The facility is unique in that it will:
\begin{itemize}
  \item Allow searches for sterile neutrinos of exquisite sensitivity
    to be carried out; and
  \item Serve future long- and short-baseline neutrino-oscillation
    programs by providing definitive measurements of
    $\parenbar{\nu}_e N$ and $\parenbar{\nu}_\mu N$ scattering cross
    sections with percent-level precision;
  \item Constitute the crucial first step in the 
    development of muon accelerators as a powerful new
    technique for particle physics.
\end{itemize}
A number of results have been reported that can be interpreted as
hints for oscillations involving sterile neutrinos
\cite{Aguilar:2001ty,AguilarArevalo:2007it,AguilarArevalo:2010wv,Mueller:2011nm,Huber:2011wv,
Mention:2011rk,Anselmann:1994ar,Hampel:1997fc,Abdurashitov:1996dp,Abdurashitov:1998ne,Abdurashitov:2005tb}
(for a recent review see \cite{Abazajian:2012ys}).
Taken together, these hints warrant a systematically different and
definitive search for sterile neutrinos.
A magnetized iron neutrino detector at a distance of 
$\simeq 2\,000$\,m from the storage ring combined with a near
detector placed at a distance of 20\,m, identical to it in all respects
but fiducial mass, will allow searches for active/sterile
neutrino oscillations in both the appearance and disappearance
channels.
Simulations of the $\nu_e \rightarrow \nu_\mu$ appearance channel show
that the presently allowed region can be excluded at the $10\sigma$
level (see Fig.~\ref{fig:TenSig}) while in the $\nu_e$ disappearance channel, nuSTORM has the
statistical power to exclude the presently allowed parameter space.
Furthermore, the definitive studies of $\parenbar{\nu}_e N$ 
($\parenbar{\nu}_\mu N$) scattering that can be done at nuSTORM
will allow backgrounds to be quantified precisely.
The race to discover CP-invariance violation in the lepton sector and
to determine the neutrino mass-hierarchy has begun with the recent
discovery that $\theta_{13} \ne 0$ 
\cite{An:2012eh,Ahn:2012nd,Abe:2011fz,Abe:2011sj,Adamson:2011qu}.
The measured value of $\theta_{13}$ is large 
($\sin^2 2 \theta_{13} \sim 0.1$), so measurements of oscillation
probabilities with uncertainties at the percent level are required.
For future long-baseline experiments to reach their
ultimate precision requires that the $\parenbar{\nu}_e N$ and
the $\parenbar{\nu}_\mu N$ cross sections are known precisely for
neutrino energies ($E_\nu$) in the range $0.5 < E_\nu < 3$\,GeV.  
nuSTORM is therefore unique as it makes it possible to measure the
$\parenbar{\nu}_e N$ and the $\parenbar{\nu}_\mu N$ cross 
sections with a precision $\simeq 1$\% over the required
neutrino-energy range.
At nuSTORM, the flavor composition of the beam and the
neutrino-energy spectrum are both precisely known.
In addition, the storage-ring instrumentation combined with measurements at a
near detector will allow the neutrino flux to be determined to the required 
precision of 1\% or better.
 
In effect, the unique $\nu$ beam available at the nuSTORM facility has the potential to be transformational 
in our approach to $\nu$ interaction physics, offering a ``$\nu$ light source" to physicists from a number 
of disciplines.    

Finally, nuSTORM's unique capabilities offer the opportunity of providing muon beams 
(simultaneously while running the neutrino program) for future investigations into muon ionization.
Muon cooling is the key enabling technology needed for future ultra-high intensity muon accelerator facilities.
Its demonstration would be a (the) major step on the path towards realization of a multi-TeV Muon Collider.
\newpage

\clearpage
\section{Overview}
\label{sec:Overview}
The idea of using a muon storage ring to produce a neutrino beam for experiments
was first discussed by Koshkarev \cite{Koshkarev:1974my} in 1974.  A detailed description of a muon storage ring for neutrino oscillation experiments was first produced by Neuffer \cite{NeufferTelmark} in 1980.  In his paper, Neuffer studied muon decay rings 
with E$_\mu$ of 8, 4.5 and 1.5 GeV.  With his 4.5 GeV ring design, he achieved a figure of merit of $\simeq 6\times10^9$
useful neutrinos per $3\times10^{13}$ protons on target.  The facility we describe here (nuSTORM) is essentially the same 
facility proposed in 1980 and would utilize a 3.8 GeV/c muon storage ring with 10\% momentum acceptance to study eV-scale oscillation physics,
$\nu_e$ and $\nu_\mu$ interaction physics and would provide a technology test bed.  In particular the facility can:

\begin{itemize}
  \item Serve a first-rate neutrino-physics program encompassing:
    \begin{itemize}
       \item Exquisitely sensitive searches for sterile neutrinos in
        both appearance and disappearance modes; and
      \item Detailed and precise studies of electron- and
        muon-neutrino-nucleus scattering over the energy appropriate to
        future long- and short-baseline neutrino
        oscillation programs; and
    \end{itemize}
  \item Provide the technology test-bed required to carry-out the R\&D
    critical for the implementation of the next step in a
    muon-accelerator based particle-physics program.
\end{itemize}

Unambiguous evidence for the existence of one or more sterile
neutrinos would revolutionize the field.
nuSTORM is capable of making the measurements required to confirm
or refute the evidence for sterile neutrinos using a
technique that is both qualitatively and quantitatively new
\cite{Kyberd:2012iz}.
The nuSTORM facility has been designed to deliver beams of $\nu_e$
($\bar{\nu}_e$) and $\bar{\nu}_\mu$ ($\nu_\mu$).
A detector located at a distance $\sim 2\,000$\,m from the end of one
of the straight sections will be able to make sensitive searches for
the existence of sterile neutrinos.
If no appearance 
($\bar{\nu}_\mu \rightarrow \bar{\nu}_e$)
signal is observed, the LSND allowed region can be ruled out at the
$\sim 10 \sigma$ level.
Instrumenting the nuSTORM neutrino beam with a near detector at a
distance of $\sim 20$\,m makes it possible to search for sterile
neutrinos in the disappearance $\nu_e \rightarrow \nu_X$ and
$\nu_\mu \rightarrow \nu_X$ channels.
In the disappearance search, the absence of a signal would permit the 
presently allowed region to be excluded at the 99\% confidence level in
our current analysis (see section~\ref{subsubsec:CAnD}). For a general
discussion of optimization of disappearance searches at short baseline,
see \cite{Winter:2012sk}.

\begin{figure}[htpb]
  \centering{
    \includegraphics[width=1.0\textwidth]{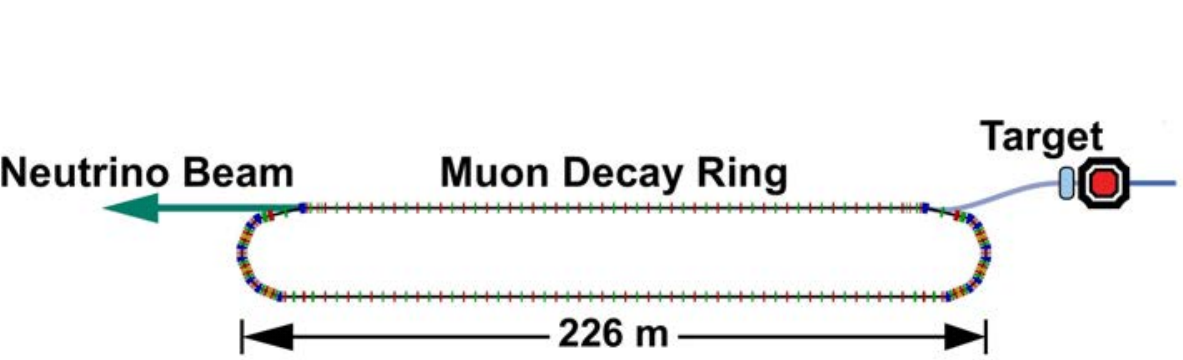}}
  \caption{Schematic of the facility}
  \label{fig:STORM}
\end{figure}

By providing an ideal technology test-bed, the nuSTORM facility
will play a pivotal role in the development of accelerator systems, 
instrumentation techniques, and neutrino detectors.
It is capable of providing a high-intensity, high-emittance,
low-energy muon beam for studies of ionization cooling and can
support the development of the high-resolution, totally-active,
magnetized neutrino detectors.
The development of the nuSTORM ring, together with the
instrumentation required for the sterile-neutrino-search and the $\nu N$-scattering programs,
will allow the next step in the
development of muon accelerators for particle physics to be defined.
Just as the Cambridge Electron Accelerator \cite{Oasis:Lib:Harvard},
built by Harvard and MIT at the end of the '50s, was the first in a
series of electron synchrotrons that culminated in LEP, nuSTORM has
the potential to establish a new technique for particle physics that
can be developed to deliver the high-energy $\nu_e$ ($\bar{\nu}_e$) 
beams required to elucidate the physics of flavor at the Neutrino
Factory and to provide the enabling technologies for a multi-TeV
$\mu^+ \mu^-$ collider.

nuSTORM itself represents the simplest implementation of the Neutrino Factory concept \cite{Geer:1997iz}. In our case,
120 GeV/c protons are used to produce pions off a conventional solid target.  The pions are collected with a magnetic horn and quadrupole magnets
and are then transported to, and injected into, a storage ring.  The pions that decay in the first straight of the ring can yield a muons that are captured in the ring.  The circulating muons then 
subsequently decay into electrons and neutrinos.  We are using a storage ring design that
is optimized for 3.8 GeV/c muon momentum. This momentum was selected to maximize the physics reach for both 
$\nu$ oscillation and the cross section physics.  See Fig.~\ref{fig:STORM} for a schematic of the facility.

Muon decay yields a neutrino beam of precisely known flavor content and energy.  For example 
for positive muons:  $\mu^+ \rightarrow e^+$ + $\bar{\nu}_\mu$ + $\nu_e$.
In addition, if the circulating muon flux in the ring is measured
accurately (with beam-current transformers, for example), then the neutrino beam flux is also accurately known.
Near and far detectors are placed along the line of one of the straight sections of the racetrack decay ring.
The near detector can be placed as close as 20 meters from
the end of the straight.  A near detector for disappearance measurements will be identical to the 
far detector, but only about one tenth the fiducial mass.
Additional purpose-specific near detectors can also be located in the near hall and will measure neutrino-nucleon cross sections 
and can provide the first precision measurements of $\nu_e$ and $\bar{\nu}_e$ cross sections. 
A far detector at $\simeq$ 2000 m would study neutrino oscillation physics and would be capable
of performing searches in both appearance and disappearance channels.
The experiment will take advantage of the ``golden channel" of oscillation appearance $\nu_e \rightarrow \nu_\mu$, where the resulting 
final state has a muon of the wrong-sign from interactions of the $\bar{\nu}_\mu$ in the beam.
In the case of $\mu^+$ stored in the ring, this would mean the observation of an event with a $\mu^-$.
This detector would need to be magnetized for the
wrong-sign muon appearance channel, as is the case for the current baseline Neutrino Factory detector \cite{NF:2011aa}.
A number of possibilities for the far detector exist.  However, a magnetized iron detector similar to that used
in MINOS is seen to be the most straight forward and cost effective approach.
For the purposes of the nuSTORM oscillation physics, a detector inspired by MINOS, but with 
thinner plates and much larger excitation current (larger B field) is assumed.
\newpage

\section{Motivation}
\label{Sect:Motivation}

The case for the nuSTORM facility rests on three themes.
First, the neutrino beam, instrumented with a pair of
magnetized detectors, near and far, will allow searches for sterile
neutrinos of unprecedented  sensitivity to be carried out.
The signal to background ratio for this combination is of the order ten
and is much larger than that in other accelerator-based projects.
Second, the uniquely well-known neutrino beam generated in muon decay
may be exploited to make detailed studies of neutrino-nucleus
scattering over the neutrino-energy range of interest to present and
future long and short-baseline neutrino oscillation experiments.
In long-baseline experiments, these measurements are required to break
the correlation between the cross-section and flux uncertainties and
to reduce the overall systematic uncertainty to a level that reinforces
the investment in super-beam experiments such
as T2HK, LBNE, LBNO and SPL-Frejus.
The nuSTORM $\parenbar{\nu} N$ scattering program is no less
important for the next generation of short-baseline experiments for
which uncertainties in the magnitude and shape of backgrounds to the
sterile-neutrino searches will become critically important.
Third, the storage ring itself, and the muon beam it contains, can
be used to carry out the R\&D program required to implement the next
step in the incremental development of muon accelerators for particle
physics.
The muon accelerator program has the potential to provide elucidation of the
physics of flavor at the Neutrino Factory and then to lead to the technological 
foundation for multi-TeV $\mu^+ \mu^-$ collisions at the Muon Collider.
The three individually-compelling themes that make up the case for
nuSTORM constitute a uniquely robust case for a facility that will
be at once immensely productive scientifically and seminal in the
creation of a new technique for particle physics. 

\subsection{Sterile neutrinos}
\label{sec:sterile-motivation}

\subsubsection{Sterile neutrinos in extensions of the Standard Model}
\label{sec:sterile-bsm}

Sterile neutrinos---fermions that are uncharged under the $SU(3) \times SU(2)
\times U(1)$ gauge group---are a generic ingredient of many extensions of the
Standard Model. Even in models that do not contain sterile neutrinos, they can
usually be added easily. For a review of models with sterile neutrinos and the
associated phenomenology see~\cite{Abazajian:2012ys}.

One important class of sterile neutrino theories are models explaining the
smallness of neutrino masses by means of a seesaw mechanism. In its simplest
form, the seesaw mechanism requires at least two heavy ($\sim 10^{14}$~GeV)
sterile neutrinos that would have very small mixings ($\sim 10^{-12}$) with the
active neutrinos. However, in slightly non-minimal models, at least some
sterile neutrinos can have much smaller masses and much larger mixing angles.
Examples for such non-minimal scenarios include the split seesaw
scenario~\cite{Kusenko:2010ik}, seesaw models with additional flavor symmetries
(see e.g.~\cite{Mohapatra:2001ns}), models with a Froggatt-Nielsen
mechanism~\cite{Froggatt:1978nt, Barry:2011fp}, and extended seesaw models that
augment the mechanism by introducing more than three singlet fermions, as well
as additional symmetries~\cite{Mohapatra:2005wk, Fong:2011xh, Zhang:2011vh}.

Also Grand Unified Theories (GUTs) can contain Standard Model singlet fermions.
In GUTs, fermions are grouped into multiplets of an extended gauge group, for
instance $SU(5)$ or $SO(10)$, which has $SU(3) \times SU(2) \times U(1)$ as a
subgroup. If the multiplets are larger than needed to accommodate the Standard
Model fermions, there will be extra states that will behave like gauge singlets
after the GUT symmetry is broken (see for instance~\cite{Bando:1998ww,
Ma:1995xk, Shafi:1999rm, Babu:2004mj} for GUT models with sterile neutrinos).

Finally, sterile neutrinos arise naturally in ``dark sector'' models, which
contain a group of particles that has nontrivial dynamics of its own, but is
only very weakly coupled to the Standard Model particles.  If this dark sector
is similar to the visible sector---as is the case, for instance, in
string-inspired $E_8 \times E_8$ models---it is natural to assume that it also
contains neutrinos~\cite{Berezhiani:1995yi, Foot:1995pa, Berezinsky:2002fa},
which would appear sterile from a Standard Model point of view.

\subsubsection{Experimental hints for light sterile neutrinos}
\label{sec:sterile-hints}

Besides their generic appearance in extended theoretical models, much of the
current interest in sterile neutrinos is motivated by experimental results.
In fact, several neutrino oscillation experiments have observed deviations from
the best available Standard Model predictions and can be interpreted as
hints for oscillations involving light sterile neutrinos with masses of order eV.

The most long-standing of these hints is the result from the {\bf LSND}
experiment, which studied short-distance $\bar\nu_\mu \to \bar\nu_e$
oscillations at a baseline of $\sim 30$~m~\cite{Aguilar:2001ty}.  A flux of
low-energy ($\lesssim 50$~MeV) muon antineutrinos was produced in the decay of
stopped pions, $\pi^+ \to \mu^+ \nu_\mu$, followed by muon decay $\mu^+ \to e^+
\bar\nu_\mu \nu_e$. In the Neutrino Standard Model, neutrinos of this energy are not expected
to oscillate over distances as short as the LSND baseline.  The detection
process for electron antineutrinos in LSND was inverse beta decay, $\bar\nu_e p
\to e^+ n$, in a liquid scintillator detector.  Backgrounds to the neutrino
oscillation search arise from the decay chain $\pi^- \to \bar\nu_\mu + (\mu^-
\to \nu_\mu \bar\nu_e e^-)$, if negative pions produced in the target decay
before they are captured by a nucleus, and from the reaction $\bar\nu_\mu p \to
\mu^+ n$, which is only allowed for the small fraction of muon antineutrinos
produced by pion decay \emph{in flight} rather than stopped pion decay.  LSND
reports an excess of $\bar\nu_e$ candidate events above this background with
a significance of more than $3\sigma$. When interpreted as $\bar\nu_\mu \to
\bar\nu_e$ oscillations through an intermediate sterile state $\bar\nu_s$, this
result is best explained by sterile neutrinos with an effective mass squared
splitting $\Delta m^2 \gtrsim 0.1$~eV$^2$ relative to the active neutrinos, and
with an effective sterile sector-induced $\bar\nu_\mu$--$\bar\nu_e$ mixing
angle $\sin^2 2\theta_{\mu e} \gtrsim 2 \times 10^{-3}$, depending on
$\Delta m^2$.

A second possible hint for oscillations involving sterile neutrinos is provided
by the {\bf MiniBooNE experiment}~\cite{AguilarArevalo:2012va}, which was
designed to test the neutrino oscillation interpretation of the LSND result.
MiniBooNE uses higher-energy (200~MeV--few~GeV) neutrinos from a horn-focused
pion beam.  By focusing either positive or negative pions, MiniBooNE can switch
between a beam dominated by $\nu_\mu$ and a beam dominated by $\bar\nu_\mu$. In
both modes, the experiment observed an excess of electron-like events at
sub-GeV energies. The excess has a significance above $3\sigma$ and can be
interpreted in terms of $\parenbar\nu_\mu \to \parenbar\nu_e$ oscillations
consistent with the LSND observation~\cite{AguilarArevalo:2012va}.

While the $\nu_\mu \to \nu_e$ and $\bar\nu_\mu \to \bar\nu_e$ oscillation channels
in which LSND and MiniBooNE observe excesses are sensitive to transitions between
active neutrino flavors mediated by sterile states, sterile neutrinos can also
manifest themselves in the disappearance of active neutrinos into sterile states.
In fact, such anomalous disappearance may have been observed in reactor experiments.
A larger number of such experiments has been carried out in the past, and while
their results were consistent with the flux predictions available at the
time~\cite{Schreckenbach:1985ep}, they are in conflict with more accurate recent
re-evaluations of the reactor antineutrino flux~\cite{Mueller:2011nm,Huber:2011wv}.
In particular, the average $\bar\nu_e$ event rate is about 6\% lower than
the current prediction~\cite{Mention:2011rk,Abazajian:2012ys}, which has a stated uncertainty of 2-3 \%.
The significance of the
deficit depends crucially on the systematic uncertainties associated with the
theoretical prediction, some of which are difficult to estimate.  If
the {\bf reactor antineutrino anomaly} is interpreted as $\bar\nu_e \to \bar\nu_s$
disappearance via oscillation, the required 2-flavor oscillation parameters
are $\Delta m^2 \gtrsim 0.3$~eV$^2$ and $\sin^2 2\theta_{ee,\rm eff} \sim 0.1$.

Interestingly, active-to-sterile neutrino oscillations with these parameters can also
explain another experimental result, often referred to as the {\bf gallium
anomaly}.  It is based on measurements by the GALLEX and SAGE solar neutrino
experiments which have used $\nu_e$ from intense artificial radioactive sources to
demonstrate the feasibility of their radiochemical detection
method~\cite{Anselmann:1994ar, Hampel:1997fc, Abdurashitov:1996dp,
Abdurashitov:1998ne, Abdurashitov:2005tb}. Both experiments observed fewer
neutrinos from the source than expected. The statistical significance of the
deficit is around $3\sigma$ and can be interpreted in terms of short-baseline
$\bar\nu_e \to \bar\nu_s$ disappearance with $\Delta m^2 \gtrsim 1$~eV$^2$ and
$\sin^2 2\theta_{ee,\rm eff} \sim 0.1$--$0.8$.~\cite{Acero:2007su,
Giunti:2012tn}.

\subsubsection{Constraints and global fit}
\label{sec:sterile-global-fit}

\begin{figure}
  \begin{center}
    \includegraphics[width=0.48\textwidth]{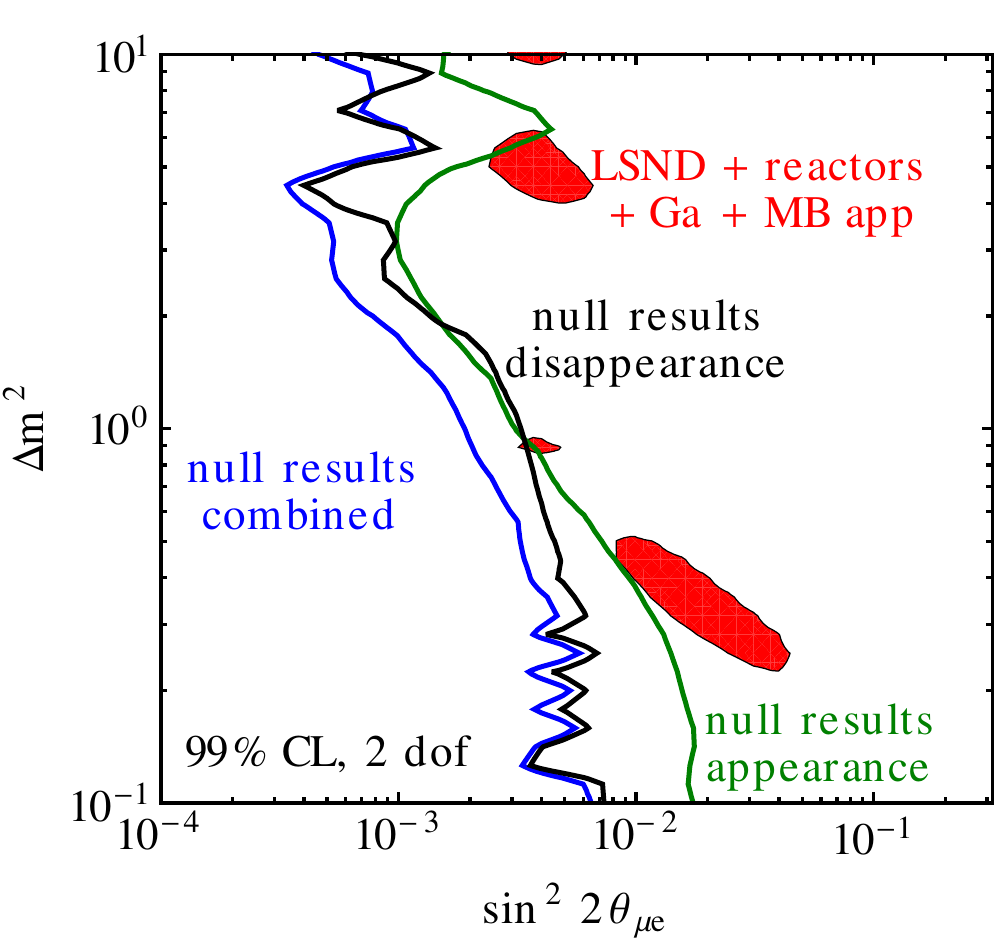}
    \includegraphics[width=0.48\textwidth]{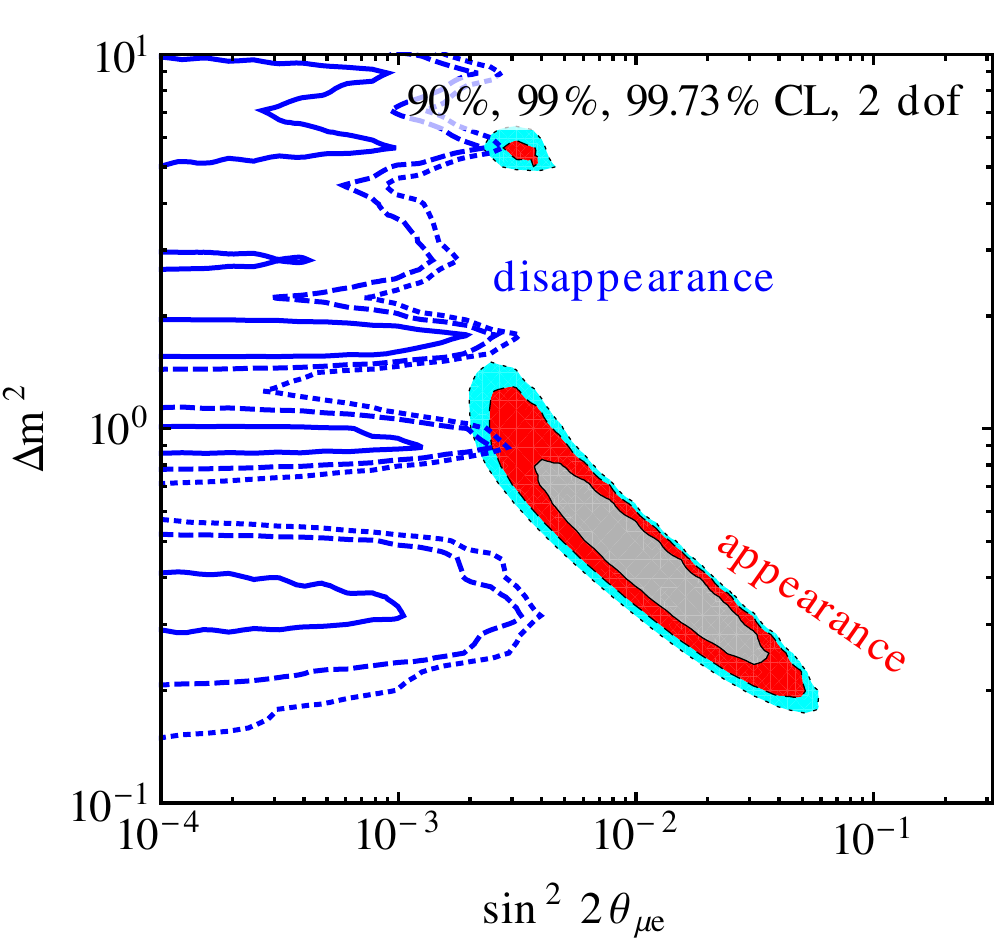}
  \end{center}
  \caption{Global constraints on sterile neutrinos in a 3+1 model. In the left panel,
    we show that $\protect\parenbar{\nu}_e$ appearance data (colored region:
    LSND~\cite{Aguilar:2001ty}, MiniBooNE~\cite{AguilarArevalo:2012va},
    KARMEN~\cite{Armbruster:2002mp}, NOMAD~\cite{Astier:2001yj},
    E776~\cite{Borodovsky:1992pn}, ICARUS~\cite{Antonello:2012fu}) is only
    marginally consistent with disappearance data (blue contours: atmospheric
    neutrinos~\cite{Ashie:2005ik}, solar neutrinos~\cite{Cleveland:1998nv,
    Kaether:2010ag, Abdurashitov:2009tn, Hosaka:2005um, Aharmim:2007nv,
    Aharmim:2005gt, Aharmim:2008kc, Aharmim:2011vm, Bellini:2011rx,
    Bellini:2008mr}, MiniBooNE/SciBooNE~\cite{AguilarArevalo:2009yj,
    Cheng:2012yy} MINOS~\cite{Adamson:2010wi, Adamson:2011ku}, reactor
    experiments~\cite{Declais:1994su, Declais:1994ma, Kuvshinnikov:1990ry,
    Vidyakin:1987ue, Kwon:1981ua, Zacek:1986cu, Apollonio:2002gd, Boehm:2001ik,
    DBneutrino, Abe:2012tg, Ahn:2012nd, Gando:2010aa},
    CDHS~\cite{Dydak:1983zq}, KARMEN~\cite{Reichenbacher:2005nc} and
    LSND~\cite{Auerbach:2001hz} $\nu_e$--${}^{12} \text{C}$ scattering data and
    gallium experiments~\cite{Hampel:1997fc, Abdurashitov:1998ne,
    Kaether:2010ag, Abdurashitov:2005tb}). In the right panel, we compare
    those experiments which see unexplained signals (LSND,
    MiniBooNE appearance measurements, reactor experiments, gallium
    experiments) to those which do not. For the analysis of reactor data, we
    have used the new reactor flux predictions from~\cite{Mueller:2011nm}, but
    we have checked that the results, especially regarding consistency with
    LSND and MiniBooNE $\bar\nu$ data, are qualitatively unchanged when the old
    reactor fluxes are used.  Fits have been carried out in the GLoBES
    framework~\cite{Huber:2004ka, Huber:2007ji} using external modules
    discussed in~\cite{GonzalezGarcia:2007ib, Maltoni:2007zf,
    Akhmedov:2010vy, Kopp:2013vaa}. Plots taken from ref.~\cite{Kopp:2013vaa}.}
  \label{fig:sterile-regions-3p1}
\end{figure}

In spite of the interesting hints from four short baseline oscillation
experiments, the existence of light sterile neutrinos is far from established.
The most important reason is that other short baseline experiments did not
observe a signal and place strong constraints on the available sterile neutrino
parameter space. The compatibility of the signals from LSND, MiniBooNE, reactor
and gallium experiments with null results from a large number of other
experiments is assessed in global fits~\cite{Kopp:2011qd, Giunti:2011cp,
Karagiorgi:2011ut, Giunti:2011hn, Giunti:2011gz, Abazajian:2012ys,
Giunti:2012tn, Archidiacono:2013xxa, Kopp:2013vaa}.  The results from one
of these fits~\cite{Kopp:2013vaa} are shown in
Fig.~\ref{fig:sterile-regions-3p1}.

\begin{figure}
  \begin{center}
    \includegraphics[width=0.44\textwidth]{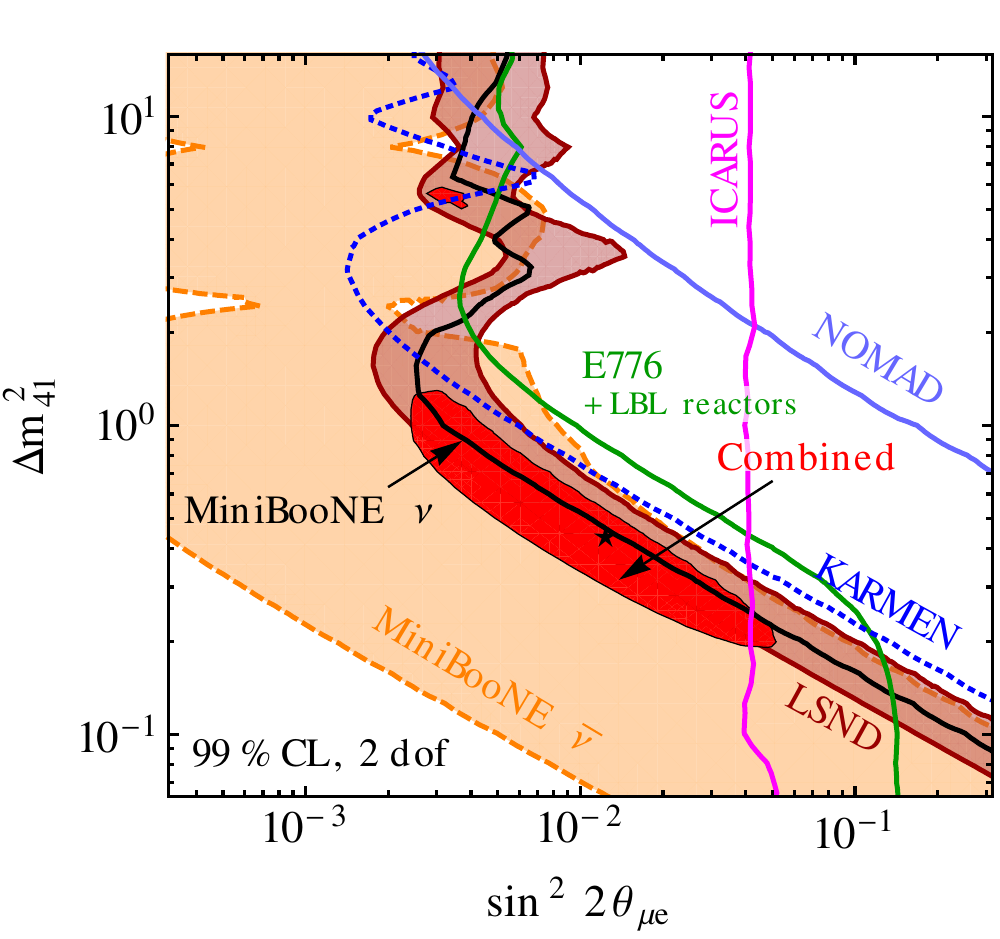}
    \raisebox{-0.3cm}{\includegraphics[width=0.5\textwidth]{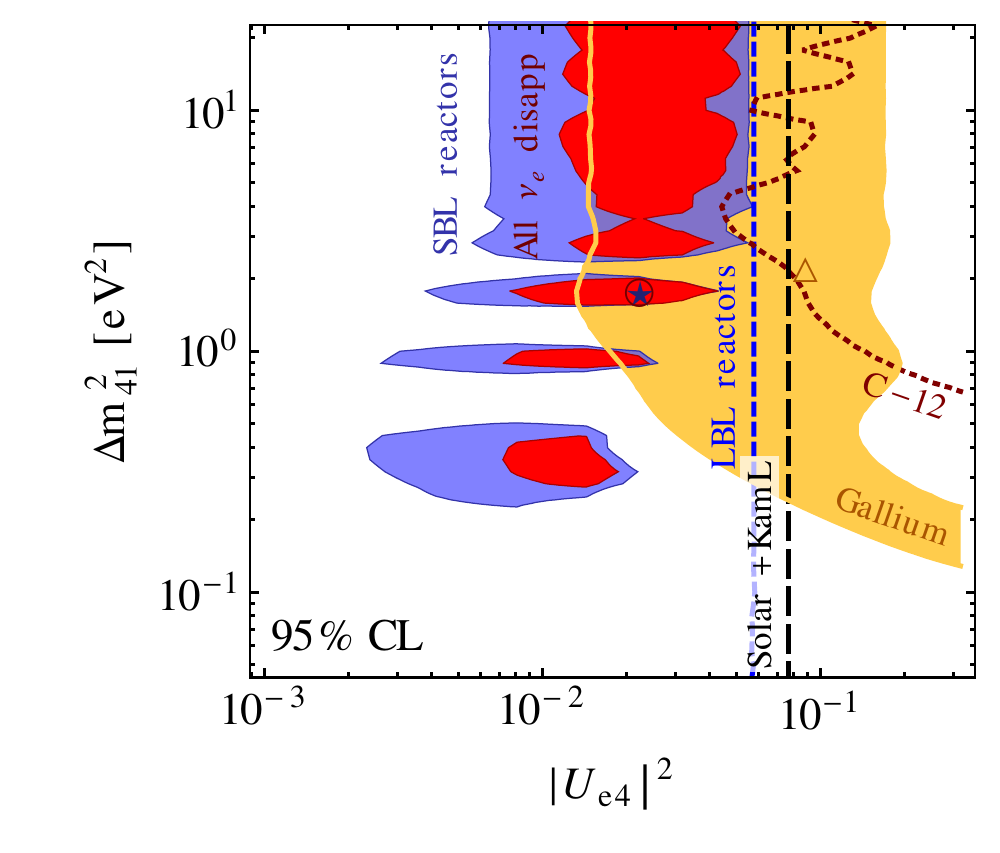}}
  \end{center}
  \caption{Global constraints on $\protect\parenbar{\nu}_\mu \to \protect\parenbar{\nu}_e$
    oscillations (left) and on $\protect\parenbar{\nu}_e$ disappearance (right).
    These subsets of the globally available data are found to be consistent,
    and it is only when they are combined, and when, in addition, exclusion limits on
    $\protect\parenbar{\nu}_\mu$ disappearance are included, that tension appears.
    Plots taken from ref.~\cite{Kopp:2013vaa}.}
  \label{fig:sterile-subsets}
\end{figure}

We see that there is significant tension in the global data set: The parameter
region favored by LSND, MiniBooNE, reactor and gallium experiments is
incompatible, at the 99\% confidence level, with exclusion limits from other
experiments (left panel of figure~\ref{fig:sterile-regions-3p1}).  Similarly, the
combination of all $\parenbar\nu_e$ appearance data is in conflict with the
$\parenbar\nu_\mu$ and $\parenbar\nu_e$ disappearance data (right panel of
figure~\ref{fig:sterile-regions-3p1}).  Quantifying this disagreement with a parameter
goodness of fit test~\cite{Maltoni:2003cu} yields p-values of order
$10^{-4}$~\cite{Kopp:2013vaa}.  (Note that different authors find different
levels of tension between appearance and disappearance
data~\cite{Karagiorgi:2011ut, Archidiacono:2013xxa, Kopp:2013vaa}. The reason
for the differences between global fits lie in the choice of data sets included
and in the details of the fits to individual experimental data sets.) It is
difficult to resolve this tension even in models with more than one sterile
neutrino.

It is important to note that in spite of the tension in the global fit, the different
appearance data sets are compatible with each other (Fig.~\ref{fig:sterile-subsets},
left), as are the disappearance
data sets among themselves (Fig.~\ref{fig:sterile-subsets},
right). This implies in particular that an explanation of
the reactor and gallium anomalies in terms of sterile neutrinos is perfectly viable.
Also, $\parenbar\nu_\mu \to \parenbar\nu_e$ transitions with parameters
suitable for explaining LSND and MiniBooNE are not directly ruled out by
other experiments measuring the same oscillation channels. It is only when
appearance and disappearance signals are related through the equation
\begin{align}
  \sin^2 2\theta_{\mu e} = 4 |U_{e4}|^2 |U_{\mu 4}|^2
\end{align}
that the tension appears. Here, $\theta_{\mu e}$ is the effective mixing angle
governing $\parenbar\nu_\mu \to \parenbar\nu_e$ transition in LSND and MiniBooNE,
and $U_{e4}$, $U_{\mu 4}$ are the elements of the leptonic mixing matrix describing
mixing between electron neutrinos and sterile neutrinos, and between muon neutrinos
and sterile neutrinos, respectively.

Possible solutions to the tension between positive hints and null results are:
\begin{enumerate}
  \item One or several of the apparent deviations from the standard
    three neutrino oscillation framework discussed in
    Section~\ref{sec:sterile-hints} have explanations not related to
    sterile neutrinos.
  \item One or several of the null results that favor the no-oscillation
    hypothesis are in error.
  \item There are sterile neutrinos plus some other kind of new physics
    at the eV scale. (See for instance~\cite{Akhmedov:2010vy,
    Karagiorgi:2012kw} for an attempt in this direction.)
\end{enumerate}

\subsubsection{Sterile neutrinos in cosmology}
\label{sec:sterile-cosmology}

Important constraints on sterile neutrino models come from cosmology, which is
sensitive to the effective number $N_\text{eff}$ of thermalized relativistic
species in the Universe at the time of Big Bang Nucleosynthesis (BBN) and at
the time of recombination.  Moreover, cosmological observations constrain the
sum of neutrino masses.  The most recent Planck data~\cite{Ade:2013lta} yields
$N_\text{eff} = 3.30^{+0.54}_{-0.51}$ when combined with polarization data from
WMAP~\cite{Hinshaw:2012fq}, high-multipole measurements from
ACT~\cite{Das:2013zf} and SPT~\cite{Reichardt:2011yv, Story:2012wx} and data on
baryon acoustic oscillations (BAO)~\cite{Percival:2009xn, Padmanabhan:2012hf,
Blake:2011en, Anderson:2012sa}.  The sum of neutrino masses is constrained by
the same data to be below 0.23~eV.  While the exact strength of exclusion
depends on which cosmological data sets are analyzed, the general conclusion
remains the same: cosmology after Planck is consistent with only the standard
three neutrino flavors, but the existence of additional thermalized neutrino
species is not ruled out. On the other hand, these extra species should have
masses significantly below 1~eV, too low to explain the short-baseline
anomalies.  However, the Planck limit on the effective relativistic degrees of 
freedom are weakened if one includes the current best fit data for the Hubble constant.

It is important to keep in mind that cosmological bounds apply only
to neutrino species that come into thermal equilibrium in the early Universe,
usually through oscillations.  Hence, if oscillations between active and
sterile neutrinos are suppressed at early times, these limits are avoided (see
\cite{Kristiansen:2013mza, Archidiacono:2013xxa} for recent combined fits of
cosmology and short baseline oscillation data).  A suppression of
active--sterile neutrino oscillations can occur if there is a very large
primordial lepton asymmetry (see for instance~\cite{Mirizzi:2012we,
Saviano:2013ktj} for recent studies) or in Majoron models~\cite{Bento:2001xi,
Dolgov:2004jw}. Other scenarios in which eV scale sterile neutrinos can be
consistent with cosmology include phenomenological non-standard extensions of
$\Lambda$CDM cosmology~\cite{Hamann:2011ge}, scenarios with very low reheating
temperature~\cite{Gelmini:2004ah}, certain types of $f(R)$
gravity~\cite{Motohashi:2012wc}, and models with additional heating mechanisms
for the primordial plasma after neutrino decoupling (see for instance
\cite{Ho:2012br}).

We conclude that, given the current experimental situation, it is impossible to
draw firm conclusions regarding the existence of light sterile neutrinos.
There is on the one hand an intriguing accumulation of experimental anomalies
that could be interpreted in the context of sterile neutrino models, while on
the other hand, it seems difficult to accommodate all these hints
simultaneously, given strong constraints from experiments with null results.
An experiment searching for short-baseline neutrino oscillations with good
sensitivity and well-controlled systematic uncertainties has great potential to
clarify the situation by either finding a new type of neutrino oscillation or
by deriving a strong and robust constraint on any such oscillation. While the
former outcome would constitute a major discovery, the latter one would also
certainly receive a lot of attention since it would provide the world's
strongest constraints on a large variety of theoretical models postulating
``new physics'' in the neutrino sector at the eV scale.  
A truly definitive experiment for both the muon appearance and muon disappearance channels 
is required to reach a convincing conclusion on the existence of light, sterile neutrinos.


\def\parenbar#1{{\null\!                        
   \mathop#1\limits^{\hbox{\tiny (--)}}       	
   \!\null}}                                    

\newcommand{\nustorm}{\mbox{$\nu$\hbox{STORM} }}

\subsection{Neutrino-nucleus scattering}
\subsubsection{Introduction}
\label{sec:motivation:physics}

Recent interest in neutrino interactions in the few GeV energy region comes from neutrino oscillation
experiments and their need to reduce systematic errors.  Cross section measurements have been performed by neutrino oscillation collaborations in the past, with T2K currently carrying on this tradition.  
Importantly, there is a dedicated cross section experiment (MINERvA) currently underway and 
others using LAr detectors (MicroBooNE) planned for the near future at Fermilab.

Even with this degree of activity, the precision with which the basic neutrino-{\em nucleon} cross 
sections are known 
is still not better than $20-30$\%. There are two main reasons for this: the poor knowledge of neutrino
fluxes and the fact that all the recent cross section measurements have been performed on nuclear targets.  It is important to recall that what current neutrino experiments are measuring are events that are generated from 
a convolution of energy-dependent neutrino flux $\otimes$  energy-dependent cross section $\otimes$ energy-dependent nuclear effects.  The experiments have, for example, then measured an effective neutrino-carbon cross section.  To extract a neutrino-{\em nucleon} cross section from these measurements requires separation of 
nuclear physics effects, which can only be done with limited precision. 
For many oscillation experiments, using the same nuclear targets for their near and far detectors is a good start.  However, even with the same nuclear target near-and-far, that there are different near and far neutrino energy spectra due to beam geometry and oscillations means there is a different convolution of cross section $\otimes$ nuclear effects near-and-far and there is no automatic cancellation between the near-and-far detectors.  
For a thorough comparison of measured neutrino-nucleon cross sections with theoretical models, 
these convoluted effects have to be understood. For further details please see~\cite{Morfin:2012kn}.
This section will summarize the current status of 
both theoretical and experimental studies of neutrino nucleus scattering with an emphasis on what 
nuSTORM, with its superior knowledge of the neutrino flux and its high-intensity source of electron-neutrinos, 
can contribute.

\begin{figure}[htb]
  \centering
  \mbox{
    \subfigure{
      \includegraphics[width=0.45\textwidth]%
        {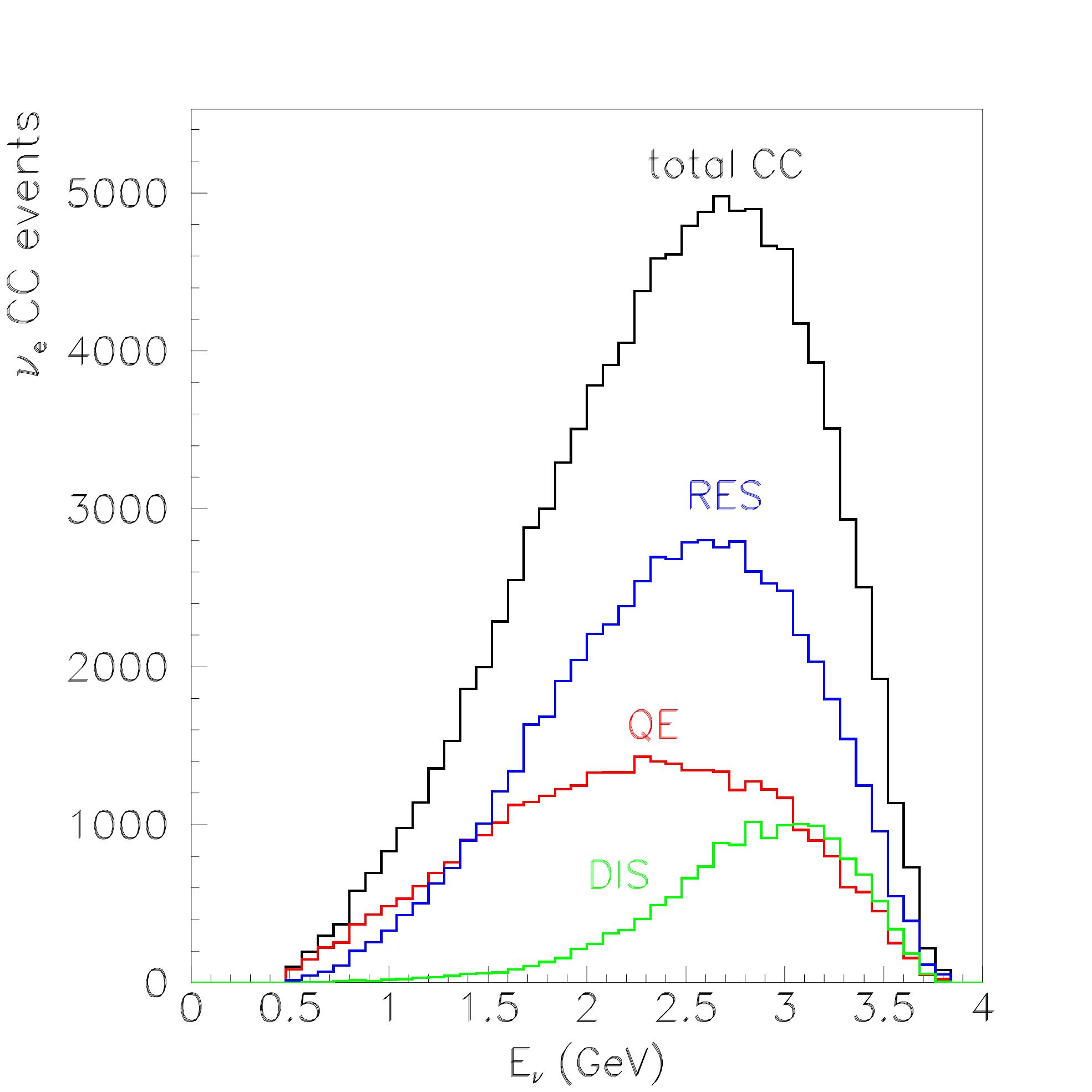}}\quad
    \subfigure{
      \includegraphics[width=0.3\textwidth]%
        {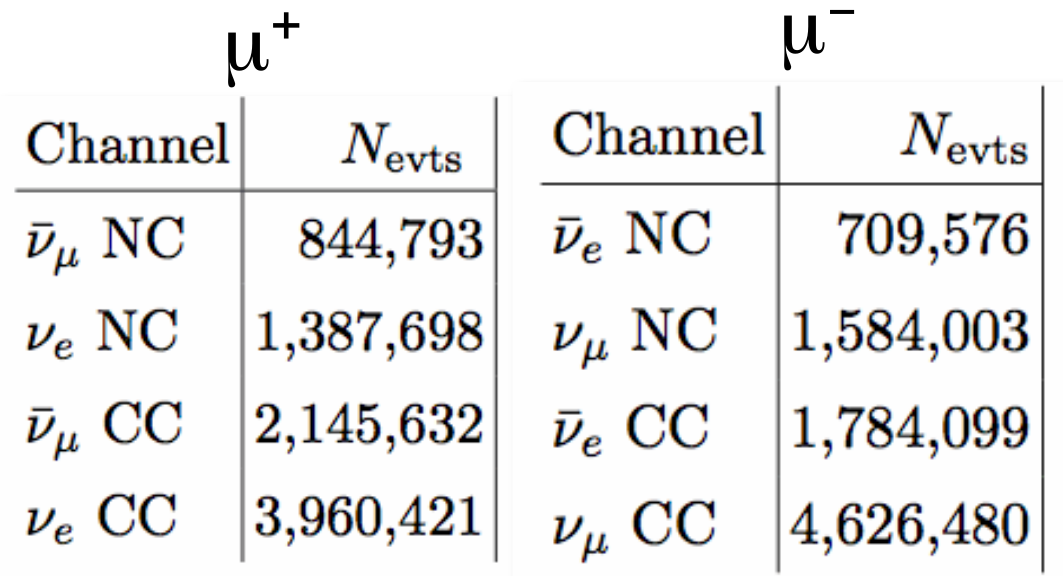}}
  }
  \caption{\footnotesize Left: The breakdown of scattering channels for a nuSTORM $\nu_{\mu}$ or $\nu_{e}$ beam circulating 3.8 GeV/c muons.
Right: The expected event rates for $\nu$ and $\overline{\nu}$ for both $\mu^+$ and $\mu^-$ circulating beams and a 100 fiducial ton detector located 50 m from the straight.  The exposure is 10$^{21}$ POT.} 
  \label{fig:ScattChan}
\end{figure}

For neutrino-{\it nucleon} interactions one can distinguish: Charged Current quasi-elastic (CCQE), Neutral Current elastic (NCEl), Resonance production (RES) and more inelastic reactions involving pion production from the $\Delta$ through the transition region up to the deep-inelastic (a rather misleading "DIS" term is often used to describe all the interactions which are neither CCQE/NCEl nor RES) domain.  Quite different theoretical tools are used to model each of them.   The expected distribution of nuSTORM events among the various scattering channels as well as the expected event rates are shown in  Fig.~\ref{fig:ScattChan}.  For neutrinos the expected distribution is 56\% resonant, 32\% 
quasi-elastic and 12 \% DIS  with event energy peaking at about 2.5 GeV. For 
anti-neutrinos the breakdown is 52\% resonant, 40\% quasi-elastic and 8 \% DIS.  

From the experimental point of view it is most natural to speak about events in terms of the visible final
state topology; that is, events with no pions in the final state or with only one pion above a given
momentum threshold, and so on. In fact, in several recent experimental measurements that 
investigated quantities  defined in this way, the dependence on assumptions of Monte Carlo event 
generators were minimal.  To compare with the experimental data given in this format, one must add 
contributions from various dynamical mechanisms and also model Final State Interactions (FSI) effects. 
Several ingredients of the theoretical models are verified simultaneously. 
It is clear that in order to validate a model one needs many samples of 
precise neutrino-nucleus scattering measurements on a variety of nuclear targets with various neutrino fluxes.

\subsubsection{Charged Current quasi-elastic}
\label{sec:motivation:ccqe}

The simplest neutrino hadronic reaction is the charge current quasi-elastic (CCQE) interaction: 
$ \nu_\ell + n \rightarrow \ell^- + p$  with two particles: a charged lepton and proton in the produced 
state and the antineutrino analog state that involves a neutron instead of a proton in the produced state.  
We need to extend this definition to the neutrino-nucleus interaction occurring on bound nucleons.  
The ejected proton is not necessarily seen in a detector because quite often its momentum is below 
the acceptance threshold. However, events with a single reconstructed  charged lepton track 
can result from a variety of initial interactions e.g. from a two body charged current interaction or 
from real pion production and its subsequent absorption. Similar problems arise in other type of 
interactions. It is becoming clear that interpretation of neutrino-nucleus interactions must rely 
on a careful data/Monte Carlo (MC) comparison done with reliable MC neutrino event generators. 

In the case of neutrino nucleus scattering we also use the term CCQE-like reaction to define 
one in which there are no pions in the final state. It then includes events with real pion 
production followed by absorption. It also includes interactions on bound-nucleon 
systems (np-nh or meson-exhange current) which will be discussed shortly.

A theoretical description of the free nucleon target CCQE reaction is based on the 
conserved vector current (CVC) and the partially conserved axial current (PCAC) hypotheses. 
The only unknown quantity is the nucleon axial form-factor $G_A(Q^2)$ for which one typically 
assumes a dipole form $G_A(0)(1+\frac{Q^2}{M_A^2})^{-2}$ with one free parameter, 
the axial mass $M_A$. This dipole form is an assumption which need not hold. Non-dipole
form factors are being investigated in \cite{Bhattacharya:2011ah}. 

In the past, several measurements of $M_A$ were performed on a deuterium target for which 
most of nuclear physics complications are minimal and it seemed that the results converged to
 a value of the order of $1.03$~GeV \cite{Bodek:2007vi}. There is an additional  argument 
in favor of a similar value of $M_A$ coming from the weak pion-production at low $Q^2$. 
A PCAC based evaluation gives an axial mass value of  $1.077\pm 0.039$~GeV~\cite{Bernard:2001rs}. 
On the other hand, all of the more recent high statistics measurements of  $M_A$, with the 
exception of the NOMAD higher-energy experiment, reported larger values, with the  
MiniBooNE  (carbon, $Q^2>0$~GeV$^2$) determination of $1.35\pm 0.17$~\cite{AguilarArevalo:2010zc} 
compared to the  NOMAD (carbon, $Q^2>0$~GeV$^2$) value of $1.07\pm 0.07$~\cite{Lyubushkin:2008pe}).  
The most recent MINERvA preliminary results for CCQE antineutrino reaction are still subject to 
large flux normalization uncertainties but they seem to be consistent with 
$M_A=0.99$~GeV \cite{minerva_ccqe-Neutrino2012}


\paragraph{ Theoretical approaches to CCQE}
\label{sec:motivation:ccqe:theory}

Several approaches have been followed to describe the CCQE-like process.  For
moderate and intermediate neutrino energies, in the few GeV region, the most relevant 
ones are: the involvement of
one nucleon, or a pair of nucleons or even three nucleon
mechanisms and the excitation of $\Delta$ or higher resonances. 

A review of theoretical model results can be found in~\cite{Boyd:2009zz}. Almost all 
approaches used at intermediate neutrino energies rely on the impulse approximation (IA)
and neutrino-nucleus CCQE interactions and are viewed as
a two step process: primary interaction and Final State Interactions (FSI), and then the
propagation of resulting hadrons through the nucleus.
In addition they consider several nuclear effects
such as the Random Phase Approximation (RPA) or Short Range Correlations (SRC). 
In the neutrino-nucleus cross section measurements, a goal is to 
learn about neutrino free nucleon target scattering parameters (an obvious exception is coherent pion
production). {\it Effective} parameters, like the
sometimes discussed quasi-elastic axial mass $M_A^{eff}$, are of little use as their values 
can depend on the neutrino
flux, target and perhaps also on the detection technique/acceptance. 

The simplest model, commonly used in Monte Carlo event generators, is the relativistic 
Fermi gas (RFG) model  proposed by Smith and
Moniz more than 40 years ago~\cite{Smith:1972xh} . The model combines the bare nucleon physics
with a model to account for Fermi motion and nucleon binding within the specific
nucleus. 
The model can be made more realistic in many ways to achieve
better agreement with a broad range of electron scattering data. For
example, the inclusion of a realistic joint distribution of target
nucleon momenta and binding energies based on short range correlation
effects leads to the spectral function (SF) approach. 
Spectral functions for
  nuclei, ranging from carbon (A = 12) to iron (A = 56)
  have been modeled~\cite{Benhar:1994hw}.  Calculations by Benhar {\em et
  al.},~\cite{Benhar:2006nr} and Ankowski {\em et
  al.},~\cite{Ankowski:2007uy} show that the SF effects only moderately
  modify the muon neutrino differential cross sections, leading
  to reductions of the order of 15\% in the total cross sections. 
  Inclusion of nucleon-nucleon long-range correlations leads to RPA effects
which improves predictions at lower momentum
transfers (and also low $Q^2$).  

\paragraph{Multinucleon mechanisms}
\label{sec:motivation:mnm}

A plausible solution to the large axial mass puzzle was first pointed out by
M. Martini\footnote{The papers of Martini {\em et al.}  are based on the older investigation by 
Marteau {\em et al.},\cite{Marteau:1999kt}. The relevant features of the model were known already at
the end of 1990s and at that time the goal was to understand better
SuperKamiokande atmospheric neutrino oscillation signal.
} {\em et al.},~\cite{Martini:2009uj,Martini:2010ex}, and later
corroborated by the IFIC group~\cite{Nieves:2011yp,Nieves:2011pp}. 
In the MiniBooNE measurement of Ref.~\cite{AguilarArevalo:2010zc}, QE is
related to processes in which only a muon is detected in the final
state. As was already discussed above, besides genuine QE events, this
definition includes multinucleon processes. The MiniBooNE analysis of the data
attempts to correct (through a Monte Carlo estimate) for real pion production that
escapes detection through absorption in the nucleus leading to
multinucleon emission. But, it seems clear that to describe the data
of Ref.~\cite{AguilarArevalo:2010zc}, it is necessary to consider, at
least, the sum of the genuine QE
(absorption by just one nucleon), and the multinucleon contributions,
respectively. The sum of these two
contributions contribute to the CCQE-like cross section.

The inclusion of the 2p2h (multinucleon) contributions enables ~\cite{Nieves:2011yp,Martini:2011wp} 
the double differential cross section \linebreak
$d^2\sigma/dE_\mu d\cos\theta_\mu$ and the integrated flux unfolded cross section measured by MiniBooNE, to be described with
values of $M_A$ (nucleon axial mass) around $1.03\pm 0.02$ GeV~\cite{Bernard:2001rs,Lyubushkin:2008pe}. This is re-assuring
from the theoretical point of view and more satisfactory than the
situation envisaged by some other works that described the MiniBooNE
data in terms of a larger value of $M_A$ of around 1.3--1.4 GeV, as
mentioned above. 
 
\paragraph{Neutrino energy reconstruction}
\label{sec:motivation:ereco}

Neutrino oscillation probabilities depend
on the neutrino energy, unknown for broad fluxes and, for CCQE,
often estimated from the measured angle and energy
of the outgoing charged lepton $\ell$ only. It is common to define a 
reconstructed neutrino energy $E_{\rm rec}$ (neglecting binding energy and the difference of proton and 
neutron masses) as:
\begin{equation}
E_{\rm rec} = \frac{M
  E_\ell-m_\ell^2/2}{M-E_\ell+|\vec{p}_\ell|\cos\theta_\ell}\label{eq-23:defereco}
\end{equation}
which would correspond to the energy of a neutrino that emits a lepton,
of energy $E_\ell$ and three-momentum $\vec{p}_\ell$, with a gauge boson
$W$ being absorbed by a free nucleon of mass $M$ {\em at rest} in a CCQE event. 
The
actual (``true'') energy, $E$, of the neutrino that has produced the
event will not be exactly $E_{\rm rec}$.  Actually, for each $E_{\rm rec}$,
there exists a distribution of true neutrino energies that give
rise to events whose muon kinematics would lead to the given value of
$E_{\rm rec}$. In the case of genuine QE events, this distribution is sufficiently 
peaked (the Fermi motion broadens the peak and binding energy shifts it a little) around the
true neutrino energy to make the algorithm in 
Eq.~(\ref{eq-23:defereco}) accurate enough to study the neutrino
oscillation phenomenon~\cite{Meloni:2012fq}. 
However, due to the presence of multinucleon events, there is a long tail in the
distribution of true energies associated to each $E_{\rm rec}$ that makes
the use of Eq.~(\ref{eq-23:defereco}) unreliable. The effects of the inclusion of multinucleon
processes on the energy reconstruction have been noticed in \cite{Sobczyk:2012ms} and investigated in
Ref.~\cite{Martini:2012fa}, within the Lyon 2p2h model and also estimated
in Ref.~\cite{Mosel:2012hr}, using the simplified model of
Ref.~\cite{Lalakulich:2012ac} for the multinucleon mechanisms. Further discussion of this
effect and its implications is given in Section~\ref{sec:motivation:lbl} with Figure \ref{fig:erec}
giving a good indication of the size of this effect.

\paragraph{Monte Carlo event generators}
\label{sec:motivation:ccqe:mc}

Monte Carlo codes (GENIE, NuWro, Neut,
Nuance, etc) describe CCQE events using a simple RFG
model, with FSI effects implemented by means of
a semi-classical intranuclear cascade. NuWro also offers a possibility to run simulations with
a spectral function and an effective momentum dependent nuclear potential. It is also currently the only
MC generator with implementation of MEC dynamics. Since the primary interaction and 
the final state effects are effectively decoupled, FSI do not change
the total and outgoing lepton differential cross sections. 

\subsubsection{The Pion-production Region}
\label{sec:motivation:pion}

In the so-called RES region the channels of interest are mainly hadronic resonances, 
with the most important being the $\Delta (1232)$. 
Typical final states are those with a single pion. During the last five years several new 
pion production measurements
have been performed. In all of them, the targets were nuclei (most often carbon) and 
interpretation of the data 
in terms of the
neutrino-nucleon cross section needed to account for nuclear effects, impossible to 
do in a model independent manner. 
On the other hand, there has been a lot of activity in the area of coherent 
pion production and this subject will be discussed
separately.

\paragraph{Experimental Results}
\label{sec:motivation:pion:exp}

\subparagraph{{\small NC $\pi^0$}}
\label{sec:motivation:pion:exp:ncpi0}

\begin{figure}[htb]
\begin{center}
\makebox[0pt]{\includegraphics[height=6.0cm]{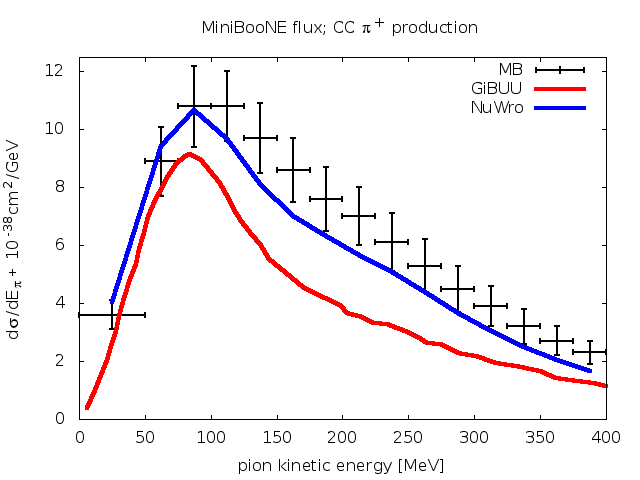}\includegraphics[height=6.0cm]{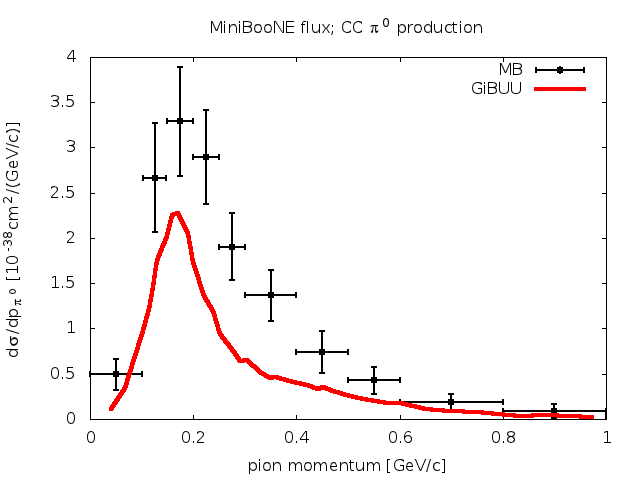}}
\end{center}
\caption{\footnotesize Left: Differential cross section for CC$1\pi^+$ production in the final state (all the FSI effects are 
included). MiniBooNE measurement  \cite{AguilarArevalo:2010bm} is compared to GiBUU \cite{Lalakulich:2011ne} and NuWro computations.
Right: the same for CC$\pi^0$ production, but only GIBUU results are shown. }\label{fig-23:pions}
\end{figure}

Neutral current $\pi^0$ production (NC$\pi^0$) is a background to the $\nu_e$ appearance 
oscillation signal.  One is interested in a $\pi^0$ leaving the nucleus and 
recent experimental data are given in this format with all the FSI effects included. 
Signal events originate mostly from: NC$1\pi^0$ primary interaction with a $\pi^0$ 
not being affected by FSI and
NC$1\pi^+$ primary interaction with the $\pi^+$ being transformed into $\pi^0$ 
in a charge exchange FSI reaction.
There are four recent measurements of NC$\pi^0$ production (K2K \cite{Nakayama:2004dp}, 
MiniBooNE neutrinos, MiniBooNE 
antineutrinos \cite{AguilarArevalo:2009ww} and
SciBooNE \cite{Kurimoto:2009wq}) that use three different fluxes: (K2K, Fermilab Booster 
neutrinos and anti-neutrinos) and 
three targets: $H_2O$ (K2K), $CH_2$ (MiniBooNE) and $C_8H_8$ (SciBooNE). 
Final results are presented as flux averaged distributions of events 
as a function of the $\pi^0$ momentum 
and, in the case of MiniBooNE and SciBooNE, also as a function of the $\pi^0$ production angle. 
\subparagraph{{\small CC $\pi^+$}}
\label{sec:motivation:pion:exp:ccpip}
MiniBooNE measured CC $1\pi^+$ production cross sections, where the signal is 
defined as exactly one $\pi^+$ in the final
state with no other mesons \cite{AguilarArevalo:2010bm}.
A variety of flux integrated differential and doubly differential cross sections, were
reported in $Q^2$ and the final state particles' momenta. The cross section results are much 
larger than NUANCE MC predictions
and the difference is on average $23\%$. In Fig.~\ref{fig-23:pions} on the 
left GiBUU and NuWro predictions for CC$\pi^+$ are compared to the MiniBooNE data. 
\subparagraph{{\small CC $\pi^0$}}
\label{sec:motivation:pion:exp:ccpi0}
MiniBooNE also measured CC $1\pi^0$ production cross sections. As before, the signal 
is defined as exactly one $\pi^0$ in the final
state \cite{AguilarArevalo:2010xt}. Various differential distributions are available. 
There is a dramatic discrepancy 
between the measured CC $1\pi^0$
production cross section as a function of neutrino energy and NUANCE MC predictions 
in the region of lower 
energies.
On average the data is larger by $56\pm 20$\%, but for $E_\nu<1$~GeV the disagreement 
is as large as a factor of $2$.
In Fig.~\ref{fig-23:pions} on the right, GiBUU predictions for CC$\pi^+$ are compared to the MiniBooNE
data. 
\paragraph{Theoretical Considerations}
\label{sec:motivation:pion:theory}
Due to nuclear effects, a comparison to the new data is possible only for MC event generators,
sophisticated computation tools like GiBUU and also a few theoretical groups which are able to evaluate FSI
effects.

Most of the interesting work was done within GiBUU. It turned out to be very difficult to reproduce 
the MiniBooNE CC$1\pi^+$ and CC$1\pi^0$ results: the measured cross section is much larger than theoretical
computations. In the case of CC $1\pi^+$ production, the discrepancy is as large as 100\%. It was also noted
that the reported shape of the distribution of $\pi^+$ kinetic energies is different 
from theoretical calculations
and does not show a strong decrease at $T_{\pi^+}>120$~MeV located in the region of maximal probability for 
pion absorption. 

The authors of \cite{Lalakulich:2011ne} mention three possible reasons for the 
data/GiBUU predictions discrepancy: 
(i) the fact that 
$\Delta$ excitation axial form factor was chosen to agree with the ANL data only, 
neglecting the larger cross section measured in the
BNL experiment; (ii) hypothetical 2p-2h-1$\pi$ pion production contribution analogous to 2p-2h discussed
in Section~\ref{sec:motivation:mnm}; (iii)
flux underestimation in the MiniBooNE experiment. For the last point, the argument gets 
support from the better 
data/theory agreement found for the ratio, as discussed below.

In the case of NC$\pi^0$ production, a systematic comparison was performed with NuWro MC predictions 
with an updated FSI model
for pions \cite{Golan:2012wx}. The overall agreement is satisfactory. Shapes of 
the distributions of final state $\pi^0$'s
are affected by an interplay between pion FSI such as absorption and {\it formation time} effects, 
understood here as an effect
of a finite $\Delta$ life-time. It is argued that NC$\pi^0$ production data 
can be very useful for benchmarking 
neutrino MC event generators.

In addition, it has been known since ANL and BNL pion production measurements 
that although being a dominant mechanism, $\Delta$ excitation alone cannot reproduce 
the data and that non-resonant background terms must be included in the theoretical models. 
There were many attempts in the past to develop suitable models
but usually they were not very well justified from the theoretical point of view. 
\paragraph{Coherent pion production}
\label{sec:motivation:pion:coh}
In coherent pion production (COH) the target nucleus remains in the ground state. 
There are four possible channels: CC and NC reactions using neutrinos or anti-neutrinos.
A clear experimental signal for the COH reaction for high energies was observed and 
the aim of recent measurements 
was to fill a gap in the 
knowledge of a region around $\sim 1$~GeV COH cross sections. At larger neutrino energies a recent 
measurement was made by 
MINOS which reported a 
NC reaction cross section at $<E_\nu>=4.9$~GeV to be consistent with the predictions 
of the Berger-Sehgal model 
(see below).
\subparagraph{{\small Experimental Results}}
\label{sec:motivation:pion:coh:exp}
In the case of the NC reaction, MiniBooNE \cite{AguilarArevalo:2008xs} and SciBooNE \cite{Kurimoto:2010rc} 
searched for the COH component. SciBooNE \cite{Kurimoto:2010rc} evaluated the ratio 
of the COH NC$\pi^0$ 
production to the total CC cross section as $(1.16\pm 0.24)\%$. 

MiniBooNE evaluated the NC COH component (plus possible hydrogen diffractive contribution about which 
little is known) in 
the NC$\pi^0$ production as 19.5\% (at $<E_\nu>\sim 1$~GeV) and then the overall 
flux averaged overall NC$1\pi^0$ 
cross section as 
$(4.76\pm0.05\pm0.76)\cdot 10^{-40}$cm$^2$/nucleon. Unfortunately, it is difficult 
to translate both measurements 
into the 
absolutely normalized value of the NC COH cross section because of strong
dependence on the NUANCE MC generator used in the data analysis.  

In the case of the CC reaction, K2K \cite{Hasegawa:2005td} and SciBooNE 
\cite{Hiraide:2008eu} reported 
no evidence for the COH component. 
For the K2K analysis, 
the 90\% confidence limit upper bound for the COH cross sections on carbon was estimated 
to be $0.6\%$ of the inclusive 
CC cross section.
The SciBooNE upper limits (also for the carbon target) are: 
$0.67\%$ at $<E_\nu>\sim 1.1$~GeV, and $1.36\%$ at 
$<E_\nu>\sim 2.2$~GeV.
SciBooNE also reported the measurement of the ratio of CC COH $\pi^+$ to NC COH $\pi^0$ 
production and estimated it 
as $0.14^{+0.30}_{-0.28}$. This is a surprisingly low value, which disagrees with results from the 
theoretical models which, at SciBooNE energies, typically predict values somewhat smaller than $2$. For massless
charged leptons, isospin symmetry implies the value of $2$ for this ratio and the finite mass corrections 
make the predicted 
ratio smaller.
\subparagraph{{\small Theoretical developments}}
\label{sec:motivation:pion:coh:theory}
Higher neutrino energy ($E_\nu >\sim 2$~GeV) COH production data 
(including recent NOMAD measurements) were successfully 
explained with 
a PCAC based model \cite{Rein:1982pf}. Adler's theorem relates 
$\sigma_{COH}(\nu + X\rightarrow \nu + X + \pi^0)$ 
at $Q^2\rightarrow 0$ 
to $\sigma(\pi^0 + X \rightarrow \pi^0 + X)$. Subsequently, the model 
for the CC reaction, has been upgraded
\cite{Berger:2007rq} 
to include lepton mass effects important for low $E_{\nu}$ studies. 
The new model predicts the $\sigma_{COH} (\pi^+)$/$\sigma_{COH} (\pi^0)$ 
ratio at $E_\nu=1$~GeV 
to be $1.45$ rather than $2$. Another important improvement is to use a better model for 
$d\sigma(\pi +\ {}^{12}C \rightarrow \pi +\ {}^{12}C)$/$dt$ in the region of pion kinematic energy 
$100$~MeV$<T_\pi<900$~MeV.  
As a result, the predicted COH cross section  from the model became reduced 
by a factor of 2-3 \cite{Berger:2009nc}. The PCAC based approach 
is also discussed in \cite{Paschos:2009dn} and critically re-derived in Ref.~\cite{Hernandez:2009vm}.
At lower energies the microscopic $\Delta$ dominance models for the COH reaction 
\cite{Nakamura:2009iq,Valverde:2009cy,Hernandez:2010jf,AlvarezRuso:2007sq}  are believed to be more reliable.

\paragraph{MC generators}
\label{sec:motivation:pion:coh:mc}
Almost all MC events generators rely on the 
Rein-Sehgal resonance model for pion resonance production \cite{Rein:1980wg}. 
The model is based on the quark resonance model and includes 
contributions from 18 resonances covering the region $W<2$~GeV. 
The model is easily implementable in MC generators and it has only one set of vector and
axial form factors. 
In the original model, the charged 
lepton is assumed 
to be massless and prescriptions to cope with this problem 
were proposed in Refs. \cite{Graczyk:2007xk}. 
It was also realized that 
the Rein-Sehgal  model can be improved in the $\Delta$ region
by modifying both vector and axial form factors using either old deuterium or new 
MiniBooNE pion production data \cite{Nowak:2009se,Graczyk:2007bc} . 

For coherent pion production, all the MCs use the Rein-Sehgal COH model~\cite{Rein:1982pf} analysis of 
of MC event generators and theoretical models described in \cite{Boyd:2009zz} show 
that in the $1-2$~GeV energy region, the Rein Sehgal COH model predictions disagree significantly
with all the recent theoretical computations and experimental results. None of the microscopic
models, which are believed to be more reliable in the $1$~GeV region, have been implemented in Monte
Carlo codes yet.

\paragraph{Duality}
\label{sec:motivation:duality}
Bridging the region between RES and DIS (where interactions occur on quarks, to a good approximation)
dynamics is a practical problem which must be resolved in all
MC event generators.  In MC event generators ``DIS'' is defined as ``anything but QE and RES''. This is
usually expressed as a condition on a lower limit for the invariant hadronic mass, such as $W>1.6$~GeV
for example. 
Notice, however, that such a definition of ``DIS'' contains a contribution from the
kinematical region $Q^2<1$~GeV$^2$ which is beyond the applicability
of the genuine DIS formalism.  
The RES/DIS transition region is not only a 
a matter of an arbitrary choice, but is closely 
connected with the hypothesis of quark-hadron duality. 

Investigation of structure functions introduced in the formalism of the
inclusive electron-nucleon 
scattering led Bloom and Gilman to the observation that the average over resonances 
is approximately equal to the leading twist contribution measured in the completely 
different DIS region. One can
distinguish two aspects of duality: (i) resonant structure functions oscillate around a DIS scaling 
curve; (ii) the resonant structure functions for varying values of $Q^2$ 
slide along the DIS curve evaluated at fixed $Q^2_{DIS}$.

As a practical procedure for addressing this region, Bodek and Yang~\cite{Bodek:2010zz}  
have introduced and refined a model that is used by many contemporary neutrino event generators,
such as NEUGEN and its successor GENIE, to bridge the kinematic region between the Delta and full DIS.
The model has been developed for both neutrino- and electron-nucleon inelastic scattering 
cross sections using leading order parton distribution functions and introducing a new 
scaling variable they call $\xi_w$. 

At the juncture with the DIS region, the Bodek-Yang model incorporates the GRV98~\cite{Gluck:1998xa} 
LO parton distribution functions replacing the variable x with $\xi_w$.  
They introduce ``K-factors", different for sea and valence quarks, to multiply the PDFs so that 
they are correct at the low $Q^2$ photo-production limit.  A possible criticism of the 
model is the requirement of using the rather dated GRV98 parton distribution functions in the DIS region 
to make the bridge to the lower W kinematic region seamless.

\subsubsection{$\nu$-A Deep-inelastic Scattering}
\label{sec:motivation:dis}
Although deep-inelastic scattering (DIS) is normally considered to be a topic for much higher energy neutrinos, 
wide-band beams such as the Fermilab NuMI, the planned LBNE and the nuSTORM beams do have real 
contributions from DIS that are 
particularly important in feed-down to the background and thus must be carefully considered.  
In addition, there are 
x-dependent nuclear effects that should be taken into account when comparing results 
from detectors with different nuclei and even when comparing results from ``identical" near 
and far detectors when the neutrino spectra entering the near and far detectors are different.

Nonetheless, the DIS fraction of events for nuSTORM are on the order of 10\% or less.  
The contribution of nuSTORM to the the study of DIS is expected to be of similar precision.
For a full review of neutrino DIS scattering results see~\cite{Morfin:2012kn} .

\paragraph{nuSTORM Contributions to the Study of Neutrino-Nucleus Scattering}
\label{sec:motivation:nustorm}

\begin{figure}
  \centering
  \includegraphics[width=0.9\textwidth]{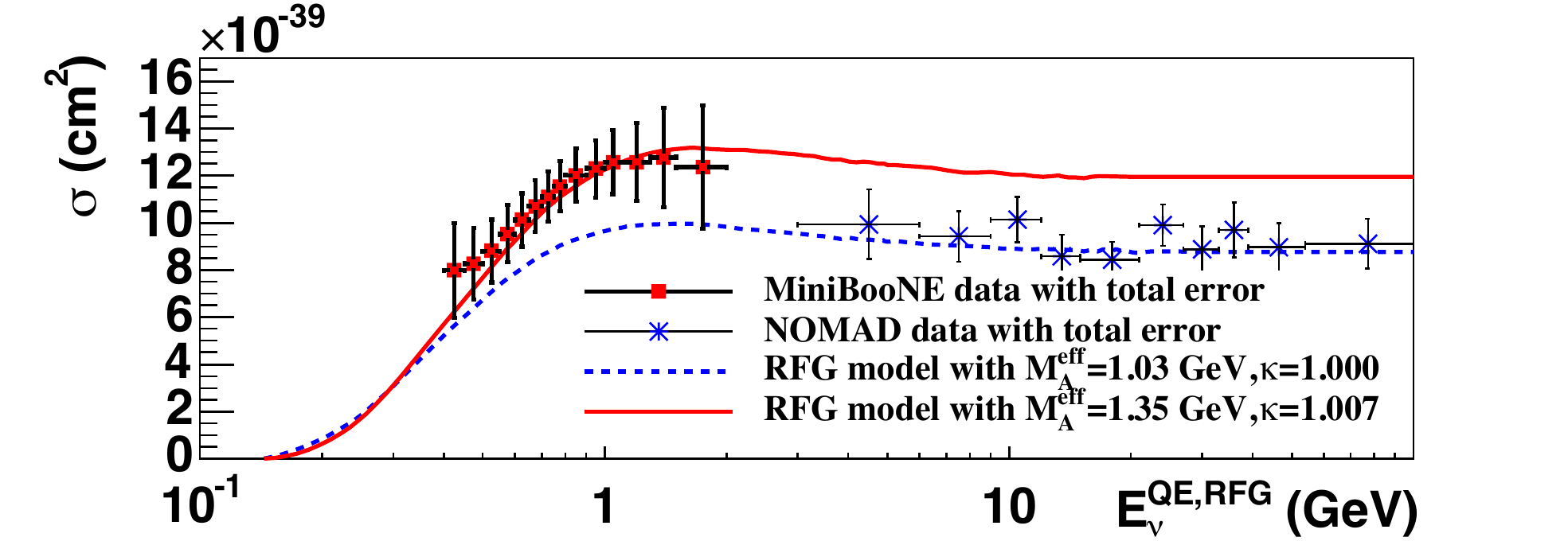}
  \caption{
    The total charged-current quasi-elastic cross section for
    $\nu_{\mu}$ as measured by the MiniBooNE and NOMAD experiments.
  }
  \label{fig:MiniNOMAD}
\end{figure}

As an example of how nuSTORM will directly contribute to the study of neutrino-nucleus scattering, 
a  comparison of the MiniBooNE and NOMAD results is shown in Fig.~\ref{fig:MiniNOMAD}.  
Certainly one explanation for the difference between these two measured quasi-elastic cross sections
is a mis-estimation of the absolute neutrino flux of one or both of the experiments.
The claimed accuracy of the MiniBooNE flux is 8\% while the NOMAD flux has a similar claim of 
8\% accuracy\cite{Astier:2003rj}.  Therefore, the difference between the MiniBooNE and 
NOMAD results just from flux normalization spread alone is less than two standard deviations. With 
the nuSTORM flux resolution of $\leq$ 1\%, the actual quasi-elastic cross section off carbon could 
be measured with an error nearly an order of magnitude smaller than either the MiniBoNE or NOMAD experiments.  

It is important to point out that no other neutrino experiment now being planned can match the power 
of nuSTORM to answer the many challenges of neutrino-nucleus scattering.  All use the 
conventional pion-decay source neutrino beams and, as such, can not expect to know 
the actual flux of neutrinos entering their detectors to better than (5-7) \% at best compared with 1\% or better for the nuSTORM beam.

\subsubsection{Differences in the energy-dependent cross sections of $\nu_{\mu}$- and $\nu_e$-nucleus interactions}
\label{sec:motivation:mue}

To determine the mass hierarchy of neutrinos and to search for
CP-invariance violation in the neutrino sector, current and upcoming
accelerator-based neutrino-oscillation experiments such as T2K
\cite{Abe:2011ks} and NOvA \cite{Ayres:2004js} as well as future
proposed experiments such as LBNE and the Neutrino
Factory \cite{Akiri:2011dv} plan to make precision measurements of the
neutrino flavor oscillations $\parenbar{\nu}_\mu \to \parenbar{\nu}_e$ or
$\parenbar{\nu}_e \to \parenbar{\nu}_\mu$.
An important factor in the
ability to fit the difference in observed event rates between the near
and far detectors will be an accurate understanding of the cross
section of $\nu_{\mu}$- and $\nu_e$-nucleus interactions.
Uncertainties on differences in expected event rates due to
differences between these cross-sections will contribute to
experimental uncertainties in these flavor-oscillation measurements.

\begin{figure}
  \centering
  \includegraphics[width=0.65\textwidth]{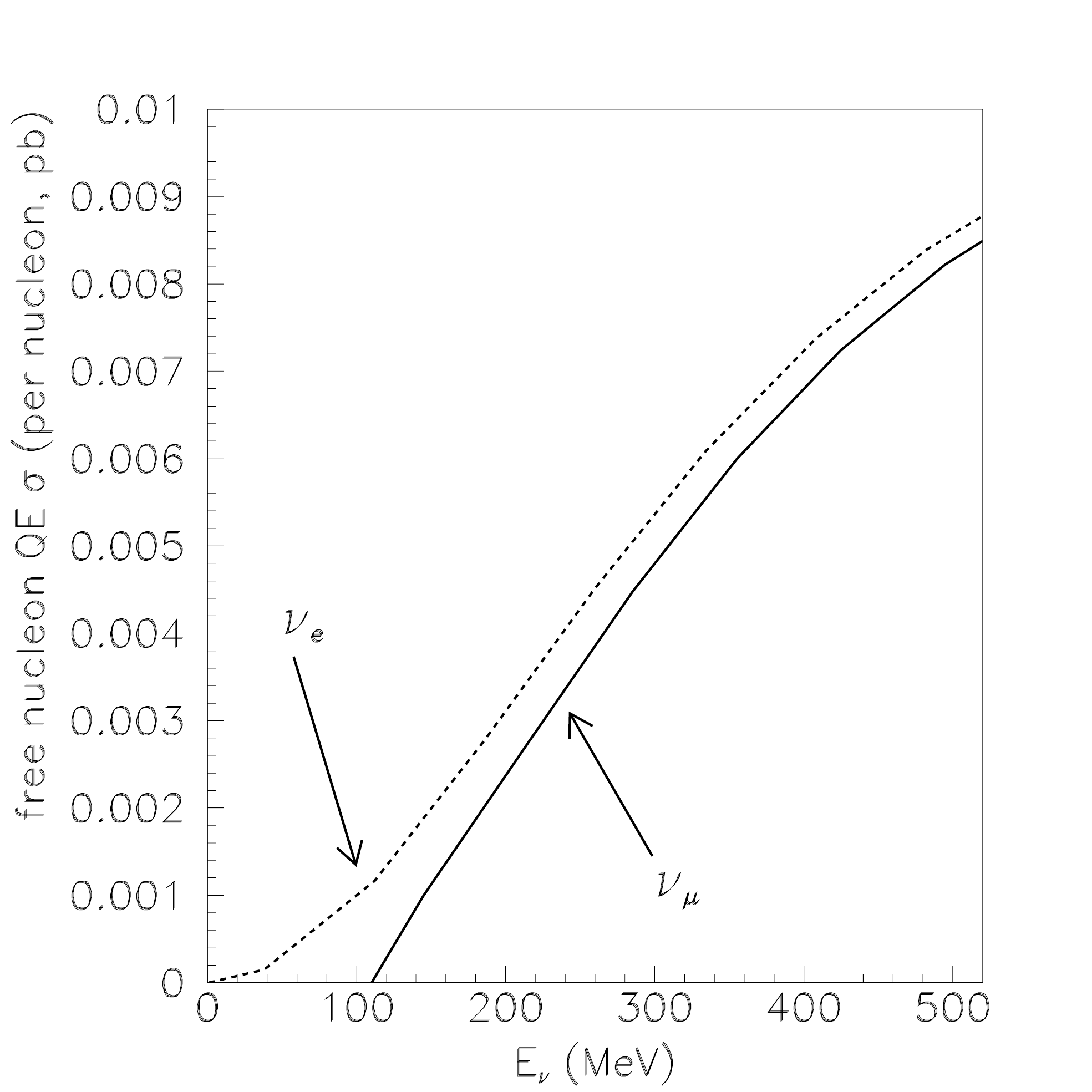}
  \caption{
    The total charged-current quasi-elastic cross-section for
    $\nu_{\mu}$ and $\nu_e$ neutrinos.
  }
  \label{fig:qe}
\end{figure}

There are obvious differences in the cross sections due to the
difference in mass of the outgoing lepton.
These can be calculated by including the lepton-mass term in the
cross-section expression.
Fig.~\ref{fig:qe} \cite{Day:2012gb}, shows these expected
differences in the cross sections as a function of neutrino energy.
Another such calculable difference occurs because of radiative
corrections.
Radiative corrections from a particle of mass $m$ are proportional to
$\log(1/m)$, which implies a significant difference since the muon is
$\sim 200$ times heavier than the electron \cite{De_Rujula:1979jj}.
This turns into a difference of $\sim 10$\% in the cross sections.
In addition to these differences, there are other more subtle
differences due to the coupling of poorly-known or unknown form
factors to the lepton tensor that reflect the differences in the
outgoing lepton mass.
These effects have been investigated in some detail~\cite{Day:2012gb}
but must be probed experimentally.

Regarding nuclear effects, while there are no differences expected in
the final-state interactions, there are expected differences in the
initial reaction cross-sections between $\nu_{\mu}$- and
$\nu_e$-nucleus interactions.
Since the lepton tensor, reflecting the mass of the outgoing lepton,
couples to the hadron-response functions, there is a difference in
nuclear effects at the interaction vertex due to the $\mu$-to-$e$ mass
difference.
The expected difference in the $\nu_{\mu}$- and $\nu_e$-nucleus
cross-section ratio is around 5\% when using a spectral-function model
\cite{Benhar:1994hw} for the initial nucleon momentum compared to the
relativistic Fermi gas model \cite{Smith:1972xh,Bodek:1980ar}.
There is another 5\% difference expected for the multi-nucleon ($n$p-$n$h)
contributions \cite{Martini:2012:private}.
These differences in cross sections extend up into the resonance
region with the low-$Q^2$ behavior of $\Delta$ production, exhibiting
10\% differences at values of $Q^2$ where the cross section has
leveled off.

While each of the individual effects outlined above may not be large
compared to current neutrino-interaction uncertainties, they are large
compared to the assumed precision of oscillation measurements in the
future LBL program.
Moreover, the sum of these effects could be quite significant and the
uncertainty in our knowledge of the size of these effects will
contribute directly to uncertainties in the neutrino-oscillation
parameters determined from these experiments---and these uncertainties
can only be reduced with good quality $\parenbar{\nu}_e$ scattering
data.
nuSTORM is the only source of a well-understood and well-controlled
$\parenbar{\nu}_e$ neutrino beam (and $\parenbar{\nu}_\mu$  )with which these cross-section
differences can be studied systematically.

\subsubsection{Systematics of Long Baseline Oscillation Measurements}
\label{sec:motivation:lbl}

\begin{figure}
  \centering
  \mbox{
      \includegraphics[width=0.6\textwidth]{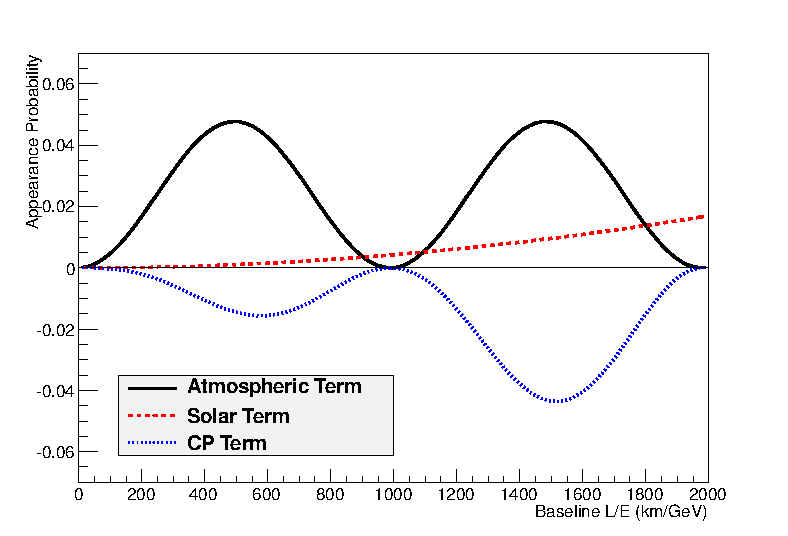}
  }
\caption{The $\nu_{\mu} \rightarrow \nu_{e}$ appearance probability as a function of baseline.
The probability contains a term that controls oscillations at the atmospheric mass scale, a term
that controls oscillations at the solar mass scale, and an interference term which contains the
CP phase.}
\label{fig:LBAppearance}
\end{figure}

The recent measurement of a large value for the previously unknown mixing angle, $\theta_{13}$, has
raised the hope that the next generation of long baseline neutrino oscillation experiments could observe
CP-violation in the lepton sector. Such a result would have
far-reaching consequences in our understanding of the flavor and baryon asymmetry problems and
could open a window to studying physics at the GUT scale. CP violation is introduced in
the neutrino flavor oscillation mechanism through a CP-violating phase, $\delta_{CP}$, in the
PMNS mixing matrix. It can be accessed experimentally through two methods : a non-zero phase
will lead to a difference between the $\nu_{\mu} \rightarrow \nu_{e}$ and $\overline{\nu_{\mu}}
\rightarrow \overline{\nu_{e}}$ appearance probabilities as well as a variation in the relative value of
the second oscillation maximum with respect to the first oscillation maximum in $\nu_{\mu} \rightarrow \nu_{e}$
appearance measurements. Measurement of this CP-violating phase is challenging, regardless of
the method used. Fig.~\ref{fig:LBAppearance} shows the components of the $\nu_{\mu} \rightarrow \nu_{e}$
appearance probability as a function of the experiment baseline calculated with oscillation parameters
equal to our best-known estimates and assuming maximal CP violation ($\delta_{CP} = \frac{\pi}{2}$).
The maximum appearance probability at any baseline is less than 6\%, and
in the baseline regions being considered by the next generation of experiments 
($L/E \approx 400-500~\mbox{km/GeV}$), the component of
the probability that is sensitive to the CP-phase is subleading. Any experiment that is attempting
to measure CP-violation using neutrino oscillations must therefore be capable of measuring a small
difference between small number of events, and in this context it is imperative that all systematic errors
be well controlled, either by external measurements or by measurements at near detectors.

\begin{figure}
  \centering
  \mbox{
      \includegraphics[width=0.6\textwidth]{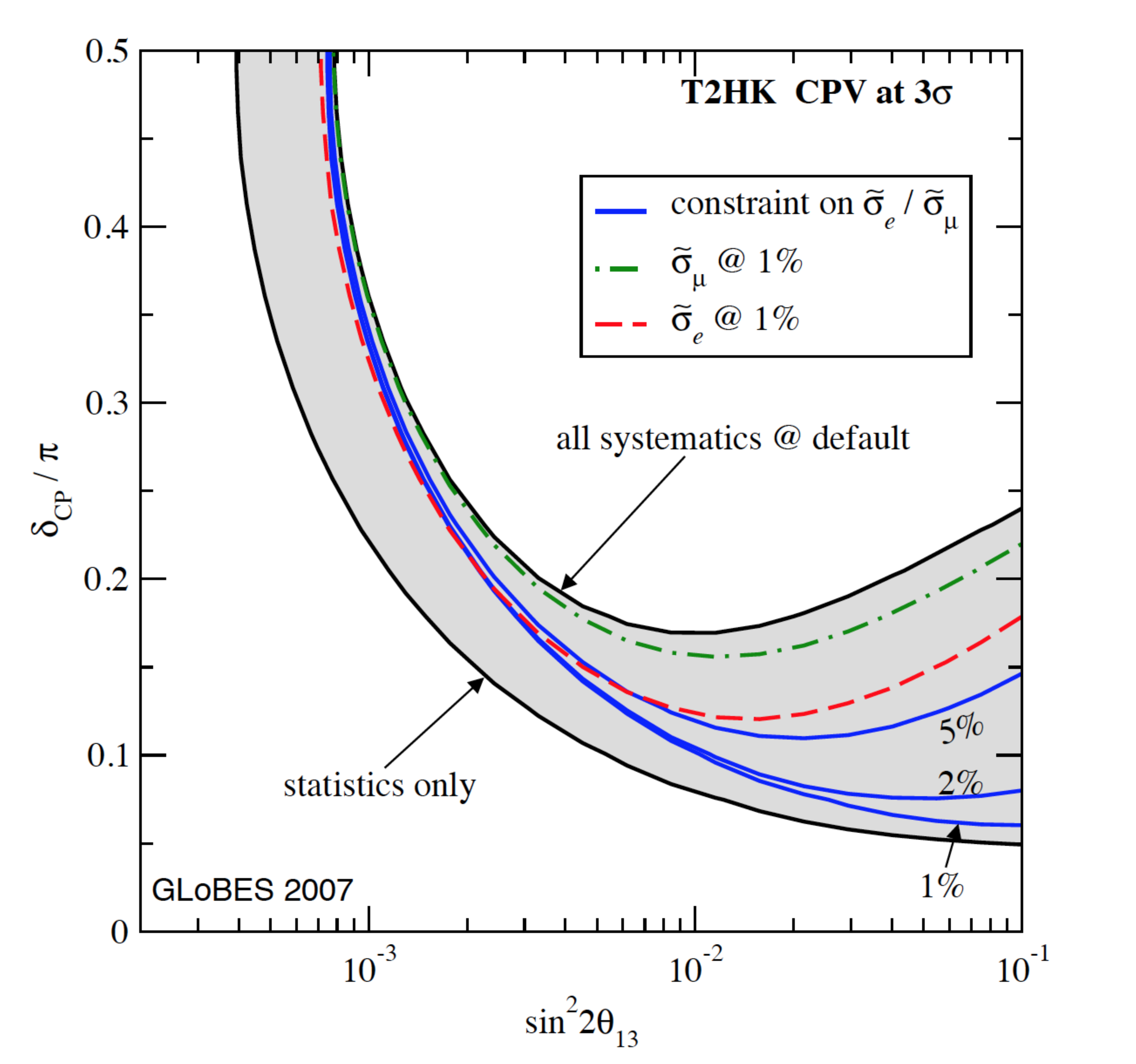}
  }
\caption{The T2HK $3\sigma$ CP phase coverage for a 2 year neutrino, 6 year antineutrino
run in the 400 kW JPARC beam using a 500 kt far detector. Details can be found in \cite{Huber:2007em}.}
\label{fig:huber2007em}
\end{figure}

One of the largest systematic errors come from poor knowledge of the interaction cross sections. 
Fig.~\ref{fig:huber2007em} shows the $3 \sigma$ coverage of $\delta_{CP}$ as a function of 
$sin^{2}(2\theta_{13})$ for the proposed experimental setup of T2HK. The recent measurements
of $sin^{2}(2\theta_{13})$ : $[0.0944 \pm 0.007 \mbox{(stat)} \pm 0.003 \mbox{(sys)}]$ from
Daya Bay\cite{Dwyer:2013wqa}, $[0.113 \pm 0.013 \mbox{(stat)} \pm 0.014 \mbox{(sys)}]$ from
RENO\cite{Ahn:2012nd} and $[0.097 \pm 0.034 \mbox{(stat)} \pm 0.034 \mbox{(sys)}]$\cite{Abe:2013sxa}
from Double CHOOZ locate the next experiments at the
extreme right-edge of the plot. If only statistical errors are considered, then T2HK has
$3 \sigma$ sensitivity over almost the entire range of $\delta_{CP}$. However, the coverage range decreases
significantly if the $\nu_{\mu}$ and $\nu_{e}$ interaction cross-section ratio is assumed to have
a systematic uncertainty of 10\%, as shown by the ''default'' line in Fig.~\ref{fig:huber2007em}. One can see
that in order to have the largest $\delta_{CP}$ coverage, the ratio of the $\nu_{e}$ to the
$\nu_{\mu}$ cross sections need to be understood at the few percent level. Such a precise constraint
cannot be made at a conventional neutrino experiment. As an example, the uncertainty in the 
$\nu_{\mu}$ to $\nu_{e}$ flux ratio at an experiment in a conventional neutrino beam
is on the order of 10\%. Only
a facility with a precisely defined beam, such as nuSTORM, is capable of delivering $\nu_{\mu}$
and $\nu_{e}$ cross section measurements with the required precision.

The effect of cross section uncertainties differs depending on the design of the experiment. T2HK 
runs with an off-axis beam tuned to a peak neutrino energy of 600 MeV. At this energy, the charged
current cross sections are dominated by the quasi-elastic scattering process which has significant
uncertainties. These uncertainties not only affect coverage, but also the precision of the measurement.
Even with the most optimistic of runtime scenarios (2 years of neutrino running followed by 6 years 
of antineutrino running), the precision to which T2HK could determine $\delta_{CP}$ never falls below
about $8^{\circ}$ \cite{Coloma:2012ji}, limiting the capability of T2HK to distinguish between different
flavor models. Other experimental setups, such LBNE, operate at higher energy and in wide-band beams. 
Coloma {\em et al.} show that these wide-band beam experiments are less affected by systematic
errors on the $\sigma_{e}/\sigma_{\mu}$ cross section ratio\cite{Coloma:2012ji}, although the study assumes an
uncertainty of about 2\% on the $\nu_{e}$ to $\nu_{\mu}$ cross section ratio in the resonance
energy regime. It should be noted that this uncertainty is purely theoretical. No measurements of this
ratio exist, but it is expected that the uncertainty on any measurement using current conventional 
neutrino beams would be dominated by the knowledge of the neutrino flux and would not be much less than
10\%.

\begin{figure}[htb]
\begin{center}
\makebox[0pt]{\includegraphics[height=6.0cm]{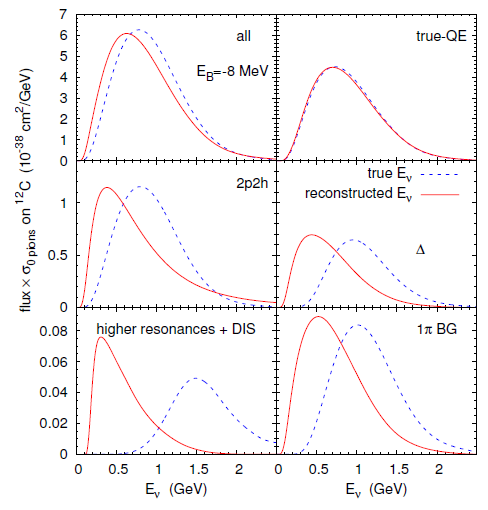}\includegraphics[height=6.0cm]{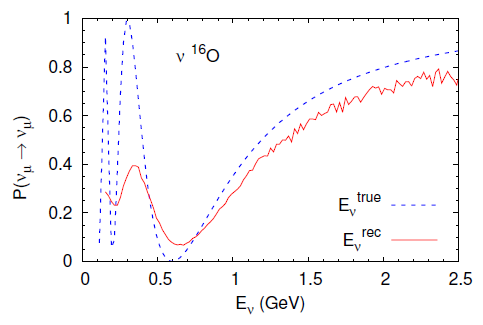}}
\end{center}
\caption{\footnotesize Left: Effect on the reconstructed neutrino energy spectrum of assuming a
quasi-elastic topology in MiniBooNE. The neutrino energy spectrum generally underestimates the true spectrum for
non-elastic event classes. Right: The muon neutrino disappearance probably for a T2K-like experiment shows
a systematic shift in the oscillation dip due to the mis-reconstruction of the neutrino energy spectrum.}\label{fig:erec}
\end{figure}

All of these experimental setups, moreover, suffer from the problem that they must estimate the neutrino energy
using the outgoing charged lepton partner from interactions on complex nuclear targets.
Imperfect knowledge of the micro-physics of the nuclear environment can generate systematic
differences in the mapping between the charged-lepton momentum and the neutrino energy between
the model and the data. Fig.~\ref{fig:erec} (left)  shows the effect of assuming a quasi-elastic like
neutrino energy reconstruction (which is a reasonably standard procedure) for different types of
event classes in the MiniBooNE experiment. The top right panel shows that the procedure reconstructs
the neutrino energy in true quasi-elastic events very well, but that for other event classes (such as
the 2p2h event class in the middle left panel), the reconstructed energy significantly underestimates
the true neutrino energy. 
Such uncertainties cannot be mitigated by a near detector 
unless and until the model calculations are sufficiently detailed to allow falsification with final-state
particle data. Systematic effects like this, which are unknown or poorly modeled, have been shown to 
lead to a bias in oscillation parameters which are obtained from fits to the reconstructed neutrino energy
spectrum \cite{Lalakulich:2012hs,Nieves:2012yz}. For example, Fig.~\ref{fig:erec} (right) shows the $\nu_{\mu}$
disappearance probability for a T2K-like experiment. The position of the first oscillation maximum is
shifted slightly from its true position due to the influence of non-quasi-elastic event classes. If these
event classes are included in the model, then a correction can be made, but the danger lies in an event
class not being included in a model where it exists in the data. Such a situation has the potential to
return a biased estimate of the oscillation parameters.
Hence, understanding and incorporating these effects 
into the models is a high-priority task. This
can only be done using dedicated experiments with high-resolution detectors, which can provide detailed information
about the final-state of neutrino interactions on different target types. These detectors must be
 paired with a neutrino
beam which is precisely known. Currently only the nuSTORM facility design is capable of making these measurements.

\subsubsection{Detector Options}
\label{sec:motivation:detectors}

The physics program outlined in Section \ref{sec:motivation:physics}
cannot be performed without the precision beam delivered by the nuSTORM
facility. This beam must be coupled with a detector or detectors which 
are capable of making the measurements required to understand the interaction
of neutrinos on nuclei.  A consideration of the physics topics
suggests that the required detector must have the following capabilities~:
\begin{itemize}
\item{\bf Different nuclear targets : } Much of the physics which complicates
the understanding of neutrino interactions arise from the necessary
interaction of the neutrino in a nuclear potential. To understand the effect 
of nuclear targets, we need to compare the final states of neutrino interactions on light and
heavy targets while using the same $\nu$ or $\overline{\nu}$ energy distribution. 
Targets from carbon to iron are now routinely used. New 
liquid argon experiments will provide data on a heavy noble liquid. One
issue that is important, but not currently being addressed, is the 
requirement for data on light targets such as helium or, preferably,
hydrogen or deuterium.
\item{\bf Vertex imaging : } Many of the nuclear models predict differing
numbers of low energy nucleons being ejected from the nucleus. Currently
no experiment can image the vertex with sufficient resolution to collect
data on low energy vertex activity to test these models. The only detectors
which could do this are liquid or gas argon TPCs or a bubble-chamber type imaging device.
\item{\bf Flavor identification : } The nuSTORM beam delivers an equal
mix of $\nu_{\mu}$ and $\overline{\nu_{e}}$ (or $\overline{\nu_{\mu}}$ and $\nu_{e}$). 
A detector must
be capable of distinguishing between the final state leptons in charged
current interactions. This requires a detector capable of either track/shower
discrimination or particle flavor identification.
\item{\bf Final state reconstruction : } It is extremely important to
understand the topology of the final-state hadronic system as this topology
reflects both the hadronic processes in the bare interactions and the
secondary interactions of final state particles as they pass
through the nucleus. These final state interactions are one of the largest
sources of uncertainty in the current round of oscillation experiments.
Comparison of the final state on light and heavy targets will help constrain
the final state models. 
\item{\bf Sign selection : } The capability for determining the charge of the final
state particles emitted from the interaction is very desirable. The inclusion of
a magnet into the detector design is considered for all stand-alone detector designs
described below. However, it may not be possible to include this capability fully
in the initial measurement phase. This is, however, important for both neutrino
oscillation physics, which relies on determining the sign of the outgoing charged lepton, and for
neutrino interaction physics which would benefit substantially by separation of
final state pions by charge.
\end{itemize}

It would be difficult to design a single detector that could embody
all of these requirements without compromising on performance. We envision
the nuSTORM beam to be a facility serving a number of different detectors
for different types of studies. nuSTORM could run as a phased research
program, initially with existing detectors, and then adding new experiments optimized
for different measurements.

The first measurements could involve using existing, possibly refurbished, 
well-understood detectors. The MINER$\nu$A experiment \cite{Osmanov:2011ig}, currently operating
in the NuMI beam in the energy range that the nuSTORM beam would deliver, 
could be moved to the nuSTORM facility.  This can
be accomplished with relatively minimal expenditure although the photosensors,
currently multi-anode PMTs, may need to be replaced with silicon 
photomultipliers. MINER$\nu$A can provide data on interactions on
carbon, lead and iron with reasonable tracking and calorimetry. It has 
some capacity to image details of the neutrino interaction vertex, but a
better detector for this purpose would be one of the liquid argon
detectors that are currently being developed, or are envisioned as part
of the LBNE research program. 
One problem with these choices of detectors is that neither 
MINER$\nu$A, nor any of the planned liquid argon systems, can
measure the charge or momentum of muons from charged current interactions. 
A new muon ranger would have to
be provided in the nuSTORM facility. 
The near-detector companion to SuperBIND for the short-baseline $\nu$ oscillation component of
nuSTORM could be used as the muon ranger (see section~\ref{sec:far}).  It is currently specified 
with a mass of 200 T so a more detailed analysis of its use at the muon ranger for the $\nu$ interaction
component of nuSTORM will have to be done.  This would be the only new 
detector needed for the initial phase of cross-section measurement program.

An alternate possibility could be a completely new, high-resolution, magnetized
tracking detector. Two concepts are being studied: 
\begin{enumerate}
\item The first is the HIRESMNU
detector, proposed and fully costed as a near detector for the LBNE experiment \cite{HIRESMNU1,HIRESMNU2}.
Fig.~\ref{fig-det-schematic} shows a schematic of the HIRESMNU
design.
This detector embeds a $4 \times 4 \times 7$\,m$^3$ Straw-tube tracker (STT),
surrounded by a 4$\pi$ electromagnetic calorimeter (ECAL) in a  dipole
magnet with  $B \simeq 0.4$\,T.
Downstream of the magnet, and within the magnet yoke, are detectors
for muon identification.
The STT will have a low average density similar to liquid hydrogen,
about 0.1\,gm/cm$^3$, which is essential for momentum determination
and the identification of electrons, protons, and pions.
The foil layers, up- and down-stream of the straw tubes, provide
the transition-radiation and constitute most of the 7\,ton fiducial
mass.
The foil layers serve both as the mass on which the neutrinos will
interact and as generators of transition radiation (TR), which
provides electron identification.
\begin{figure}
  \begin{center}
    \includegraphics[width=0.8\textwidth]{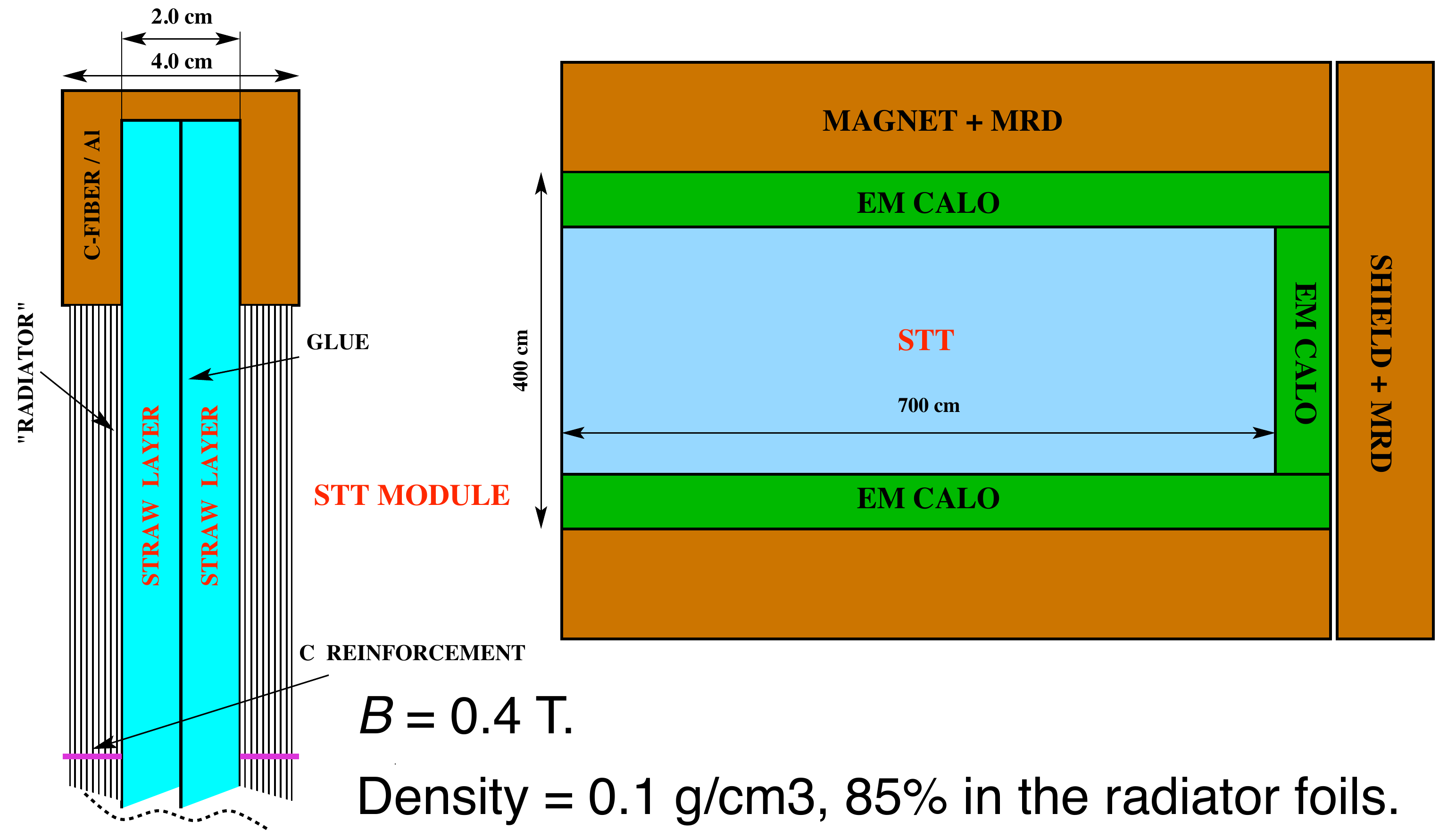}
  \end{center}
  \caption{
    Schematic of the HIRESMNU concept showing the straw tube tracker (STT), the
    electromagnetic calorimeter (ECAL) and the magnet with the muon
    range detector (MRD).
    The STT is based upon ATLAS \cite{ATLAS-TRT1,ATLAS-TRT2,ATLAS-TRT3}
    and COMPASS \cite{COMPASS1,COMPASS2} trackers.
    Also shown is one module of the proposed straw tube tracker (STT).
    Interleaved with the straw tube layers are plastic foil radiators,
    which provide 85\% of the mass of the STT.
    At the upstream end of the STT are layers of nuclear-target for
    cross section measurements.
  }
  \label{fig-det-schematic}
\end{figure}
\item A versatile detector option has been proposed as a near detector for LBNO,
the Gas Argon Modular Apparatus for Neutrino Detection
($\gamma$$\nu$det) \cite{LBNO:EoI2CERN}.
It is based on a pressurized argon time projection chamber (TPC) located in
a large 5 m diameter pressure vessel, as shown in Fig.~\ref{fig:GAMAND}.
A magnetic field is applied to the full volume of the pressure vessel.
A magnet design similar to that of the UA1/NOMAD spectrometer
dipole magnet provides a field with characteristics close to those
required for this detector.
The pressure vessel can accommodate several layers of a
scintillator-based calorimeter, such as the SuperBIND plastic
scintillator material, between the TPC and the pressure vessel inner
surface.
One advantage of this design, compared with a more compact pressure
vessel enclosing only the TPC, is that there is less redundant/passive
material between the TPC and first layers of the scintillator detector
and fewer blind spots.
The outer layers of the embedded scintillating material can be
interleaved with radiators to improve the containment within the
pressure vessel of more energetic secondaries from events of interest
occurring in the TPC.
High energy muons ($>$ 1 GeV) of both signs  from neutrino events
originating in the TPC would be measured downstream in the muon
ranger/spectrometer described above.

\begin{figure}
  \begin{center}
    \includegraphics[width=0.7\textwidth]{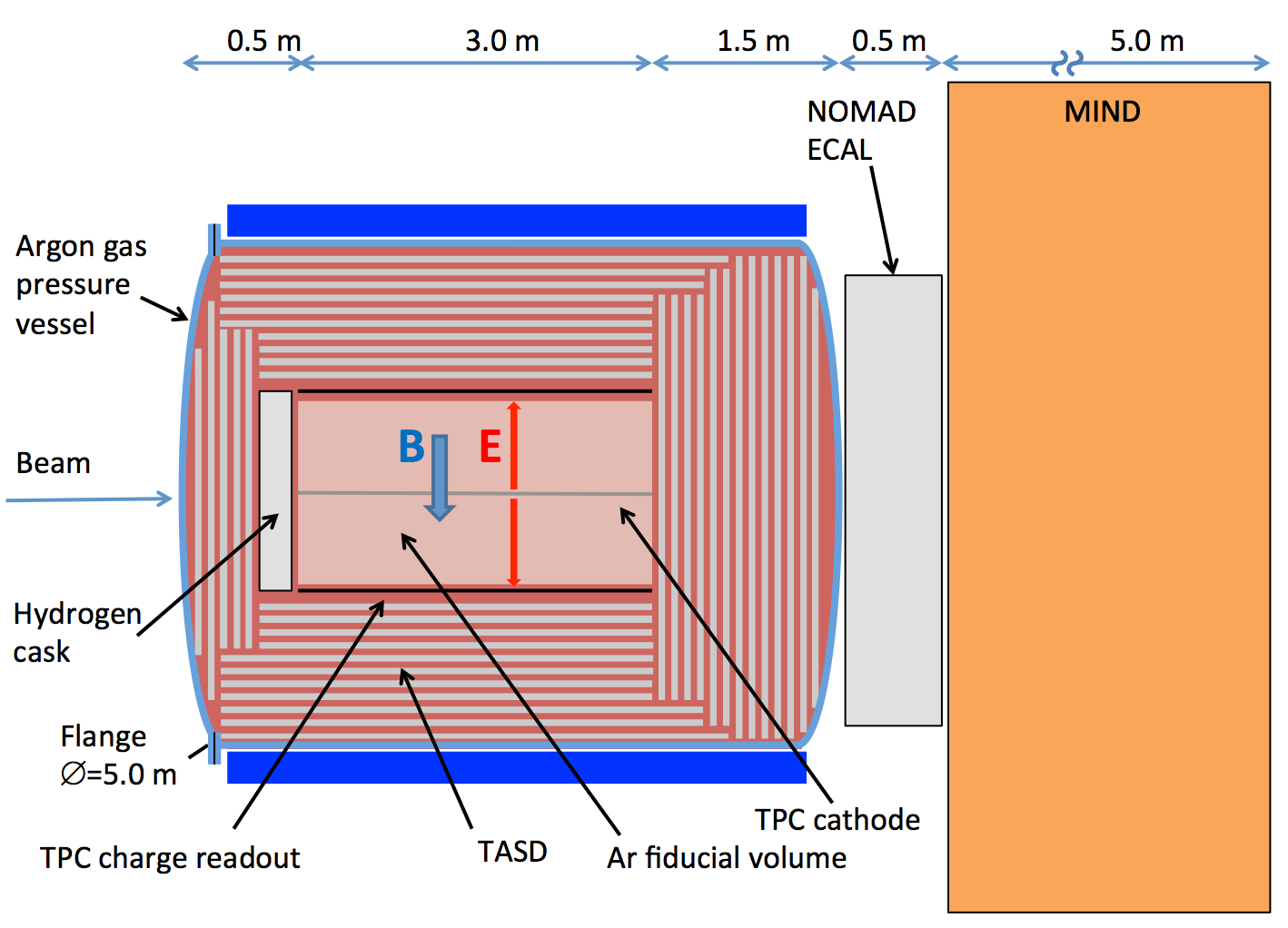}
  \end{center}
  \caption{
    Schematic of the pressurized argon gas-based TPC detector.
    Both the TPC and scintillator calorimeter layers surrounding it
    are enclosed in a pressure vessel.
    A 0.5\,T magnetic field is applied to the pressure vessel volume.
    Downstream of the TPC are also an electromagnetic calorimeter
    (ECAL) and a magnetized iron neutrino detector (MIND/SuperBIND).
    The latter acts as a muon spectrometer for neutrino interactions
    occurring in the TPC and as an independent near detector for the
    sterile neutrino program.
  }
  \label{fig:GAMAND}
\end{figure}
\end{enumerate}

The details and costing of each of these options is currently under study.

One limitation that is apparent in these detector options is 
that they all involve interactions on heavy targets. It is
becoming clear, however, that much of the current confusion around neutrino interaction
physics arises from the low statistics, sometimes conflicting, data on 
light targets that are the foundation of many of the models. It is possible that a 
light-nuclei target could be integrated with either HIRESMNU or the gaseous TPC option
in the form of a solid target inside the magnet of HIRESMNU or the pressure vessel
of the gaseous TPC detector. However, the constraints of the proposed designs make
implementation of such a system difficult and sub-optimal. An alternate possibility
in a later phase of 
nuSTORM would involve repeating, or in many cases performing the first,  measurements
of interactions on hydrogen or deuterium nuclei. The techniques of the 
bubble chamber experiments could be revived, complemented by modern readout technology
in the form of fast CCD or CMOS imaging systems and fast, accurate image analysis
using techniques adapted from the liquid argon TPC program.

It is clear that the measurements envisioned at the nuSTORM facility cannot
be accomplished with just one experiment. A careful study of the performance of each
of the detector options, in conjunction with a prioritization of the most important
measurements by the community, will define a phased research program that will, finally,
clear up the many areas of poor knowledge of neutrino interaction cross sections.

\subsection{Technology test-bed}
\label{SubSect:RnD}

\subsubsection{Muon beam for ionization cooing studies}
\label{SubSect:MuBm}

Muon ionization cooling improves by a factor $\sim 2$ the stored-muon
flux at the Neutrino Factory and is absolutely crucial for a Muon
Collider of any center-of-mass energy in order to achieve the required
luminosity.
The Muon Ionisation Cooling Experiment (MICE) \cite{MICE:2005zz} will
study four-dimensional ionization cooling and work is underway to
specify the scope of a follow-on six-dimensional (6D) cooling
experiment.
MICE is a ``single-particle'' experiment; the four-momenta of single
muons are measured before and after the cooling cell and then input and
output beam emittances are reconstructed from an ensemble of
single-muon events.
A 6D cooling experiment could be done in the same fashion, but doing
the experiment with a high-intensity pulsed muon beam is preferred.  
One feature of nuSTORM is that an appropriate low-energy muon beam
with these characteristics can be provided in a straightforward
fashion, see Section~\ref{SubSect:6DICE}.
\subsubsection{Neutrino cross-section measurements for Super Beams}
\label{SubSubSect:nuXSectfor SB}

\begin{figure}
  \begin{center}
    \includegraphics[width=0.85\textwidth]%
    {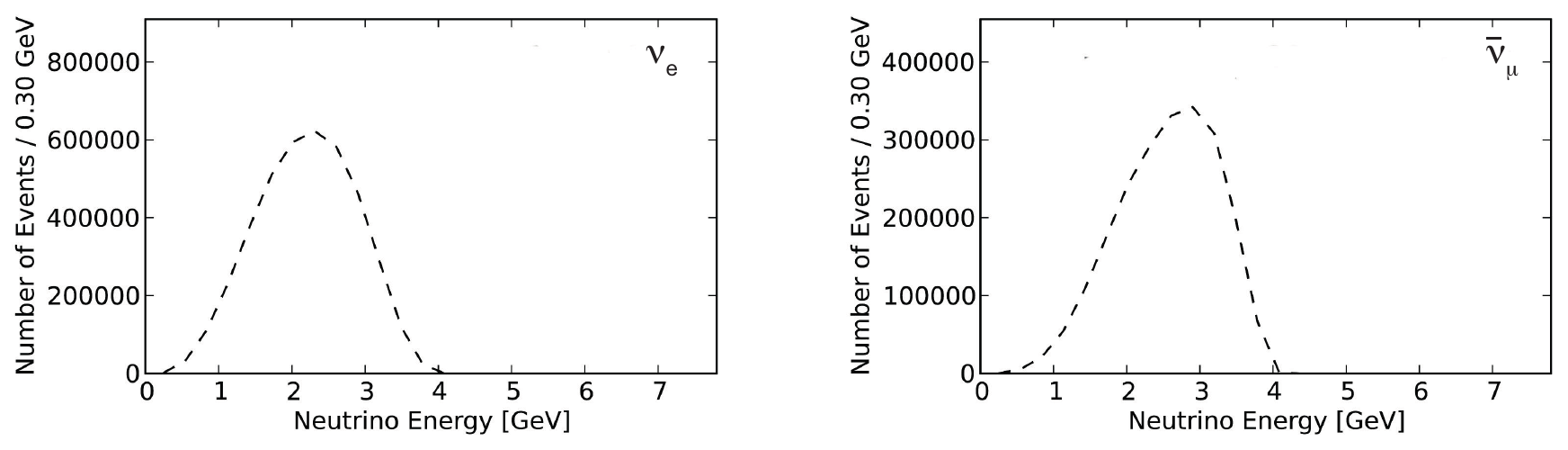}  \\
    \centering{
      \includegraphics[width=0.35\textwidth]%
      {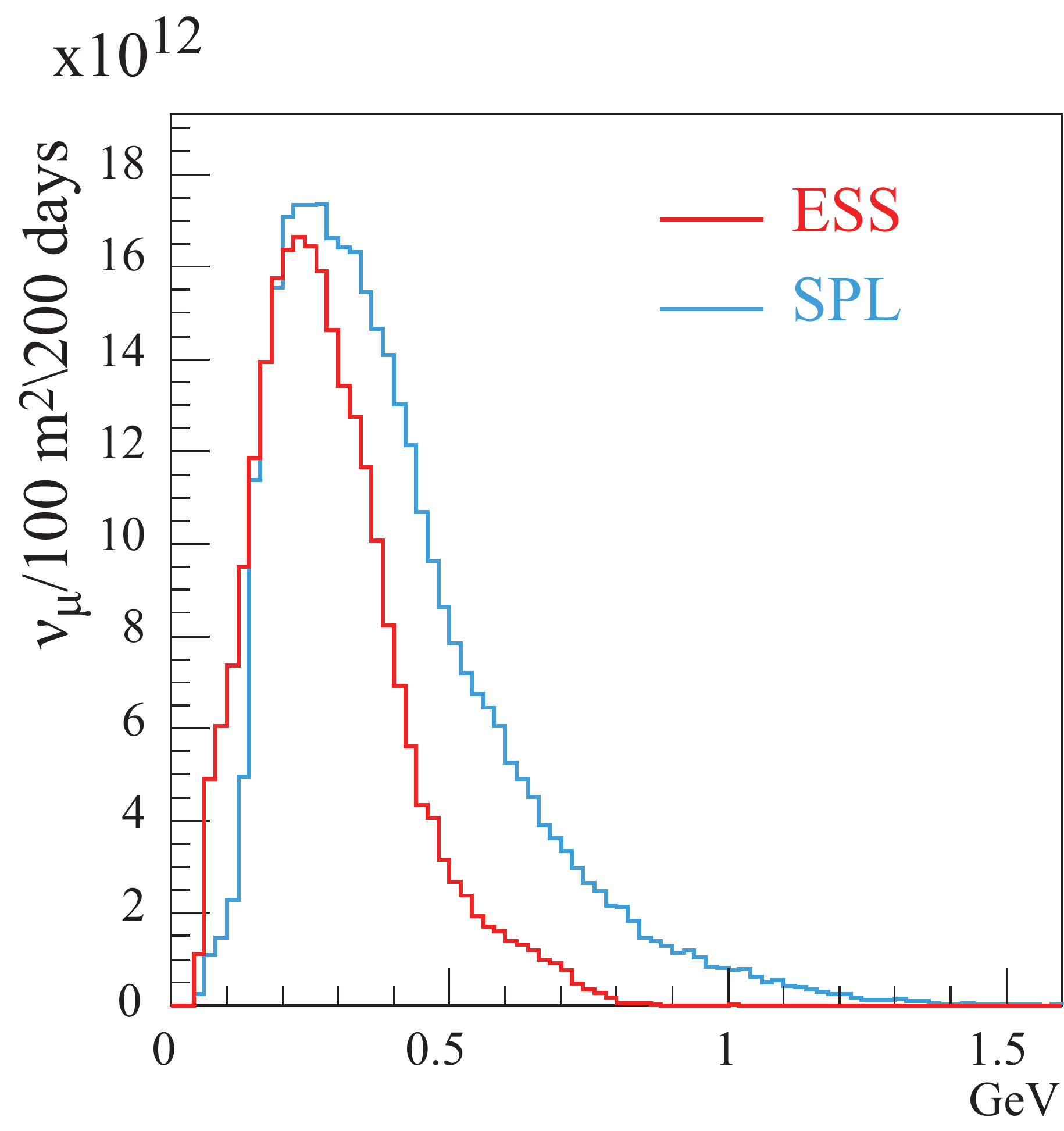}
    }
  \end{center}
  \caption{
    Top row: $\nu_e$ (left) and $\bar{\nu_\mu}$ (right) event rates per
    100\,T in a detector placed $\sim 50$\,m from the end of one of
    the straight sections (for a stored $\mu^+$ beam) at nuSTORM.
    Bottom row: neutrino energy spectrum for the SPL (with proton
    energy of 4.5\,GeV) and ESS (with proton energy of 2.5\,GeV) based 
    super beams.
  } 
  \label{Fig:Fluxes}
\end{figure}

The neutrino spectrum produced by the nuSTORM 3.8\,GeV/c stored muon
beam is shown in Fig.~\ref{Fig:Fluxes}.
The nuSTORM flux at low neutrino energy ($< 0.5$\,GeV) is
relatively low.
The neutrino energy spectrum that would be produced at a low-energy
super-beam such as the SPL-based beam studied in
\cite{Baussan:2012wf} or the recent proposed super beam at the
European Spallation Source (ESS) \cite{Baussan:2012cw} is also
shown in Fig.~\ref{Fig:Fluxes}.
Both the SPL and ESS based super beams propose to use the MEMPHYS
water Cherenkov detector \cite{deBellefon:2006vq,Agostino:2012fd}.
To enhance the event rate in the low neutrino-energy region of
importance to such facilities, the possibility of capturing muons with
a central momentum below 3.8\,GeV/c is being studied.

The ``degrader'' introduced in Section~\ref{SubSect:6DICE} provides an alternative technique by
which the required low-energy neutrino beam could be produced.
The muons exiting the degrader with P$_\mu >$ 400 MeV/c (see Fig.~\ref{fig:lowE}) could be used to
produce neutrinos with
energies of around 300\,MeV. 
A detector placed a few tens of meters behind the degrader would make
it possible to measure the $\parenbar{\nu}_e N$ and
$\parenbar{\nu}_\mu N$ cross sections required for the SPL or ESS based super beams.
\clearpage

\section{Facility}
\label{sec:Facility}
The basic concept for the facility was presented in Fig.~\ref{fig:STORM}.  A high-intensity proton 
source places beam on a target, 
producing a large spectrum of secondary pions.  Forward pions are focused by a horn into a capture and transport channel. 
Pion decays within the first straight of the decay ring can yield a muon that is stored in the ring.
Muon decay within the straight 
sections will produce $\nu$ beams of known flux and flavor via:
$\mu^+ \rightarrow e^+$ + $\bar{\nu}_\mu$ + $\nu_e$ or $\mu^- \rightarrow e^-$ + $\nu_\mu$ + $\bar{\nu}_e$.
For the implementation which is described here, we choose a 3.8 GeV/c storage ring to obtain the desired spectrum 
of $\simeq$ 2 GeV neutrinos (see Fig.~\ref{Fig:Fluxes}). This 
means that we must capture pions at a momentum of approximately 5 GeV/c.
\subsection{Primary proton beam}
This section describes the reference design for the nuSTORM primary (proton) beamline. This system will extract protons from 
the Fermilab MI (MI) synchrotron, using a single-turn single-batch extraction method and then transport them to the target in the nuSTORM target hall. The nominal range of operation will be for protons from 60 to 120 GeV/c.

The principal components of the primary beamline include the standard magnets in the MI-40 abort line to capture protons in the synchrotron and redirect them to the nuSTORM beamline. This beamline is a series of dipoles and quadrupole magnets to transport the protons to the target. All of the nuSTORM primary-beam technical systems are being designed to support sustained, robust and precision beam operation.

\begin{figure}[h]
  \centering{
    \includegraphics[width=0.9\textwidth]{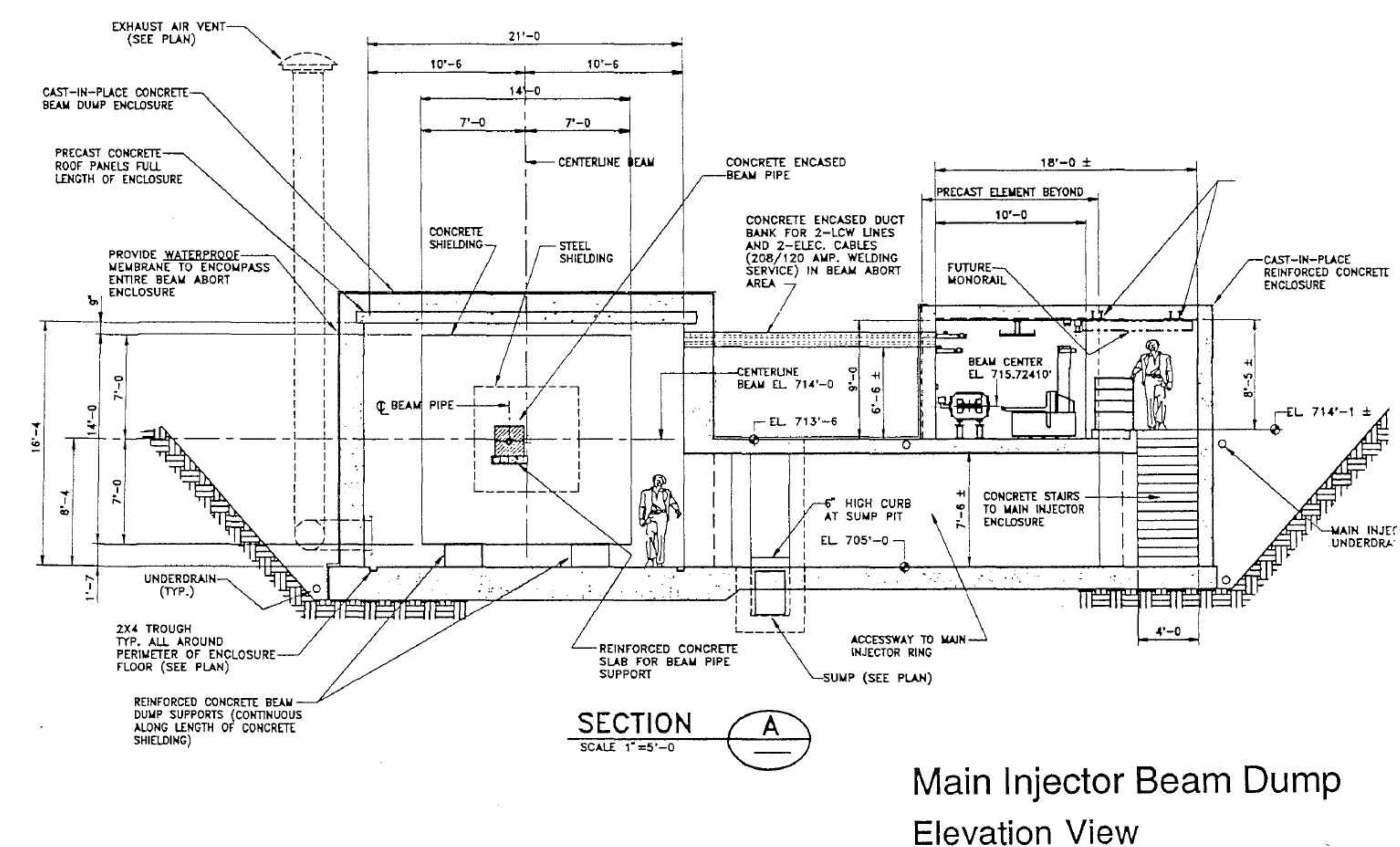}}
  \caption{MI absorber room cross section.}
  \label{fig:MIabRoom}
\end{figure}

In 1994, the NuMI Project Definition Report originally designed the extraction line from the MI-40 area. The MI absorber was built with the transport beam pipe installed at an estimated elevation of 714 feet. The nuSTORM facility will extract protons from the MI through this channel and continue through two enclosures towards the nuSTORM target hall \cite{MITDH:1994z}.
In Fig.~\ref{fig:MIabRoom}, this channel is indicated by the shaded-in square offset from the center of the MI Absorber.
The nuSTORM primary beam is extracted using single-turn, or ``fast" extraction, in which a portion of the protons accelerated in the MI synchrotron ring, will be diverted to the nuSTORM beamline within one revolution after each acceleration cycle. The train of bunches dedicated for the nuSTORM target hall extends one seventh of the MI circumference or for one ``Booster Batch."  The remaining bunches are dedicated to additional operational beamlines running concurrently, i.e. NuMI, NoVA, or LBNE \cite{NUMITDH:2002z,Papadimitriou:2011fz}. 
After extraction, the beam is controlled by a series of dipoles (bending) and quadrupole (focusing) magnets. 

The nuSTORM primary-beamline is designed to direct the beam towards the nuSTORM target hall and collection channel, with a spot size appropriate for maximizing pion production (see Section~\ref{subsec:PiP}). nuSTORM will implement a point to point focusing design in two sections. This is allowed due to the relatively short distance between THE MI and the nuSTORM target hall (600') and also because a large beam spot is required for the MI absorber line (upstream beamline  common to MI abort line and nuSTORM). The first section focuses the proton beam after the MI absorber line. The second and final focus of the beam will be used for controlling the beam spot size on the target.
\subsubsection{Extraction line}
In order to achieve this focused beam, the MI absorber line will be modified with respect to placement of existing quadrupole magnets and replacement of their HV cables from the MI quad focusing buss onto separate individual power supplies. All absorber line quadrupoles, Q001 through Q003, will need independent power supplies. Q002 and Q003 will need to be moved upstream of their current locations by 23 and 30 feet respectfully. This relocation will provide adequate space for nuSTORM's extraction switch magnet and allow for proper optics in the remaining beamline. The two large bending magnets will remain in their current location, and continue to operate on the MI bend buss.

Once MI beam has achieved its flattop energy, three new kicker magnets in the MI-40 area apply a horizontal kick to the beam to the outside of the ring. This beam passes through a set of quadrupoles, Q401 and Q402, to continue the horizontal trajectory at elevation 715 feet, 9 inches. Then a series of three specialized magnets called Lambertsons \cite{ILA}, L001 L002 and L003, vertically extract the beam from the MI. These extraction Lambertsons are unique in that this set bend the beam downward for extraction, while in other Lambertson areas, such as MI-52 and MI-60, they bend the beam upward. 
	
\begin{figure}[h]
  \centering{
    \includegraphics[width=0.55\textwidth]{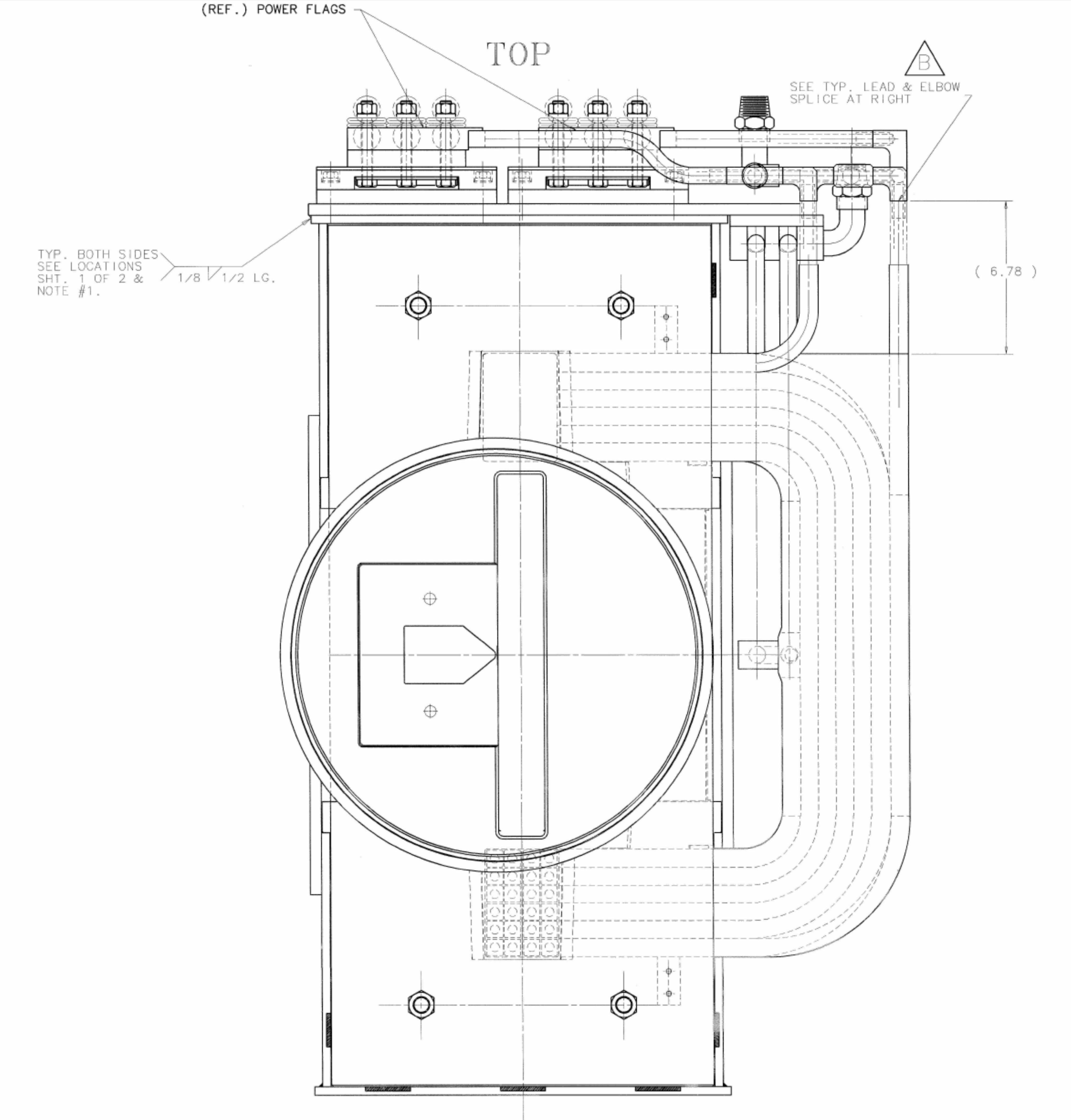}
    \includegraphics[width=0.40\textwidth]{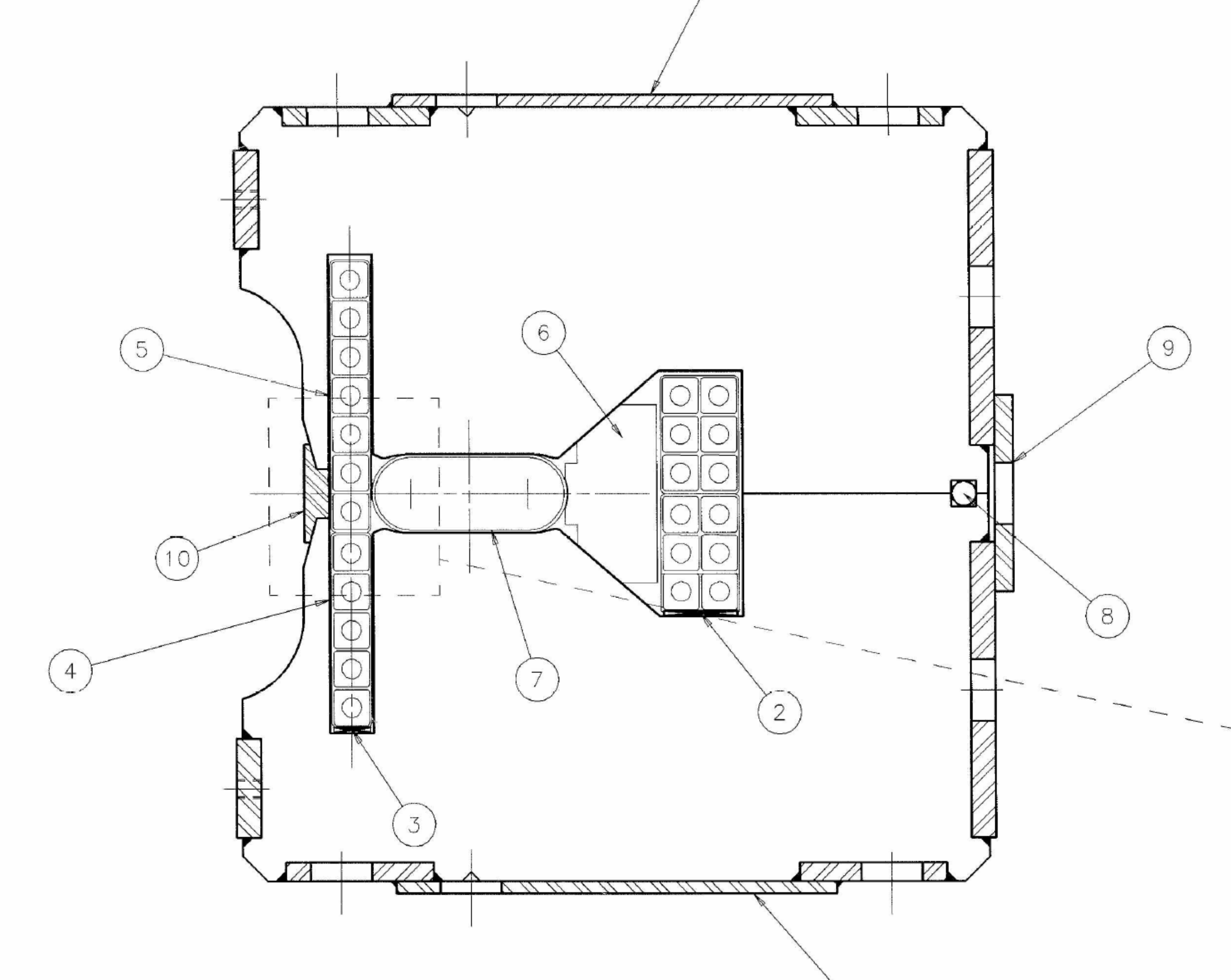}}
  \caption{MI Lambertson (left) and C magnet drawings (right).}
  \label{fig:Lambert}
\end{figure}

The Lambertsons sit in the path of the beam both when beam circulates and when it is being extracted, so they must accommodate both paths. The circulating beam passes through a field-free region in the magnet, and the extracted beam passes through the region of magnetic field and is bent downward from the circulating MI trajectory by 18.24 milliradians. Fig.~\ref{fig:Lambert} shows a cross section of the MI Lambertson.
Each Lambertson in the line bends the beam, such that after passing through the string of Lambertsons the extracted beam is sufficiently separated from the MI orbit to pass through the first bending magnet external to the MI, a C-magnet \cite{ICA}. The C-magnet clears the MI beam tube downstream of the third Lambertson and provides an additional downward bend, 9.34 milliradians, enough so that the extracted beam can pass below the outside of the next quadrupole in the MI lattice. This also accommodates clearance for the first quadrupole, Q001, in the absorber line. Fig.~\ref{fig:Lambert} also shows a cross section of the MI C-magnet.

Once the nuSTORM primary beam is extracted from the MI, it shares the beam line components with the MI aborted beam. With its trajectory, the nuSTORM primary beam will be directed and focused exactly like the MI aborted beam by a series of three quadrupoles, two dipoles and several trim magnets. However, once past the last quadrupole, Q003, the beam will be bent towards the transport beam pipe in the MI absorber using a pulsed EDB (extended dipole bender) dipole named NSHV1 \cite{EDB}.  This horizontal pulsed magnet is rolled by 0.37 radians clockwise to necessitate the horizontal and vertical differences between the MI absorber and the transport beam pipe aperture, as shown in Fig.~\ref{fig:MIabRoom}. This results in the beam being bent 4.97 milliradians horizontally away from the MI ring and 1.93 milliradians vertically downward.

\begin{figure}[h]
  \centering{
    \includegraphics[width=0.9\textwidth]{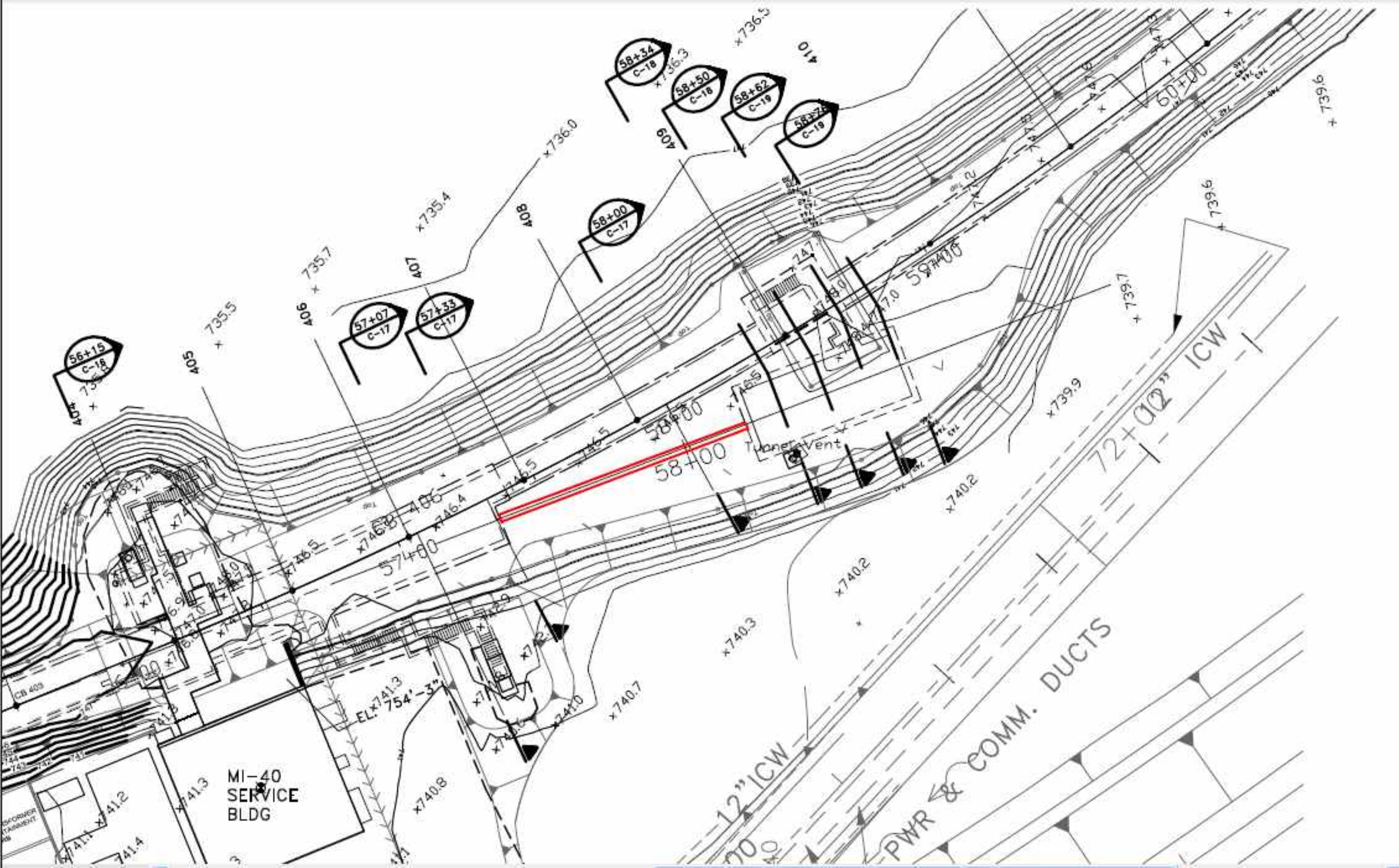}}
  \caption{MI absorber berm pipe.}
  \label{fig:BermP}
\end{figure}

Between the MI enclosure and the MI absorber room is a distance of 88 feet. This stretch of beamline is transitioned into an 24 inch wide ``bermpipe" commonly used at Fermilab. This vacuum isolated stretch of bermpipe \cite{MIBP} will contain three beamlines, two of them converging towards the MI Absorber, while the nuSTORM trajectory is towards the Transport Beam Pipe.  Fig.~\ref{fig:BermP} shows this berm pipe relative to the MI enclosure and absorber room.
\subsubsection{Beam line}
From the back end of the MI Absorber Room, the nuSTORM beamline continues into new construction enclosures. The first section of optics is to match the MI Abort line optics to the final focusing optics of the nuSTORM beamline. This is achieved by using only two 3Q120 quadrupole magnets, NS1Q1 and NS1Q2, at the beginning of the first enclosure NS1 \cite{NS1}. Once the beam has passed the first quadrupoles in the NS1 enclosure, the beam is bent upwards by NS1V1 to correct for the downward trajectory created naturally by the MI Absorber line and the NSHV1. Here the beam line is at its lowest point 713 feet, 3 inches.

After the beam has had the necessary drift space to be focussed by NS1Q1 and NS1Q2, and be placed on the correct vertical trajectory by NS1V1, the first series of horizontal dipoles, NS1H1, bends the beam 0.0934 radians. This horizontal string of magnets is comprised of four B2 magnets \cite{NSV1} with a 1.54 Tesla field. After this string of dipoles the beam is then transported to the next enclosure, NS2.

\begin{figure}[h]
  \centering{
    \includegraphics[width=0.9\textwidth]{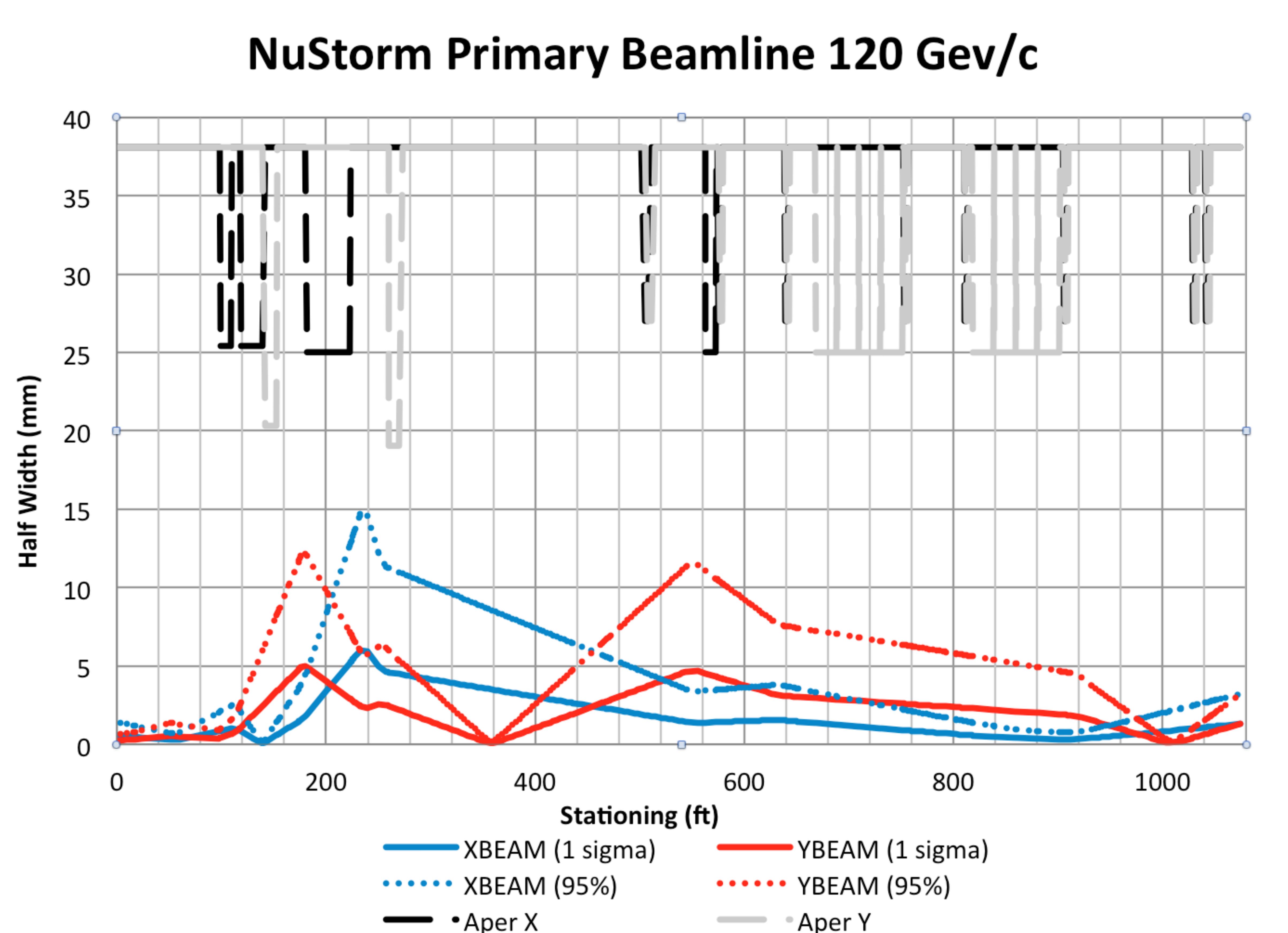}}
  \caption{nuSTORM beamline half width sizes and apertures for 120 Gev/c protons.}
  \label{fig:BeamLine120}
\end{figure}

\begin{figure}[h]
  \centering{
    \includegraphics[width=0.9\textwidth]{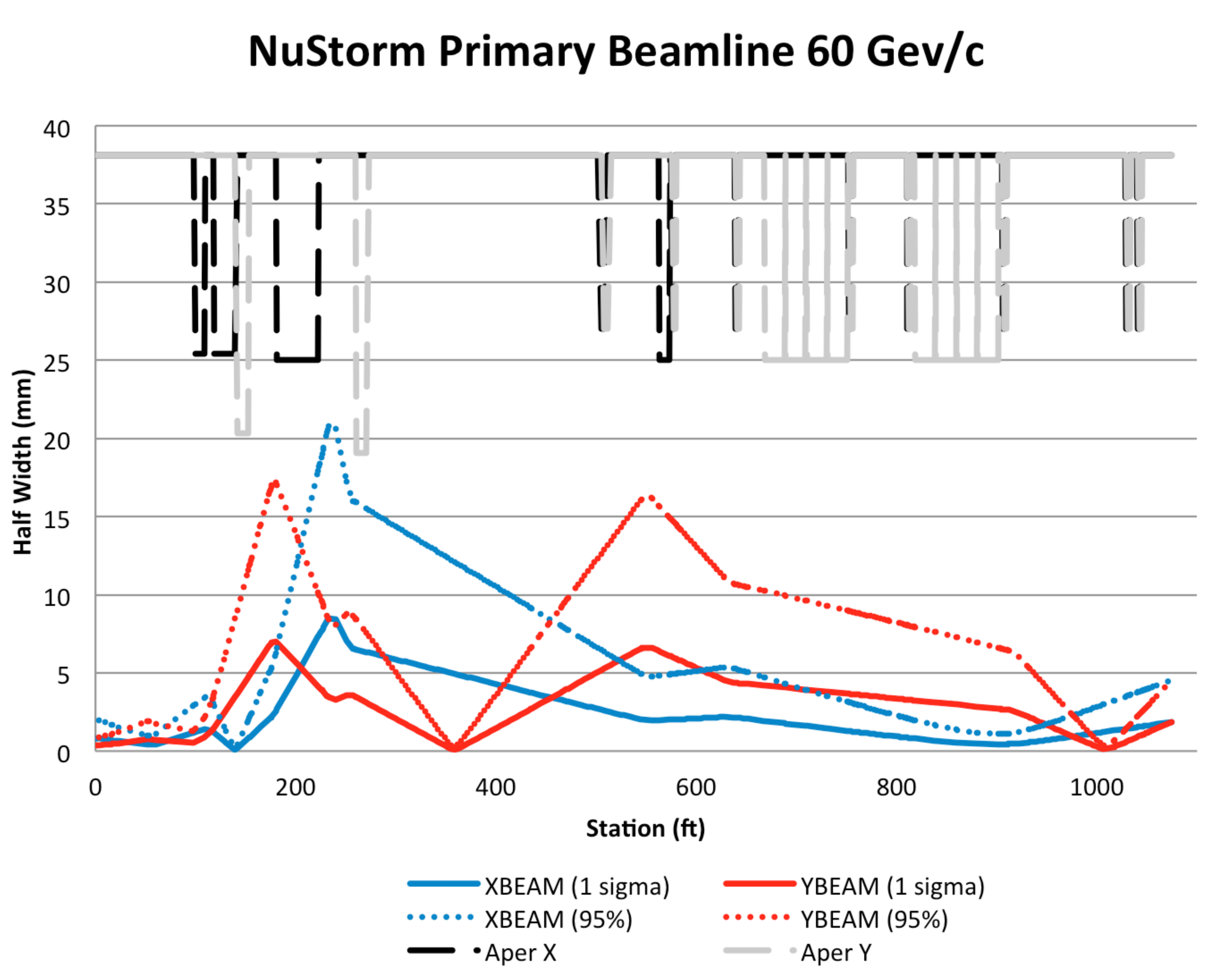}}
  \caption{nuSTORM beamline half width sizes and apertures for 60 Gev/c protons.}
  \label{fig:BeamLine60}
\end{figure}

In NS2, the beam is bent with another series of four B2 dipoles, NS2H1. This string bends the beam an additional 0.0934 radians, with a bend field of 1.54 T. After the last bend string, a single quadrupole magnet NS2Q1 focuses the beam vertically. This allows the beam to be focused after the bending by the NS2 horizontal magnet string. After the last bend, the beam is focused onto the nuSTORM target via NS2Q1. This quadrupole will be able to focus the beam on target to maximize pion production and efficiency. The beam half widths and component apertures are shown in Fig.~\ref{fig:BeamLine120} and Fig.~\ref{fig:BeamLine60}.
They include the FODO lattice of the MI through the absorber line and through the nuSTORM primary beamline. The nuSTORM primary beamline starts at station 481 ft. This beamline is also specifically planned so that it has very little impact on current operational surroundings.  The nuSTORM beamline optics share three quadrupoles with the MI absorber line, and three dedicated quadrupoles in the primary beamline. The optics and trajectory of this beamline have been designed to handle different momentum protons to be operational with concurrently running experiments utilizing the MI in the future.  Fig.~\ref{fig:BeamLine120} is designed for 120 GeV/c protons, however, Fig.~\ref{fig:BeamLine60} shows the beam line is capable of focusing the 60 GeV/c protons for future experiments. Table~\ref{tab:BLmag} contains the data for the nuSTORM quadrupoles for 120 GeV/c protons.
\begin{table}
\centering
\caption {Magnet names types and settings for 120 Gev/c protons.}
\label{tab:BLmag}
\begin{tabular}{|llrl|}
\hline
Dipoles & Type & Current (Amps) & Notes\\
\hline\noalign{\smallskip}
NSHV1 &	EDB	& 796.4	& Rolled 0.37 radians\\
NS1V1 &	EDB	& 664.7	& Critical Device \#1\\
NS1H1 (4) &  B2 & 3833.1 &  Critical Device \#2\\
NS2H1 (4) &  B2 &	3833.1 &  	\\
\hline
Quadrupoles & Type &Current (Amps) &	Notes\\
\hline
Q001 &	IQA	& 1883.2	& Vertical Focusing\\
Q002 &	IQB	& 1423.9	& Horizontal Focusing\\
Q003  &	IQB	& 853.0	& Vertical Focusing\\
NS1Q1 &	3Q120 & 22.3	& Vertical Focusing\\
NS1Q2 &	3Q120 & 11.9	& Horizontal Focusing\\
NS2Q1 &	3Q120& 19.5	& Vertical Focusing\\
\noalign{\smallskip}\hline
\end{tabular}
\end{table} 
The nuSTORM beamline contains 18 correctors or 9 pairs of horizontal and vertical trims, before and after every large angle change mentioned above. Four of these correctors are placed just prior to the nuSTORM target hall for the operational ability to change the position and angle of the beam interacting with the target.  This set of correctors will be useful for target scans and alignment. These correctors can correct a 0.25 mm transverse mis-alignment offset for all the major bending dipoles. For roll tolerances, the maximum acceptable error is 0.5 milliradians \cite{Kyle1}.

\begin{figure}[h]
  \centering{
    \includegraphics[width=0.9\textwidth]{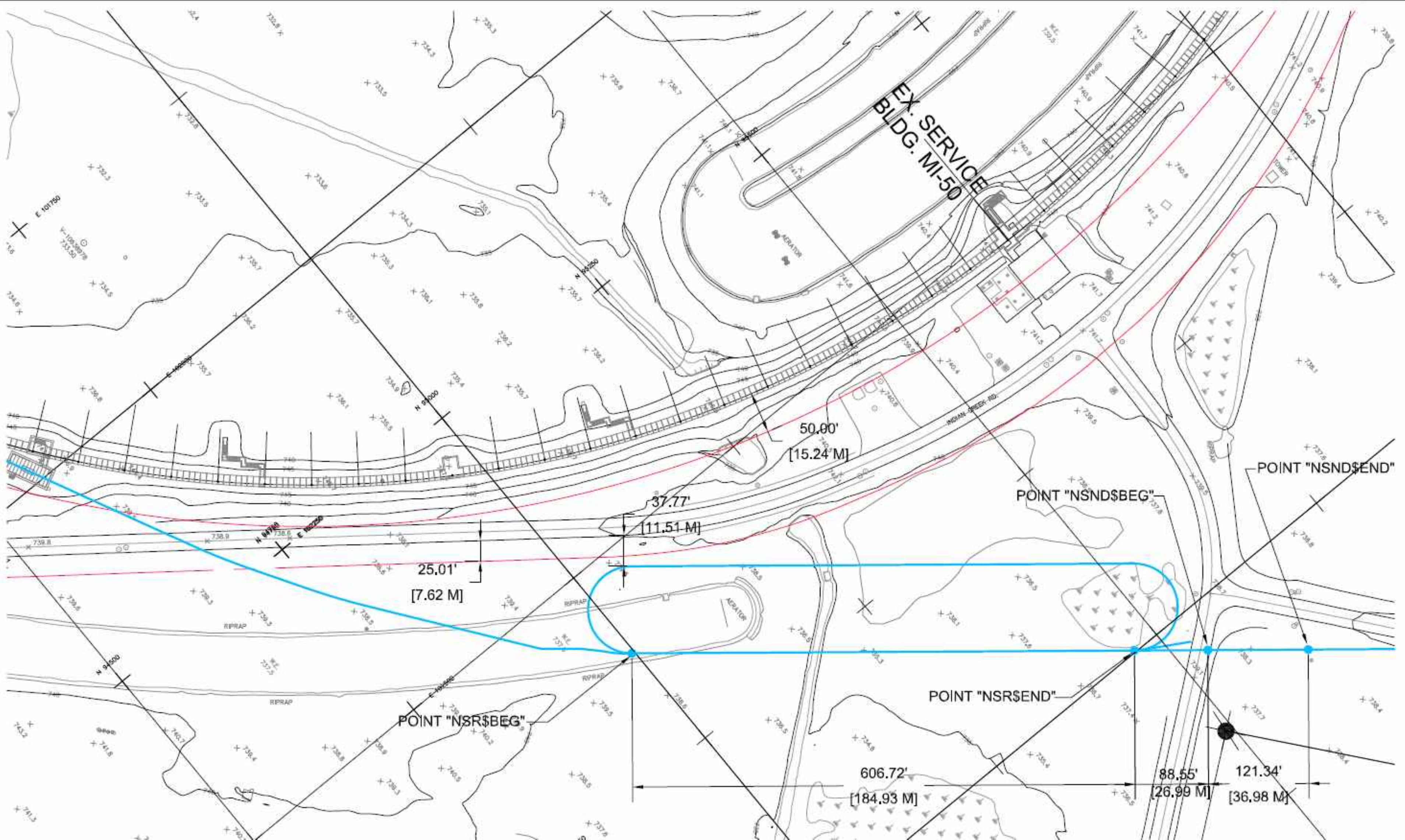}}
  \caption{Enclosure layout of nuSTORM primary beamline.}
  \label{fig:Enclosures}
\end{figure}

As mentioned above, the nuSTORM primary beamline focusing and bending elements are split between two different enclosures. These enclosures are split by a jacked pipe of 59 feet in length. Running through this jacked pipe, are individual pipes that contain the beam line and other utility supply and return lines. The primary beamline enclosure layout is shown in Fig.~\ref{fig:Enclosures} along with the proposed target hall, Pion Decay Channel and Muon Decay Ring.
nuSTORM's primary beamline will use two critical devices, as stipulated in the Fermilab Radiological Control Manual \cite{RCM}
(FRCM). 
The first critical device, NSHV1, is located inside the MI-40 area as described above. This switch magnet must be powered in order to allow beam to the  nuSTORM
target.   The second critical device will be NS1H1, as described above. 

nuSTORM's primary beamline will also include  instrumentation packages containing Loss Monitors, Toroids, Multiwires, and Beam Position Monitors (BPMs). In this current design the beamline contains three toroids, ten Multiwires, and twelve BPMs.  Loss monitors will be placed on magnets with tighter aperture constraints. Each major bending string, NS1V1, NS1H1, and NS2H1 will have loss monitors mounted on the first magnet. 

Of the three toroids, the first will be located inside the MI enclosure just prior to the berm pipe. This will be able to report an input for transfer efficiency to the nuSTORM target hall. The second toroid is positioned just after the NS1H1 magnet string in the NS1 enclosure. The last toroid will be positioned in enclosure NS2 just prior to the target hall. This will provide a comparison for transfer efficiency and protons on target measurements.

Of the ten Multiwires, enclosure NS1 has four and enclosure NS2 contains the remaining six. In each Multiwire, the wire spacing is assembled with 1 mm pitch, a common spacing at Fermilab, and each plane will contain forty wires. Each BPM unit contains horizontal and vertical plates to provide horizontal and vertical positions, saving on longitudinal space in the beamline. nuSTORM will adopt the LBNE style button BPMs. In enclosure NS1, five BPMs will be installed and the NS2 enclosure will have six BPMs.

nuSTORM's two large bending strings of NS1H1 and NS2H1 have the same operating current and bend field. For power supplies to be used by this line, it is cost efficient to use one power supply to power both the first and the second string of B2 magnets. Other magnets such as the correctors, quadrupoles, and the first critical device NSHV1, will need to use independent power supplies .
\subsection{Target Station}
\label{subsec:42}
\subsubsection{Target station overview and conceptual layout}
\label{subsubsec:421}
The nuSTORM target station conceptual layout addresses the requirement of providing a reliable pion production facility while providing a platform for component maintenance and replacement.  Due to the severe service environment encountered in target halls, it is expected that the beamline elements will require replacement on the order of every few years.

The general operation of the target station utilizes a primary proton beam extracted from the MI and transported to interact with a target to produce pions (along with other short-lived hadrons) which are subsequently focused toward a set of capture quadrupoles by a single magnetic focusing horn.  The design parameters of beam spot size, target material and interaction length, and focusing horn current and geometry are considerations for providing reliable and sufficient pion production to support the experimental requirements.  The following parameters for the proton beam on target are specified for the baseline in Table~\ref{tab:POT}.
\begin{table}
\centering
\caption {Proton beam parameters}
\label{tab:POT}
\begin{tabular}{|l|r|}
\hline
Energy			&  120 GeV\\
Protons per pulse	&  $8\times10^{12}$\\
Pulse width		&  1.6 $\mu$s\\
Beam $\sigma$		&  1.1 mm\\
Cycle time			&  1.33 sec.\\
\hline
\noalign{\smallskip}
\end{tabular}
\end{table} 
Several considerations must be adequately understood and addressed to ensure safe and reliable operation of the target station.  The target station is one area where nuSTORM can draw heavily on the successful operation and experience gained in NuMI target hall operations.  The nuSTORM target station baseline conceptual layout will employ a very similar design approach to that used in the NuMI target hall complex.  The following key elements address the primary requirements for safe and reliable target station operation:
\begin{itemize}
\item Provide adequate shielding to safely accommodate a maximum beam power of 400kW.
\item Utilize economical shielding and target pile design methodology to minimize cost impact.
\item Provide for the installation and operation of the following active beamline components:
	\begin{enumerate} 
	  \item Production target
	  \item Focusing horn, electrical stripline bus, and pulsed power supply
	  \item Pair of capture quadrupole magnets and related power supplies and utilities
	  \item Water-cooled collimators for quadrupole secondary beam spray protection
	\end{enumerate}
\item Provide target pile shield with air cooling while considering humidity control to minimize component corrosion and provide the ability for safe and effective tritium management.
\item Provide a target, horn, and capture quadrupole-positioning module support platform including shielding, hardware support provisions, remote hot handling capability with rigid, stable alignment capability.
\item Provide for hot handling operations such as failed component replacement and cool-down storage for highly radioactivated components.
\item Provide the infrastructure layout for constructing and operating the facility including radioactive water cooling (RAW) systems, component power supplies, air handling systems, adequate floor space for placement of shielding blocks during maintenance periods, and a component handling interface including remote pick functionality, cameras to facilitate hot handling, and a suitable overhead crane for material handling.
\end{itemize}
Many of the above elements will be addressed by using NuMI-style components and general target station layout as the design costs and operational characteristics are well understood for NuMI.  In addition, manpower is a significant cost driver for new projects and utilizing or slightly modifying existing design concepts minimizes cost impact.

\begin{figure}[h]
  \centering{
    \includegraphics[width=0.9\textwidth]{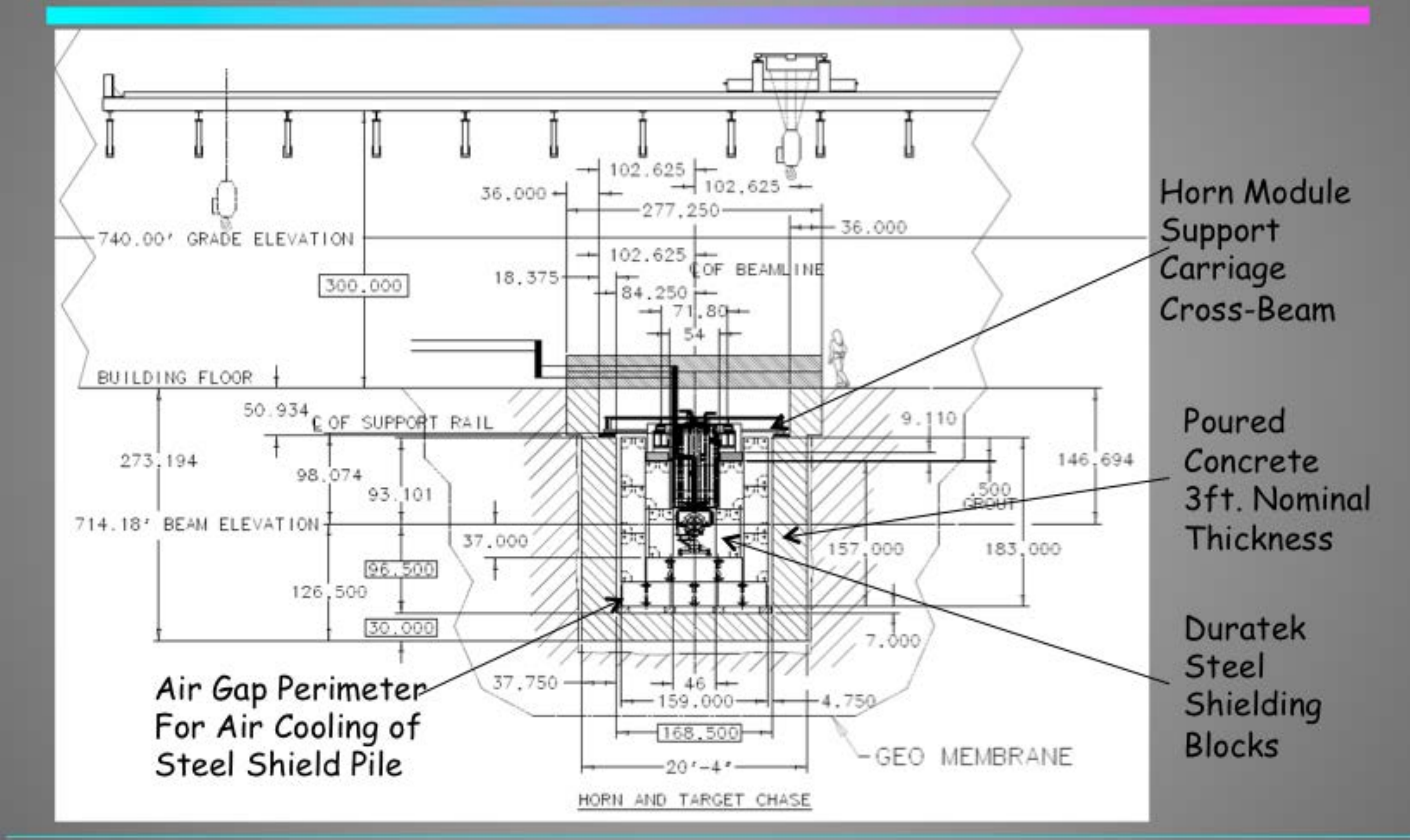}}
  \caption{nuSTORM target pile and beamline chase cross-section downstream of horn looking in beam-upstream direction.}
  \label{fig:TP}
\end{figure}

The general target chase (i.e., beam space) shielding requirements will be met by utilizing a shield pile layout nearly identical to that of NuMI for 400kW operation.  The nuSTORM target station shielding specifies 4 feet of steel surrounded by 3 feet of concrete, with vertical shielding requirements for sky-shine requiring 9 feet of steel with 6 in of borated polyethylene for prompt neutron shielding. Fig.~\ref{fig:TP} shows a cross-section of the target pile just downstream of the horn position.
Several key elements appear in Fig.~\ref{fig:TP}, including the Duratek steel (now Energy Solutions) shielding block stack around the chase perimeter.  These shielding blocks measure 26x52x52 in. ~with a corresponding weight of 10 tons each.  The current nuSTORM conceptual layout requires approximately 140 Duratek blocks.  As a cost benchmark, NuMI purchased a quantity of 500 Duratek blocks at a price of \$226 per block (1.13 cents per pound).  This underscores the benefit of using Duratek blocks, as they represent a significant source of very inexpensive steel.

\begin{figure}[h]
  \centering{
    \includegraphics[width=0.7\textwidth]{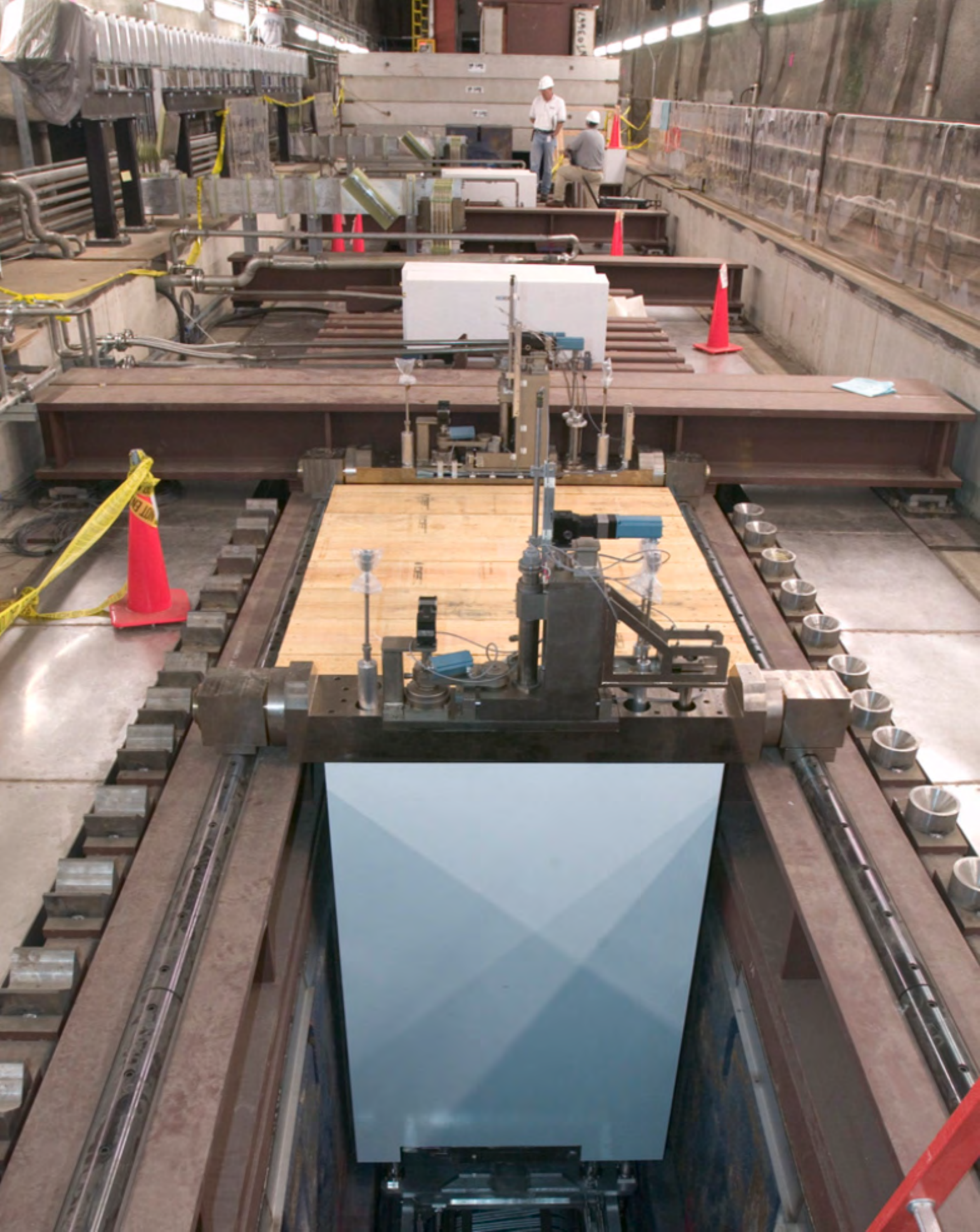}}
  \caption{NuMI target pile chase, target module, and carriage I-beam support structure depicting target module installed in Duratek-
		lined beamline chase (view is upstream of target module looking downstream).}
 \label{fig:NumiTP}
\end{figure}

Fig.~\ref{fig:TP} also depicts the horn-positioning module supported by a rigid I-beam carriage structure that is supported on an isothermal surface for accurate alignment during beam operation.  Fig.~\ref{fig:NumiTP} of this shielding configuration, as seen during NuMI construction, highlights additional detail.
It is envisioned that this component mounting scheme and shield pile arrangement will be used for the target, horn, and capture quadrupoles.

\begin{figure}[h]
  \centering{
    \includegraphics[width=0.9\textwidth]{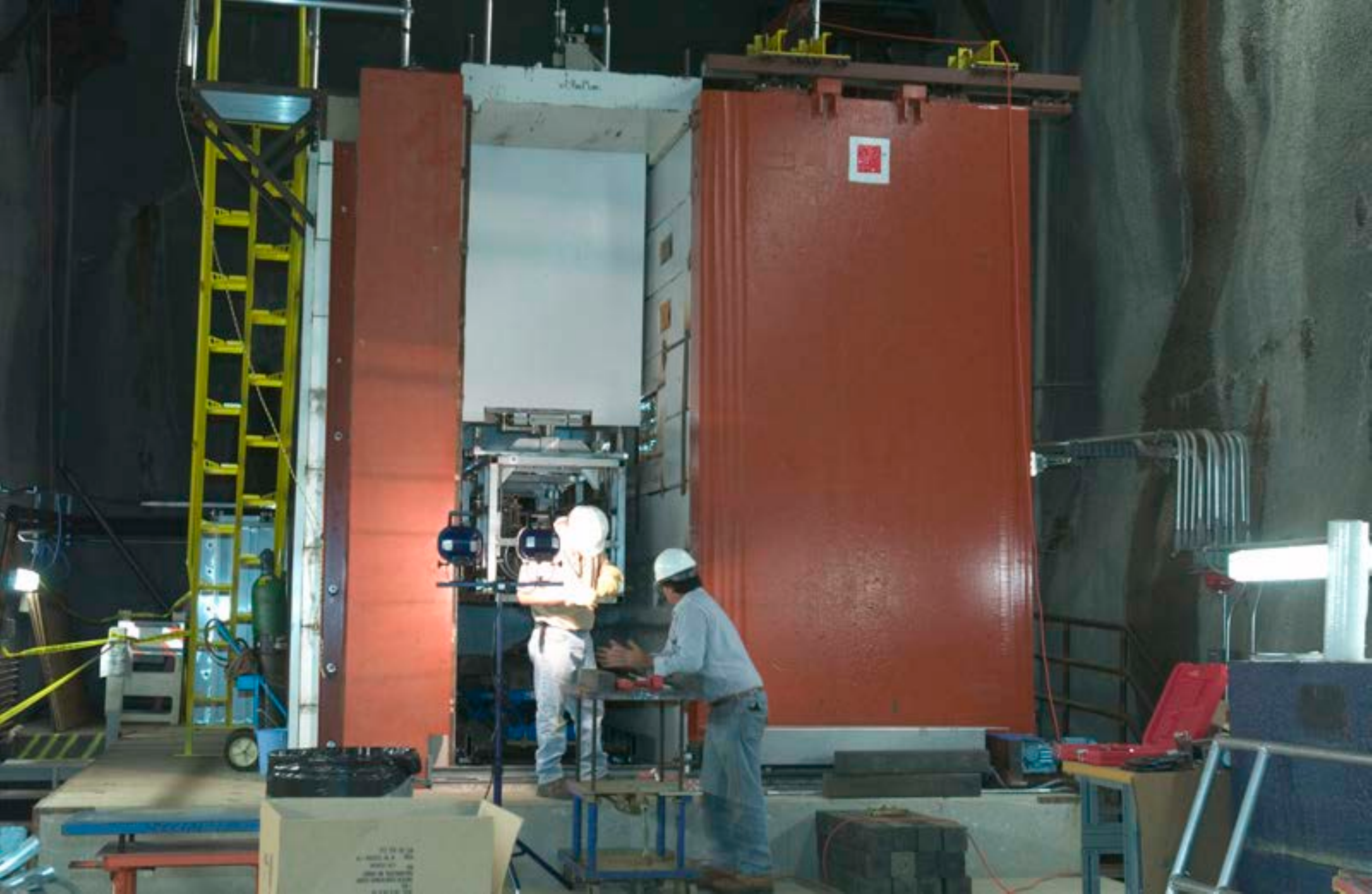}}
  \caption{NuMI Workcell for hot handling operations and failed component replacement.  Image shows NuMI target module in the workcell 
			for a hot handling procedure dry run during NuMI construction}
  \label{fig:NumiWC}
\end{figure}

\begin{figure}[h]
  \centering{
    \includegraphics[width=0.9\textwidth]{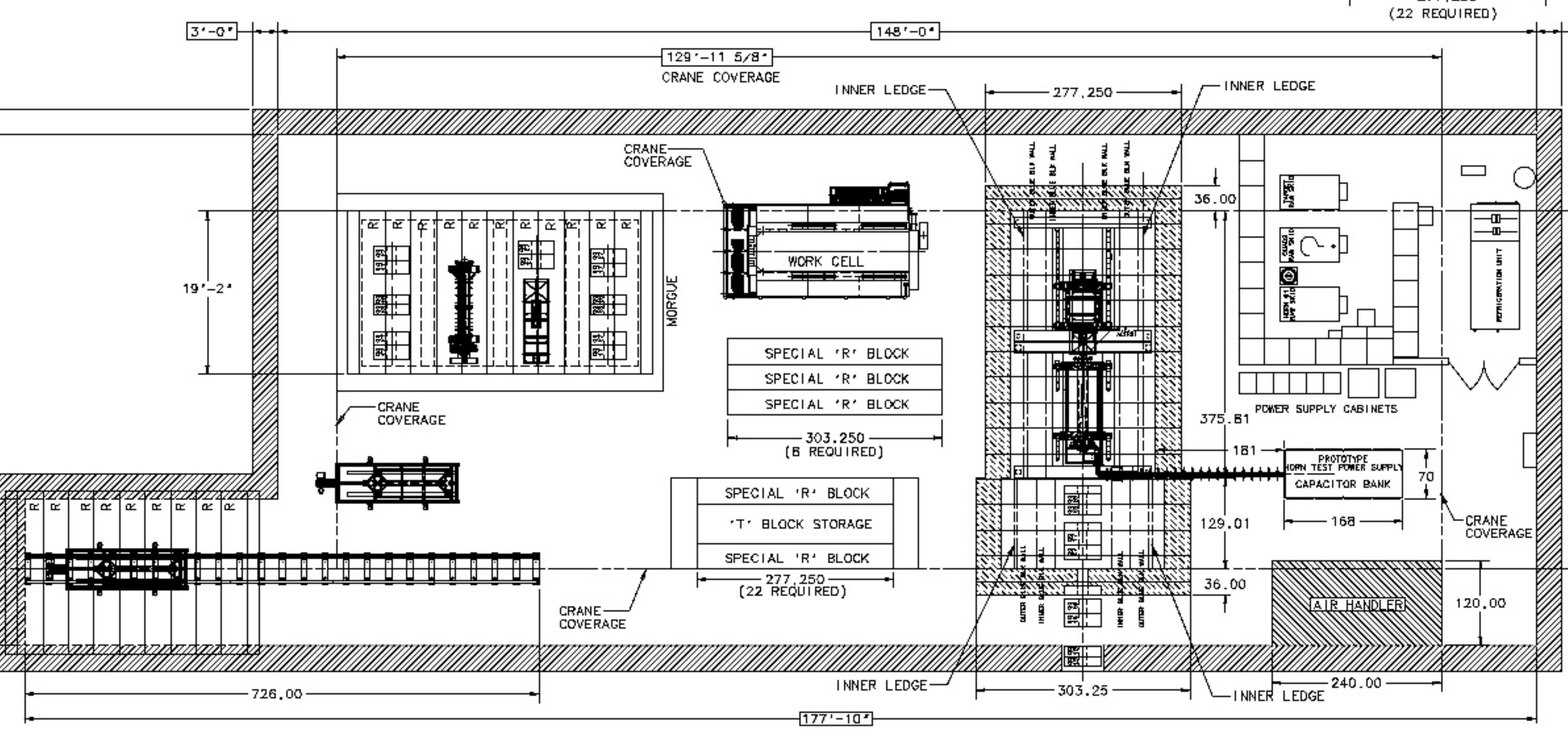}}
  \caption{nuSTORM target station main level plan view.}
  \label{fig:TSPV}
\end{figure}

Hot component handling and failed component replacement is addressed in the inherent design of the NuMI-style module that provides for remote handling capability.  Component replacement is addressed by the use of a NuMI-style workcell that has a rail landing area identical to the beamline chase and allows for remote placement of the module into the workcell, after which a remotely operated door is closed and shielding hatch covers placed over the top-center of the module space to allow personnel access to the module top for remote disconnects and component servicing.  Fig.~\ref{fig:NumiWC} provides a sense of scale of the NuMI workcell during construction with the target module inside for trial hot handling practice.
The above noted elements are incorporated into the target station conceptual layout as an input to the civil facility construction.   Fig.~\ref{fig:TSPV} gives a plan view of the current version of the target station layout.  Note that the layout incorporates a drop hatch and rail cart for rigging material into the complex for construction and removing failed components in a shielded transportation cask for long-term storage.  A morgue is included in the layout for temporary storage of radioactive elements and failed component cool-down.  The active functional area includes the target shield pile and beamline chase and an area that will be likely separated by a masonry wall to house power supplies, RAW water skids and a chase air handling system.
\subsubsection{Pion Production Target}
The nuSTORM  baseline target is very similar to the NuMI low energy target used in the MINOS and MINER$\nu$A run from CY2005 thru CY2012.  This style of target has undergone significant analysis and optimization during that time period.  When properly constructed and qualified by sufficient QA steps during construction, the target has been successfully operated for $6\times10^{20}$ protons on target (NuMI target NT-02) at beam powers near 400kW and $4.4\times10^{13}$ protons per pulse.  Target replacement is, therefore, not expected for the $10^{21}$ POT exposure for nuSTORM. 

Since the energy deposition is proportional to atomic number, Z, this style of low-Z target has been specified and successfully used in several neutrino experiments when coupled with a large depth of field focusing device, such as the magnetic focusing horn, to achieve high pion yields at a range of energies dependent upon the relative placement of target to horn.  For the NuMI low energy beamline with $4.4\times10^{13}$ protons on target, the energy deposition per pulse was calculated at 5.1 kJ per pulse.  For the 400kW case cycle time of 1.87 seconds, the average power into the graphite is approximately 2.7kW.  Scaling to nuSTORM with $0.8\times10^{13}$ protons per pulse and a cycle time of 1.33 seconds, one expects a total average power of 770W.  Note that these targets have successfully operated at much higher power levels.

Graphite possesses favorable material properties for pulsed-beam operation to minimize dynamic stress waves and thermal stress gradients including high thermal conductivity, relatively high specific heat capacity, relatively low thermal expansion and Young's Modulus of elasticity, possesses reasonable strength when considering the above properties, and is able to survive at high temperatures in the absence of an oxidizing (i.e., oxygen) environment.  Scaling from the NuMI experience to the nuSTORM expected power levels and the integrated yearly proton on target value, the graphite should not experience significant material irradiation damage for the specified beam parameters for several years of operation.

Physically, the nuSTORM current baseline design will consist of a 2-interaction length POCO graphite grade ZXF-5Q fin-style target.  The target construction consists of 47 graphite target segments each 2 cm in length, 15mm in vertical height, and 6.4mm in width.  The nominal target overall length is 95 cm, with the fins being brazed to a cooling pipe on the top and bottom surfaces of the graphite fin.  Cooling of the target fins from beam interaction is accomplished via conduction of heat through the graphite fin to the cooling pipe water.  The target core assembly is encapsulated in a thin-wall (0.4mm) aluminum tube to minimize pion absorption and capped at each end by a 0.25mm thick beryllium window.  This allows the target canister to be evacuated and back-filled with helium to minimize graphite oxidation and provides a conduction path for cooling the outer aluminum tube at higher beam powers (note that for the NuMI low energy configuration the target inserts 60cm into the horn and not much external airflow is available for convection cooling on the outer surface of the aluminum tube).  In addition, the thin-wall aluminum tube provides containment to mitigate the spread of radioactive contamination from target material degradation due to large amounts of beam exposure.

Colleagues from IHEP in Protvino, Moscow Region fabricated all NuMI low energy targets for the MINOS/MINER$\nu$A running that ended in CY2012.  Towards the last 1$\frac{1}{2}$ years of that run period, enhancements to the targets were made at Fermilab, including a prototype core fabrication of a complete ZXF-5Q graphite fin target brazed to a titanium cooling pipe.  Analysis has shown this combination to be more robust than the prior targets which used a Russian proprietary grade of high-chromium steel cooling pipe.  It is likely that some amount of engineering effort will be required to repackage and slightly redesign the low energy target for nuSTORM use, but given our experience and knowledge base, this effort is viewed as low risk and not requiring large amounts of manpower for redesign.

Efforts are ongoing to investigate the use of a medium-Z target material (e.g., Inconel 718) to enhance pion yield (see Section~\ref{sec:Inconel}),
but this effort will need to provide a thorough analysis of target material energy deposition and thermal-mechanical response as well as understanding the additional heat load input to the horn inner conductor from a higher-z target material.

\begin{figure}[hb]
  \centering{
    \includegraphics[width=0.9\textwidth]{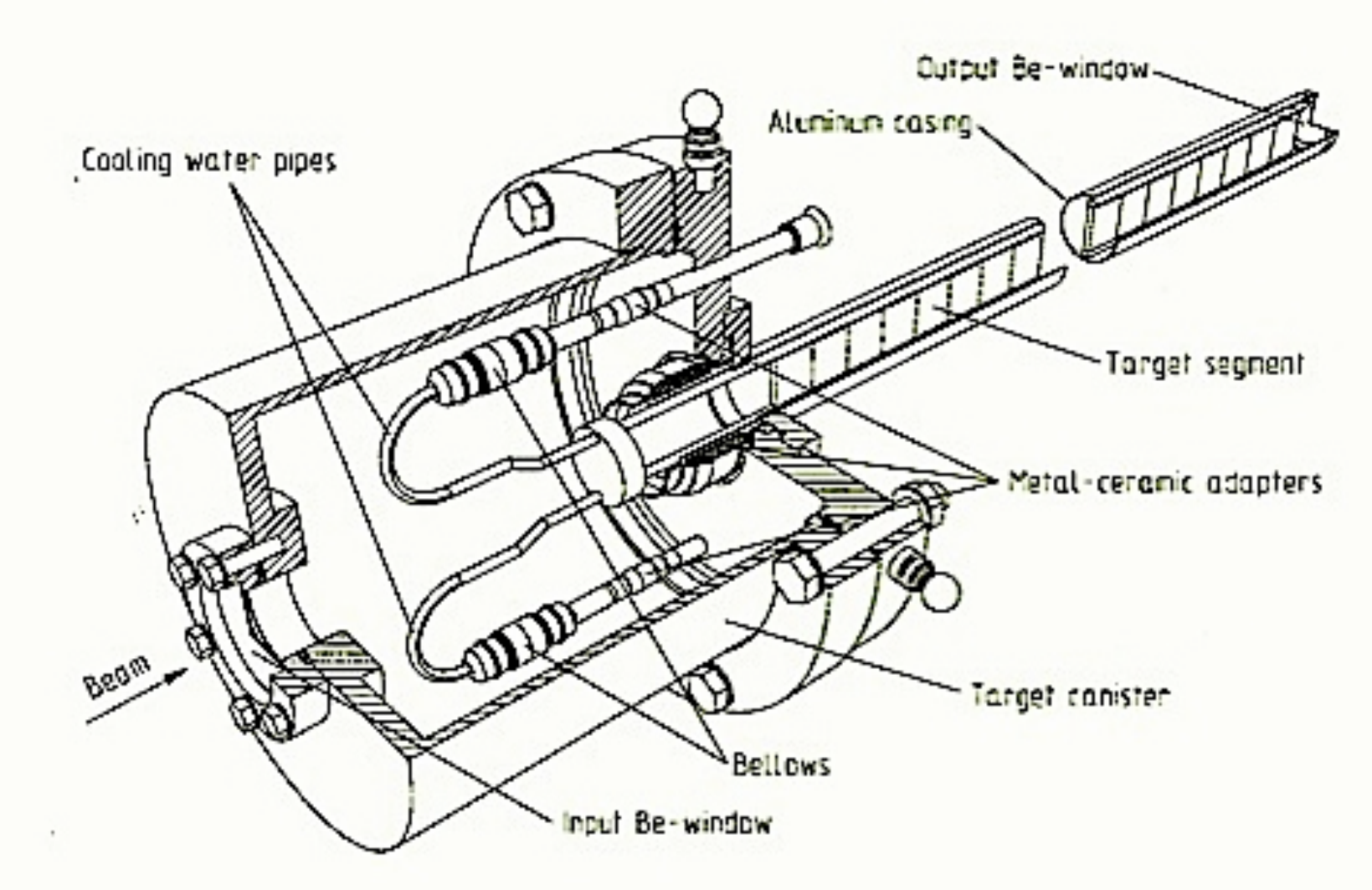}}
  \caption{NuMI-style low energy graphite fin target.}
  \label{fig:Target}
\end{figure}

\begin{figure}[h]
  \centering{
    \includegraphics[width=0.9\textwidth]{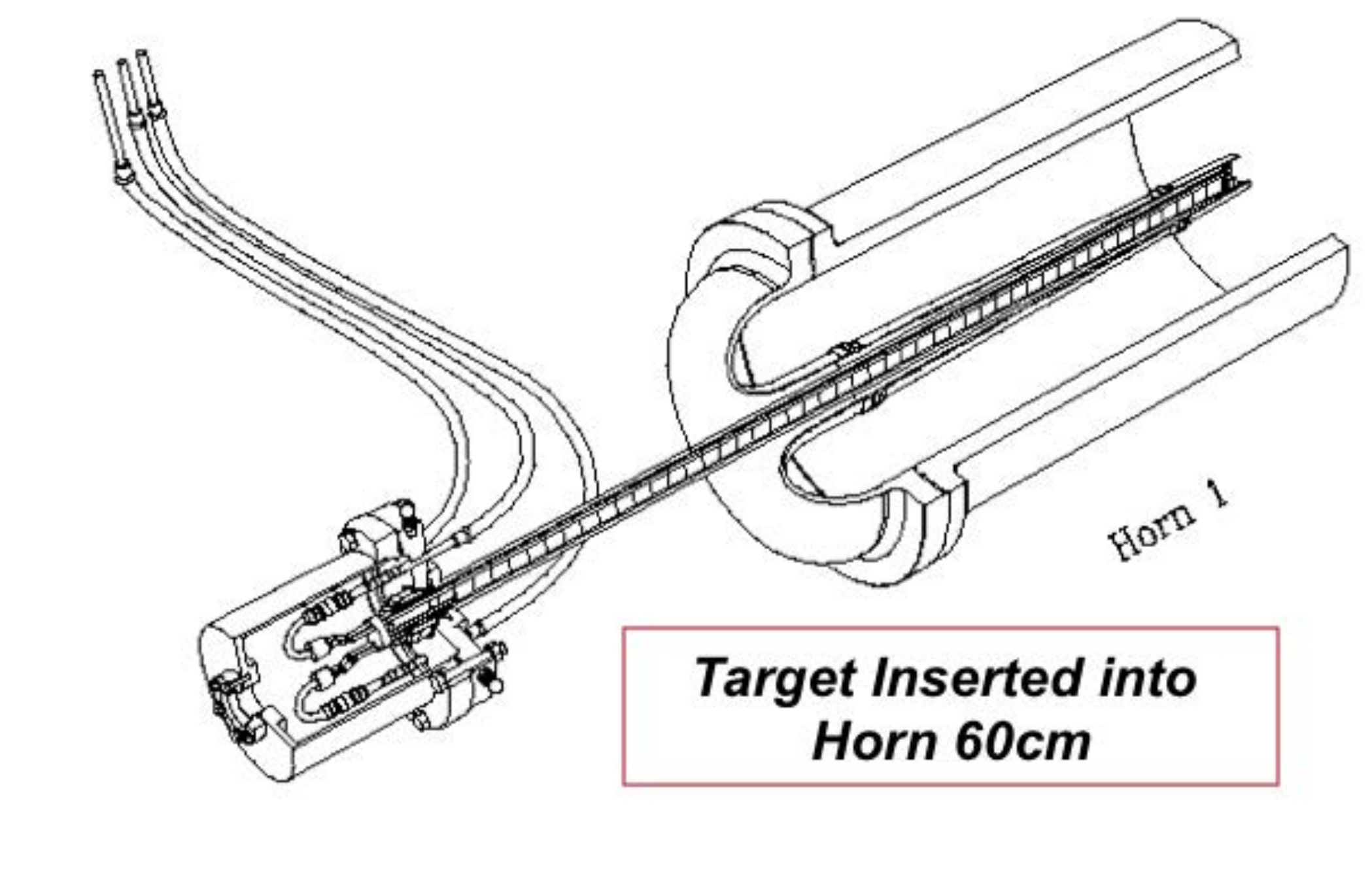}}
  \caption{NuMI-style graphite fin target in low energy position (inserted into horn 60 cm).}
  \label{fig:TarHorn}
\end{figure}

Fig.~\ref{fig:Target} delineates the basic design of the nuSTORM target, which in the baseline configuration is very similar to the NuMI low energy targets.
As previously noted, the NuMI target is designed in conjunction with the horn inner conductor shape to optimize neutrino production in the 1 to 3 GeV region.  This low energy range is achieved by inserting the target 60cm into the field free bore-region of the horn.  The target is mounted on a target carrier frame that allows for target z-motion of 2.5m along the beamline.  This entire carrier frame is mounted to the positioning module and allows for different target to horn spacing or simply provides the means to insert the target into the horn for normal low energy running.   Fig.~\ref{fig:TarHorn} demonstrates the target to horn relationship.
\clearpage
\paragraph{INCONEL target option}
\label{sec:Inconel}
Although the baseline nuSTORM production target material is graphite, as can be seen in Section~\ref{subsec:PiP}, the use of Inconel gives up to a 40\% increase in pion yield.  We have, therefore,  considered a cylindrical Inconel 718 target with a 7 mm diameter and 25cm length.  We have performed an energy deposition study
with MARS and have determined the resulting thermal load.  Based upon this thermal loading, the resulting temperature profiles for a variety of cooling methods were evaluated, with forced Helium cooling being the leading candidate  for further analysis.  Basic structural analyses were also carried out, showing that further work on the structural design is necessary due to plastic deformation.  Further analysis will be necessary to look into target geometry optimization, more realistic temperature estimates via CFD, dynamic stress effects, and resonant effects on the structural analyses.
\subparagraph{{\small MARS 15 Model and Results}}
MARS15(2013) was used to calculate the energy deposition into the target from the incoming proton beam.  The simulation parameters used are listed in Table~\ref{tab:MARSI}.
\begin{table}
\centering
\caption {MARS simulation parameters.}
\label{tab:MARSI}
\begin{tabular}{|l|rl|}
\hline
Parameter		& Value	& Unit\\
\hline
Material		& Inconel 718		& \\
Diameter		& 0.7 	& cm\\
Length		& 25 	& cm\\
Beam Energy	& 120	& GeV\\
Beam $\sigma_{x,y}$	 & 1.1	& mm\\
Protons/pulse	& $8\times10^{12}$		& \\
Pulse Length	& 1.6 	& $\mu$sec\\
Rep Rate		& 0.752		& Hz\\
\hline\noalign{\smallskip}
\noalign{\smallskip}
\end{tabular}
\end{table} 
Based on these inputs, the energy deposition was compiled and converted to a format suitable for input into ANSYS.  The total heat loading on the target is about 4.8kW.  A graphical, axisymmetric plot of energy deposition along the target is shown below in Fig.~\ref{fig:Edep}.  The beam travels from left to right starting at the (0,0) point.
\begin{figure}[h]
  \centering{
    \includegraphics[width=0.9\textwidth]{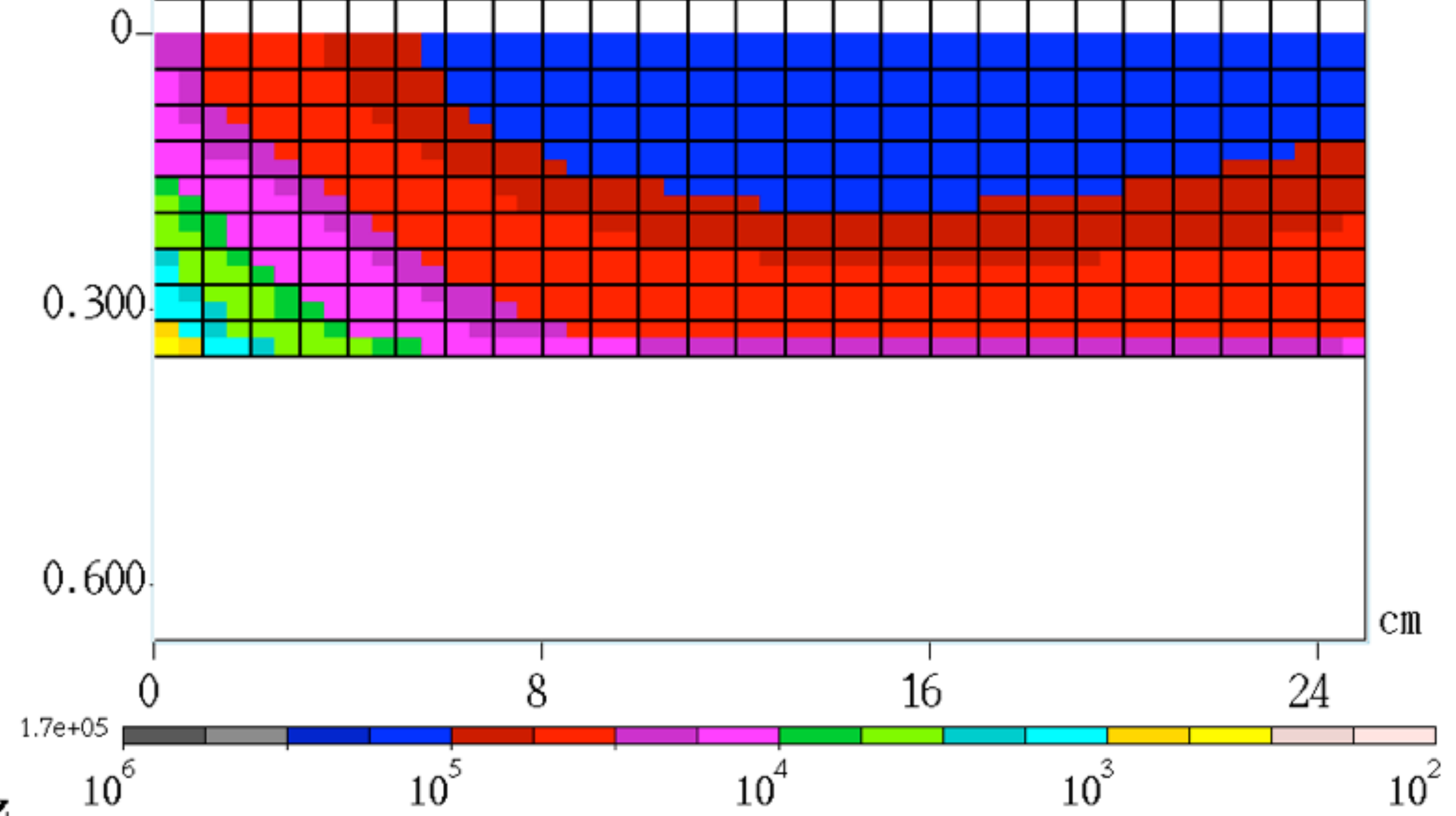}}
  \caption{Energy deposition [mW/g] into Inconel target.}
  \label{fig:Edep}
\end{figure}
It should also be noted that MARS15 predicts high DPA rates in the radial center of the target, with a peak of about 40DPA/yr. assuming continuous beam.  Material degradation from radiation damage must be considered in the final design and target lifetime calculations.  A plot of DPA/yr is shown in Fig.~\ref{fig:DPA}.
\begin{figure}[h]
  \centering{
    \includegraphics[width=0.9\textwidth]{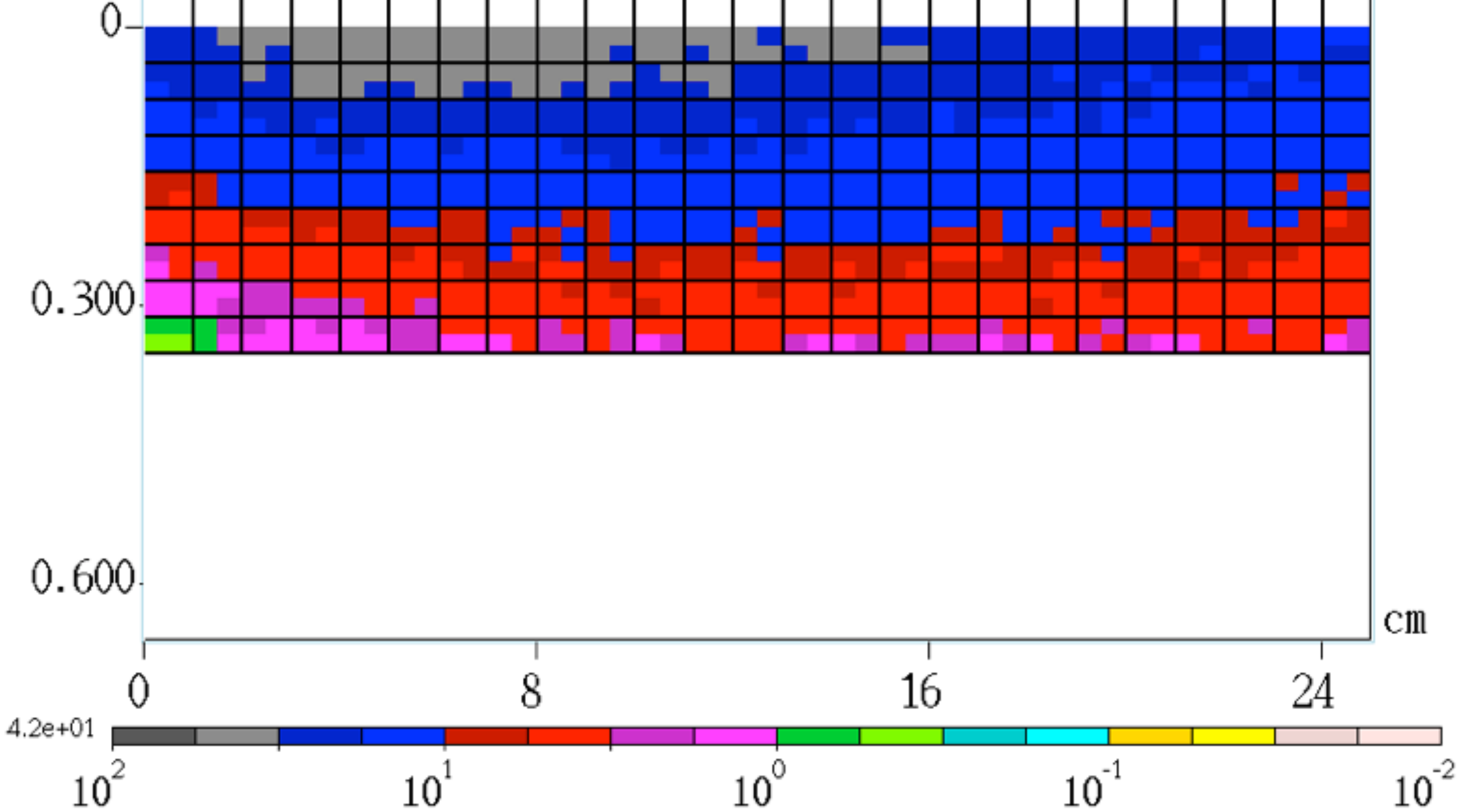}}
  \caption{DPA/yr in Inconel target.}
  \label{fig:DPA}
\end{figure}
\subparagraph{{\small ANSYS Simulations}}
ANSYS 14 was used to calculate the thermal and structural effects from the energy deposition.  Temperature dependent material properties for Inconel 718 were used for all analyses, containing information up to 1200C.  A simple axisymmetric model was used for initial scoping to minimize simulation time.  PLANE77 (thermal) and PLANE183 (structural) elements were used with a size of 0.4mm to accurately capture the maximum temperatures ($\simeq$ 3 elements per beam sigma) and resulting stress effects.\\ \\
{\bf Thermal analysis - Steady state}\\
A convection boundary condition was placed on the outer surfaces of the target and was varied according to different cooling methods as shown in Table~\ref{tab:Thermal} with a reference temperature of $22^\circ$C.  It should be noted that this model does not include heating of the fluid from flow along the surface.  The fluid is also assumed to be in full contact with the outer wall of the target.  A full CFD model will be necessary for accurate predictions of all temperatures.
\begin{table}
\centering
\caption {Steady state target temperatures with varying convection coefficients.}
\label{tab:Thermal}
\begin{tabular}{|l|rll|}
\hline
Method	& Convection 	& Peak Steady-State 	& Peak Fluid Interface \\
		& Coefficient	& Temperature			& Temperature\\
\hline		
	        & [W/(m$^2$-K)]		& $^\circ$C			& $^\circ$C\\
\hline
Forced Helium (1 bar, mach 0.3)	& 1500		& 1065		& 950\\
Forced Air (10 bar, mach 0.3)		& 3500		& 580		& 420\\
Forced Helium (10 bar, mach 0.3)	& 10000		& 370		& 160\\
Water (5 m/s)					& 25000		& 310		& 80\\
Water (10 m/s)					& 45000		& 285		& 55\\
\hline\noalign{\smallskip}
\noalign{\smallskip}
\end{tabular}
\end{table} 
Pressurized Helium and Air both appear to be possible candidates for further study, and water cooling will need further analysis to determine the fluid interface temperature during the dynamic effect of a beam pulse.  Optimizations of target geometry would also be helpful to reduce peak steady state temperatures.\\ \\
{\bf Thermal analysis - Steady state + transient effects}\\
To understand the peak temperatures at both the radial center of the target and the cooling fluid interface immediately following a pulse, a transient thermal analysis was run.  This run uses a steady state heat generation profile until the steady state temperature is reached (~5-15s), then the heat generation from a pulse is applied for 
$1.6\mu$s..  A typical plot of temperature vs. time is shown 
in Fig.~\ref{fig:trise}.
\begin{figure}[h]
  \centering{
    \includegraphics[width=0.9\textwidth]{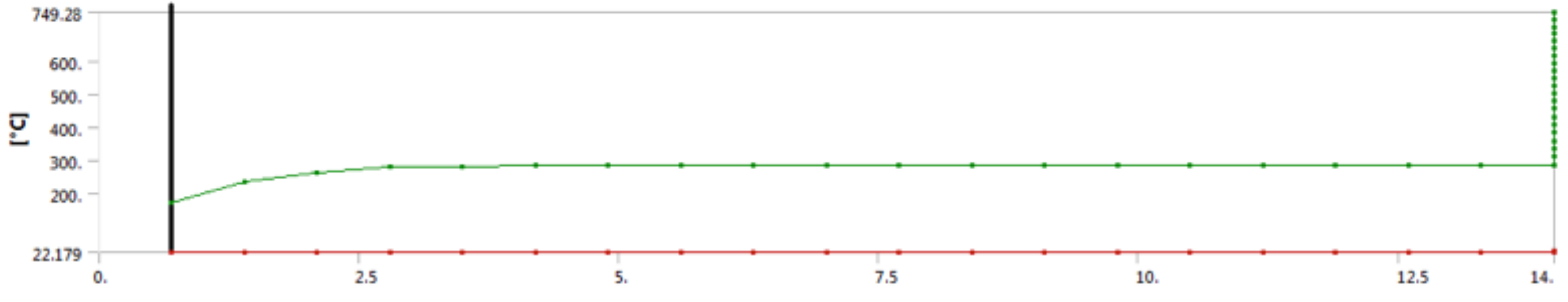}}
  \caption{Typical temperature vs time profile for transient case.}
  \label{fig:trise}
\end{figure}
Using the same boundary conditions and convection coefficients as in the previous analysis, peak steady state temperature, peak temperature in the target, and peak temperature at the fluid interface were found and are shown in Table~\ref{tab:PeakT}.  Temperatures listed are obtained immediately after the pulse is applied.
\begin{table}
\centering
\caption {Steady state + transient target temperatures with varying convection coefficients.}
\label{tab:PeakT}
\begin{tabular}{|l|rlll|}
\hline
Method	& Convection 	& Peak  			& Peak Temp.	       & Peak Fluid\\
	        &			& Steady-State		&			       & Interface\\
		& Coefficient	& Temp.  			& After Pulse 		& Temp.\\
\hline				
	        & [W/(m$^2$-K)]			& $^\circ$C	& $^\circ$C			& $^\circ$C\\
\hline
Forced Air (10 bar)		& 1500		& -		& -		& \\
Forced Helium (1 bar)	& 3500		& 580	& 1010	& 585\\
Forced Helium (10 bar)	& 10000		& 370	& 820	& 350\\
Forced Water (5 m/s)	& 25000		& 310	& 770	& {\bf 270}\\
Forced Water (10 m/s)	& 45000		& 285	& 750	& {\bf 245}\\
\hline\noalign{\smallskip}
\noalign{\smallskip}
\end{tabular}
\end{table} 
Water cooling can be ruled out since the interface temperature is above the boiling point of water and could lead to accelerated corrosion and pitting.  Pressurized Helium appears to be the leading candidate for cooling based on this analysis.\\ \\
{\bf Structural analysis - Steady state + transient effects}\\
A structural model was constructed in the same manner as the previous transient thermal analysis to determine the stress state of the target both in the steady state condition and after a beam pulse.  Refer to Fig.~\ref{fig:trise} for a plot of temperature over time.  Instead of modeling multiple pulses of the beam and observing the ratcheting effect, this method saves some simulation time while giving accurate results.  Only two cooling cases were analyzed: the compressed air and helium cases.  A large temperature gradient develops across a very short radial dimension which induces a large stress in the material.  A typical plot of maximum principal stress is shown in Fig.~\ref{fig:stress} where blue colors are compressive and red colors are tensile.  
\begin{figure}[h]
  \centering{
    \includegraphics[width=0.9\textwidth]{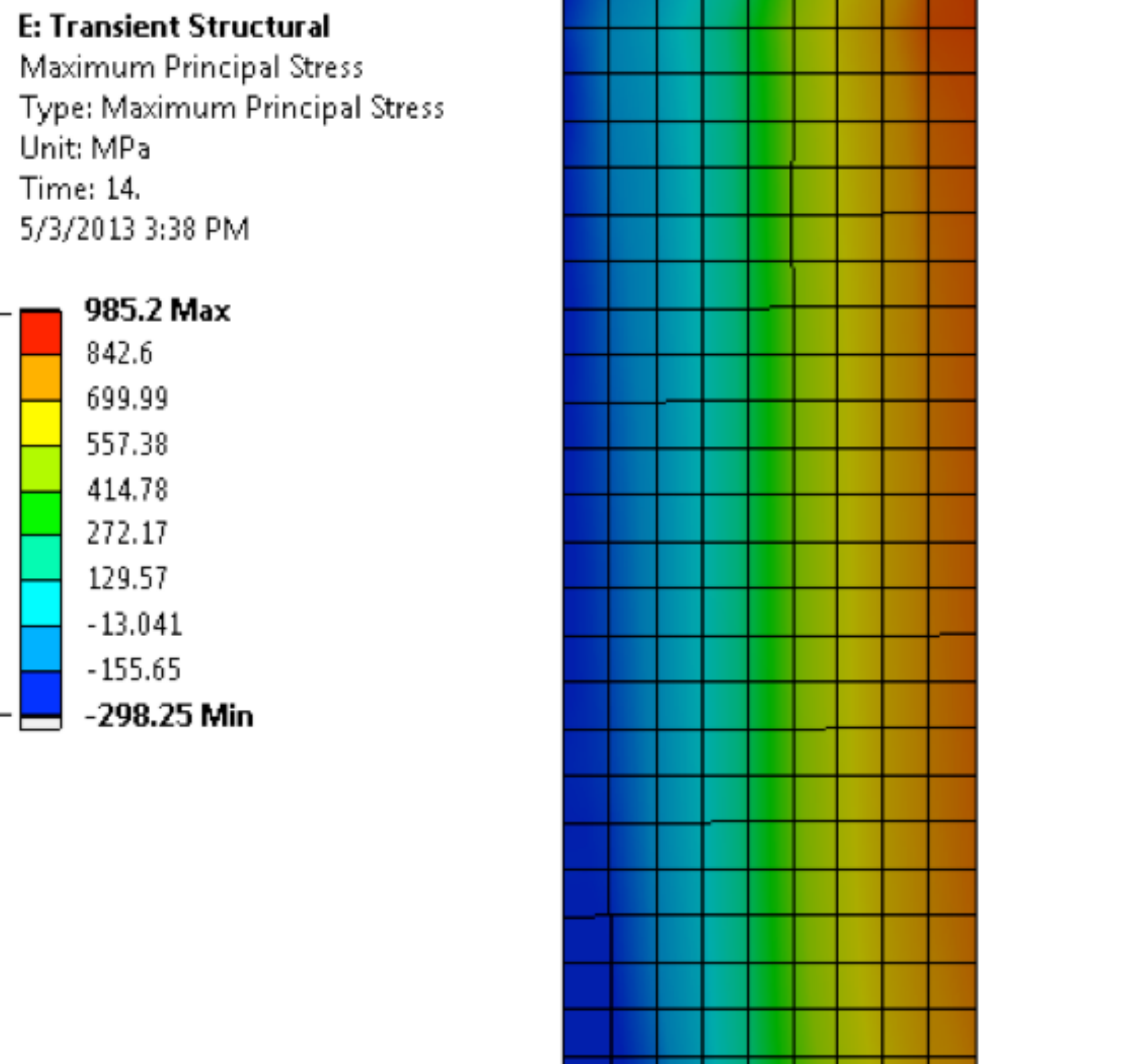}}
  \caption{Typical maximum principal stress plot.}
  \label{fig:stress}
\end{figure}
The peak steady-state Von-Mises stress for both cooling cases is listed in Table~\ref{tab:stress}.  Fig.~\ref{fig:Tyield} shows the temperature dependent tensile yield and ultimate strengths for Inconel 718.  Steady state stresses are relatively low for this material and temperature.
\begin{table}
\centering
\caption {Steady state VM stress for various cooling methods.}
\label{tab:stress}
\begin{tabular}{|rl|}
\hline
Convection Coefficient		& Peak VM Stress Steady State\\
W/(m$^2$-K)				& Mpa (ksi)\\
\hline
3500					& 300 (44)\\
10000					& 380 (55)\\
\hline\noalign{\smallskip}
\noalign{\smallskip}
\end{tabular}
\end{table} 
\begin{figure}[h]
  \centering{
    \includegraphics[width=0.9\textwidth]{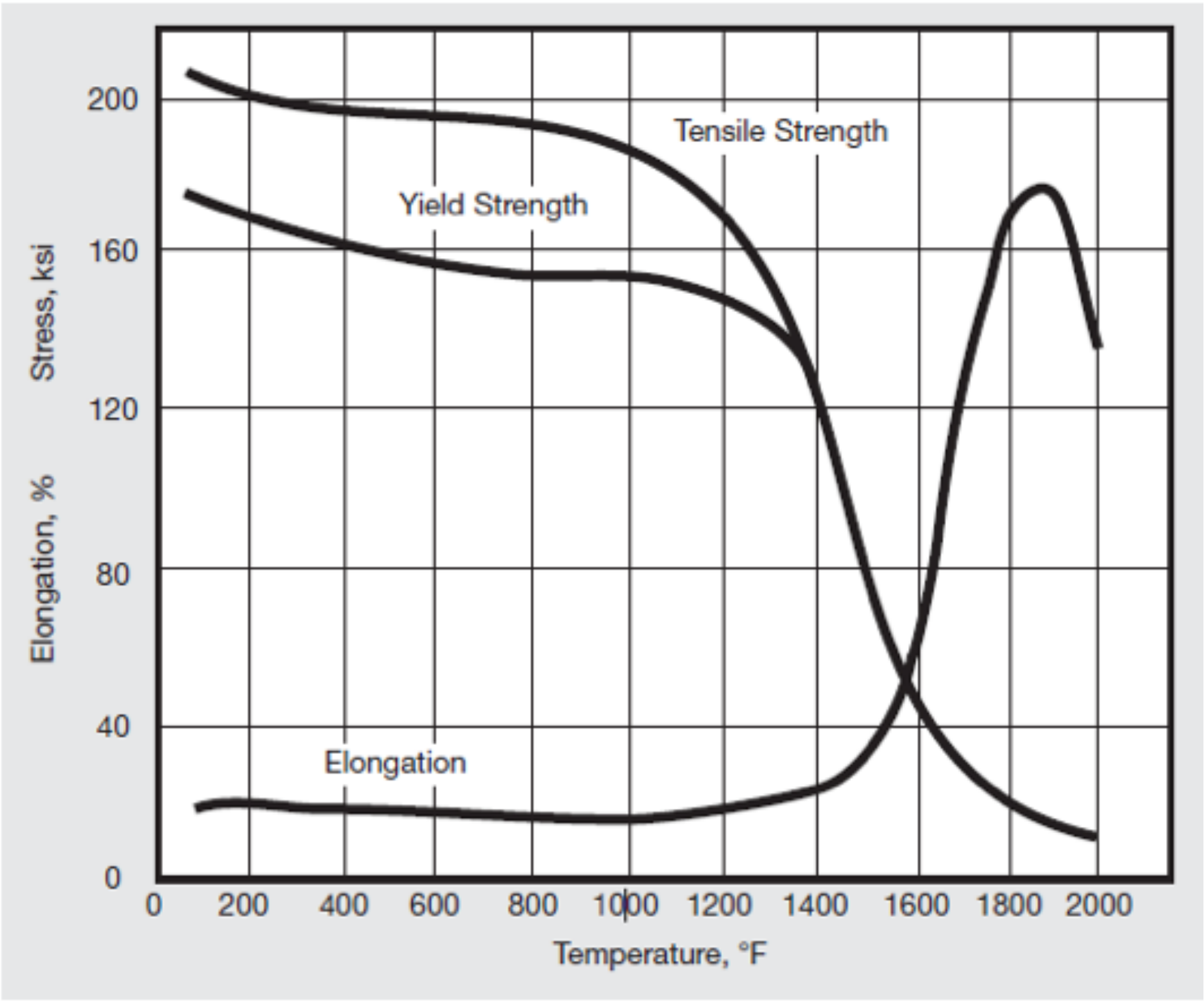}}
  \caption{Tensile yield, ultimate strength, and elongation as a function of temperature.}
  \label{fig:Tyield}
\end{figure}
Next the transient effects were evaluated with a beam pulse after the steady state temperature was obtained.  Table~\ref{tab:PeakS} shows the peak tensile and compressive temperatures and stresses.  While it is possible that there will be no yielding on the outer surface of the target, yielding will most likely occur in the radial center of the target where the compressive stress is concentrated.  Segmenting the target may help reduce the stress and should be further analyzed.
\begin{table}
\centering
\caption {Peak temperatures and stresses after pulse.}
\label{tab:PeakS}
\begin{tabular}{|rl|}
\hline
Convection Coefficient		& Peak VM Stress Steady State\\
W/(m$^2$-K)				& [Mpa (ksi)]\\
\hline
3500					& 300 (44)\\
10000					& 380 (55)\\
\hline\noalign{\smallskip}
\noalign{\smallskip}
\end{tabular}
\end{table} 
\subparagraph{{\small Next steps in the analysis for Inconel}}
Further analysis of the Inconel target is needed in at least four areas to get a better understanding of the beam response and interaction with the target geometry:
\begin{itemize}
\item Fluid flow and heating (CFD)
\item Dynamic stress effects
\item Resonant effects
\item Target segmentation
\end{itemize}
CFD analysis is needed to determine a more accurate temperature profile of the target.  ANSYS thermal calculations do not include fluid heating and this can only be evaluated with a CFD model.  Dynamic stress waves have been shown to increase stress in a target by up to 2$\times$ depending on the geometry because of the interference between stress waves.  This effect needs to be studied with a longer run after the beam pulse with short time-steps.  Resonant modes of the target will also come into play and have an effect on the dynamic stress.  A modal analysis needs to be conducted to determine the resonant modes of the target and any supporting structures.  The effect of target segmentation needs to be looked at, including all of the previously described areas for further analysis.  It has also been suggested to investigate the additional heat loading onto the horn because of the larger shower coming from the target.  This can be done with a fairly simple MARS model.
While a completely solid target will undergo plastic deformation during a beam pulse, optimization of the geometry may allow the target stresses to be reduced to an acceptable range.  Geometry optimization would also allow lower steady state temperatures and stresses.  Further analysis is needed to define an optimized target geometry.
\subsubsection{Focusing Horn}
\begin{figure}[h]
  \centering{
    \includegraphics[width=0.8\textwidth]{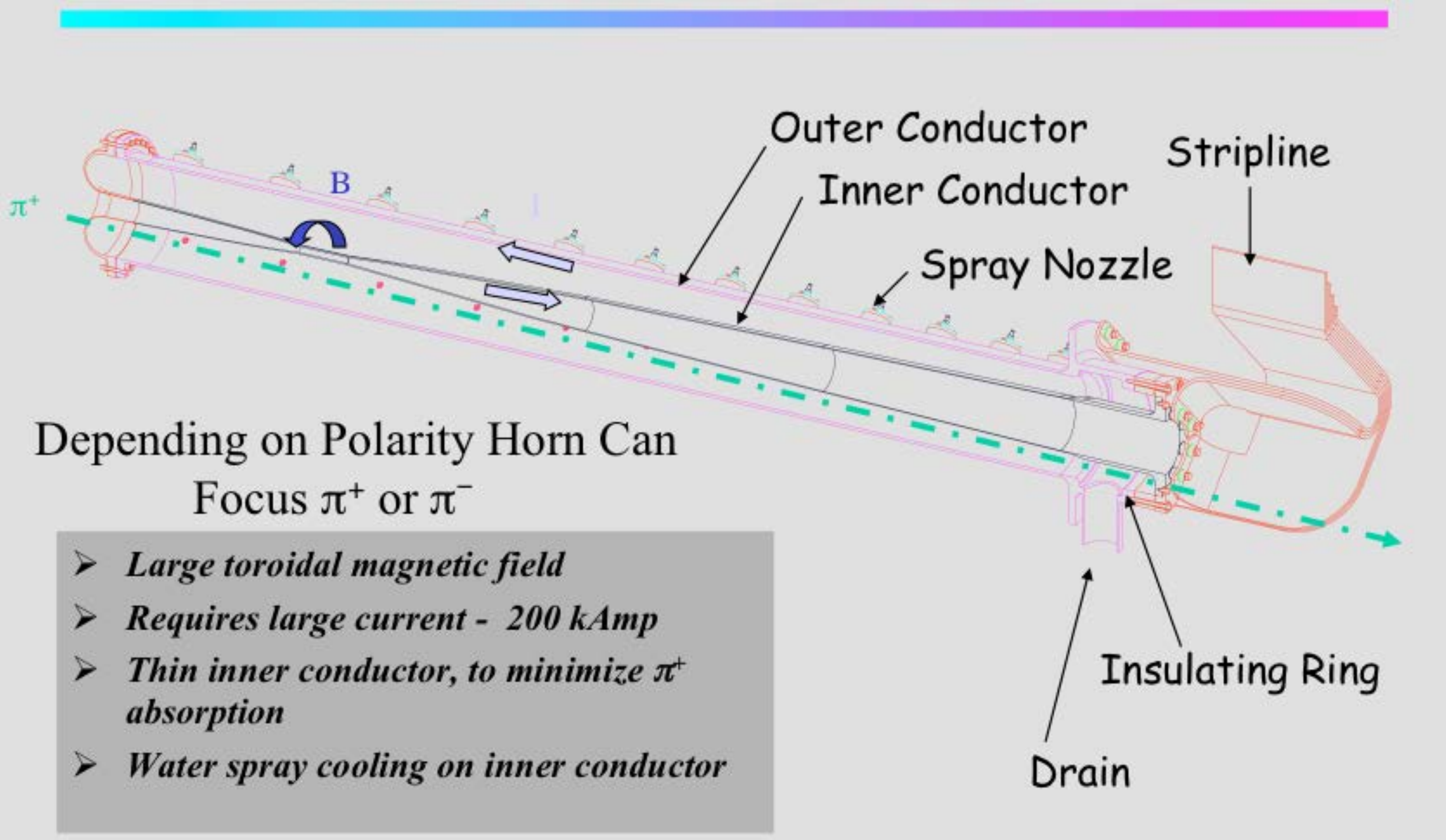}}
  \caption{NuMI Horn 1 basic elements.}
  \label{fig:Horn}
\end{figure}

The primary objective of the focusing horn is to collect the pion phase space from the target interaction with the primary beam and focus the pions such that they are stably captured by the quadrupole magnets just downstream of the horn.  This leads to several requirements of the horn for reliable and effective operation:
\begin{enumerate}
\item The horn should have the ability to operate at sufficiently high currents to maximize the pion focus to the capture quadrupoles.
\item The inner conductor of the horn should be sufficiently thin to minimize  absorption of off-axis pions produced in the target that enter the field region of the horn.
\item The horn should have sufficient structural integrity to survive a reasonable number of high current pulses with adequate safety factor to allow reliable operation for several years of service.
\item The horn should employ radiation hard structural and insulating materials.
\item The horn should employ suitable corrosion resistant materials to mitigate erosion of the thin-wall inner conductor from the water-cooling spray.
\item The horn should be designed with a fabrication tolerance that assure uniform field region.
\item The horn should have adequate cooling plus overhead for the expected ~100kW beam power and the 200kA (or higher) current pulse.
\end{enumerate}
The horn provides large depth of field focusing compatible with a low-Z target in order to maximize the pion capture rate.  This is achieved by using a target with a small cross-sectional component. allowing pions to exit the target material without further interaction.  Pions with a nearly on-axis trajectory travel through the field free region of the inner bore of the horn inner conductor, while pions with an off-axis trajectory travel through the thin-wall inner conductor of the horn into the field region and are subject to a focusing force vector which is proportional to the vector cross-product of instantaneous velocity and the magnetic field vector.  The selection of the horn inner conductor thickness is a balance between minimizing pion absorption, providing adequate thickness to allow for precision machining and fabrication, and providing adequate margin of safety relative to the operating stress that results from the thermal components of secondary particle beam heating, joule heating from current pulse resistive heating, and the magnetic forces from the interaction of current density and magnetic field.  Fig.~\ref{fig:Horn}  outlines the basic structure of the focusing horn.
nuSTORM expects to utilize the successful operational experience of NuMI and the baseline configuration consists of a NuMI style horn 1.  The NuMI horns have operated successfully for more than $3\times10^6$ pulses of operation with average beam powers as high as 375kW and maximum proton intensity of $4.4\times10^{13}$ protons per pulse.  In addition, significant analysis and understanding of horn operational characteristics and reliability exists at Fermilab.  Such analyses are well documented for the 400kW NuMI/MINOS beamline operation, the upcoming 700kW NOvA era running, and the 700kW baseline LBNE configuration.  In particular, a recent analysis for LBNE indicates that it may be possible to operate with adequate safety margin at horn currents of 230kA for the 700kW LBNE beam.  It is believed that this scenario is likely for nuSTORM, but would require some further analysis to understand the entire load history to accurately calculate the fatigue life safety factor.  Note that the fatigue life value is highly dependent upon the coupled loading of beam thermal energy, current pulse resistive heating, and magnetic loading stress value and sign (compressive vs. tensile stress), as the magnitude of alternating stress has significant effect on fatigue life.

\begin{figure}[h]
  \centering{
    \includegraphics[width=0.65\textwidth]{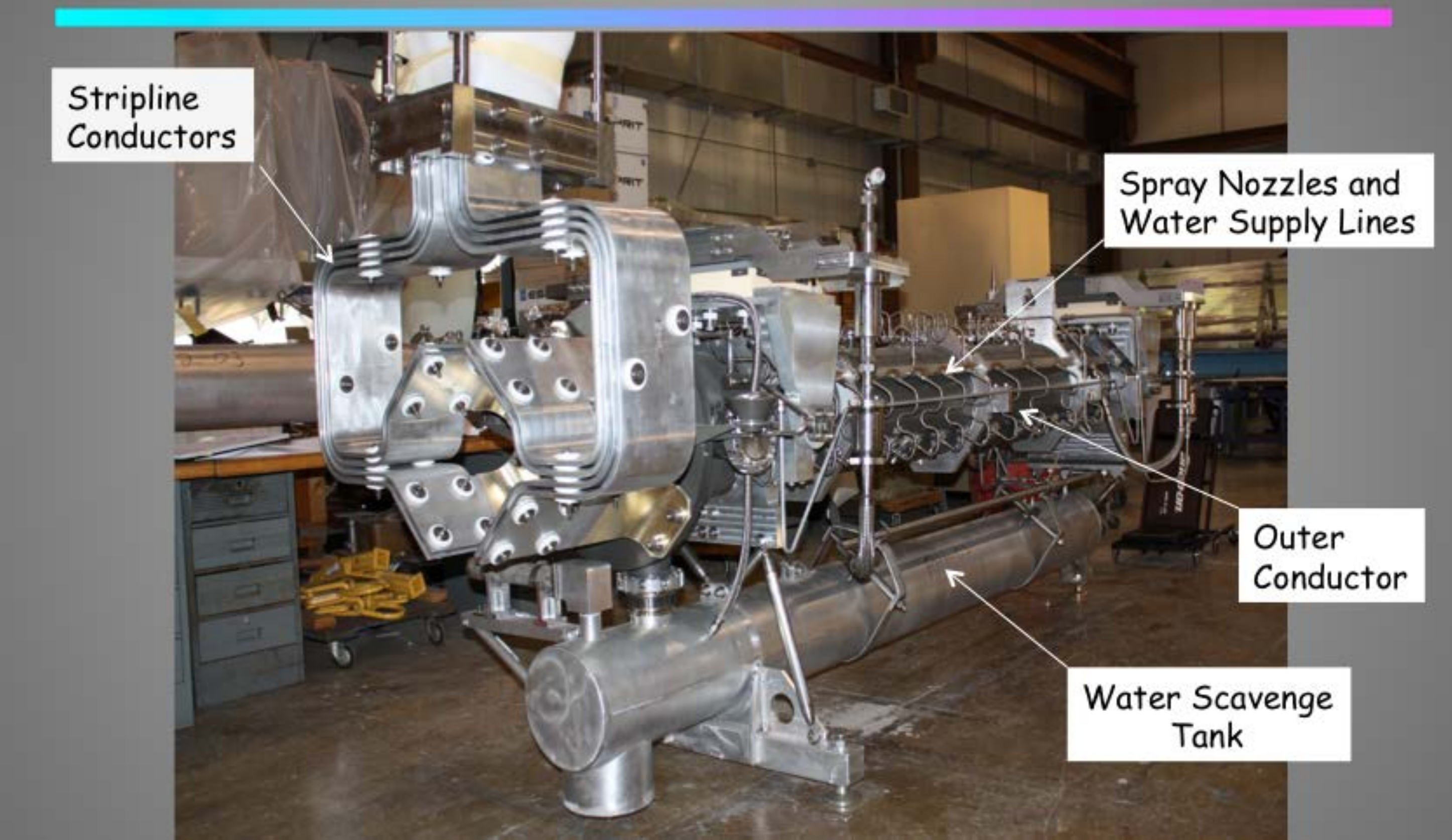}}
  \caption{NuMI Horn 1 ready spare.}
  \label{fig:NumiHorn}
\end{figure}

\begin{figure}[h]
  \centering{
    \includegraphics[width=0.75\textwidth]{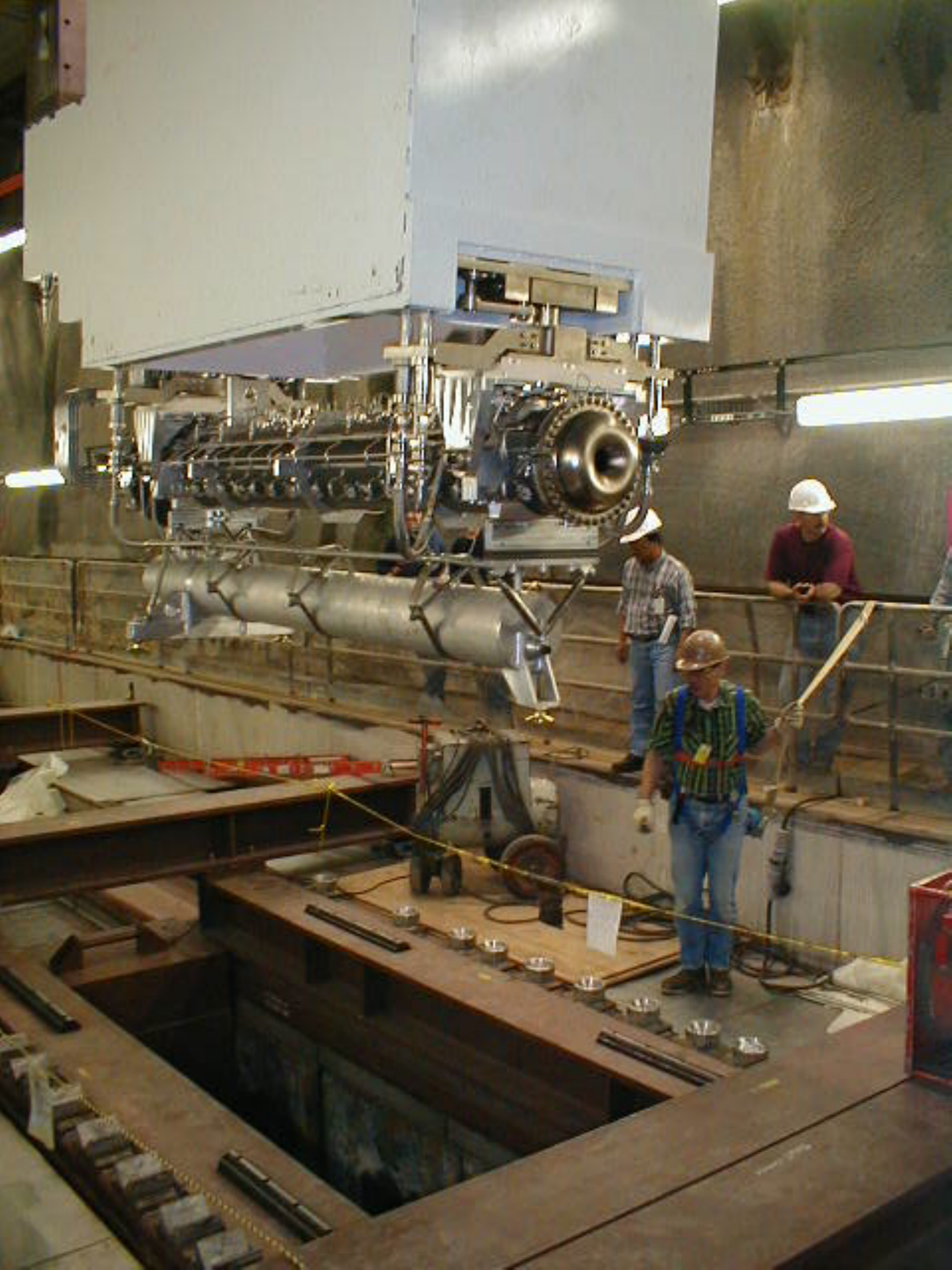}}
  \caption{NuMI Horn 1 and positioning module installation in beamline chase.}
  \label{fig:NumiHorn2}
\end{figure}

An additional consideration for using a NuMI-style horn is that Fermilab has the infrastructure for, and experience in, fabricating all NuMI horns to date.  A few examples of the benefits to such expertise in addressing the challenge of fabricating these critical devices include the in-house CNC welding and straightening of the inner conductor structure to very high tolerances ($\pm$ 0.25mm), a ready design with little modification for nuSTORM use (i.e., minimizes the cost of manpower for a new design or substantial redesign), and the infrastructure to pulse test and field map a finished horn before beamline installation.  Fig.~\ref{fig:NumiHorn} shows a NuMI horn 1 ready spare.
The design package of a new-style horn, module, and supporting hardware represents significant manpower effort and cost.  It is envisioned that nuSTORM can utilize most elements of the NuMI design.  The fabrication costs and operating envelope are well understood.  Minor design effort does exist in that the horn hangers and module for NuMI are designed for a beamline pitch of 58 milliradians.  The nuSTORM beamline pitch is level.  Such modifications to the NuMI horn and module design are viewed as relatively minor when compared to more substantial redesign efforts and are expected to pose negligible risk to design reliability.   Fig.~\ref{fig:NumiHorn2}  depicts the NuMI horn 1 package during the  installation into  the NuMI beamline.  Similar hardware is specified for the target, the first 2 capture quadrupoles immediately downstream of the horn, and beam spray protection collimators for those quadrupole magnets.
\newpage
\subsubsection{Quad capture magnets}
\label{sec:HTS-Quads}
Unlike a conventional neutrino beam where pions exiting the horn are  launched into a decay pipe, in nuSTORM we have a pion capture and transport channel that brings the pions to the decay ring.  The initial part of this section (just downstream of the horn) consists of two quadrupoles.  They are in the target chase and see a very large radiation dose and, therefore, must be radiation hard.  There are two options we have considered: MgO insulated magnets and HTS magnets following the BNL magnets designed for the Facility for Rare Isotope Beams (FRIB) \cite{Chouhan:2011}.  The current baseline design for the capture quad section of the pion capture channel (see Section~\ref{subsubsec:422}) uses magnets with a 40 cm bore.  However, recent studies indicate that a 20 cm bore can be used with little loss in flux.  This would allow nuSTORM to use the same magnets that have been designed for FRIB and would represent a substantial cost savings.  These magnets have a unique design and technology that uses  Second Generation (2G) High Temperature Superconductor (HTS).  These magnets can survive very intense heat and radiation loads (up to 10 MGy/year) and can operate at an elevated temperature of ~40 K, instead of $\simeq$ 4 K as needed in the conventional low temperature superconductors. This elevated temperature has a major impact regarding the heat load on the cryogenic system. In addition, a large temperature margin allows for robust operation by tolerating a large local increase in temperature. These HTS quadrupoles are the baseline design for the fragment separator region of FRIB which follows the production target ($\simeq$ 400 kW beam power). 

\begin{figure}[h]
  \centering{
    \includegraphics[width=0.45\textwidth]{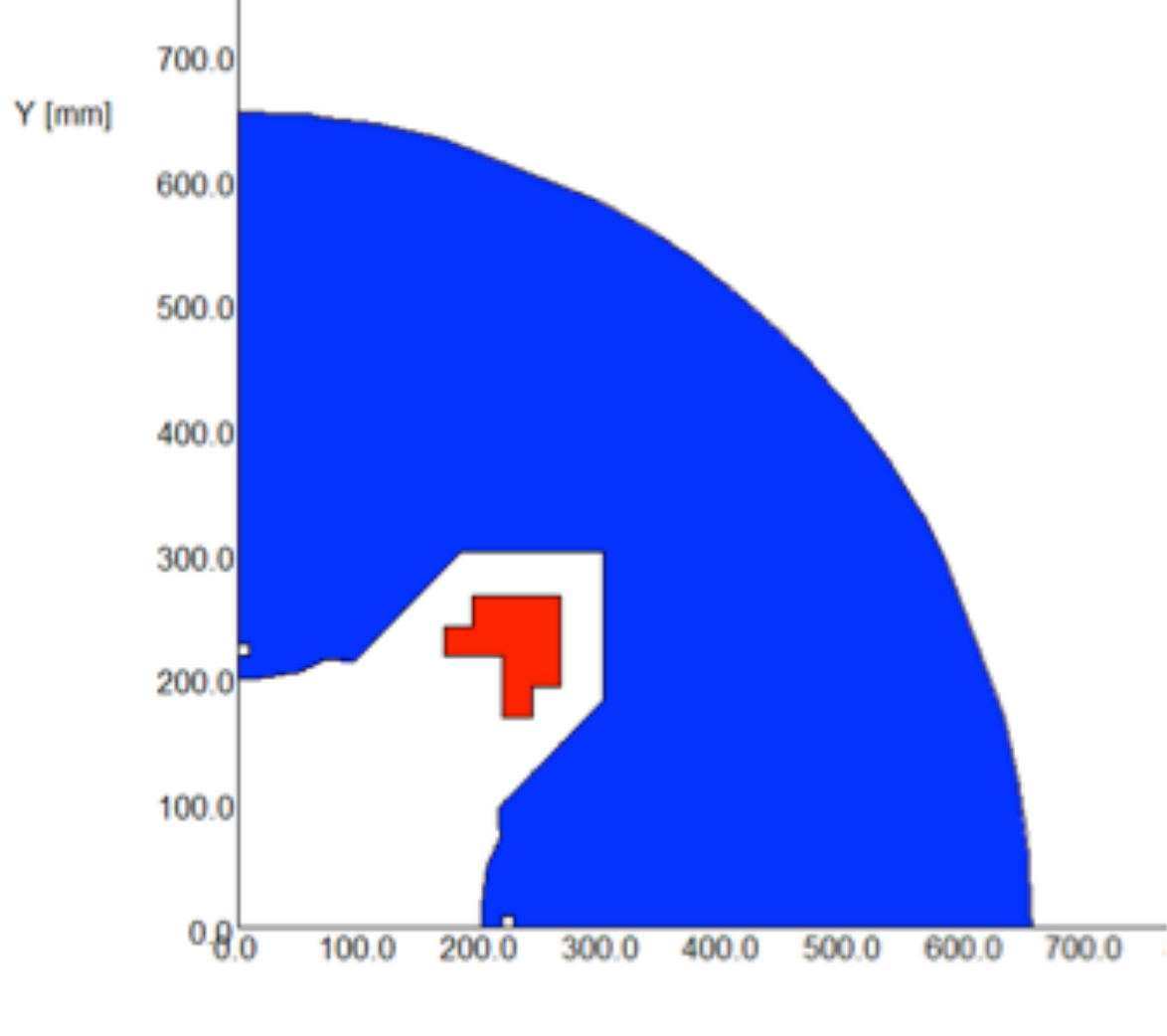}
    \includegraphics[width=0.45\textwidth]{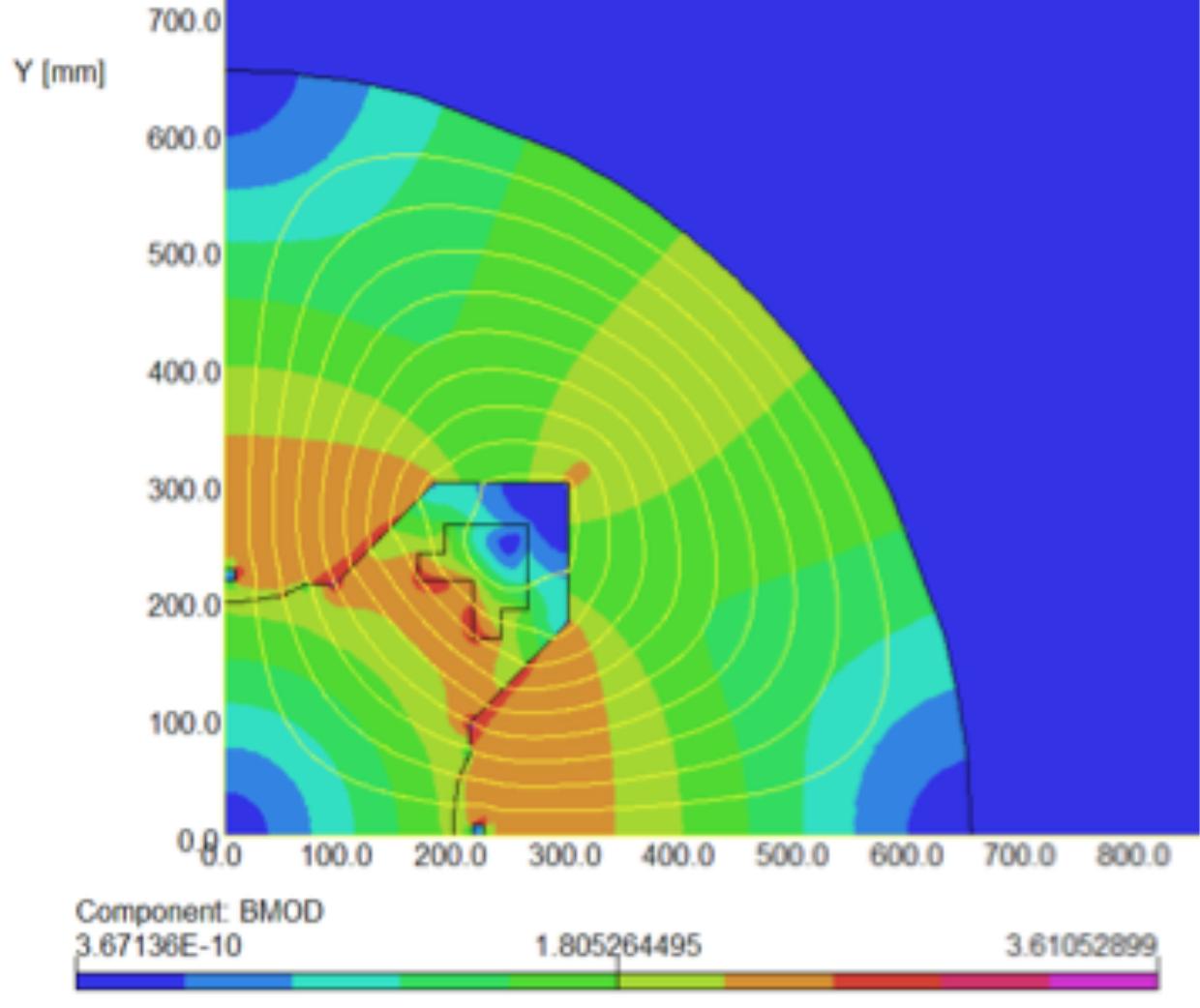}}
  \caption{A preliminary magnetic design of the HTS quadrupole for nuSTORM (model on left and field contours on right).  In the model, blue represents the warm steel and 
  the red represents the HTS conductor.}
  \label{fig:HTS-Quad}
\end{figure}

The BNL magnet group \cite{Wanderer1} has looked into the feasibility of a HTS Quad with a 40 cm bore for nuSTORM.  Fig.~\ref{fig:HTS-Quad} shows the preliminary magnetic design and field contour for this magnet.
\subsection{Pion Production}
\label{subsec:PiP}
It is known that the maximum pion yield can be achieved with a target radius $\sim$ 3 times 
the proton beam RMS size. The optimal target length depends on the target material and 
the secondary pion momentum. 
The results of our optimization study are presented in Table~\ref{tab:yield1}. 
We see that approximately 0.11 $\pi^+$/POT can be collected into $\pm$10\% momentum acceptance 
off medium to heavy targets with very small proton beam size (0.15 mm). The difference between the yield 
from heavy and light targets becomes about 2 $\times$ smaller for a more realistic beam size with a 1 mm RMS.
These numbers are for pions within the 2 mm-radian acceptance of the decay ring.

We have considered two options for pion capture, a lithium lens and a horn.
The existing Fermilab lithium lens has a working gradient of 2.6 Tesla/cm at 15 Hz. The optimal
distance between the target and lens center is about 25 cm. Pions produced into 2 mm-radian
acceptance have a wide radial distribution, however. The current lens with its 1 cm radius would only
capture 40\% of the pions within the desired $\pm$10\% momentum acceptance. With a 2 cm lens radius, 
the pion capture efficiency increases to 60\%. Further improvement could be achieved by increasing the lens
gradient, but increasing the gradient reduces the focal length. Maximal efficiency would reach 80\%
with a 4 Tesla/m gradient and a 2 cm lens radius.  These parameters are beyond the state-of-art for an operating 
lens, however, and the target's downstream end would need to be very close to the lens.  Therefore for nuSTORM,
we have abandoned consideration of a Li Lens.

The existing NuMI horn was considered as the other capture option. Different target materials, horn currents,
beam RMS sizes (target size) were studied. Results for a 60 GeV/c proton beam are presented in 
Table~\ref{tab:yield2}. We see that approximately 0.06 $\pi^+$/POT can be captured into a $\pm$10\% 
momentum acceptance for 1 mm ($\sigma_{b}^{rms}$) proton beam. Yield can be increased by using a heavy target and very small
beam transverse size. 

\begin{figure}[h]
  \centering{
    \includegraphics[width=0.7\textwidth]{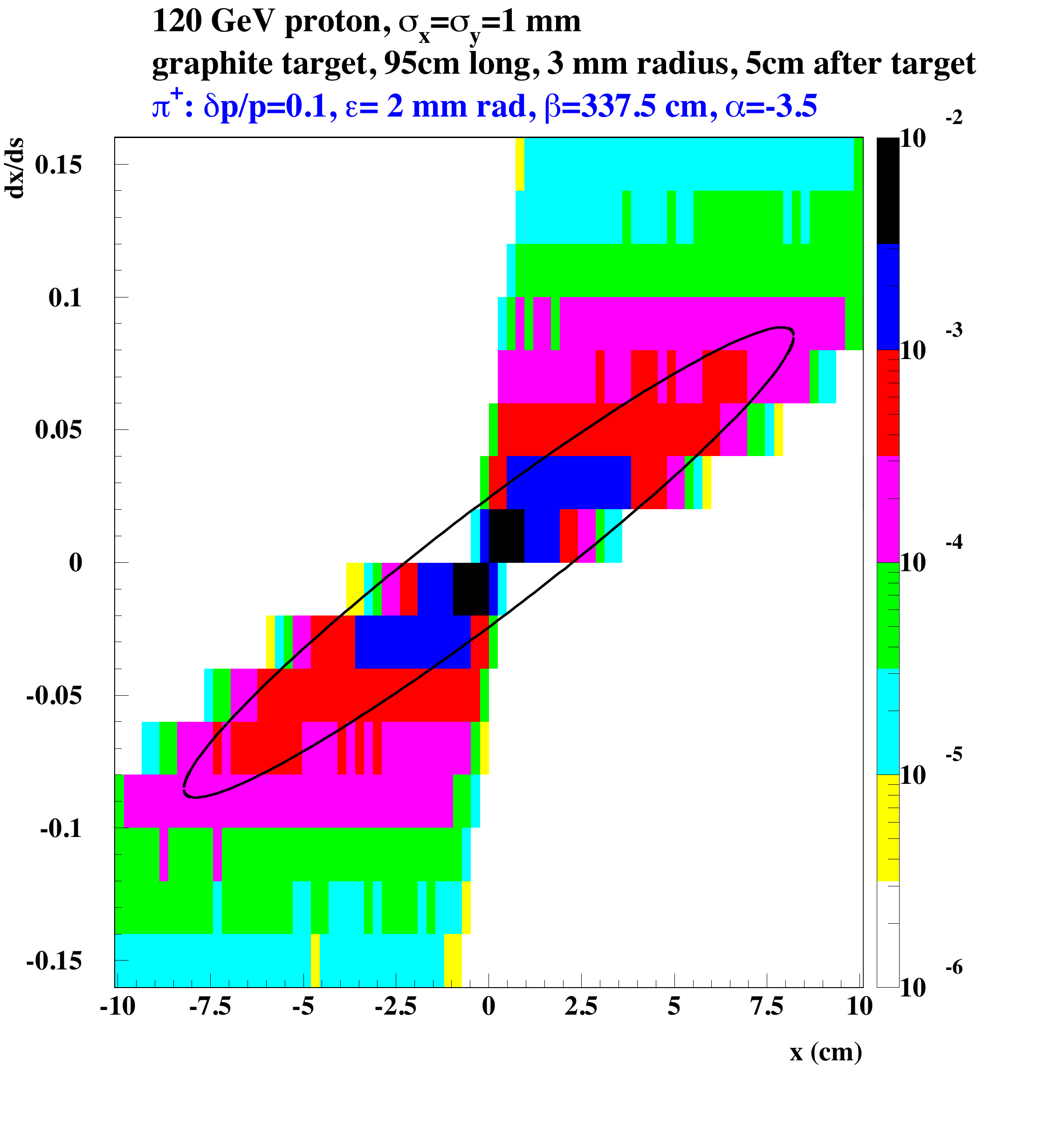}}
  \caption{Phase-space  distribution of pions 5 cm downstream of target}
  \label{fig:piTar}
\end{figure}

\begin{figure}[h]
  \centering{
    \includegraphics[width=0.7\textwidth]{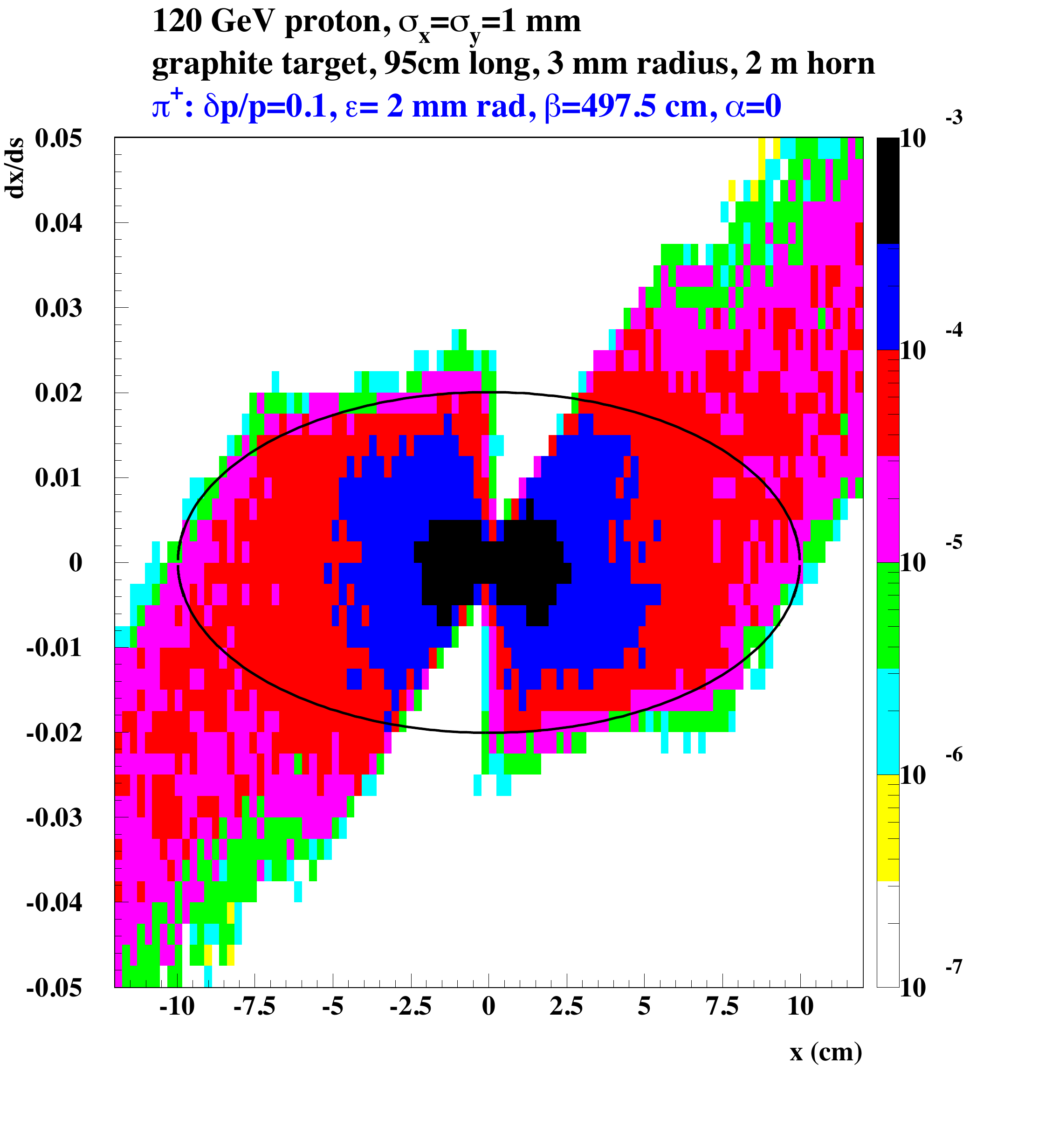}}
  \caption{Phase-space distribution of pions downstream of the horn}
  \label{fig:piHorn}
\end{figure}

Approximately twice as many pions can be collected with 120 GeV/c protons  and a NuMI-style horn. Table~\ref{tab:yield3}
presents the pion yield dependence on target material for this configuration.  Using a conventional graphite target and NuMI horn
at 230 kA, 0.094 $\pi^+$/POT are captured after the horn into the 2 mm-radian acceptance. The phase-space  distribution of pions downstream of the 
target and downstream of the horn are shown in Fig.~\ref{fig:piTar} and Fig.~\ref{fig:piHorn}, respectively.   The ellipse shown in each figure represents
the acceptance of the muon decay ring.
\begin{table}
\centering
\caption {$\pi^+$ yield/POT at 60 GeV into 2 mm radian acceptance. }
\label{tab:yield1}
\begin{tabular}{|lclllccc|}
\hline
material & momentum & $\pm 15\%$ & $\pm 10\%$    &$\pm 5\%$  & target length  & density 		& $\sigma_b$ (mm)\\
	     & (GeV/c)	  &			&			  &		       & (cm)	       & (g/cm$^3$) 	& \\
\hline\noalign{\smallskip}
Carbon   & 3   & 0.085       & 0.056  & 0.028 & 27.3 & 3.52 & 0.15 \\
Carbon   & 5   & 0.099       & 0.067  & 0.033 & 32.2 & 3.52 & 0.15  \\
Inconel  & 3   & 0.131       & 0.087  & 0.044 & 19.2 & 8.43 & 0.15  \\
Inconel  & 5   & 0.136       & 0.091  & 0.045 & 27.0 & 8.43 & 0.15  \\
Tantalum  & 3   & 0.164       & 0.109  & 0.054 & 15.3 & 16.6 & 0.15  \\
Tantalum  & 5   & 0.161       & 0.107  & 0.053 & 21.3 & 16.6 & 0.15  \\
Gold      & 3   & 0.177       & 0.118  & 0.059 & 18.0 & 19.32 & 0.15  \\
Gold      & 5   & 0.171       & 0.112  & 0.056 & 20.0 & 19.32 & 0.15  \\
Gold      & 5   & 0.143       & 0.094  & 0.047 & 20.0 & 19.32 & 1.  \\
Graphite  & 5   & 0.085       & 0.057  & 0.028 & 95.0 & 1.789 & 0.15  \\
Graphite  & 5   & 0.096       & 0.064  & 0.032 & 95.0 & 1.789 & 1.  \\
\hline
\noalign{\smallskip}
\end{tabular}
\end{table}

\begin{table}
\centering
\caption {$\pi^+$ yield/POT at 60 GeV into 2 mm radian acceptance after NuMI horn. 5 $\pm$ 0.5 GeV/c }
\label{tab:yield2}
\begin{tabular}{|lrcllll|}
\hline
material & Current (kA)  & horn length (cm) & $\sigma_b^{rms}$ (mm) & $\alpha$ & $\beta$ (cm) & yield\\
\hline\noalign{\smallskip}
Gold      & 300       & 200  & 0.15 & 0.  & 522.5 & 0.081   \\
Gold      & 300       & 300  & 0.15 & 0.  & 427.5 & 0.077   \\
Gold      & 300       & 200  & 1.   & 0.  & 542.5 & 0.063  \\
Gold      & 300       & 300  & 1.   & 0.5 & 812.5 & 0.059  \\
Graphite  & 300       & 200  & .15  & 0.5 & 257.5 & 0.049  \\
Graphite  & 300       & 300  & .15  & 0.5 & 202.5 & 0.049  \\
Graphite  & 300       & 200  & 1.   & 0.5 & 282.5 & 0.056  \\
Graphite  & 300       & 300  & 1.   & 0.  & 217.5 & 0.056  \\
Graphite  & 185       & 200  & 1.   & -1  & 682.5 & 0.056  \\
Graphite  & 185       & 300  & 1.   & -0.5 & 987.5 & 0.054  \\
Graphite  & 230       & 200  & 1.   & 0.5  & 512.5 & 0.057  \\
Graphite  & 230       & 300  & 1.   & 0.5  & 687.5 & 0.056  \\
BeO       & 230       & 200  & 1.   & 0.   & 602.5 & 0.065  \\
BeO       & 230       & 300  & 1.   & 0.5  & 782.5 & 0.066  \\
NuMI tgt  & 185       & 200  & 1.1   & -1.5 & 852.5 & 0.056  \\
NuMI tgt  & 185       & 300  & 1.1   & -0.5 & 1097.5 & 0.054  \\
\hline
\noalign{\smallskip}
\end{tabular}
\end{table}

\begin{table}
\centering
\caption {$\pi^+$ yield/POT after NuMI horn at 230 kA. 120 GeV/c proton beam with 1 mm RMS. 5$\pm$0.5 GeV/c}
\label{tab:yield3}
\begin{tabular}{|lccllcc|}
\hline
material  & target length(cm) & horn length (cm)  & $\alpha$ & $\beta$ (cm) & yield (2 mm rad) & yield (r$\le$20 cm)\\
\hline\noalign{\smallskip}
Graphite  & 95 & 200  & 0.   & 497.5 & 0.094 & 0.156 \\
Graphite  & 95 & 300  & 0.5  & 612.5 & 0.092 & 0.139  \\
BeO       & 66 & 200  & 0.   & 547.5 & 0.115 & 0.174  \\
BeO       & 66 & 300  & 0.5  & 742.5 & 0.115 & 0.154 \\
Inconel   & 38 & 200  & -0.5 & 742.5 & 0.122 & 0.179 \\
Inconel   & 38 & 300  & 0.5  & 897.5 & 0.126 & 0.157 \\
\hline
\noalign{\smallskip}
\end{tabular}
\end{table}
\subsubsection{Capture and Transport}
\label{subsubsec:422}
As described in the previous section, the pion distribution at the downstream end of the horn was simulated and generated by MARS. A shell script code converts the MARS output to G4beamline\cite{G4bl} conventions, for the simulation's input beam. Another script converts OptiM\cite{OptiM} lattice output to G4beamline lattice input.  Assuming a phase space acceptance of 2 mm$\cdot$rad for pions, it is convenient to fit the transverse distribution to a 2D Gaussian with a covariance matrix obtaining  Courant-Snyder(CS or Twiss) parameters \cite{SYLEE}.  These parameters are used for the matching condition for designing the transport line.

From the downstream end of the horn, we continue the pion transport line with several  quadrupoles. Although conventional from the magnetic field point of view, these first two to four quads still need special and careful treatment in their design in order to maximize their lifetime in this high-radiation environment.   As described in Section~\ref{sec:HTS-Quads}, the FRIB quads show promise regarding  suitability for nuSTORM. 

The injection scenario for the decay ring, that will be discussed in section \ref{subsec:43}, needs a large dispersion value at the injection point. This dispersion $\mathrm{D_x}$ and its derivative $\mathrm{D_x'}$ both need to be matched to 0 at the horn, although this is not necessarily the case in the whole transport line. Considering this constraint, along with the requirement for spacing between target station and decay ring, two bending dipoles are used in the transport line, which bend the pions to the ring injection point. Two long drift spaces are reserved before the first dipole for collimators to reduce downstream radiation. The field strength, which is roughly 4.9 T in the first dipole, requires it to be superconducting. The long distance between the two dipoles separates the pions from the high energy residual protons, which go into a separate transport line and then into a MI-style proton absorber. 

\subsubsection{Proton Absorber}
\label{subsubsec:423}
\begin{figure}[h]
  \centering{
    \includegraphics[width=0.8\textwidth]{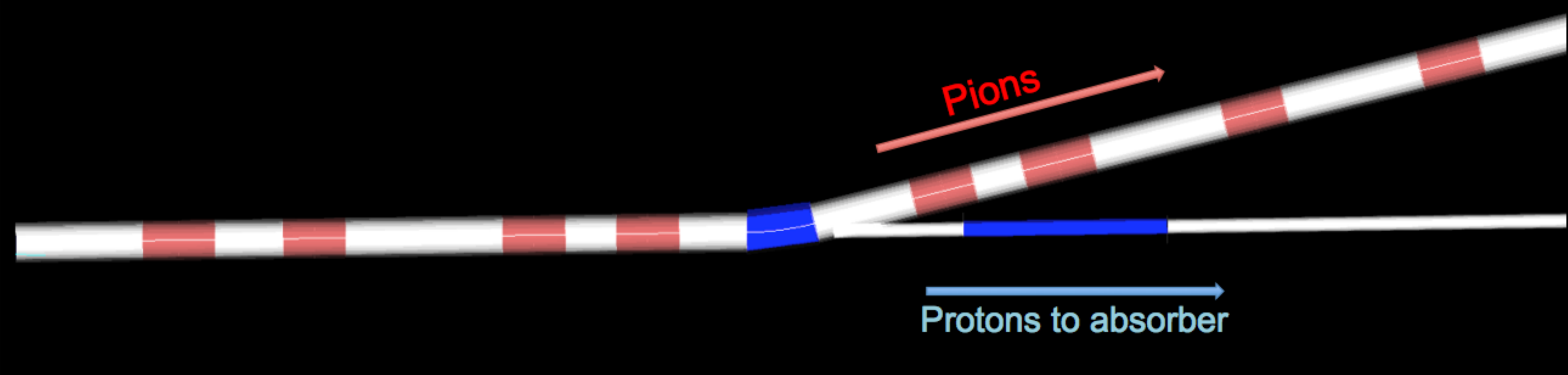}}
  \caption{Separation of proton absorber beamline and pion transport line.  Red: quadrupole, Blue: dipole, White: drift.}
  \label{fig:protonBend}
\end{figure}

The current design uses the same proton absorber size that the MI uses, which is 4.3 meters both in width and height, and 10.7 meters in length. Because the first dipole also bends the protons at the same time as the $\pi^+$ or $\pi^-$, another long dipole is needed in order to bend protons back to their initial direction. No other magnets are needed for the proton absorber beam line. The layout drawing is shown in Fig.~\ref{fig:protonBend}. The absorber, located at $\sim~$32 meters after the second dipole in Fig.~\ref{fig:protonBend} collects 43.5\% of the total energy from the residual protons remaining after the target. 
\subsection{Pion Injection}
\label{subsec:43}
The straight-section FODO cells were designed to have betatron functions $\beta_x, \beta_y$  (the Twiss parameters) optimized for beam acceptance and neutrino beam production.  Larger  betatron functions increase the beam size, following $\sigma = \sim \sqrt{\beta\epsilon_{rms}}$ and cause aperture losses.  On the other hand, smaller betatron functions increase the divergence of the muon beam and also the divergence of the resulting neutrino beams, following $\delta \theta = \sim \sqrt{ \epsilon_{rms}/ \beta}$.  In addition, the muon beam emittance is increased by the angular divergence from pion decay following $ \delta \epsilon \sim \beta_t {\theta_{decay}}^2/2$.
 
Balancing these criteria, we have chosen FODO cells with $\mathrm{\beta_{max}}$  $\sim$ 30.2 m, and $\mathrm{\beta_{min}}$  $\sim$ 23.3 m for the 3.8 GeV/c muons, which for the 5.0 GeV/c pions, implies $\sim$ 38.5 m and $\sim$ 31.6 m pion's $\mathrm{\beta_{max}}$ and $\mathrm{\beta_{min}}$, respectively.
 
\begin{figure}[h]
  \centering{
    \includegraphics[width=0.7\textwidth]{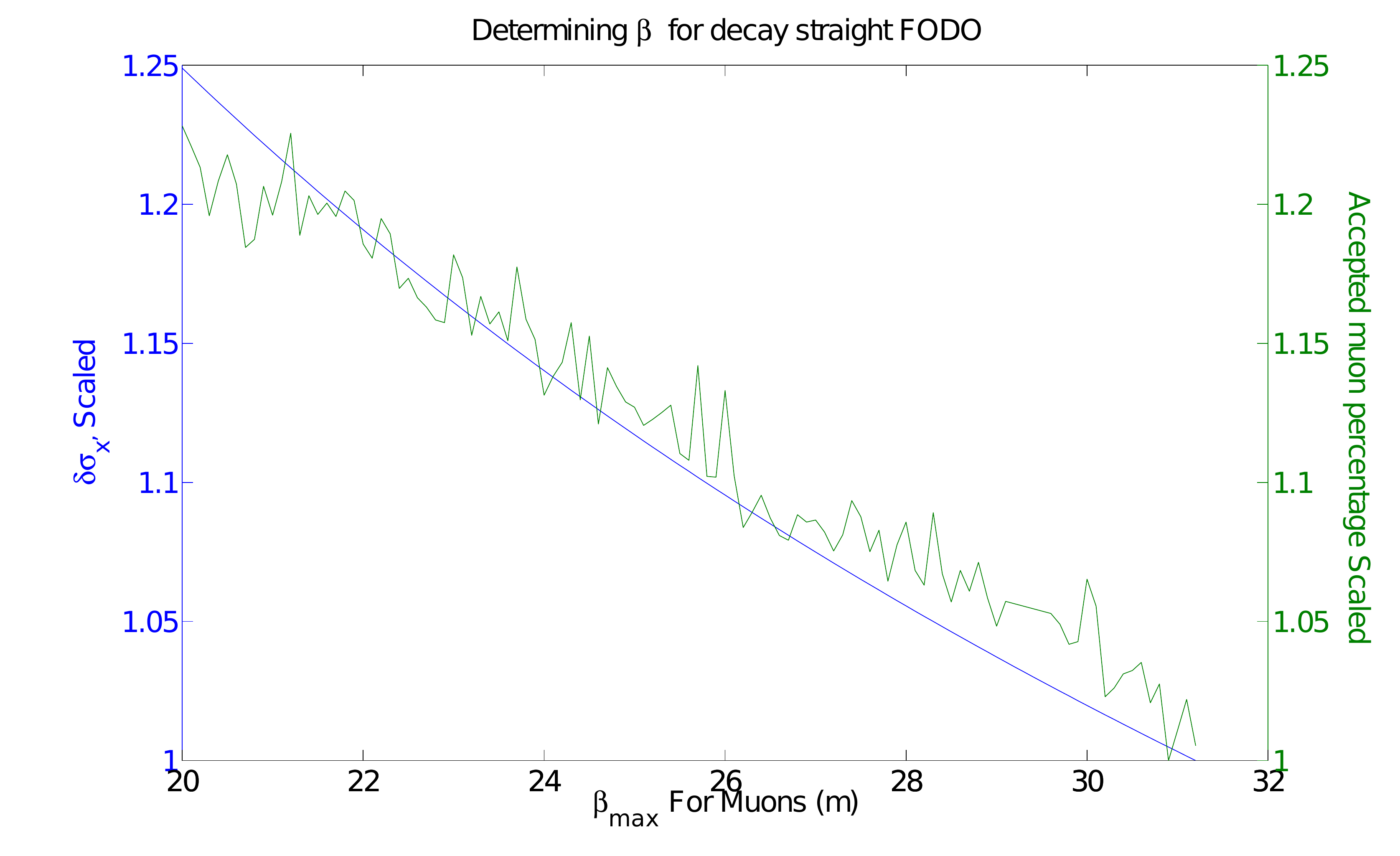}}
  \caption{Determining the Twiss for decay straight FODO cells}
  \label{fig:FODObeta}
\end{figure}

Fig.~\ref{fig:FODObeta} shows results of a set of simulations used in determining this lattice design, showing the relative increase in angular divergence obtained by reducing the FODO betatron functions.  

\begin{figure}[h]
  \centering{
    \includegraphics[width=0.8\textwidth]{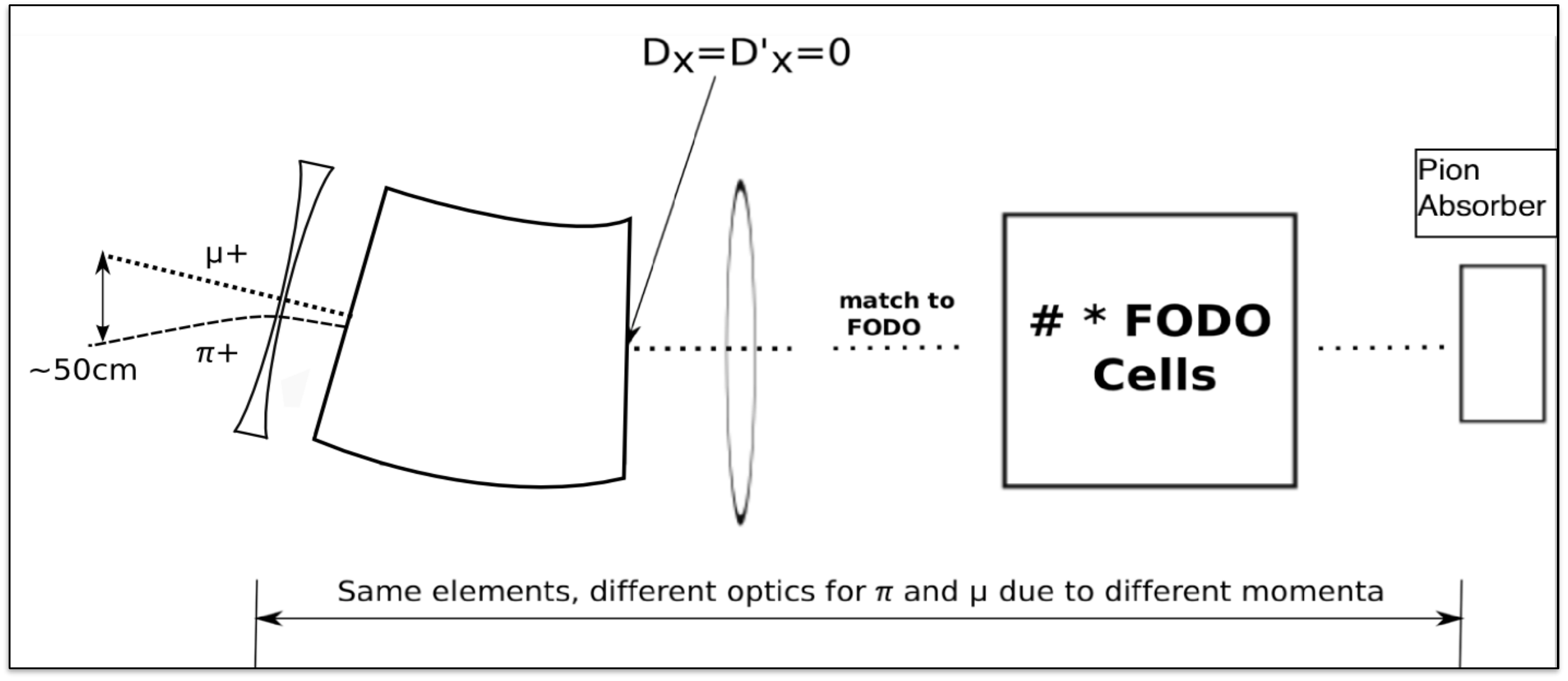}}
  \caption{The schematic drawing of the injection scenario}
  \label{fig:Sche_Stoch}
\end{figure}

\begin{figure}[h]
  \centering{
    \includegraphics[width=0.85\textwidth,height=0.45\textwidth]{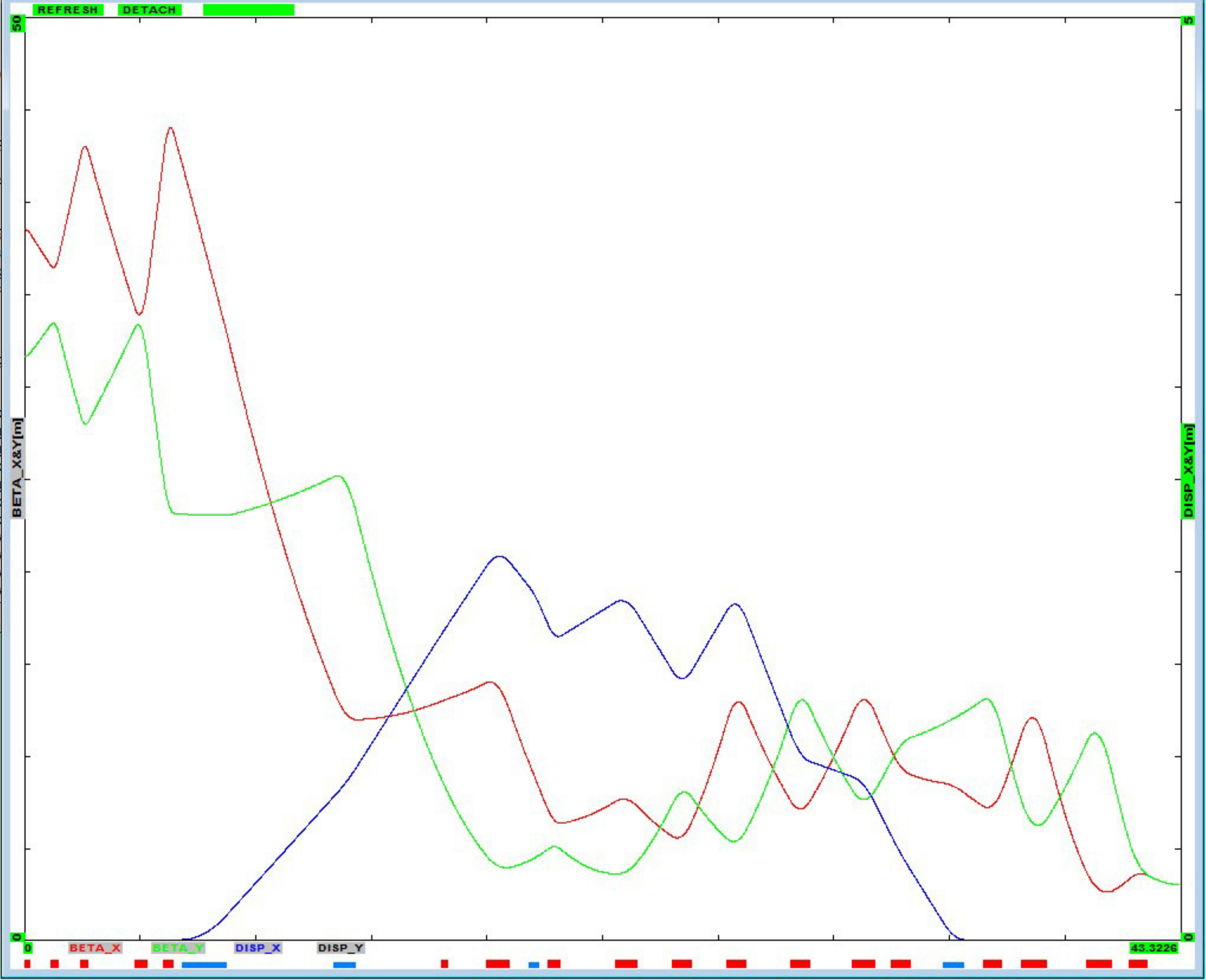}}
  \caption{The optics of $\pi$ transport and injection. The optics shown start in the decay straight and end at the downstream end of the horn (left to right).larger}
  \label{fig:optics}
\end{figure}

\begin{figure}[h]
  \centering{
    \includegraphics[width=1\textwidth]{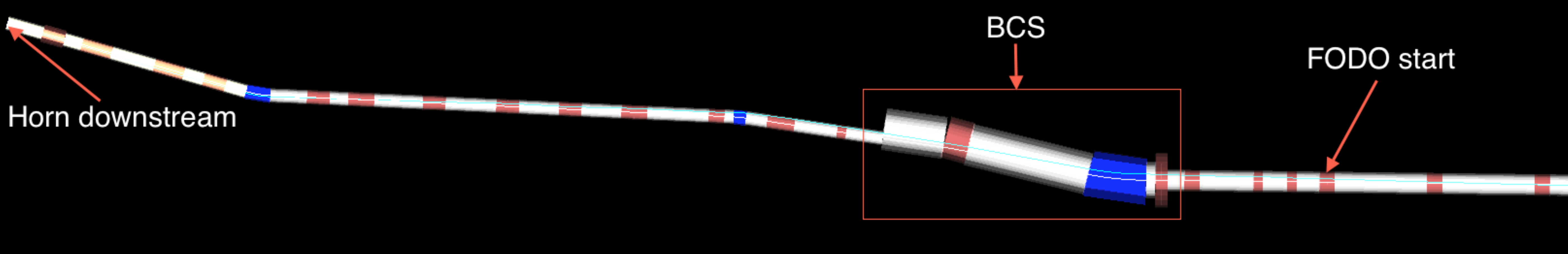}}
  \caption{The G4beamline drawing from horn downstream to the FODO cells.  Red: quadrupole, Blue: dipole, White: drift.}
  \label{fig:G4bl_layout}
\end{figure}

As discussed in section \ref{subsubsec:422}, a large dispersion $\mathrm{D_x}$ is required at the injection point, in order to achieve beam separation. A carefully designed ``beam combination section", or ``BCS", can readily reach this goal. The schematic drawing of the injection scenario is shown in Fig.~\ref{fig:Sche_Stoch}.
The pure sector dipole for muons in the BCS has an exit angle for pions that is non-perpendicular to the edge, and the pure defocusing quadrupole in the BCS  for muons is a combined-function dipole for the pions, with both entrance and exit angles non-perpendicular to the edges. The corresponding optics from OptiM are shown in Fig.~\ref{fig:optics}.  The BCS will be followed by a short matching section to the decay FODO cells (see Fig.~\ref{fig:G4bl_layout}).
At the other end of the decay straight, another BCS will be used to extract the pions which have not decayed. This BCS, which also extracts muons within the pion momentum range, will be discussed in Section~\ref{SubSect:6DICE}. 

\begin{figure}[h]
  \centering{
    \includegraphics[width=0.45\textwidth]{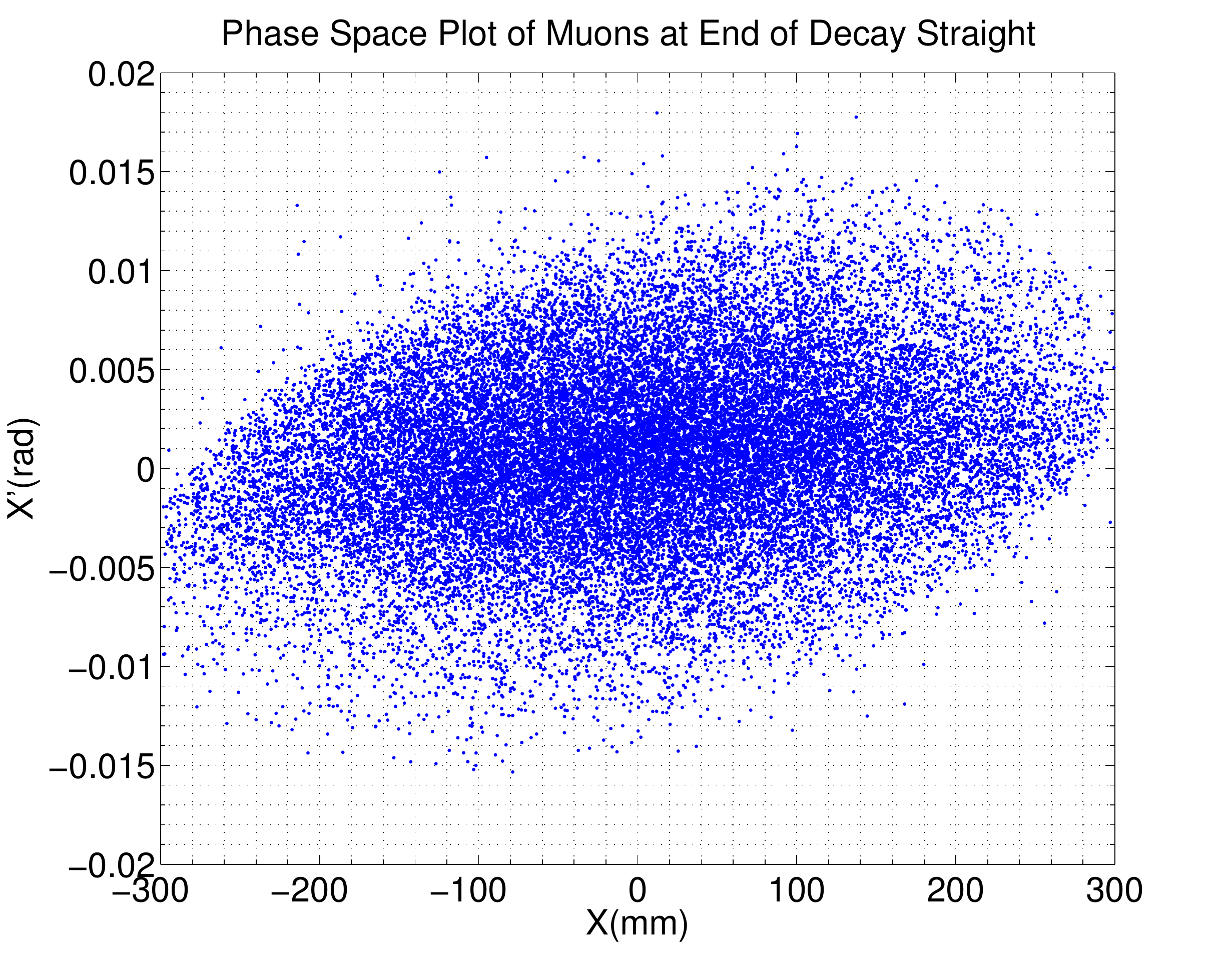}
    \includegraphics[width=0.45\textwidth]{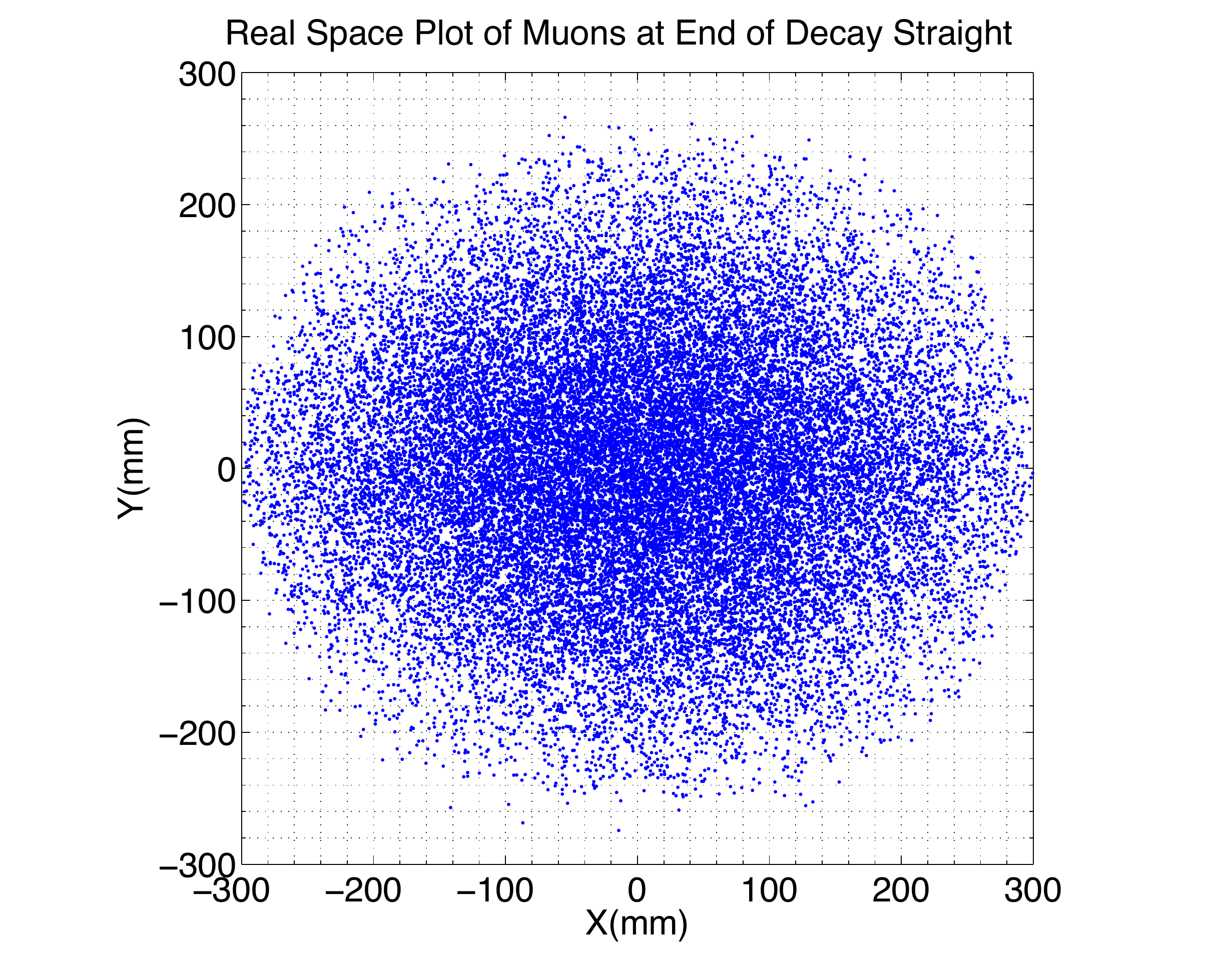}
    \includegraphics[width=1\textwidth]{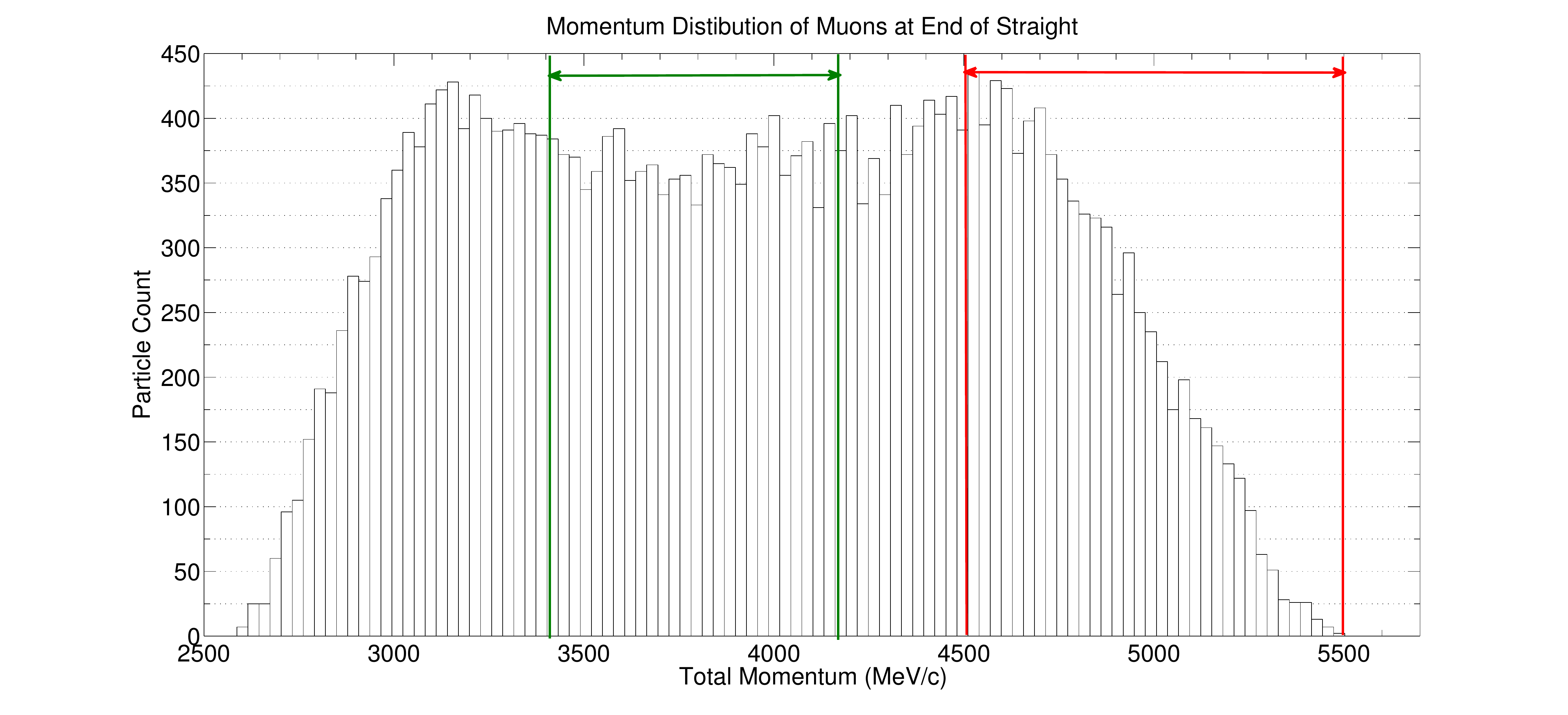}
    }
  \caption{The muon real space distribution (upper right), phase space distribution (upper left) and the muon momentum distribution (lower) at the end of decay straight.}
  \label{fig:muonMom}
\end{figure}

The performance of the injection scenario can be gauged by determining the number of muons at the end of the decay straight using G4beamline.
We were able to obtain 0.012 muons per POT (see Fig.~\ref{fig:muonMom}). These muons have a wide momentum range (beyond that which the ring can accept, 3.8 GeV/c $\pm$ 10\%) and thus will only be partly accepted by the ring. The green region in Fig.~\ref{fig:muonMom} shows the 3.8$\pm$10\% GeV/c acceptance of the ring, and the red region shows the high momentum muons which will be extracted by the BCS. This will be discussed in Section~\ref{SubSect:6DICE}. The muons also occupy a very large phase-space, which is also shown in Fig.~\ref{fig:muonMom}. 
\subsection{Decay ring}
\label{subsec:44}
We have investigated both a FODO racetrack and a FFAG racetrack for the muon decay.  The FODO ring that is described in detail below uses both normal and superconducting magnets.  A FODO lattice using only normal-conducting magnets (B $\lesssim$~1.8T) is also being studied.   The racetrack FFAG (RFFAG) is described in 
Section~\ref{subsec:RFFAG}.  Table~\ref{tab:rings} gives a comparison between the FODO and the RFFAG with regard to the ratio of the total number of useful muons stored per POT, assuming that capture off the target and injection into the ring are the same for both.  Acceptance for all the decay ring options we are considering will be studied and compared in order to obtain a cost/performance optimum but, for now, the FODO lattice is the nuSTORM baseline.
\begin{table}[h]
\centering
\caption {Relative $\mu$ yield for FODO vs. RFFAG rings}
\label{tab:rings}
\begin{tabular}{|l|l|l|}
\hline
Parameter  & FODO \hspace{5mm} & RFFAG \\
\hline\noalign{\smallskip}
L$_{straight}$ (m) & 185 & 240 \\
Circumference (m) & 480 & 606 \\
Dynamic aperture A$_{dyn}$  & 0.6 & 0.95   \\
Momentum acceptance   & $\pm~10\%$ & $\pm~16\%$ \\
$\pi$/POT within momentum acceptance \hspace{10mm} & 0.094 & 0.171 \\
Fraction of $\pi$ decaying in straight (F$_s$) & 0.52 & 0.57 \\
Ratio of L$_{straight}$ to ring circumference ($\Omega$) & .39 & .40  \\
Relative factor (A$_{dyn} \times \pi$/POT $\times$ F$_s \times \Omega$) & 0.011 & 0.037 \\
\noalign{\smallskip}\hline
\end{tabular}
\end{table} 
\subsubsection{FODO ring Lattice Design}
\label{subsubsec:441}
\begin{figure}[h]
  \centering{
    \includegraphics[width=.95\textwidth]{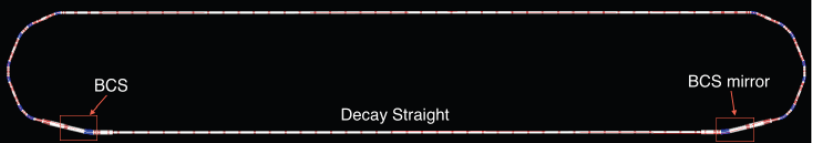}}
    \caption{Racetrack ring layout. Pions are injected into the ring at the Beam Combination Section (BCS). Similarly, extraction of pions and muons at the end of the production straight is done using a mirror image of the BCS.}
  \label{fig:ringlayout}
\end{figure}

A FODO ring with such a large phase space and momentum acceptance has not been previously developed~\cite{paper:AoiPAC13}. Here we propose a compact racetrack ring design (480 m in circumference) based on large aperture, separate function magnets (dipoles and quadrupoles). The ring is configured with FODO cells combined with DBA (Double Bend Achromat) optics. The ring layout, including pion injection/extraction points, is illustrated in Fig.~\ref{fig:ringlayout} and the current ring design parameters are given in Table \ref{tab:Dring}.
\begin{table}[h]
\centering
\caption {Decay ring specifications}
\label{tab:Dring}
\begin{tabular}{|l|r|l|}
\hline
Parameter &  Specification &   Unit\\
\hline\noalign{\smallskip}
Central momentum P$_\mu$  & 3.8 & GeV/c \\
Momentum acceptance & $\pm$ 10\%  &   \\
Circumference   & 480 & m  \\
Straight length   & 185 & m  \\
Arc length  & 50 & m \\
Arc cell &  DBA &  \\
Ring Tunes ($\nu_x, \nu_y$) &  9.72, 7.87 & \\
Number of dipoles & 16 & \\
Number of quadrupoles & 128  & \\
Number of sextupoles & 12 &   \\
\hline
\noalign{\smallskip}
\end{tabular}
\end{table}
The design goal for the ring was to maximize both the transverse and momentum acceptance (around the 3.8 GeV/c central momentum), while maintaining acceptable physical apertures of the magnets. These requirements would drive the lattice design towards strongly focusing (large transverse acceptance) and low chromaticity (large momentum acceptance) optics in the arcs. Furthermore, one side of the arc needs to accommodate pion injection/extraction sections. The stochastic injection, as described in Section~\ref{subsec:43}, drives the dispersion value to about 3 m, which puts a serious limitation on the transverse acceptance. The large dispersion at the injection point must be suppressed in the arc. To accommodate this, we have used Double-Bend Achromat (DBA) optics in the arcs, which controls the beam size. 

\begin{figure}[h]
  \centering{
    \includegraphics[width=1\textwidth]{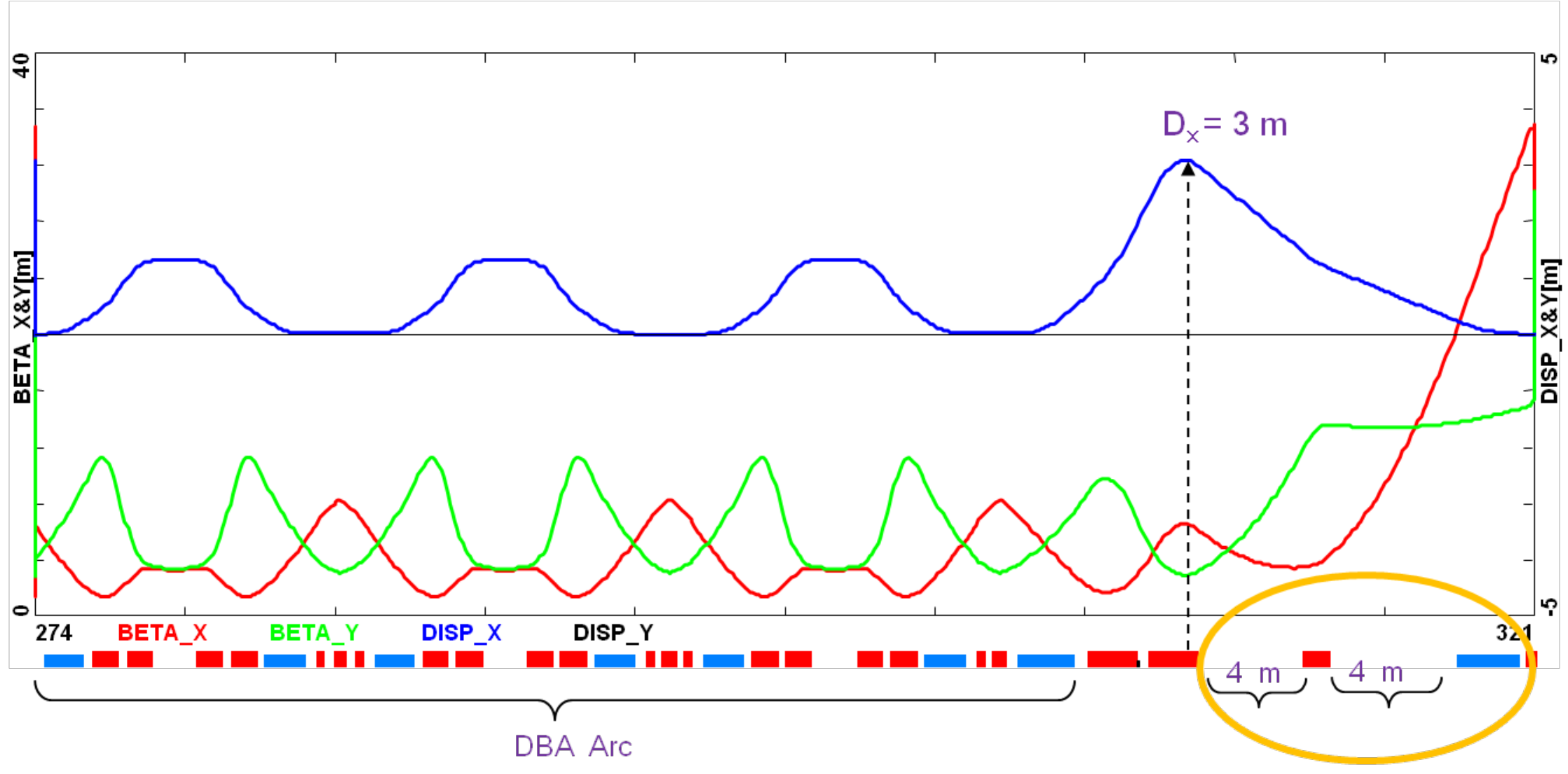}
    }
  \caption{The arc optics illustration}
  \label{fig:arc}
\end{figure}

To maintain the compactness of the arc, while accommodating adequate drift space between magnets, we limit the overall arc length to about 50 m,  keeping the dipole fields at $\simeq$ 4 Tesla. We use DBA optics, which maintains reasonably small values of the beta functions and dispersion. We limit the maximum field at the quadrupole magnet pole tip to be less than 5 Tesla. The overall arc optics are illustrated in Fig.~\ref{fig:arc}. 
The decay straight requires much larger values of the beta functions ($\simeq$ 27 m average) in order to assure small beam divergence ($\simeq$ 7 mrad). The resulting muon beam divergence is a factor of 4 smaller than the characteristic decay cone of 1/$\gamma$ (~0.029 at 3.8 GeV). The decay straight is configured with a much weaker focusing FODO lattice ($\simeq$ 15 deg. phase advance per cell). It uses normal conducting, 30 cm radius aperture quads with a modest gradient of 2 Tesla/m (0.6 Tesla at the pole tip). 

The opposite straight, which is not used for neutrino production, can be designed with much smaller beta functions. This straight also uses normal conducting quads, but with a gradient of 11 Tesla/m (1.6 Tesla at the pole tip).

\begin{figure}[h]
  \centering{
    \includegraphics[width=1\textwidth]{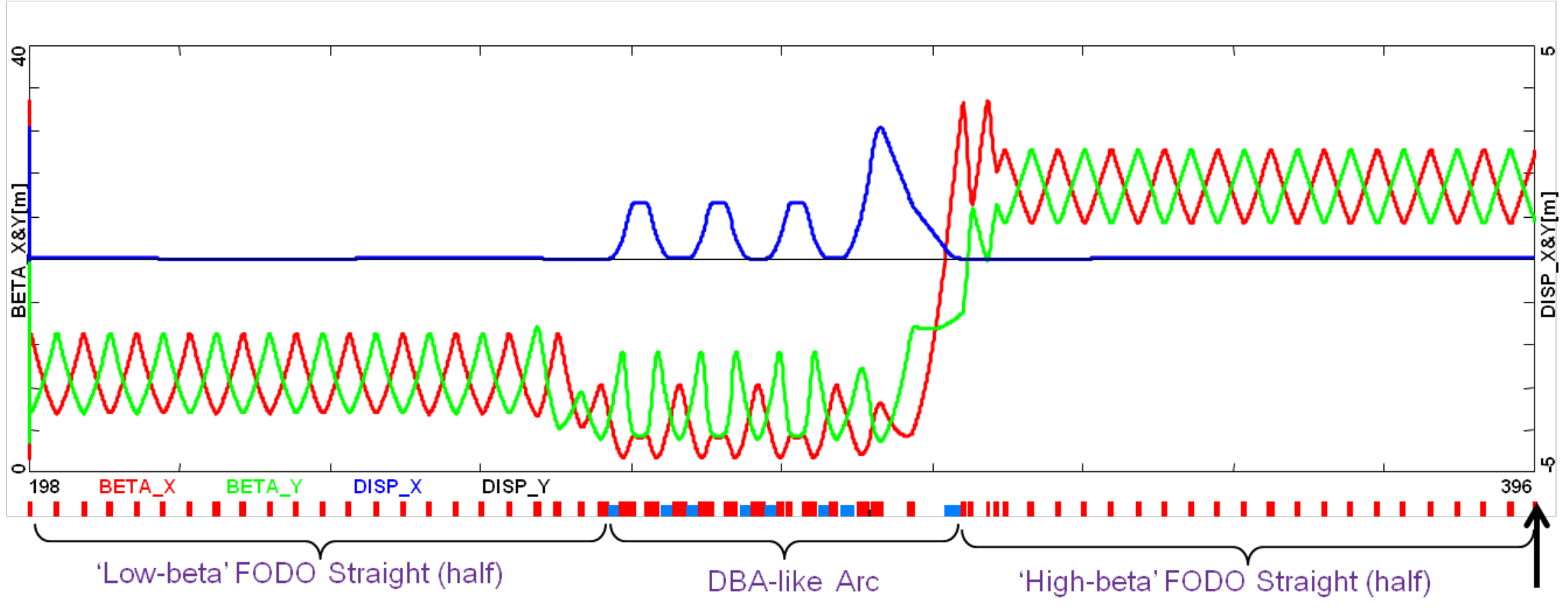}
    }
  \caption{Ring optics.}
  \label{fig:straight}
\end{figure}

\begin{figure}[h]
  \centering{
    \includegraphics[width=1\textwidth]{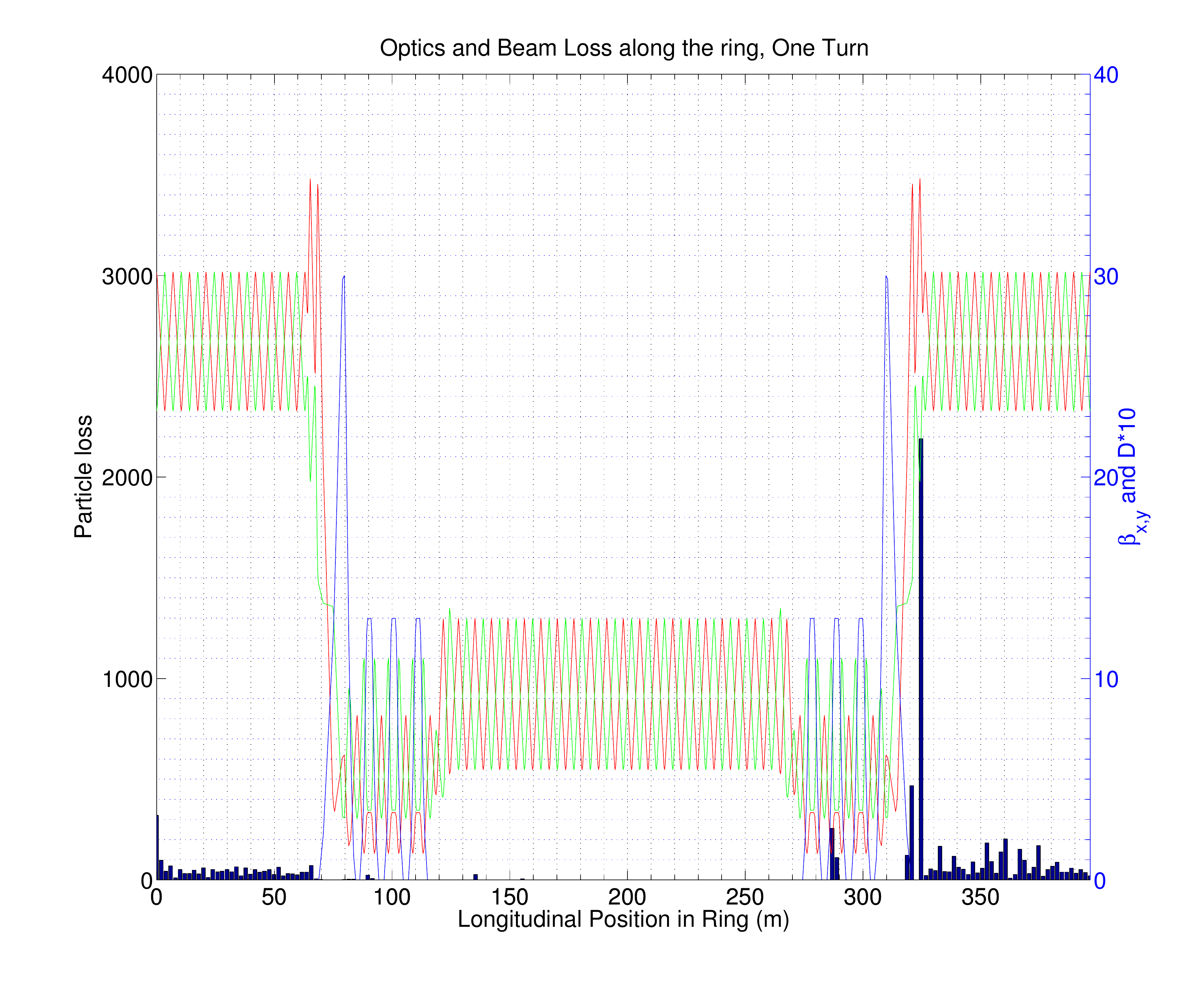}
    }
  \caption{The complete ring optics and single-turn beam loss histogram.}
  \label{fig:complete}
\end{figure}

Finally, the racetrack ring optics are illustrated in Fig.~\ref{fig:straight}.  It features a low-beta straight matched to a 180 deg. arc and is then followed by a high-beta decay straight connected to the other arc with a compact telescope insert. The complete ring optics and the single turn beam loss histogram are shown in 
Fig.~\ref{fig:complete}.

\begin{figure}[h]
  \centering{
    \includegraphics[width=.7\textwidth]{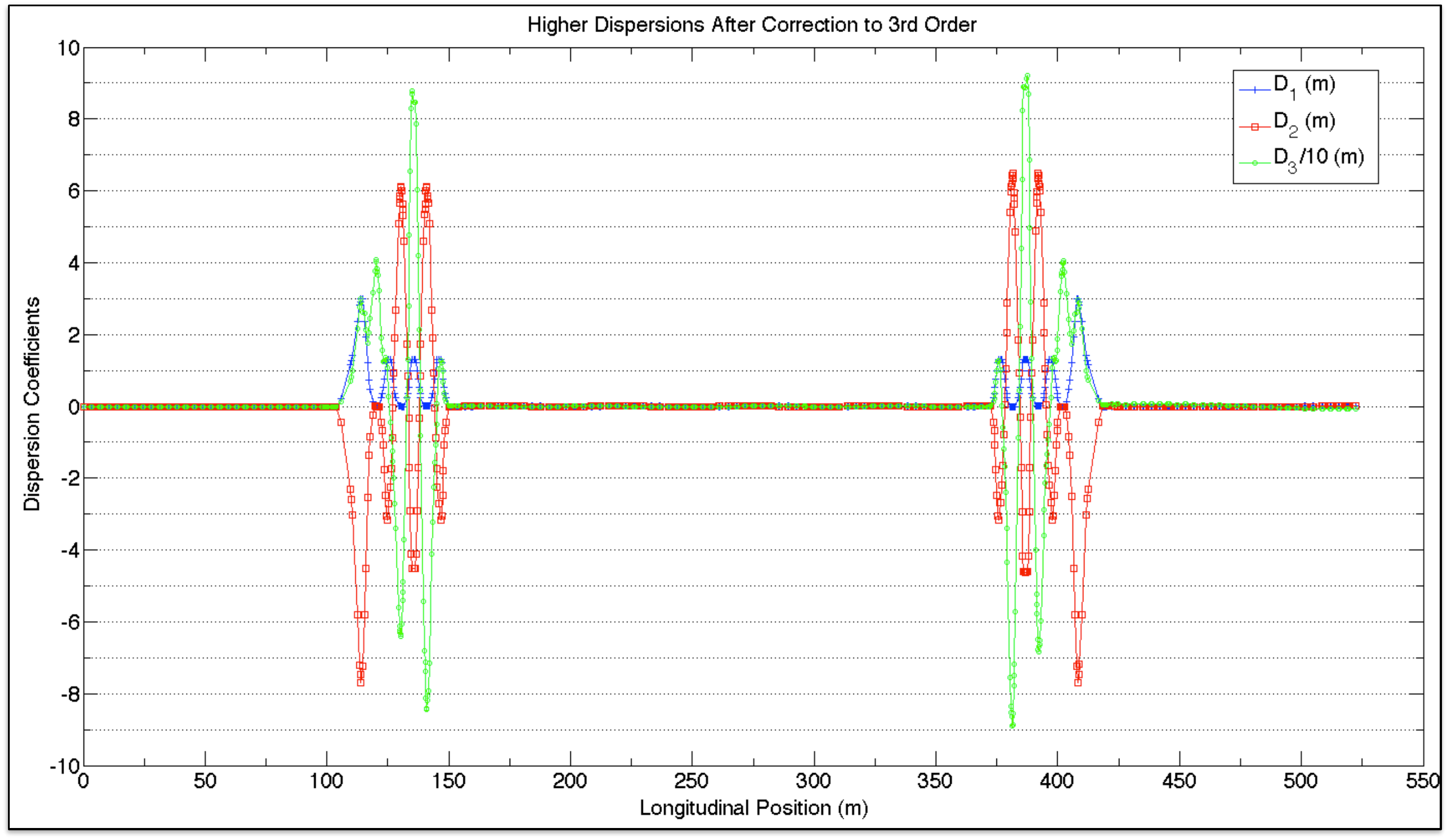}\\
    \includegraphics[width=.7\textwidth]{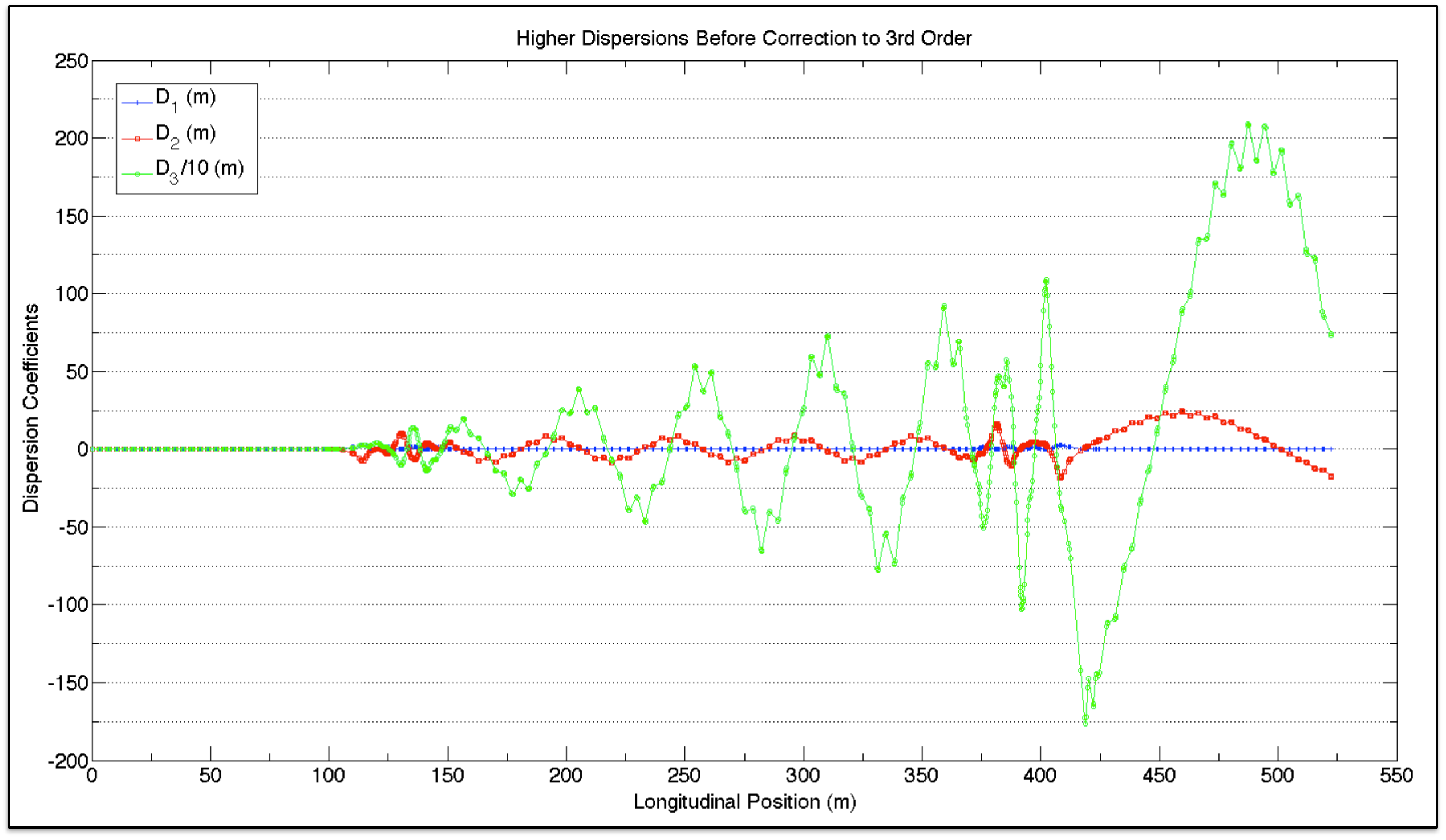}
    }
  \caption{The higher order dispersions (up to 3rd order) before(lower) and after(upper) correction}
  \label{fig:highDisp}
\end{figure}

\begin{figure}[h]
  \centering{r
    \includegraphics[width=.8\textwidth]{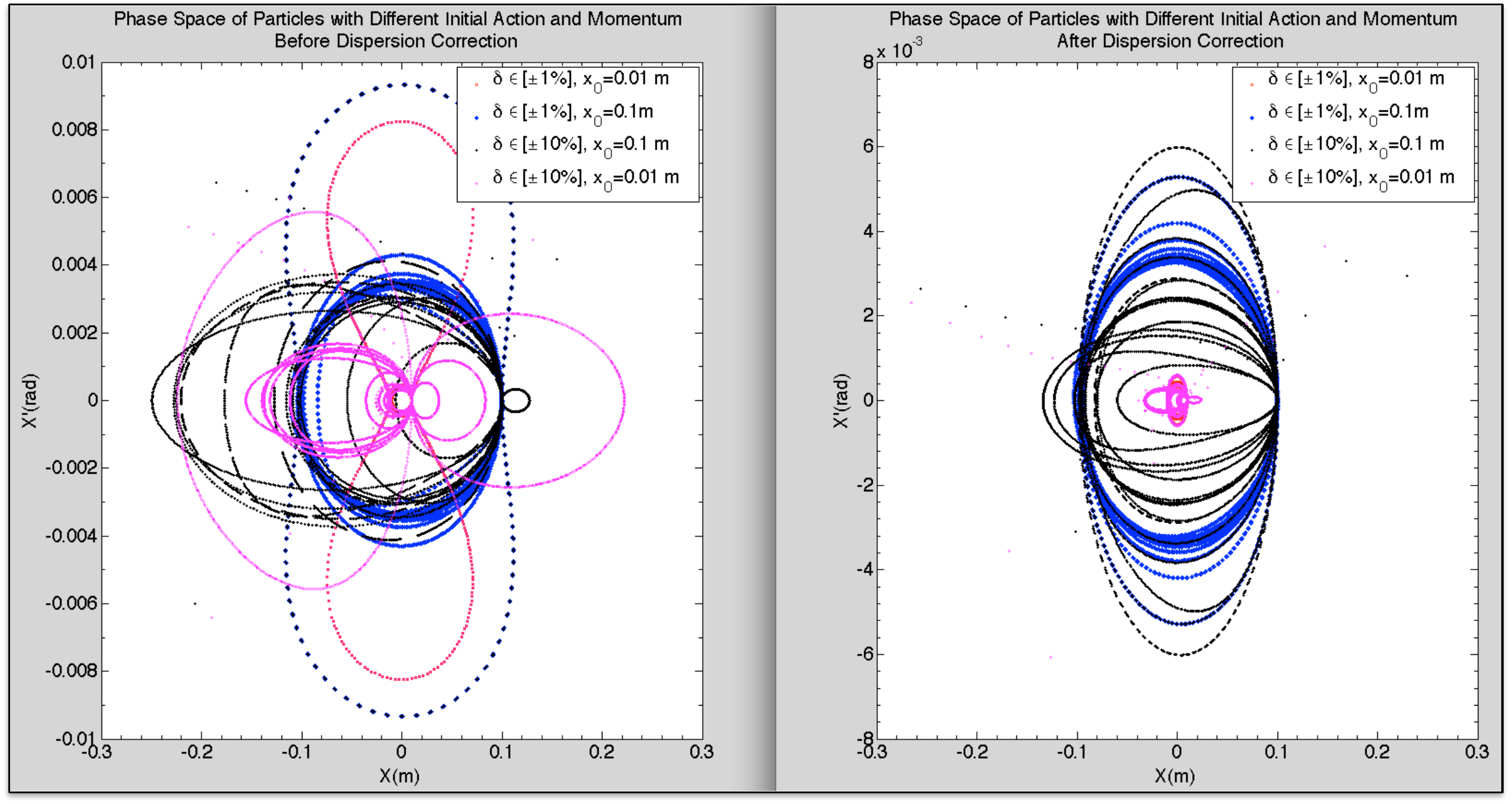}
    }
  \caption{The 100-turn tracking for off-momentum particles before(left) and after(right) correction.}
  \label{fig:orbitResp}
\end{figure}

It is very likely that the large beam loss where the beam enters the decay ring arc is caused by beta chromaticity, or ``beta beat" raised by momentum difference.  It is observed in simulations that, with sextupole and octupole correction, the orbit response of off-momentum particles can be well corrected (See Fig.~\ref{fig:highDisp} and Fig.~\ref{fig:orbitResp}). Using particle tracking in G4beamline, we are able to achieve approximately 60\% beam survival rate, for a Gaussian distributed muon beam, after 100 turns in the ring (without higher-order correctors). This number can be increased when further detailed corrections of tune chromaticity and beta chromaticity are developed.
\clearpage
\subsubsection{Advanced scaling FFAG}
\label{subsec:RFFAG}
\begin{figure}[h!]
	\begin{center}
		\includegraphics[width=\textwidth]{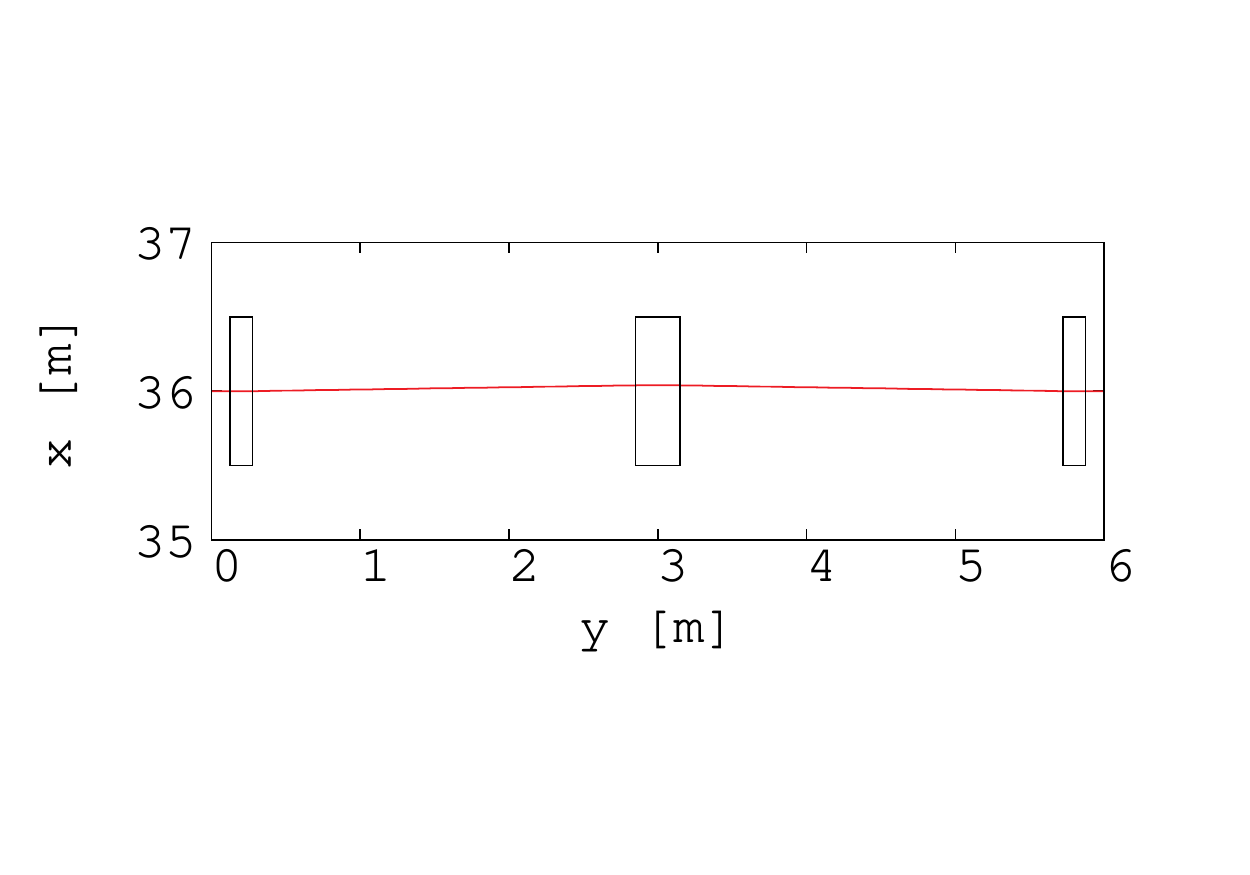}
		\caption{Top view of the straight scaling FFAG cell. The 3.8 GeV/c muon reference trajectory is shown in red. 
		Effective field boundaries with collimators are shown in black.}
		\label{str-traj}
	\end{center}
\end{figure}

\begin{figure}[h!]
   \begin{minipage}[b]{.48\linewidth}
       \includegraphics[width=8.cm]{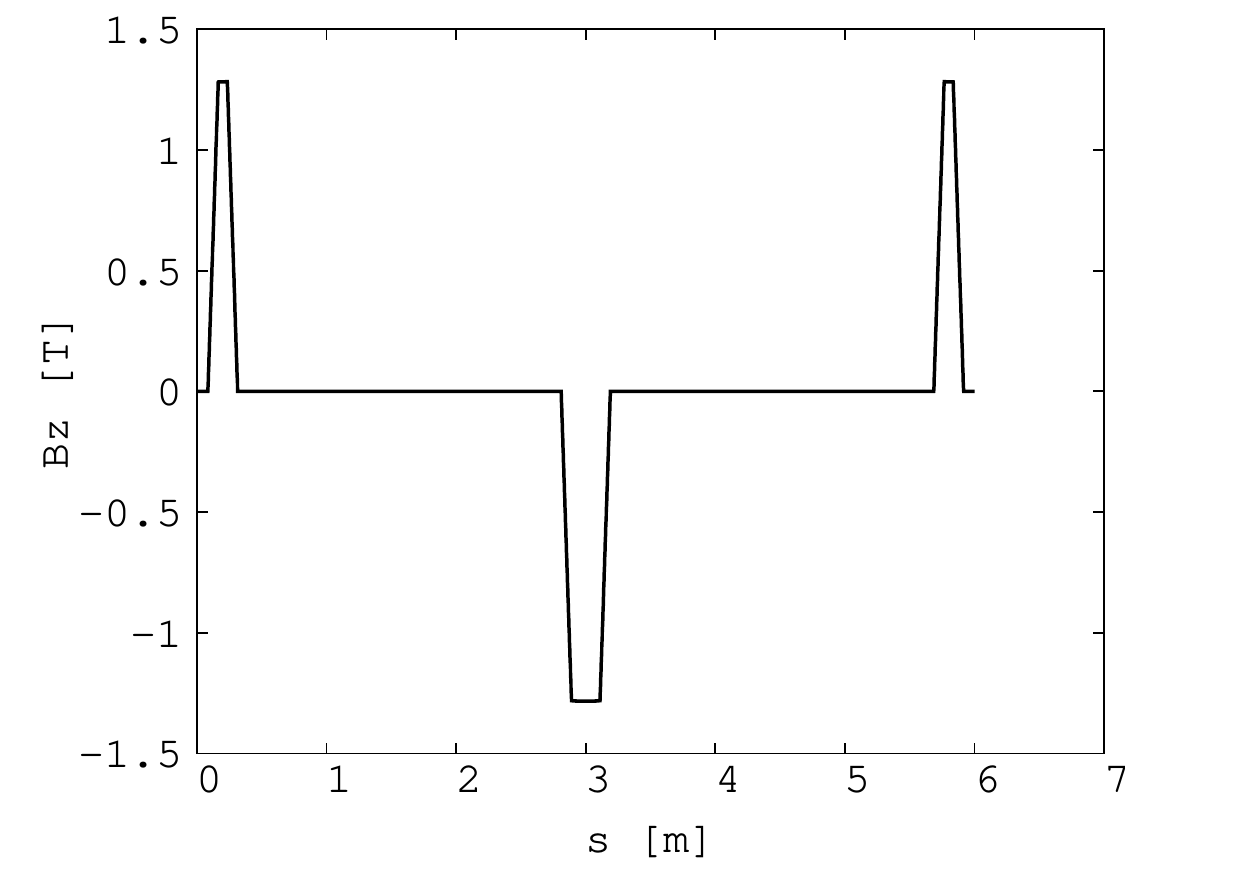}
	\caption{Vertical magnetic field for 3.8 GeV/c muon reference trajectory in the straight scaling FFAG cell.}
	\label{str-bz}
	 \end{minipage} \hfill
   \begin{minipage}[b]{.48\linewidth}
    	\includegraphics[width=8cm]{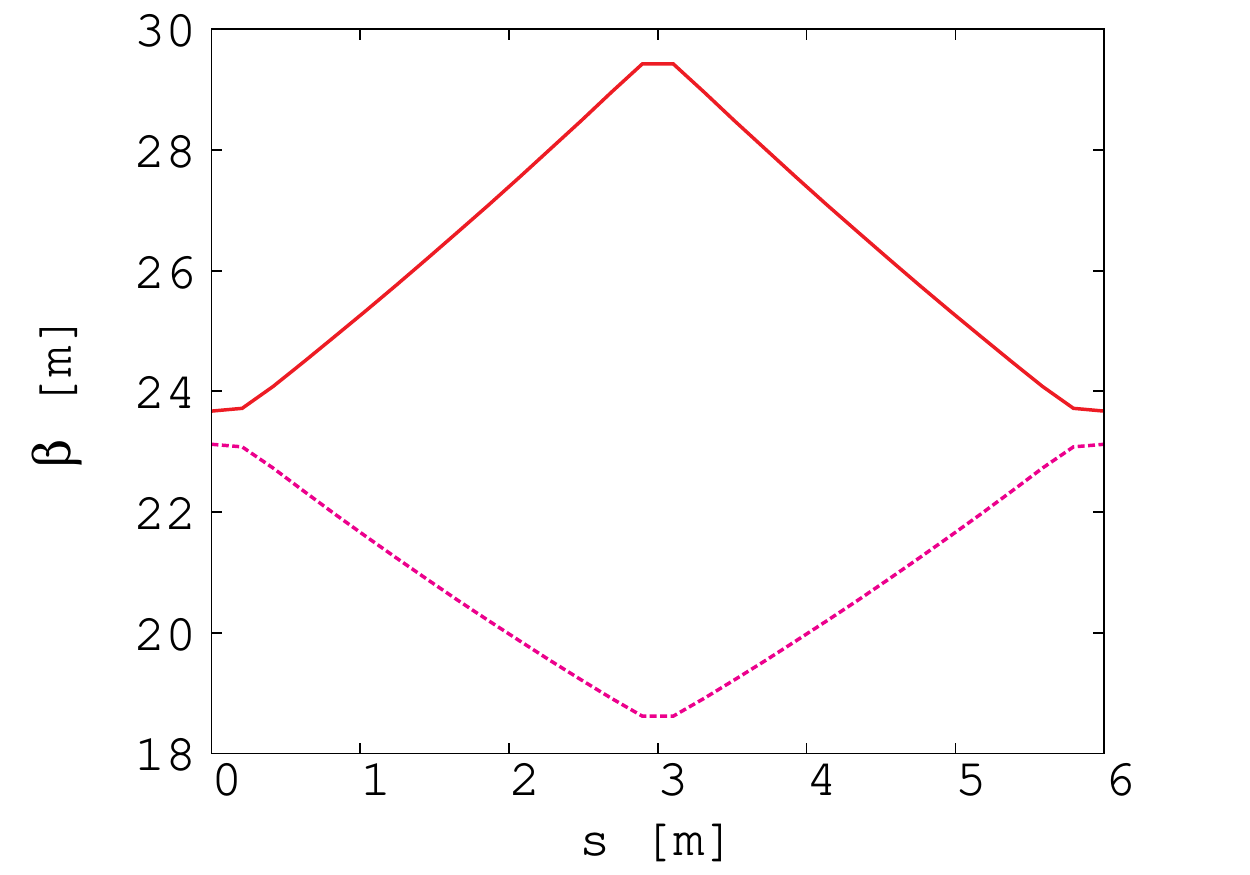}
	\caption{Horizontal (plain red) and vertical (dotted purple) periodic $\beta$ functions of the straight scaling FFAG cell.}
	\label{str-beta}
   \end{minipage}
\end{figure}

\begin{figure}[h!]
	\begin{center}
		\includegraphics[width=10cm]{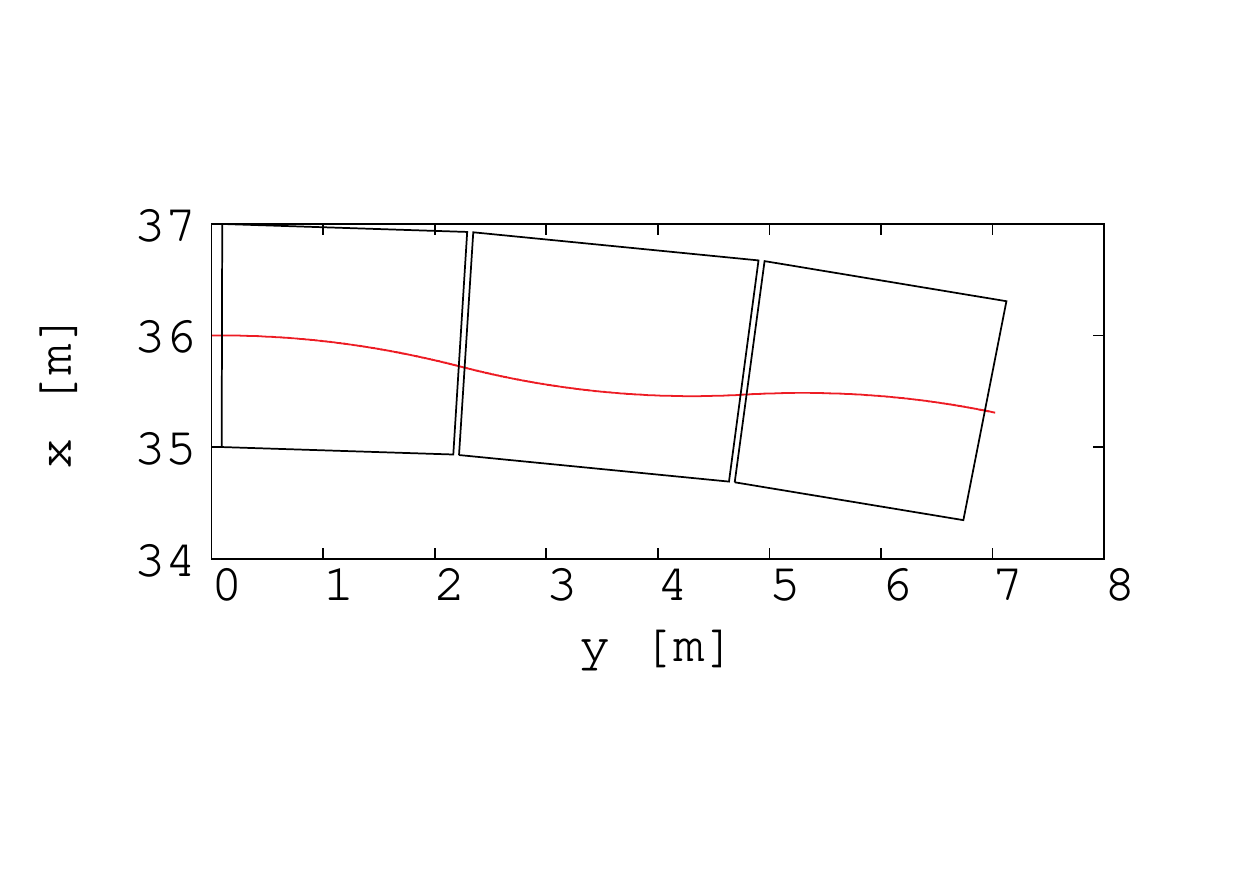}
		\caption{Top view of the circular scaling FFAG cell. The 3.8~GeV/c muon reference trajectory is shown in red. 
		Effective field boundaries with collimators are shown in black.}
		\label{circ-traj}
	\end{center}
\end{figure}

\begin{figure}[h!]
   \begin{minipage}[b]{.49\linewidth}
       \includegraphics[width=8.cm]{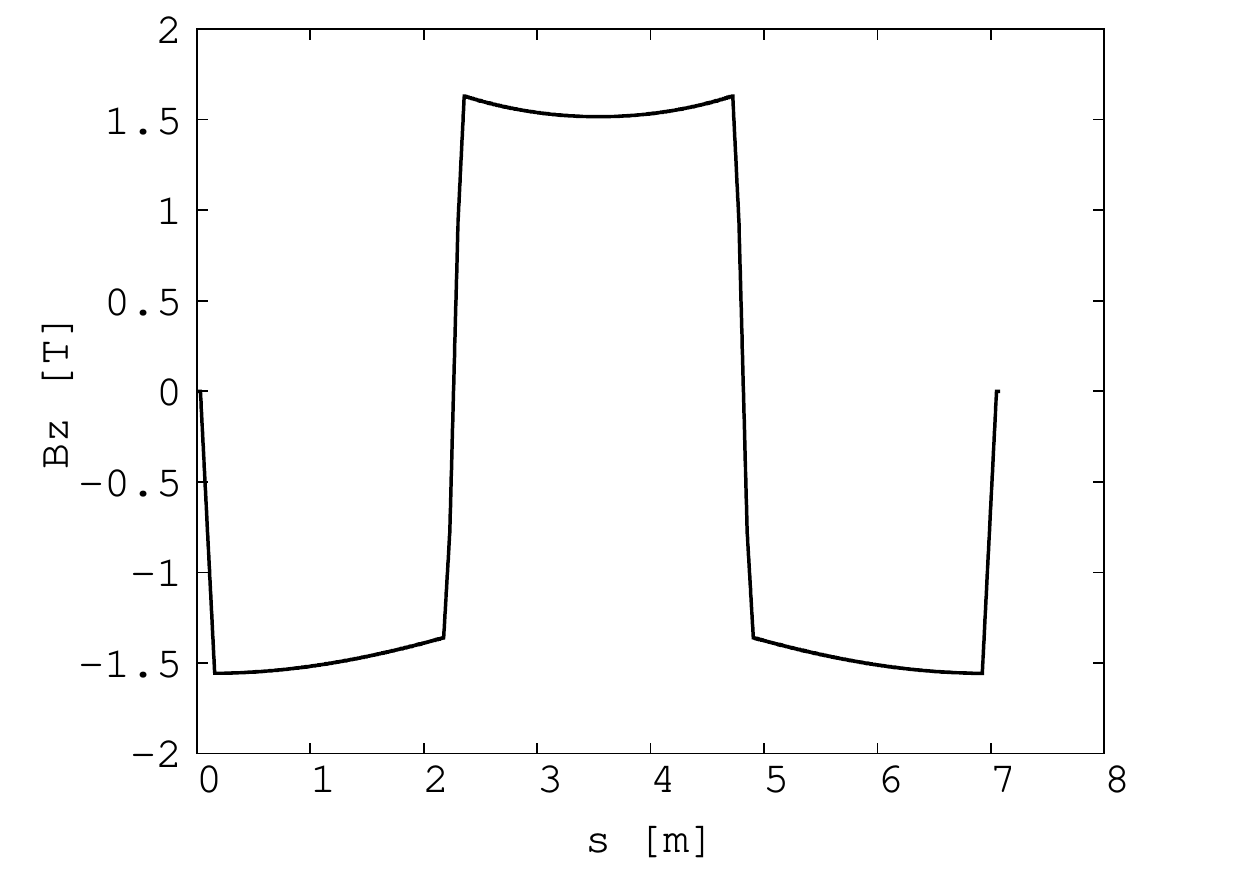}
	\caption{Vertical magnetic field for the 3.8~GeV/c muon reference trajectory in the circular scaling FFAG cell.}
	\label{circ-bz}
	 \end{minipage} \hfill
   \begin{minipage}[b]{.49\linewidth}
    	\includegraphics[width=8cm]{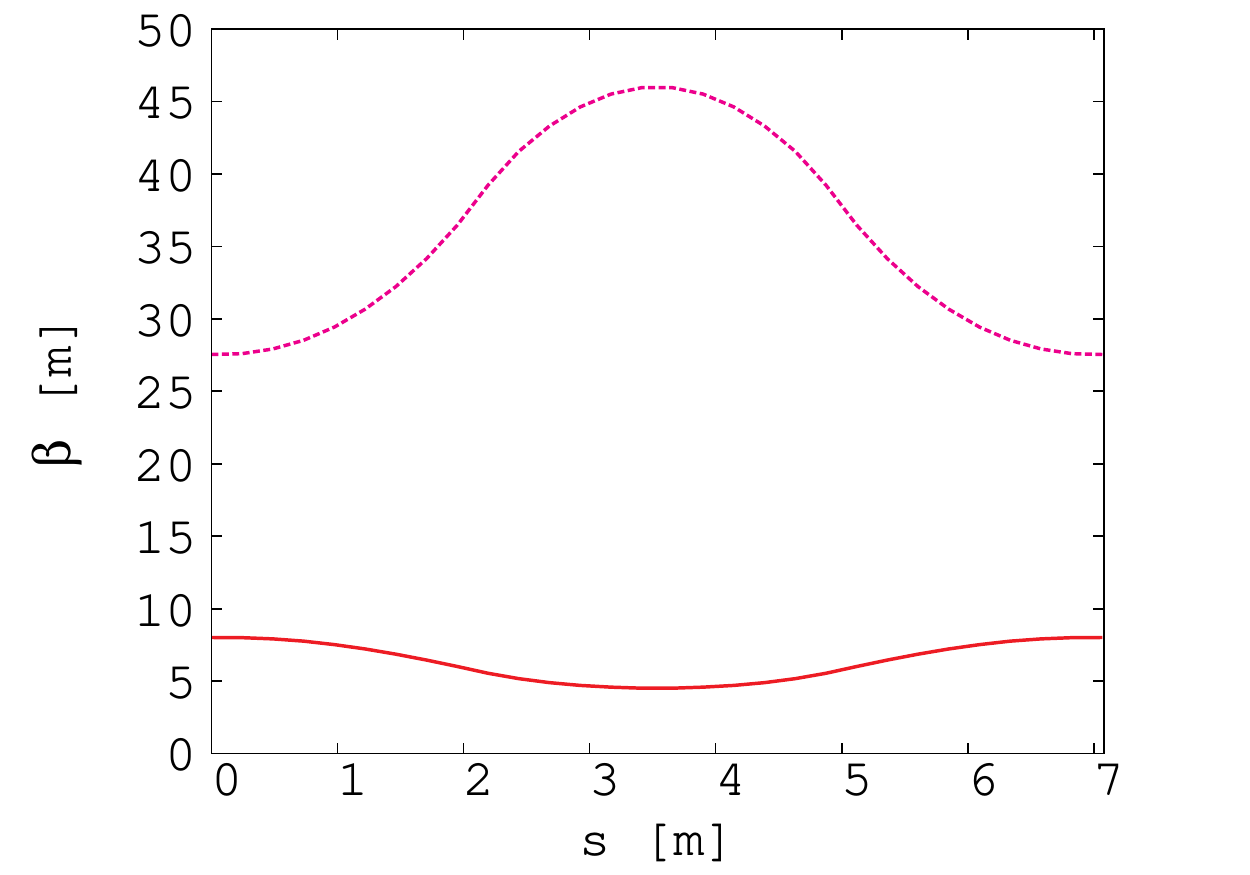}
	\caption{Horizontal (plain red) and vertical (dotted purple) periodic $\beta$ functions of the circular scaling FFAG cell.}
	\label{circ-beta}
   \end{minipage}
\end{figure}

The racetrack FFAG ring is composed of two cell types: a) a straight scaling FFAG cell and b) a
circular scaling FFAG cell.  There are 40~straight FFAG cells in each long straight section 
(80 for the  whole ring) and 16~circular FFAG cells in each of the arc sections.
\paragraph{Straight scaling FFAG cell parameters}
In the straight scaling FFAG cell, the vertical magnetic field $B_{sz}$ in the median plane follows:
\begin{displaymath}
B_{sz}=B_{0sz} e^{m(x-x_0)}  \mathcal{F},
\end{displaymath}
with $x$ the horizontal Cartesian coordinate, $m$ the normalized field gradient, $ \mathcal{F}$ an 
arbitrary function and $B_{0sz}=B_{sz}(x_0)$.
The parameters of the straight scaling FFAG cell are summarized in Table~\ref{tab-straight}.
\begin{table}[!h]
    \centering
     \caption{Parameters of the straight scaling FFAG cell.}
\begin{tabular}{|lcc|}
\hline
Cell type			&		& DFD triplet \\[-1.5mm]
Number of cells in the ring	 &	& 80 \\[-1.5mm]
Cell length	&	& 6~m\\[-1.5mm]
 $x_0$ &&	36~m\\[-1.5mm]
m-value      &         & 2.65\,m$^{-1}$         \\[-1.5mm]
         Packing factor & & 0.1 \\[-1mm]
         Collimators ($x_{min}, x_{max}, z_{max}$) && (35.5~m, 36.5~m, 0.3~m)\\[-1.5mm]
         Periodic cell dispersion  && 0.38~m\\[-1.5mm]
          Horizontal phase advance  && 13.1\,deg. \\[-1.5mm]
Vertical phase advance  && 16.7\,deg. \\[-1.mm]
\hline
D$_1$ magnet parameters &&\\[-1.5mm]
& Magnet center & 0.2~m\\[-1.5mm]
& Magnet length & 0.15~m\\[-1.5mm]
& Fringe field fall off & Linear (Length: 0.04~m)\\[-1.5mm]
& $B_0(x_0=36~m)$ & 1.28067~T\\[-1.mm]
\hline
F magnet parameters &&\\[-1.5mm]
& Magnet center & 3~m\\[-1.5mm]
& Magnet length & 0.3~m\\[-1.5mm]
& Fringe field fall off & Linear (Length: 0.04~m)\\[-1.5mm]
& $B_0(x_0=36~m)$ & -1.15037~T\\[-1.mm]
\hline
D$_2$ magnet parameters &&\\[-1.5mm]
& Magnet center & 5.8~m\\[-1.5mm]
& Magnet length & 0.15~m\\[-1.5mm]
& Fringe field fall off & Linear (Length: 0.04~m)\\[-1.5mm]
& $B_0(x_0=36~m)$ & 1.28067~T\\[-1.mm]
\hline
\end{tabular} 

 \label{tab-straight}
\end{table}
The cell is shown in Fig.~\ref{str-traj}. The red line represents the $\simeq$ 3.8 GeV/c muon reference trajectory, 
and its corresponding magnetic field is shown in Fig.~\ref{str-bz}. Periodic $\beta$ functions are shown in 
Fig.~\ref{str-beta}.
%
\paragraph{Circular scaling FFAG cell parameters}
In the circular scaling FFAG cell, the vertical magnetic field $B_{cz}$ in the median plane follows
\begin{displaymath}
B_{cz}=B_{0cz} \left( \frac{r}{r_0}\right)^k \mathcal{F},
\end{displaymath}
with $r$ the radius in polar coordinates, $k$ the geometrical field index, $ \mathcal{F}$ an 
arbitrary function and $B_{0cz}=B_{cz}(r_0)$.
The parameters of the circular scaling FFAG cell are summarized in Table~\ref{tab-circ}.
\begin{table}[h!]
    \centering
     \caption{Parameters of the circular scaling FFAG cell.}
\begin{tabular}{|lcc|}
\hline
Cell type			&		& FDF triplet \\[-1.5mm]
Number of cells in the ring	 &	& 32 \\[-1.5mm]
Cell opening angle	&	& 11.25~deg\\[-1.5mm]
 $r_0$ &&	36~m\\[-1.5mm]
k-value      &         & 10.85        \\[-1.5mm]
         Packing factor & & 0.96 \\[-1.5mm]
         Collimators ($r_{min}, r_{max}, z_{max}$) && (35~m, 37~m,  0.3~m)\\[-1.5mm]
         Periodic cell dispersion  && 1.39~m (at~ 3.8~GeV/c)\\[-1.5mm]
          Horizontal phase advance  && 67.5\,deg. \\[-1.5mm]
Vertical phase advance  && 11.25\,deg. \\[-1.mm]
\hline
F$_1$ magnet parameters &&\\[-1.5mm]
& Magnet center & 1.85~deg\\[-1.5mm]
& Magnet length & 3.4~deg\\[-1.5mm]
& Fringe field fall off & Linear (Length: 0.1~deg)\\[-1.5mm]
& $B_0(r_0=36~m)$ & -1.55684~T\\[-1.mm]
\hline
D magnet parameters &&\\[-1.5mm]
& Magnet center & 5.625~deg\\[-1.5mm]
& Magnet length & 4.0~deg\\[-1.5mm]
& Fringe field fall off & Linear (Length: 0.1~deg)\\[-1.5mm]
& $B_0(r_0=36~m)$ & 1.91025~T\\[-1.mm]
\hline
F$_2$ magnet parameters &&\\[-1.5mm]
& Magnet center & 9.4~deg\\[-1.5mm]
& Magnet length & 3.4~deg\\[-1.5mm]
& Fringe field fall off & Linear (Length: 0.1~deg)\\[-1.5mm]
& $B_0(r_0=36~m)$ & -1.55684~T\\[-1.mm]
\hline
\end{tabular} 
 \label{tab-circ}
\end{table}
The cell is shown in Fig.~\ref{circ-traj}. The red line represents the 3.8~GeV/c muon reference trajectory, and its 
corresponding magnetic field is shown in Fig.~\ref{circ-bz}. Periodic $\beta$ functions are shown in Fig.~\ref{circ-beta}.
\paragraph{Single particle tracking}
Stepwise tracking using Runge Kutta integration in a field model with linear fringe fields has been performed where 
interpolation of the magnetic field away from the mid-plane has been done to first order. Only single particle tracking has 
been done so far. We used $\mu^+$ with a central momentum, $p_0$, of 3.8~GeV/c, a minimum momentum, 
$p_{min}$, of 3.14~GeV/c and a maximum momentum, $p_{max}$, of 4.41~GeV/c.  $\Delta p/p_0$ is thus $\pm 16\%$.
The tracking step size was 1~mm. The exit boundary of a cell is the entrance boundary of the next cell.

\begin{figure}[h!]
	\begin{center}
		\includegraphics[width=10cm]{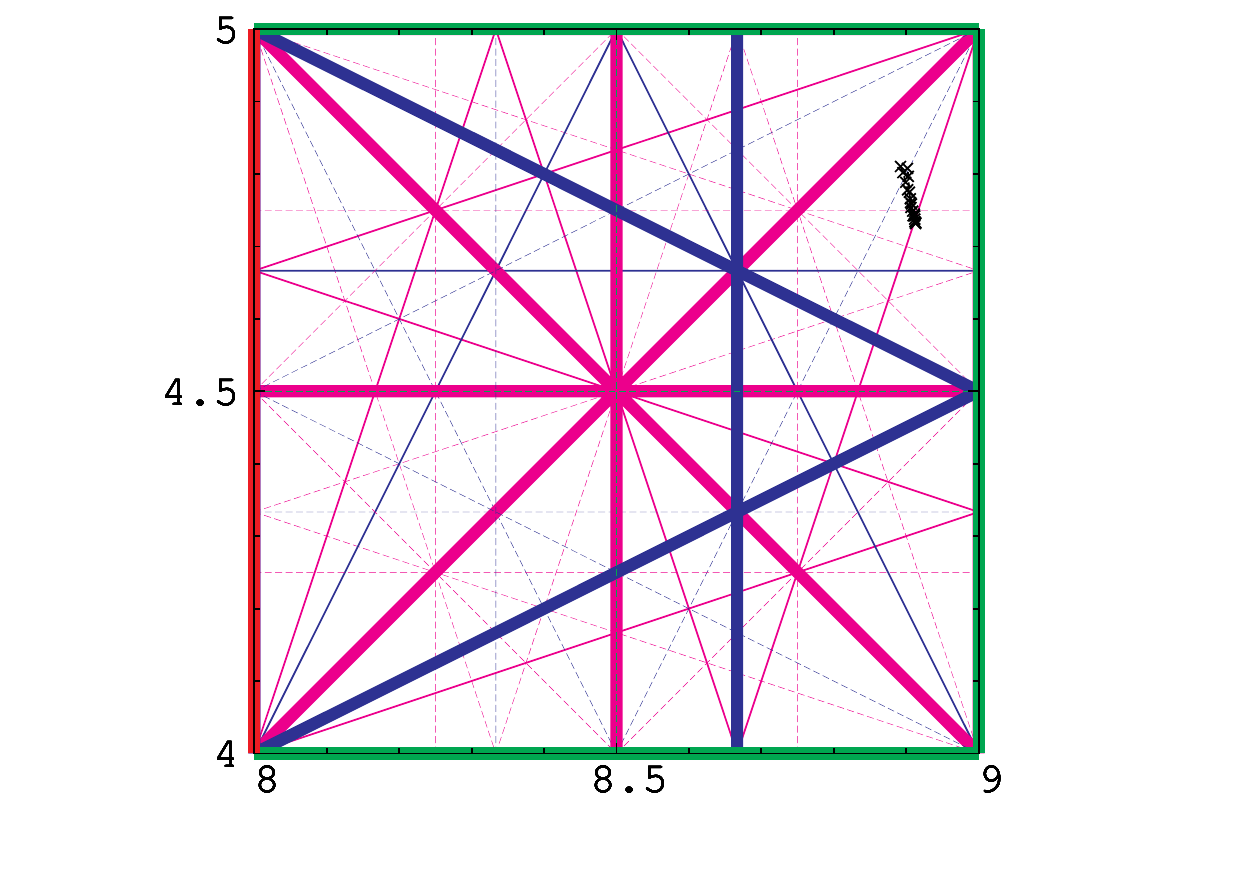}
		\caption{Tune diagram for muons from $p_{min}$ to $p_{max}$ ($\pm16\%$ in momentum around 3.8~GeV/c). Integer (red), 
		half-integer (green), third integer (blue) and fourth integer (purple) normal resonances are plotted. Structural 
		resonances are in bold.}
		\label{tunediag}
	\end{center}
\end{figure}

The ring tune point is (8.91,4.72) at the central momentum, $p_0$. Stability of the ring tune has been studied over the momentum range. The 
tune shift is presented in Fig.~\ref{tunediag}. The tune point stays within a 0.1 shift in both planes over this momentum range.
 
\begin{figure}[h!]
	\begin{center}
		\includegraphics[width=15cm]{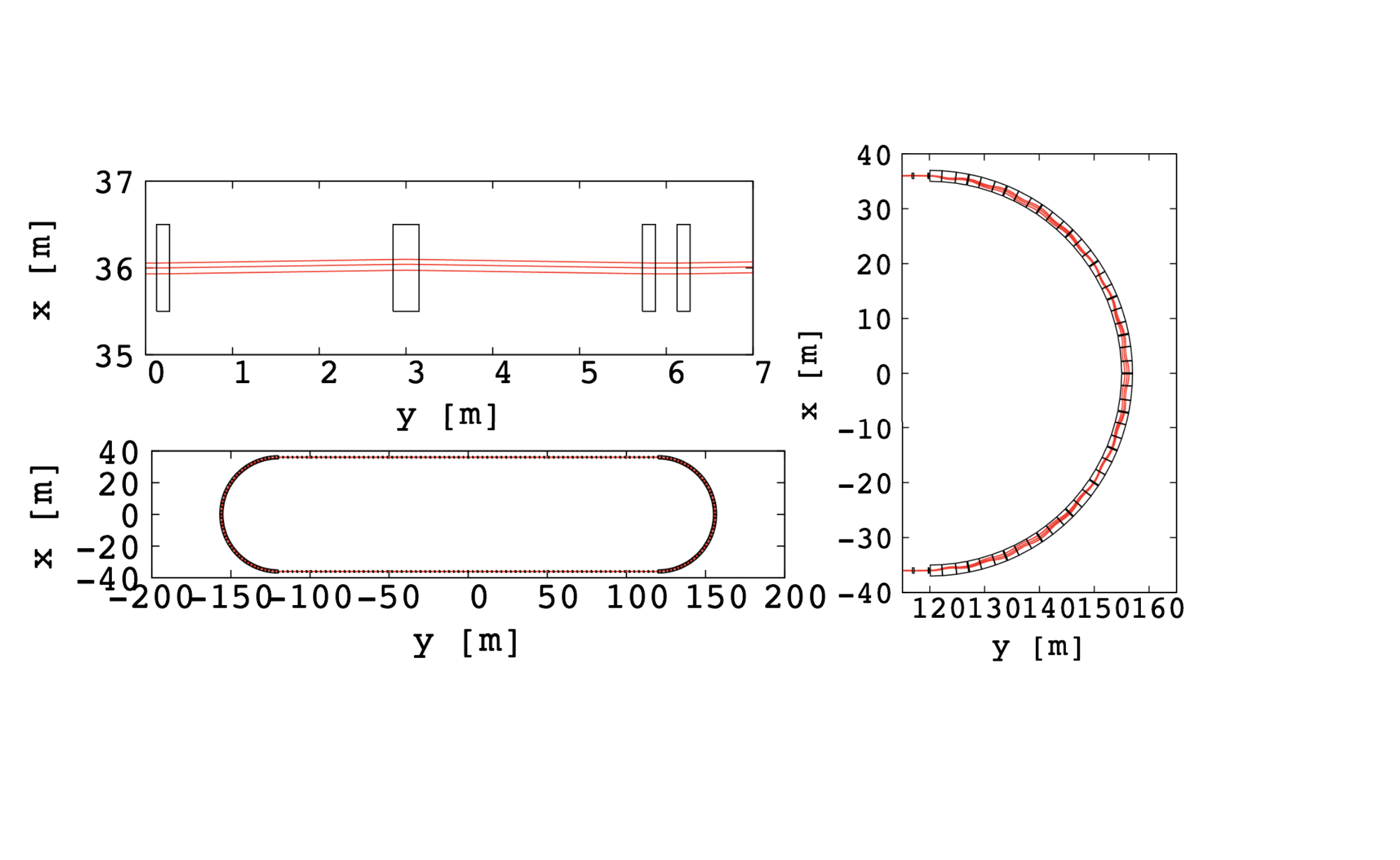}
		\caption{Top view of the racetrack FFAG lattice (bottom left scheme). The top left shows a zoom 
		of the straight section and on the right we show a zoom of the arc section. $p_0$, $p_{min}$, and $p_{max}$ 
		muon closed orbits are shown in red. Effective field boundaries with collimators are shown in black.}
		\label{single-traj}
	\end{center}
\end{figure}

\begin{figure}[h!]
	\begin{center}
		\includegraphics[width=10cm]{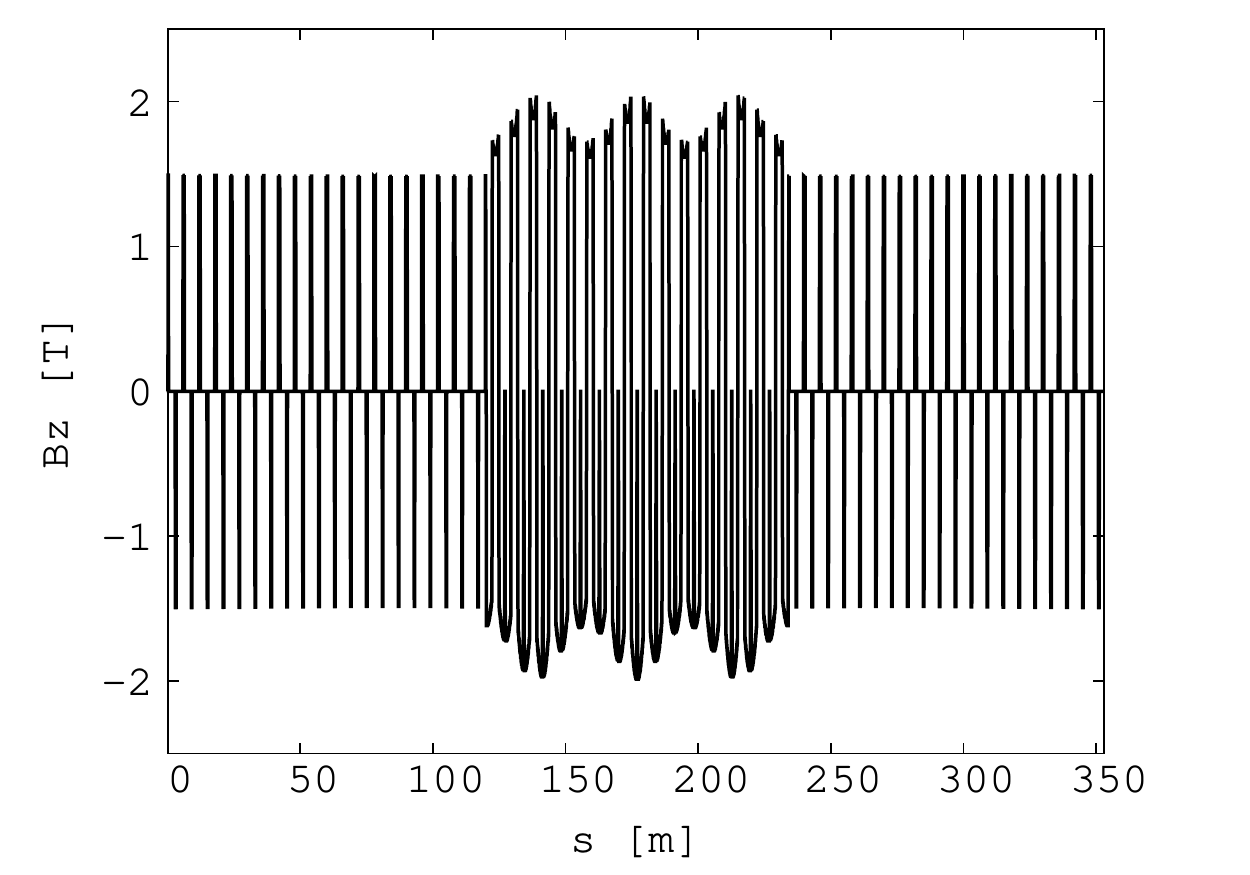}
		\caption{Vertical magnetic field for $p_{max}$ muon closed orbit in the racetrack FFAG ring.}
		\label{bz-single}
	\end{center}
\end{figure}

\begin{figure}[h!]
	\begin{center}
		\includegraphics[width=10cm]{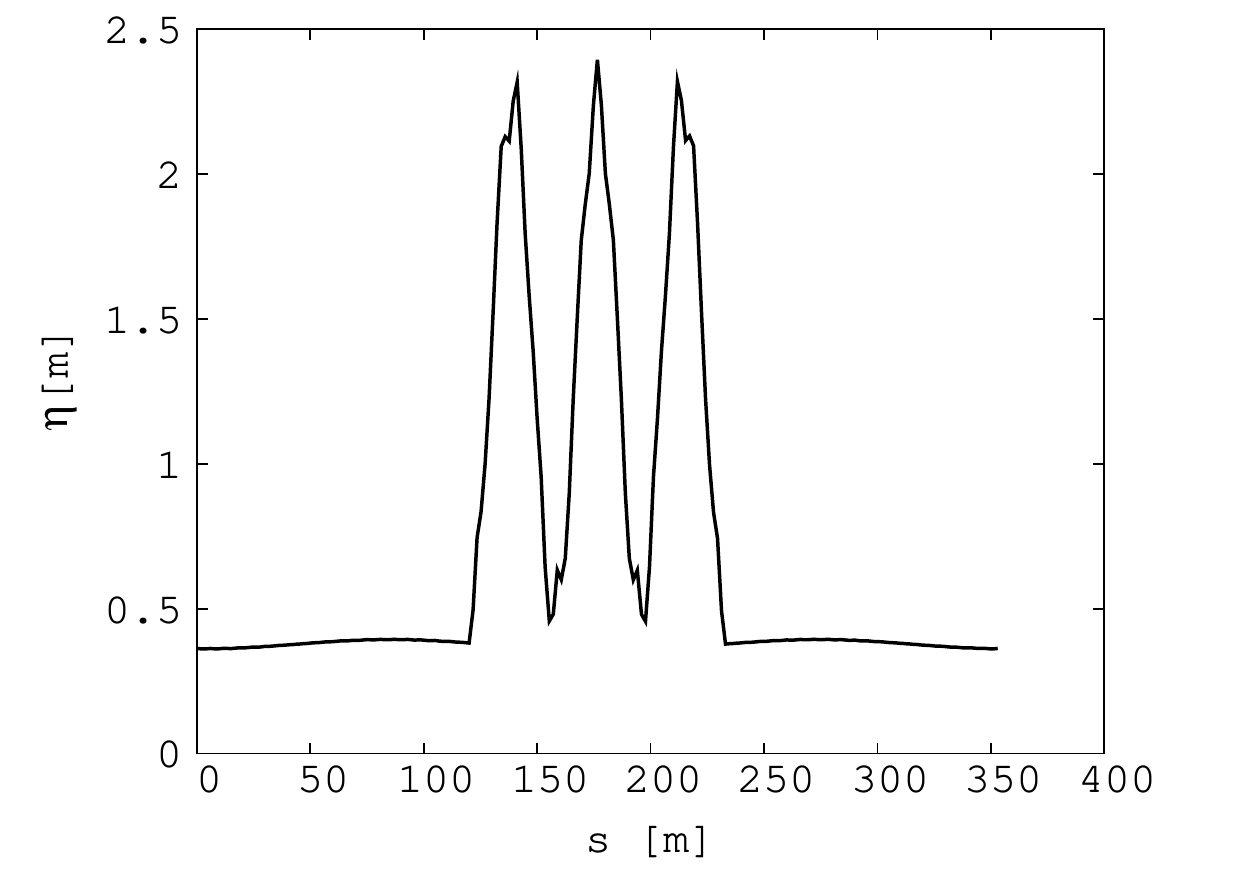}
		\caption{Dispersion function for $p_0$ in half of the ring. The plot is centered on the arc part.}
		\label{disp-single}
	\end{center}
\end{figure}

\begin{figure}[h!]
	\begin{center}
		\includegraphics[width=12cm]{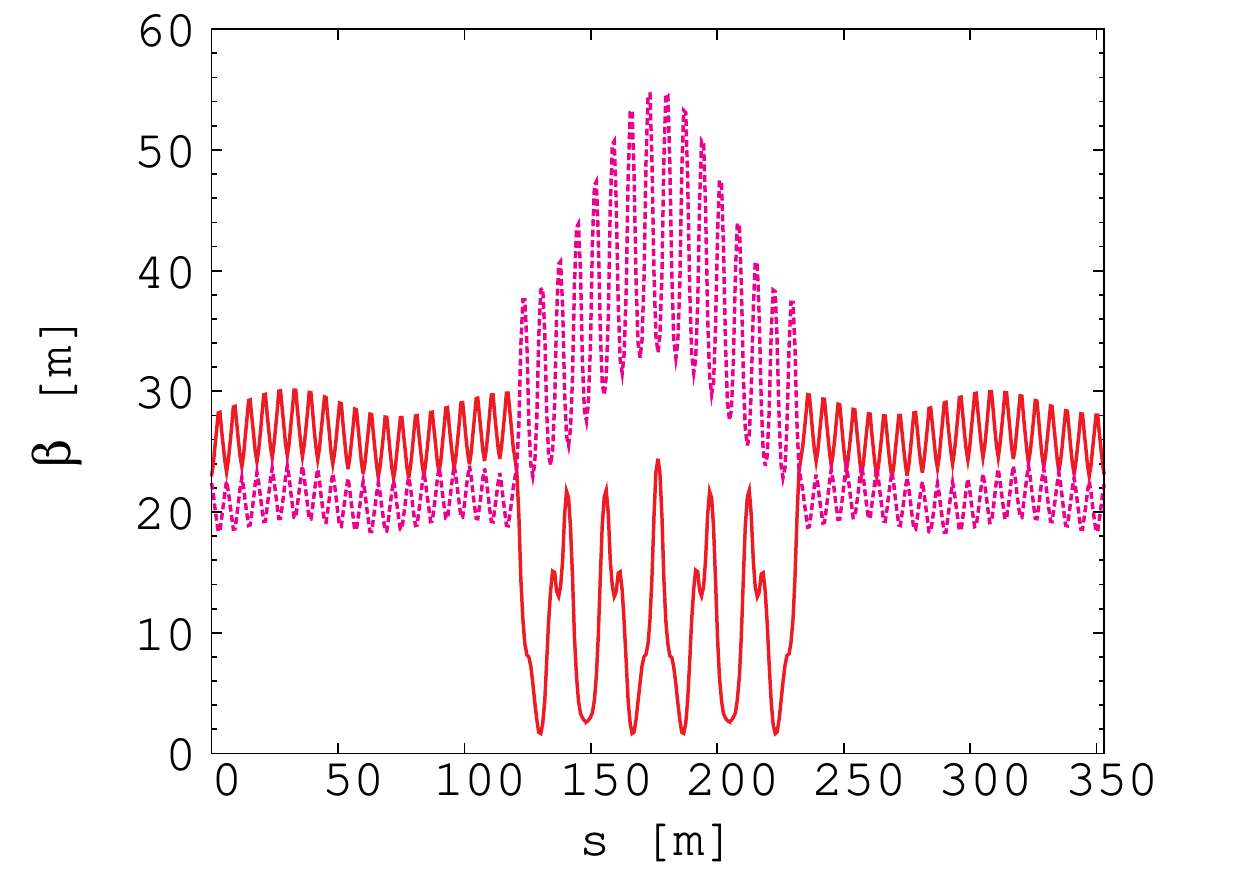}
		\caption{Horizontal (plain red) and vertical (dotted purple) periodic $\beta$ functions of half
		of the ring for $p_{0}$. The plot is centered on the arc part.}
		\label{beta-single}
	\end{center}
\end{figure}

Closed orbits of $p_0$, $p_{min}$, and $p_{max}$ particles are shown in Fig.~\ref{single-traj}. The magnetic field for the $p_{max}$ closed orbit is presented in Fig.~\ref{bz-single}. Dispersion at $p_{0}$ is shown in Fig.~\ref{disp-single}. $\beta$ functions for $p_0$, $p_{min}$, and $p_{max}$ are plotted in Fig.~\ref{beta-single}. 

\begin{figure}[h!]
   \begin{minipage}[b]{.49\linewidth}
   \centering
       \includegraphics[width=8.cm]{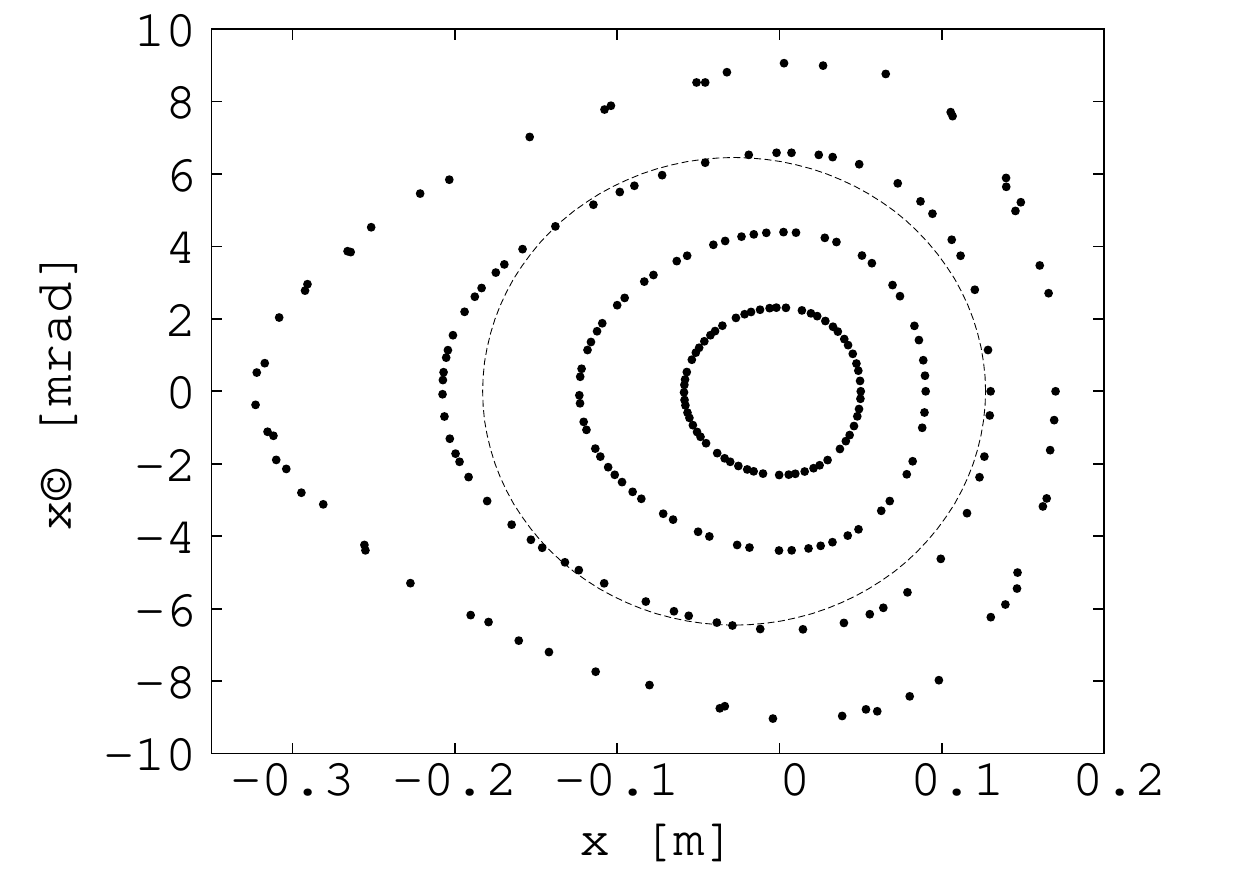}
	\caption{Stable motions in the horizontal Poincare map for different initial amplitudes (5~cm, 9~cm, 13~cm and 17~cm) over 30 turns for $p_0$. The ellipse shows a 1 mm-radian unnormalized emittance.}
	\label{poincarrex-single}
	 \end{minipage} \hfill
   \begin{minipage}[b]{.49\linewidth}
   \centering
    	\includegraphics[width=8.cm]{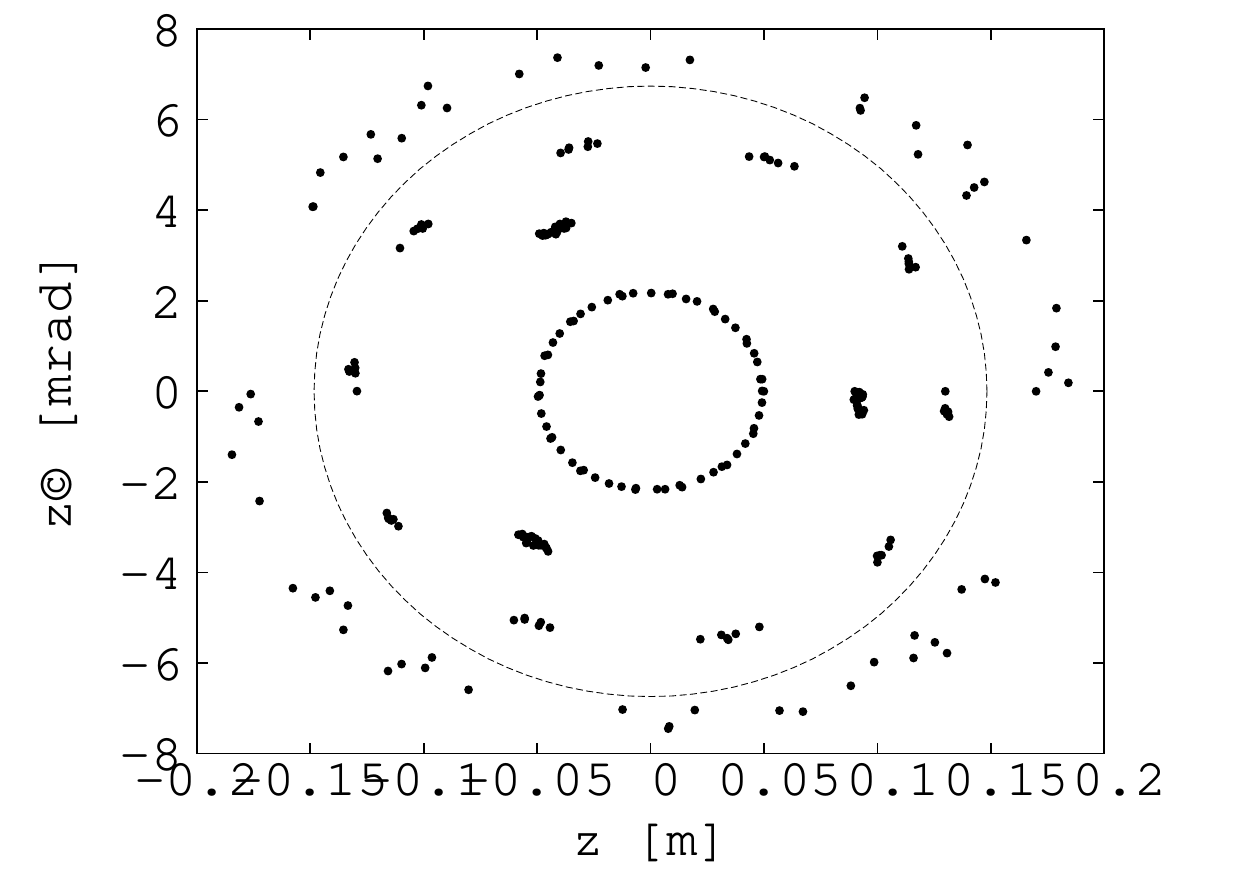}
	\caption{Stable motions in the vertical Poincare map for different initial amplitudes (5~cm, 9~cm, 13~cm and 17~cm) over 30 turns for $p_0$. The ellipse shows a 1 mm-radian unnormalized emittance.}
	\label{poincarrez-single}
   \end{minipage}
\end{figure}

\begin{figure}[h!]
   \begin{minipage}[b]{.49\linewidth}
   \centering
       \includegraphics[width=8.cm]{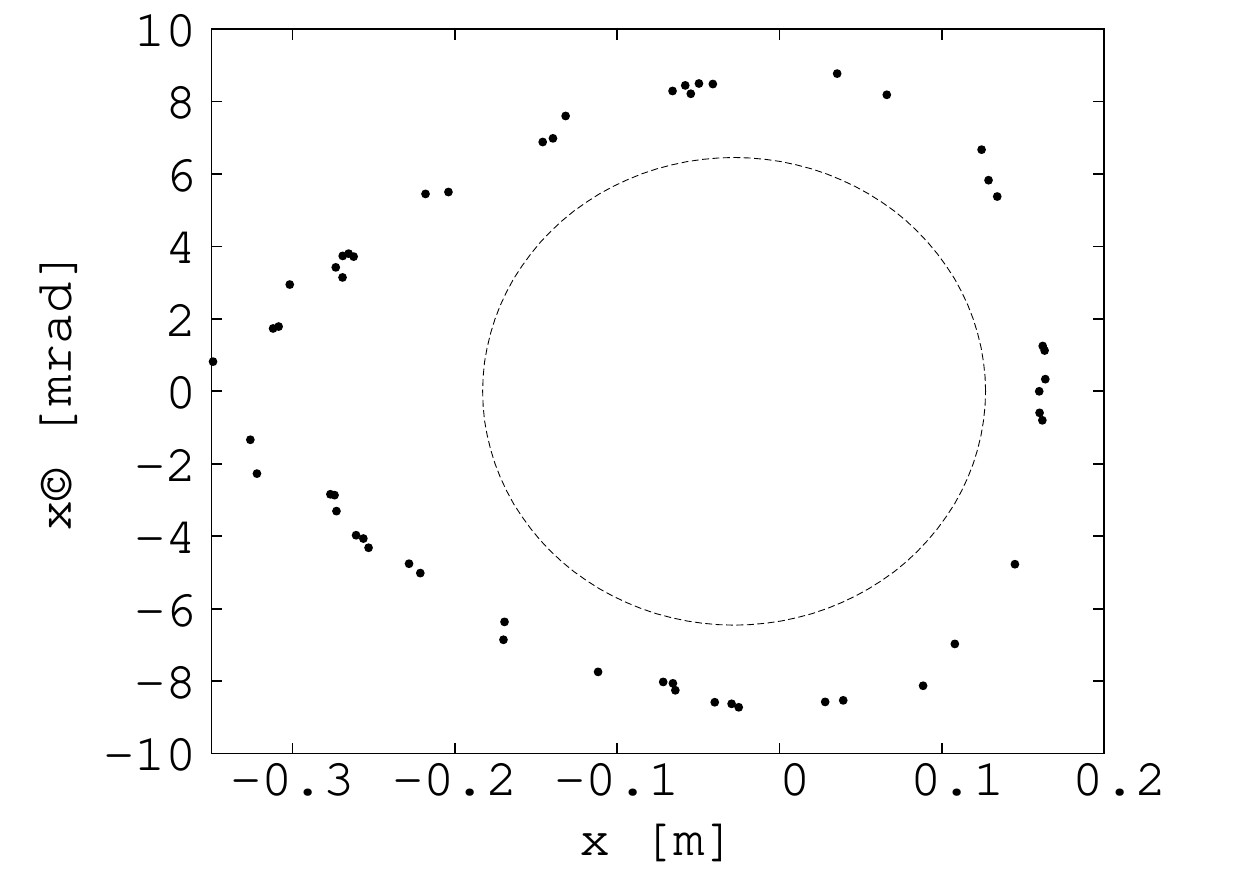}
	\caption{Horizontal Poincare map for maximum initial amplitude (16~cm) with stable motion over 30 turns for $p_{min}$. The ellipse shows a 1 mm-radian unnormalized emittance.}
	\label{poincarrex-single-pmin}
	 \end{minipage} \hfill
   \begin{minipage}[b]{.49\linewidth}
   \centering
    	\includegraphics[width=8.cm]{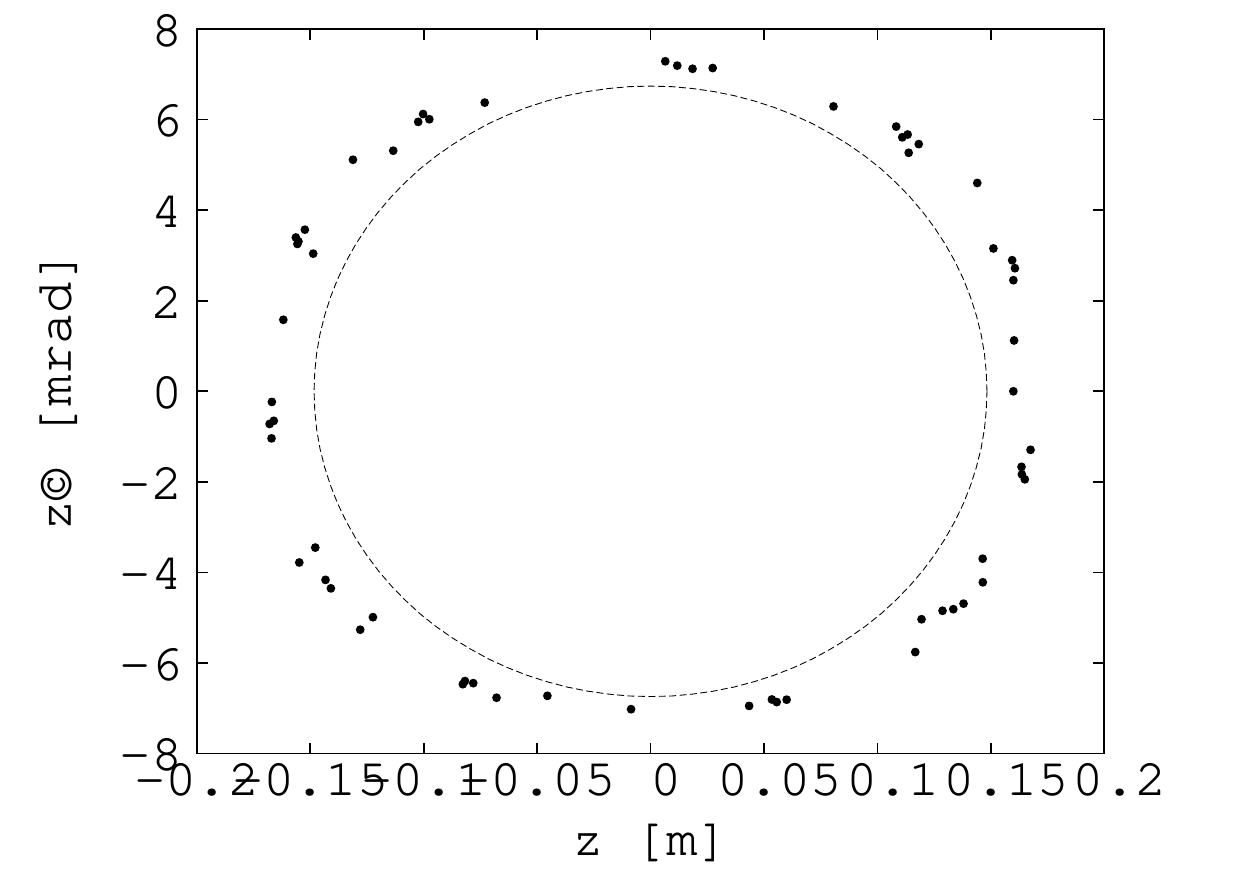}
	\caption{Vertical Poincare map for maximum initial amplitude (16~cm) with stable motion over 30 turns for $p_{min}$. The ellipse shows a 1 mm-radian unnormalized emittance.}
	\label{poincarrez-single-pmin}
   \end{minipage}
\end{figure}

\begin{figure}[h!]
   \begin{minipage}[b]{.49\linewidth}
   \centering
       \includegraphics[width=8.cm]{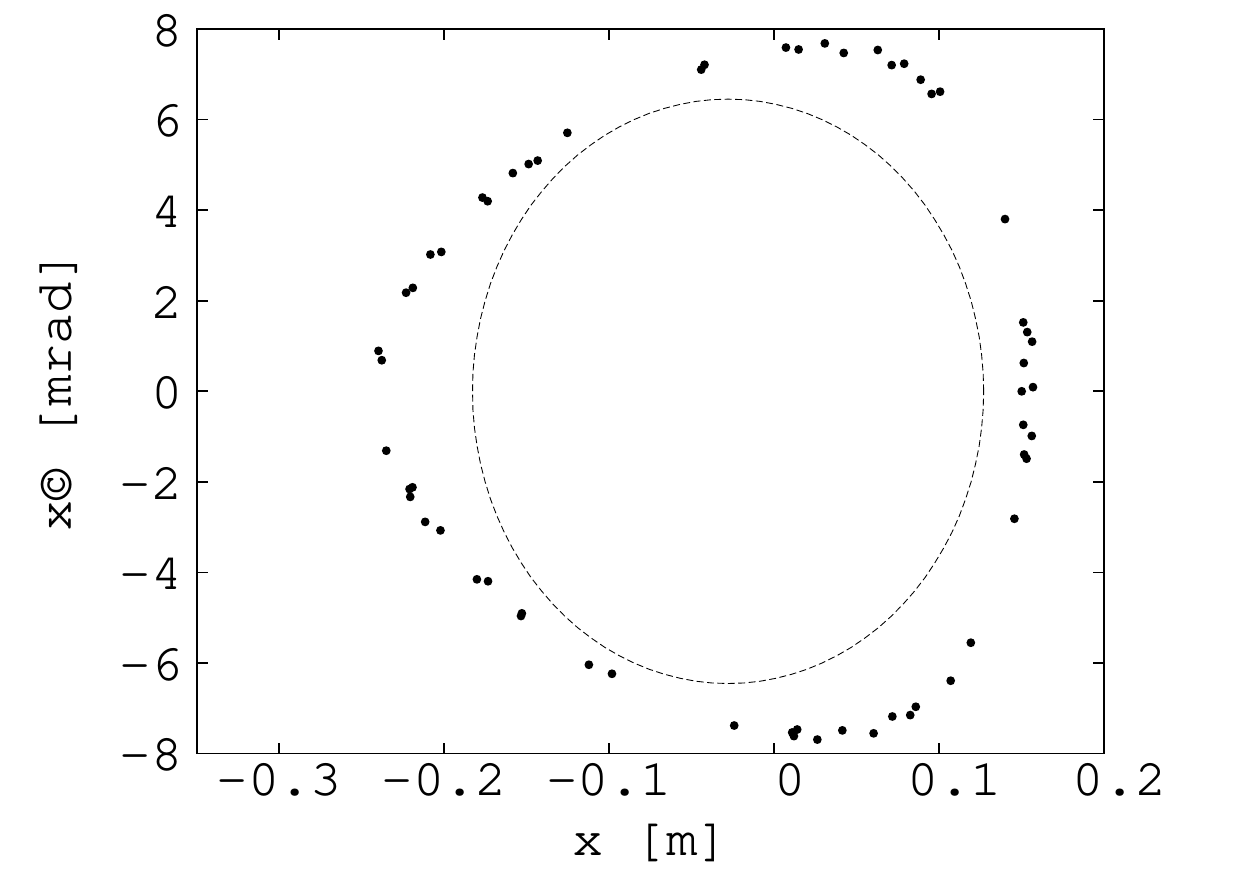}
	\caption{Horizontal Poincare map for maximum initial amplitude (15~cm) with a stable motion over 30 turns for $p_{max}$. The ellipse shows a 1 mm-radian unnormalized emittance.}
	\label{poincarrex-single-pmax}
	 \end{minipage} \hfill
   \begin{minipage}[b]{.49\linewidth}
   \centering
    	\includegraphics[width=8.cm]{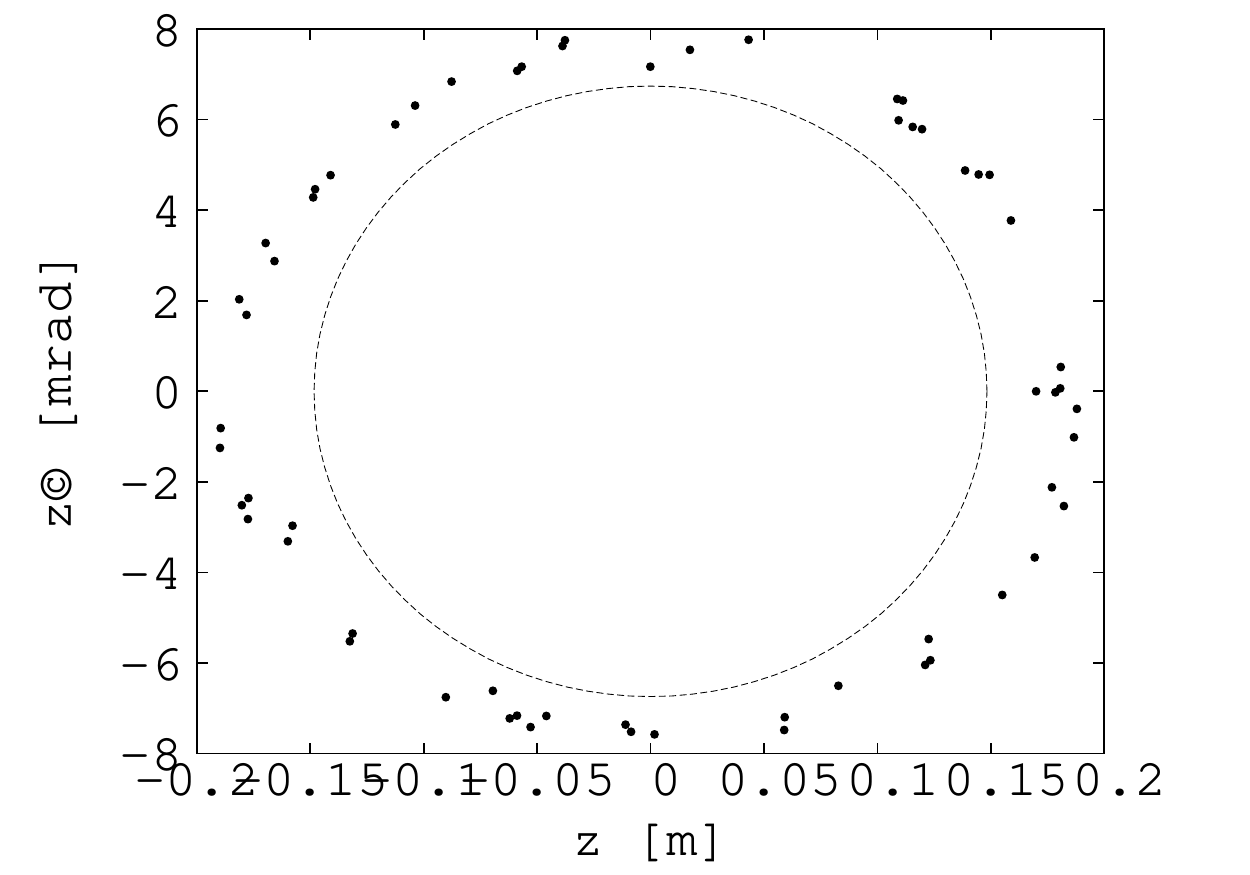}
	\caption{Vertical Poincare map for maximum initial amplitude (17~cm) with a stable motion over 30 turns for $p_{max}$. The ellipse shows a 1 mm-radian unnormalized emittance.}
	\label{poincarrez-single-pmax}
   \end{minipage}
\end{figure}

An acceptance study at fixed energy has also been done. The maximum amplitudes with stable motion at $p_0$ over 30 turns 
are shown for horizontal and vertical motion in Fig.~\ref{poincarrex-single} (left) and in Fig.~\ref{poincarrez-single} 
(right), respectively. 
The same procedure has been done for $p_{min}$ (see Fig.~\ref{poincarrex-single-pmin})
and $p_{max}$ (see Fig.~\ref{poincarrex-single-pmax}). The results are comparable. The 
unnormalized maximum emittance is more than 1 mm-radian.

\begin{figure}[h!]
   \begin{minipage}[b]{.49\linewidth}
   \centering
       \includegraphics[width=8.cm]{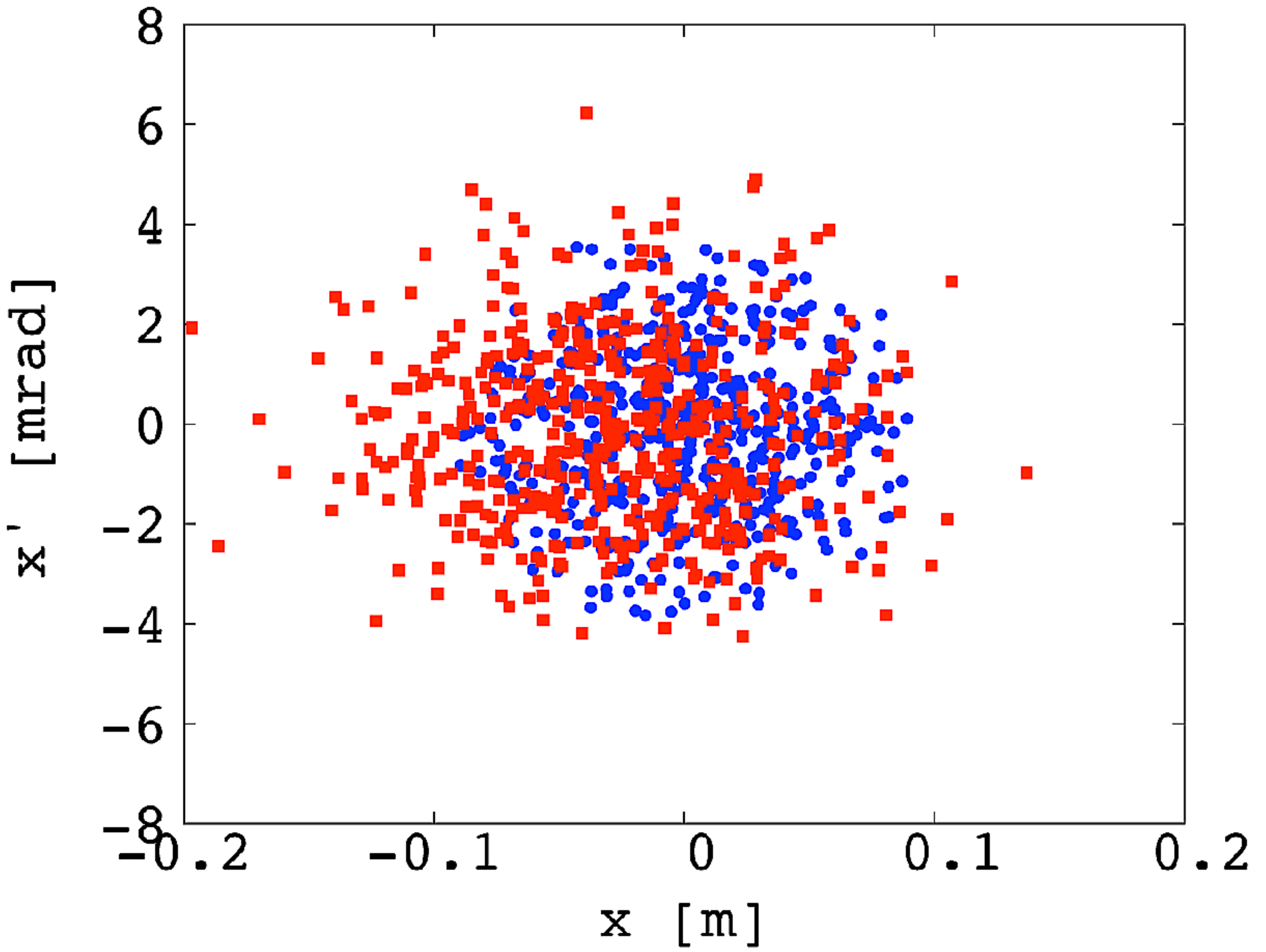}
	\caption{Beam tracking results in the horizontal phase space for a beam with $\Delta p/p_0=\pm 16\%$.   The blue shows the initial particle distribution and the red the final distribution after 60 turns.}
	\label{fig:Bthp}
	 \end{minipage} \hfill
   \begin{minipage}[b]{.49\linewidth}
   \centering
    	\includegraphics[width=8.cm]{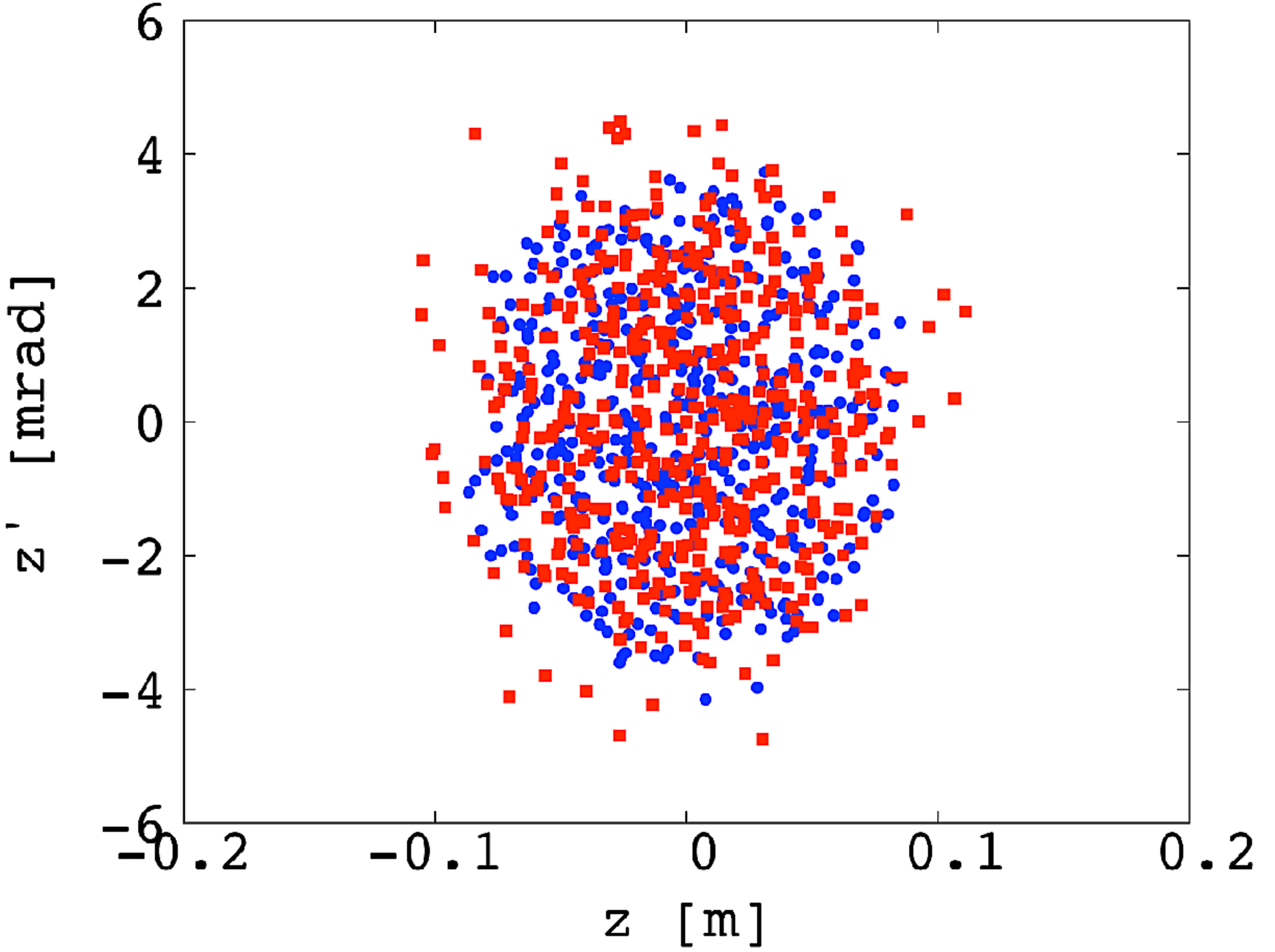}
	\caption{Beam tracking results in the vertical phase space for a beam with $\Delta p/p_0=\pm 16\%$.   The blue shows the initial particle distribution and the red the final distribution after 60 turns. }
	\label{fig:Btvp}
   \end{minipage}
\end{figure}

\paragraph{Multi-particle tracking}
 Multi-particle beam tracking in 6-D phase space has been carried out for the beam with $\Delta p/p_0=\pm 16\%$.   Fig.~\ref{fig:Bthp} and \ref{fig:Btvp} show the results of the beam tracking simulation in the horizontal and vertical directions, respectively.  A normalized emittance of 14 mm-radian in the transverse direction is assumed.    In these figures, the blue dots show the initial particle distribution and the red ones are after 60 turns.  No beam loss is observed in 60 turns.
 
\begin{figure}[h]
	\begin{center}
		\includegraphics[width=0.9\textwidth]{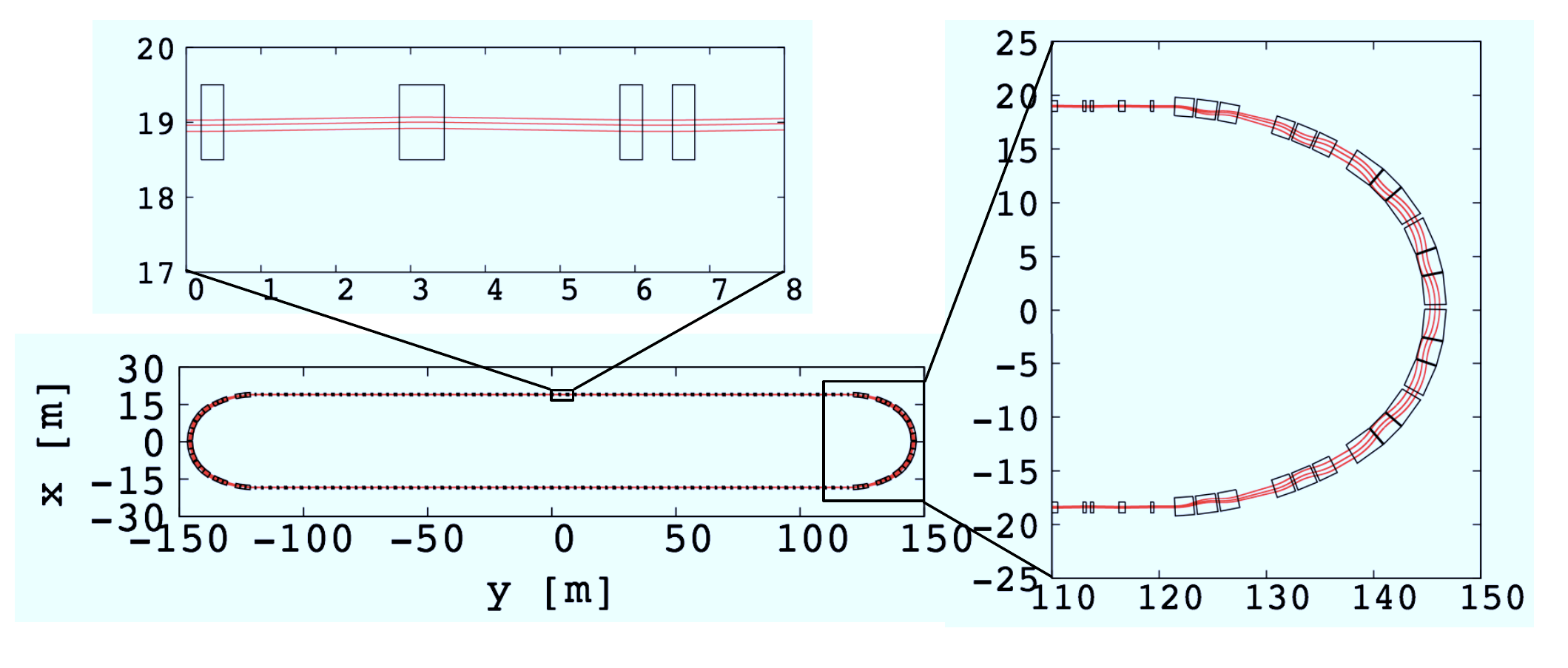}
		\caption{Top view of the racetrack FFAG lattice (bottom left figure). The top left figure shows a detail of the straight section and the right figure a detail of the arc.  Muon closed orbits for $p_0$, $p_{min}$, and $p_{max}$ are shown in red.  Effective field boundaries with collimators are shown in black.}
		\label{fig:single-traj}
	\end{center}
\end{figure}

\begin{figure}[h]
	\begin{center}
		\includegraphics[width=0.9\textwidth]{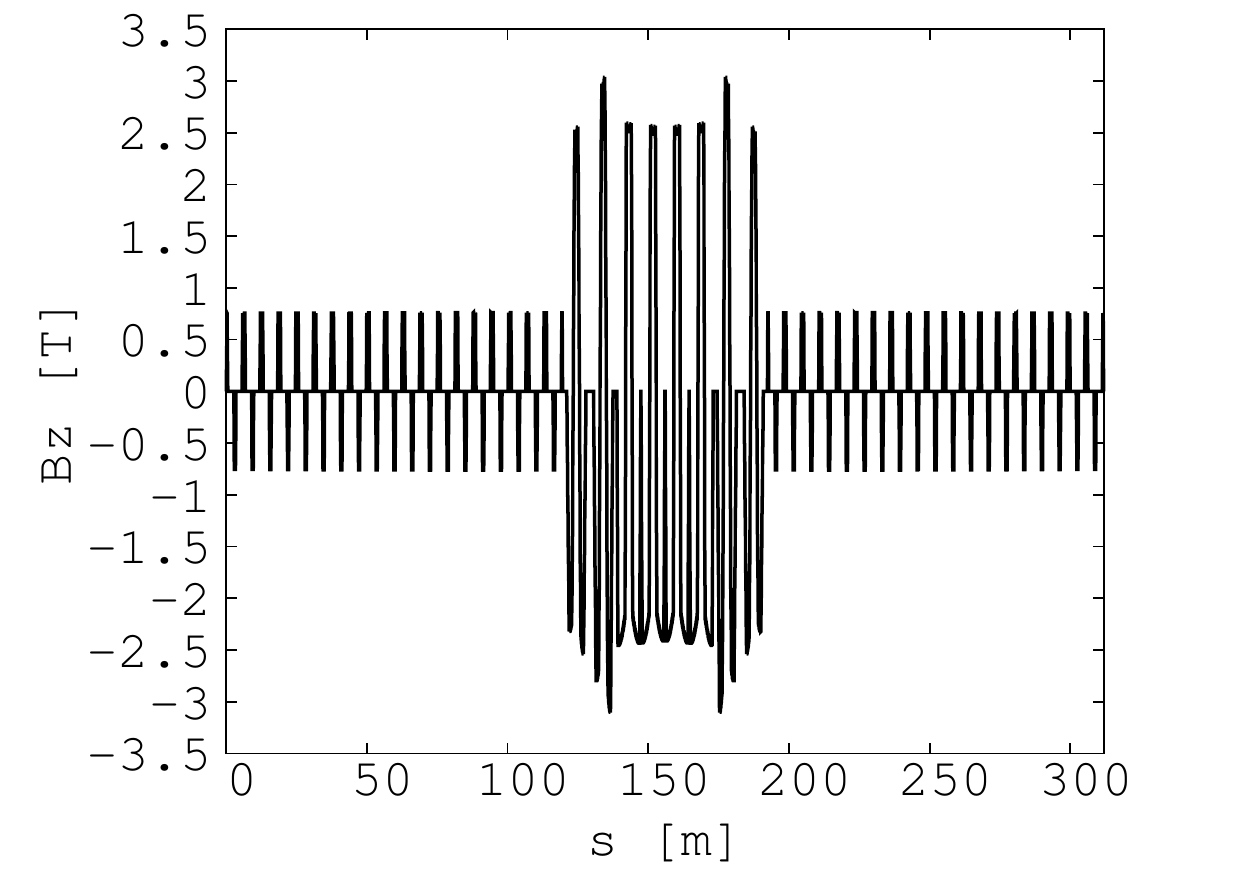}
		\caption{Vertical magnetic field for the muon closed orbit at $p_{max}$ in the racetrack FFAG ring.}
		\label{fig:bz-single}
	\end{center}
\end{figure}

\begin{figure}[h]
	\begin{center}
		\includegraphics[width=0.9\textwidth]{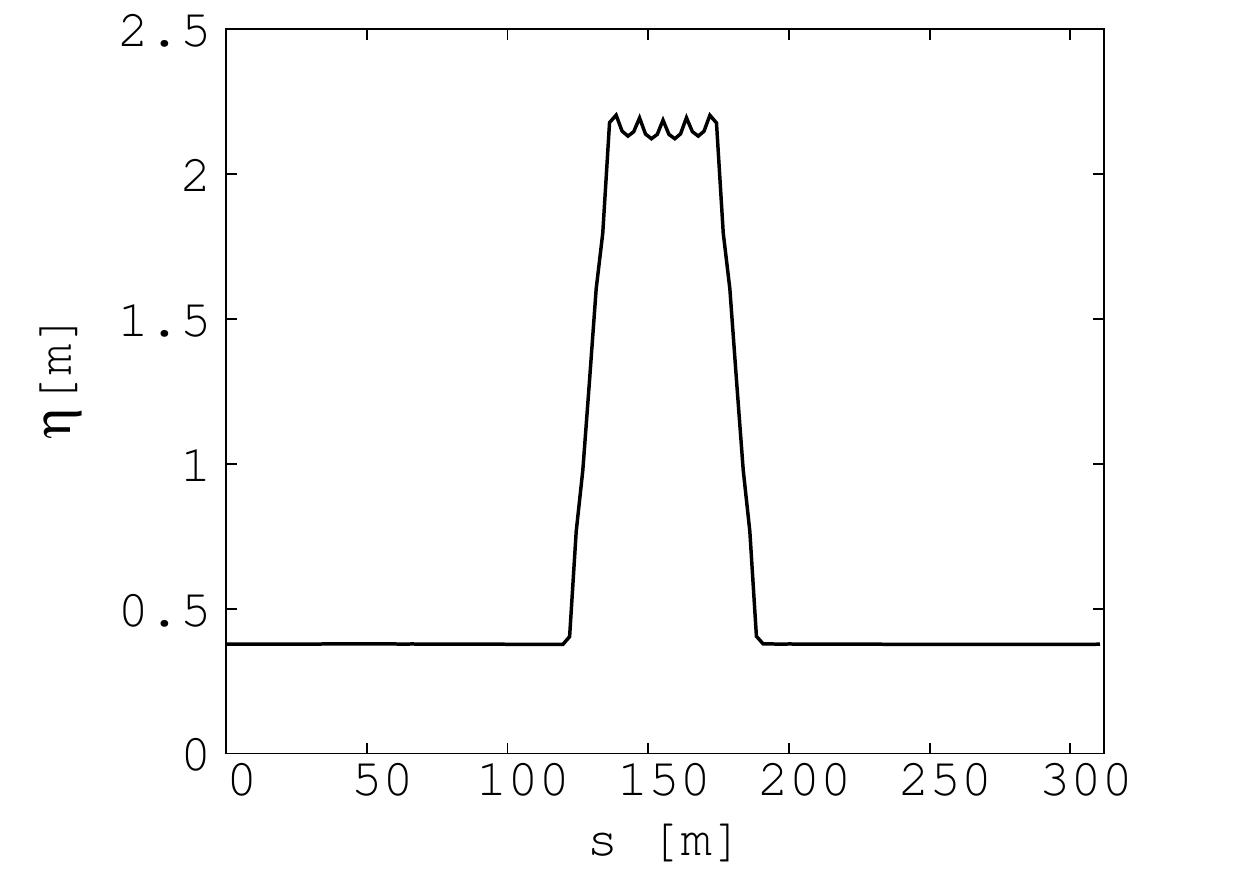}
		\caption{Dispersion function for $p_0$ in half of the ring. The plot is centered on the arc.}
		\label{fig:disp-single}
	\end{center}
\end{figure}

\begin{figure}[h]
	\begin{center}
		\includegraphics[width=0.9\textwidth]{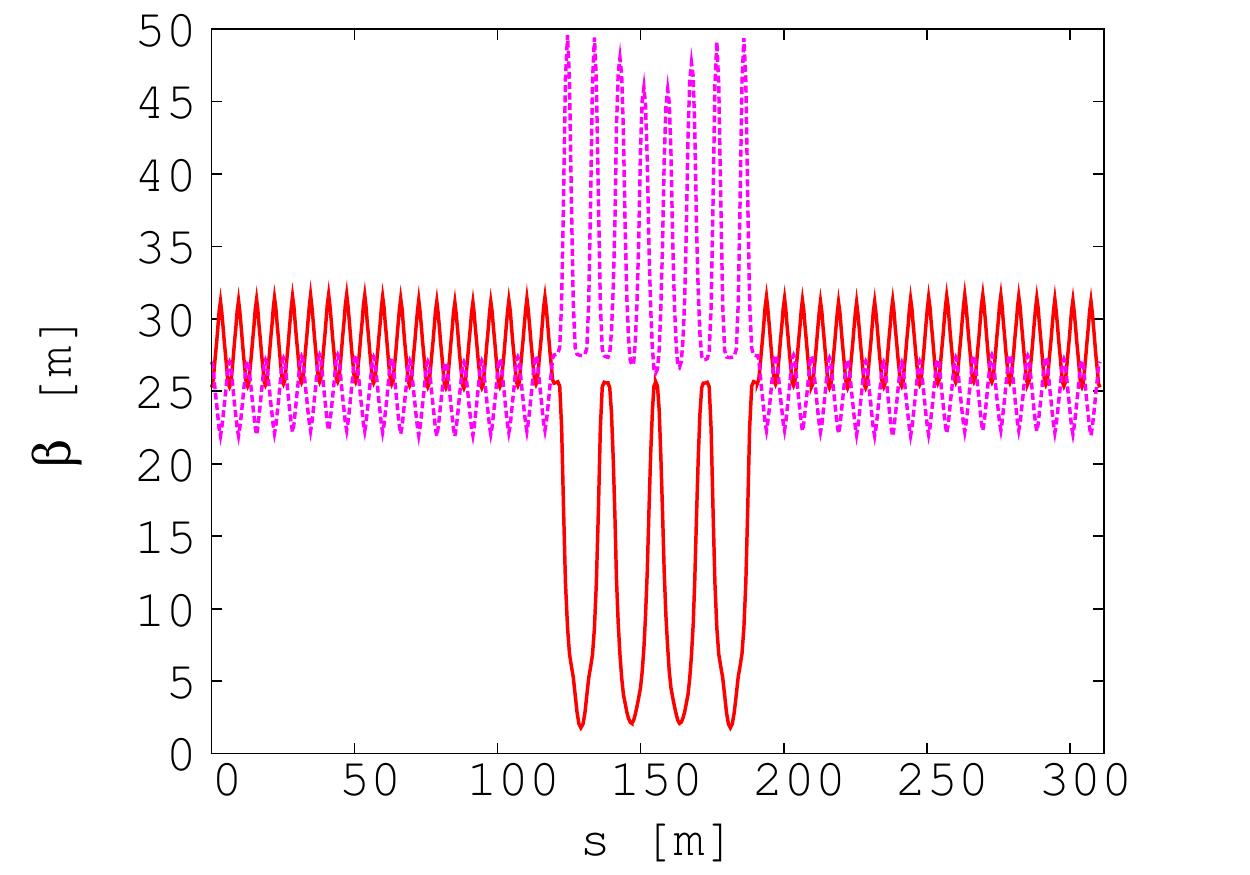}
		\caption{Horizontal (plain red) and vertical (dotted purple) periodic beta-functions of half of the ring for $p_{0}$. The plot is centered on the arc.}
		\label{fig:beta-single}
	\end{center}
\end{figure}

\clearpage
\paragraph{Compact arc design for the racetrack FFAG ring}
\label{subsubsec:4525}
In order to reduce the construction and operational cost of the racetrack FFAG decay ring, a compact arc has been designed.  It assumes the use of super-ferric combined-function magnets with magnetic field strengths up to 3~T. The arc consists of four regular FDF triplet cells in the centre and two matching FDF triplet cells on each side of the regular cells, four matching cells in total for the arc. The purpose of the matching section is to match the dispersion function between the production section, assuming straight FFAG cells with small, but non-zero dispersion of 0.38~m, and the centre of the arc.  It must also accommodate straight sections which allow for the stochastic injection of 5~GeV/c pions.  The parameters of the regular scaling FFAG arc cell are summarized in Table~\ref{tab:tab-circ}.
\begin{table}[h]
    \centering
     \caption{Parameters of the circular regular scaling FFAG arc cell.}
\begin{tabular}{|lcc|}
\hline
Cell type			&		& FDF triplet \\
Number of cells in the ring	 &	& 8 \\
Cell opening angle	&	& 30~deg\\
 $r_0$ &&	16~m\\
k-value      &         & 6.25        \\
         Packing factor & & 0.92 \\
         Collimators ($r_{min}, r_{max}, z_{max}$) && (15~m, 17~m,  0.4~m)\\
         Periodic cell dispersion  && 2.21~m (at~ 3.8~GeV/c)\\
          Horizontal phase advance  && 90\,deg. \\
Vertical phase advance  && 13.92\,deg. \\
\hline
F$_1$ magnet parameters &&\\
& Magnet center & 5.8~deg\\
& Magnet length & 10~deg\\
& Fringe field fall off & Linear (Length: 0.7~deg)\\
& $B_0(r_0=16~m)$ & -1.70382~T\\
\hline
D magnet parameters &&\\
& Magnet center & 15~deg\\
& Magnet length & 7.6~deg\\
& Fringe field fall off & Linear (Length: 0.7~deg)\\
& $B_0(r_0=16~m)$ & 2.13119~T\\
\hline
F$_2$ magnet parameters &&\\
& Magnet center & 24.2~deg\\
& Magnet length & 10~deg\\
& Fringe field fall off & Linear (Length: 0.7~deg)\\
& $B_0(r_0=16~m)$ & -1.70382~T\\
\hline
\end{tabular} 
 \label{tab:tab-circ}
\end{table}
In the circular matching scaling FFAG cell, the vertical magnetic field in the median plane follows the circular scaling law, as in the circular cell.
The parameters of the matching scaling FFAG cell are summarized in Table~\ref{tab:tab-match}.
\begin{table}[h]
    \centering
     \caption{Parameters of the matching scaling FFAG cell.}
\begin{tabular}{|lcc|}
\hline
Cell type			&		& FDF triplet \\
Number of cells in the ring	 &	& 8 \\
Cell opening angle	&	& 15~deg\\
 $r_0$ &&	36.15~m\\
k-value      &         & 26.98        \\
         Packing factor & & 0.58 \\
         Collimators ($r_{min}, r_{max}, z_{max}$) && (35.3~m, 37~m,  0.4~m)\\
         Periodic cell dispersion  && 1.29~m (at~ 3.8~GeV/c)\\
          Horizontal phase advance  && 90\,deg. \\
Vertical phase advance  && 16.95\,deg. \\
\hline
F$_1$ magnet parameters &&\\
& Magnet center & 4.2~deg\\
& Magnet length & 2.8~deg\\
& Fringe field fall off & Linear (Length: 0.7~deg)\\
& $B_0(r_0=36.15~m)$ & -2.18805~T\\
\hline
D magnet parameters &&\\
& Magnet center & 7.5~deg\\
& Magnet length & 3.0~deg\\
& Fringe field fall off & Linear (Length: 0.7~deg)\\
& $B_0(r_0=36.15~m)$ & 2.74622~T\\
\hline
F$_2$ magnet parameters &&\\
& Magnet center & 10.8~deg\\
& Magnet length & 2.8~deg\\
& Fringe field fall off & Linear (Length: 0.7~deg)\\
& $B_0(r_0=36.15~m)$ & -2.18805~T\\
\hline
\end{tabular} 
 \label{tab:tab-match}
\end{table}
The arc layout of the ring is shown in Fig.~\ref{fig:single-traj}. 
The central orbit, $p_0$ (3.8 GeV/c) and the orbits for $p_{min}$ (-16\%) and $p_{max}$ (+16\%) are also shown in Fig.~\ref{fig:single-traj}.  The magnetic field for the $p_{max}$ closed orbit is shown in Fig.~\ref{fig:bz-single}. The dispersion at $p_{0}$ is shown in Fig.~\ref{fig:disp-single} and the beta-functions at $p_0$ in Fig.~\ref{fig:beta-single}. 

Future studies on the racetrack FFAG decay ring will include a ring reconfiguration to adjust the ratio of the production straight and the arc length, detailed optimization of the working point and tracking studies. In addition, the detailed geometry of the pion injection together with the design of the super-ferric magnets must also be addressed.
\clearpage
\subsubsection{Decay ring instrumentation}
\begin{figure}[h]
  \centering{
    \includegraphics[width=.8\textwidth]{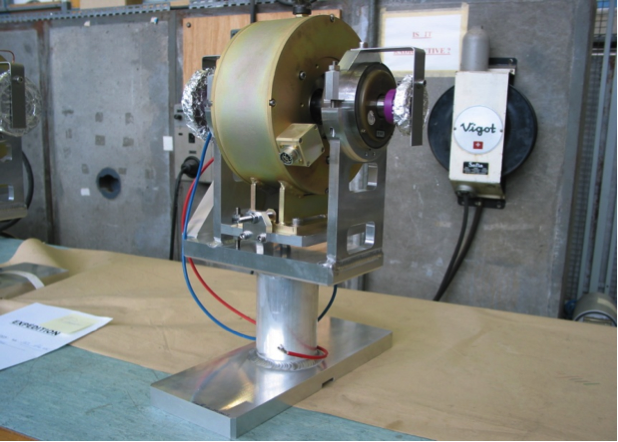}
    }
  \caption{CERN L4 beam current transformer}
  \label{fig:BCT}
\end{figure}

\begin{figure}[h]
  \centering{
    \includegraphics[width=.8\textwidth]{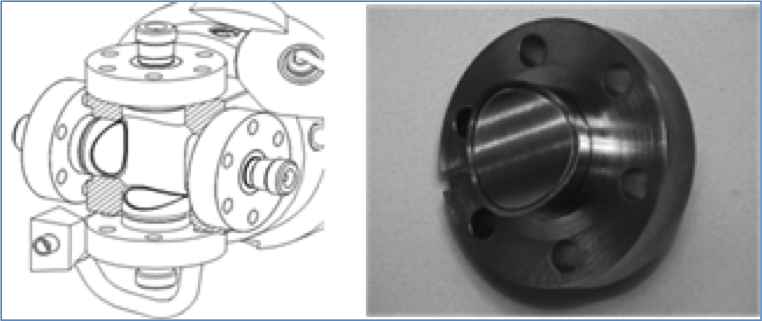}
    }
  \caption{LHC beam-position monitor button}
  \label{fig:BPM}
\end{figure}

\begin{figure}[h]
  \centering{
    \includegraphics[width=.8\textwidth]{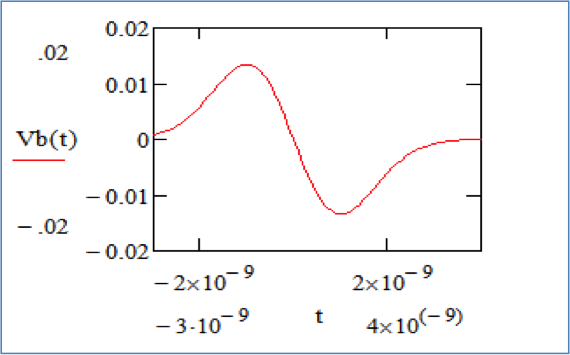}
    }
  \caption{Button impulse response}
  \label{fig:BPM_sig}
\end{figure}

The goal of the beam instrumentation for nuSTORM is twofold.  First,
the instrumentation is needed in order to determine the neutrino flux at the near and far detectors with an absolute precision of $<$ 1\%. 
Both the number of neutrinos and their energy distribution must be determined.  If both the circulating muon flux in the storage ring is known on a turn-to-turn basis, and the orbit and orbit uncertainties (uncertainty on the divergence) are known accurately, then the neutrino flux and energy spectrum can be predicted with equal precision.  
Our measurement goals for the suite of beam instrumentation diagnostics for the decay ring are summarized below:
\begin{enumerate}
\item Measure the circulating muon intensity (on a turn by turn basis) to 0.1\% absolute.
\item Measure the mean momentum to 0.1\% absolute.
\item Measure the momentum spread to 1\% (FWHM).
\item Measure the tune to 0.01.
\end{enumerate}
Second, from the accelerator standpoint, in order to commission and run the storage ring, turn-by-turn measurements of the following parameters are crucial:
trajectory, tune, beam profile and beam loss.  Our current estimate of these requirements is summarized in Table~\ref{tab:DringI} below.
\begin{table}[h]
\centering
\caption {Decay ring instrumentation specifications}
\label{tab:DringI}
\begin{tabular}{|l|c|c|}
\hline
			&  Absolute accuracy &  Resolution\\
Intensity		&  0.1\%			&  0.01\% \\
Beam position	&  5 mm			&  1 mm \\
Beam profile	&  5 mm			&  1 mm \\
Tune			&  0.01			&  0.001 \\
Beam loss		&  1\%			&  0.5\% \\
Energy		&  0.5\%			&  0.1\% \\
Energy spread	&  1\%			&  0.1\% \\
\hline
\noalign{\smallskip}
\end{tabular}
\end{table}
\paragraph{Beam intensity}
In order to measure the circulating muon intensities, one option is to use a toroid-based Fast Beam Current Transformer (FBCT), such as the one recently developed at CERN for L4~\cite{Soby}and shown in Fig.~\ref{fig:BCT}. Its specifications are given in Table~\ref{tab:BCT}. It consists of a one turn calibration winding and a 20 turn secondary winding, wound on a magnetic core and housed in a 4 layer shielding box. The mechanical dimensions will have to be adapted to the  large beam pipe of nuSTORM.  It should be noted that obtaining an absolute precision of 0.1\% will be challenging.  Problems associated with the pulsed calibration and with EMI will influence the absolute accuracy of
the FBCT.
\begin{table}[h]
\centering
\caption{FBCT parameters}
\label{tab:BCT}
\begin{tabular}{|l|c|}
\hline
Droop @ 500 $\mu$s	&  0.5\% \\
Bandwidth				&  10 MHz \\
Accuracy				&  $\simeq$ 1\% \\
Resolution				&  $\simeq 10\mu$A \\
Rise time				&  35 ns \\
\hline
\noalign{\smallskip}
\end{tabular}
\end{table}
\paragraph{Beam position}
Button beam-position monitors (BPMs) are, in general, cheap and their frequency response fits very well with that of nuSTORM. They are widely used in the LHC.  A photo of the 25mm diameter button is shown in Fig.~\ref{fig:BPM}.
Due to the large vacuum chamber size and the, at present, uncertainties regarding the circulating beam parameters, it is difficult to estimate the ultimate resolution which can be obtained. 
However, we can estimate the resolution assuming the following: a 100mm diameter button, bunch length of 4 ns, bunch intensity of $5\times10^8$ and an input noise of 
2nV/$\sqrt{Hz}$.  With these assumptions, the expected electrode signal can be simulated and is shown in Fig.~\ref{fig:BPM_sig}.   As a first order approximation the 15mV peak amplitude on the button corresponds to 50µV/mm in a 600mm diameter vacuum chamber.
With the assumed bandwidth and noise of the system (output noise $\sim30\mu$V), the expected resolution is on the order of 5 mm with a signal to noise ratio of 10. This single bunch, single turn resolution can of course be improved by averaging over all bunches, i.e., by a factor of up to 10.
\paragraph{Transverse profile measurements}
Due to the relatively low intensities (compared to primary or secondary beamlines) and very short lifetime ($100\mu$s), our first estimations indicate that using Ionization Profile Monitors (IPM) to measure the transverse profiles is not feasible.  Other detectors based on ionization (MWPC, IC, GEM) are destructive and would require a quite complicated design. (The use of low-mass MWPC is possible with further study, however.)  The use of wire scanners is not possible due to the short beam lifetime.  

\begin{figure}[h]
  \centering{
    \includegraphics[width=.8\textwidth]{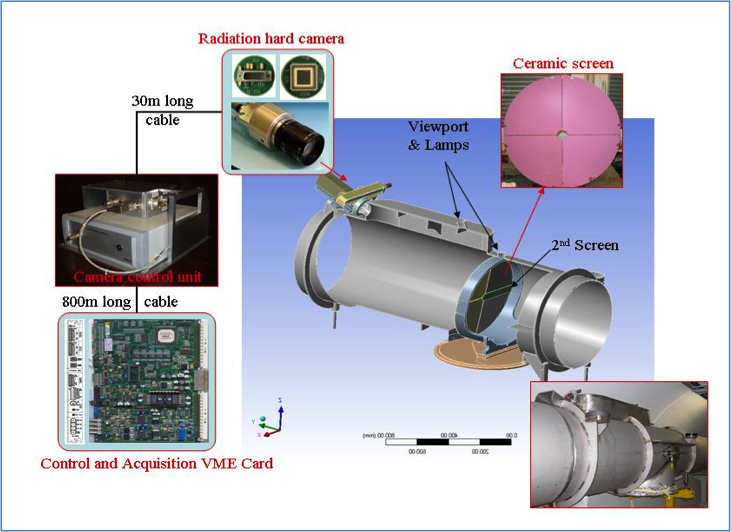}
    }
  \caption{LHC dump line BTV}
  \label{fig:BTV}
\end{figure}

\begin{figure}[h]
  \centering{
    \includegraphics[width=.8\textwidth]{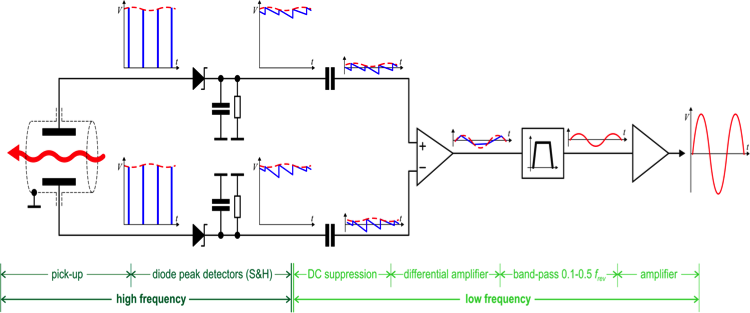}
    }
  \caption{Base-band tune measurement principle}
  \label{fig:tune}
\end{figure}

Destructive measurement techniques using either scintillation screens or SEM-strips can be utilized. One option is to adapt the LHC dump line BTV to nuSTORM.  The BTV consists of a 2 m long by 60 cm diameter vacuum tank (see Fig.~\ref{fig:BTV}).  In this design, the fixed scintillation screen is observed with a camera, which is read out using a VME based control and data acquisition card.  We estimate that this system could provide an overall position accuracy of $\sim$ 2 mm in the nuSTORM decay ring.  As designed for the LHC dump line, the system can only be used for diagnostics. It would have to be taken out of the beam during running due to the mass of the screen.  However, investigating whether a low-mass screen option is possible is worth further study.  The specification on the maximum tolerable material budget in the screen can be determined once we have the decay ring lattice fixed and have circulating beam in our G4Beamline simulation.  
\paragraph{Tune measurements}
For the measurement of the non-integer part of the nuSTORM tune we propose to use the Direct Diode Detection Base-Band Q (3D-BBQ) developed at CERN by M. Gasior \cite{Gasior:2005di} and shown schematically in Fig.~\ref{fig:tune}. This very sensitive method allows for the observation of very small amplitude modulations on high level signals. The pulses obtained from any of the beam position monitors are connected to simple diode detectors which convert the amplitude modulation of the BPM pulses to a  signal in the audio frequency range. The dominate part of the BPM signal is related to the beam intensity, and becomes a DC voltage.  This DC component is easily removed using a simple capacitor at the detector output. The two base band signals are then subtracted in a difference amplifier and digitized for tune calculations in the frequency domain.

\begin{figure}[h]
  \centering{
    \includegraphics[width=.8\textwidth]{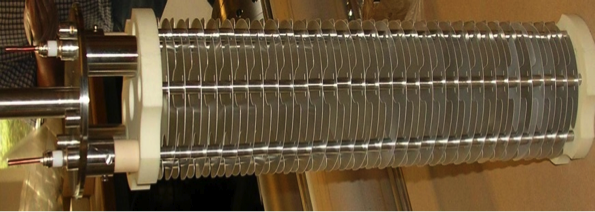}
    }
  \caption{Photograph of an ionization chamber for total beam loss (slow) measurements}
  \label{fig:BeamLoss}
\end{figure}

\begin{figure}[h]
  \centering{
    \includegraphics[width=.8\textwidth]{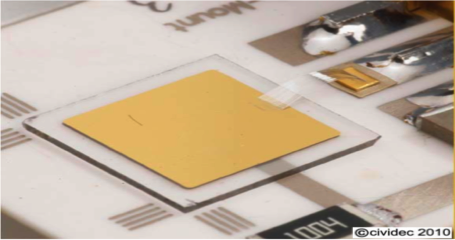}
    }
  \caption{Diamond-based secondary emission monitor}
  \label{fig:Diamond}
\end{figure}

Due to our uncertainty on the bunch intensities at this time,  it is, at present, difficult to estimate if a tune kicker is needed to enhance the betatron amplitudes to obtain the required tune and time resolution. At first estimate, we believe that connecting one or several BPMs to the system should be sufficient.
\paragraph{Beam loss measurements}
Beam loss monitors would be used mostly for diagnosing the performance of the ring with emphasis on the injection point using some monitors in the straight sections and in the arcs.  We propose to use ``slow" ionisation chambers (Fig.~\ref{fig:BeamLoss}) for integration of the total loss  around the ring and diamond-based fast secondary emission monitors (Fig.~\ref{fig:Diamond}) for observation of the ``fast" injection losses.  Both types of detectors are sensitive to charged particles only.

The ionization chamber consists of a 60 cm long stainless steel cylinder, with parallel Al electrodes separated by 0.5 cm to which a voltage of 1.5 kV is applied to every second electrode.  The entire volume is filled with Nitrogen gas.   The output current from the grounded electrodes is then proportional to the beam loss.
The diamond detector consists of a 10 mm $\times$ 10 mm $\times$  0.5 mm polycrystalline chemical vapor deposition (pCVD) diamond substrate coated on each side with 200 nm thick gold electrodes.  A biasing voltage of $\sim$ ~500 V is used.  A 40 dB broadband radiation hard current amplifier with a bandwidth 100 MHz to 2 GHz is needed to amplify the very low currents.
\paragraph{Summary}
In the above sections, we have shown how existing beam instrumentation or extrapolations from existing designs can meet the needs of nuSTORM and gives the basis for making realistic cost estimates for the instrumentation.  However, obtaining full knowledge regarding how well the instrumentation will perform in the nuSTORM decay ring will only come after the lattice design is finalized (addition of higher-order correctors).  Once this has been accomplished, we then have the tools in place within the G4Beamline simulation framework to understand, in detail, how this instrumentation will perform.
\clearpage
\subsection{A future 6D muon ionization cooling experiment}
\label{SubSect:6DICE}
\begin{figure}[ht]
  \begin{center}
    \includegraphics[width=0.85\textwidth]%
      {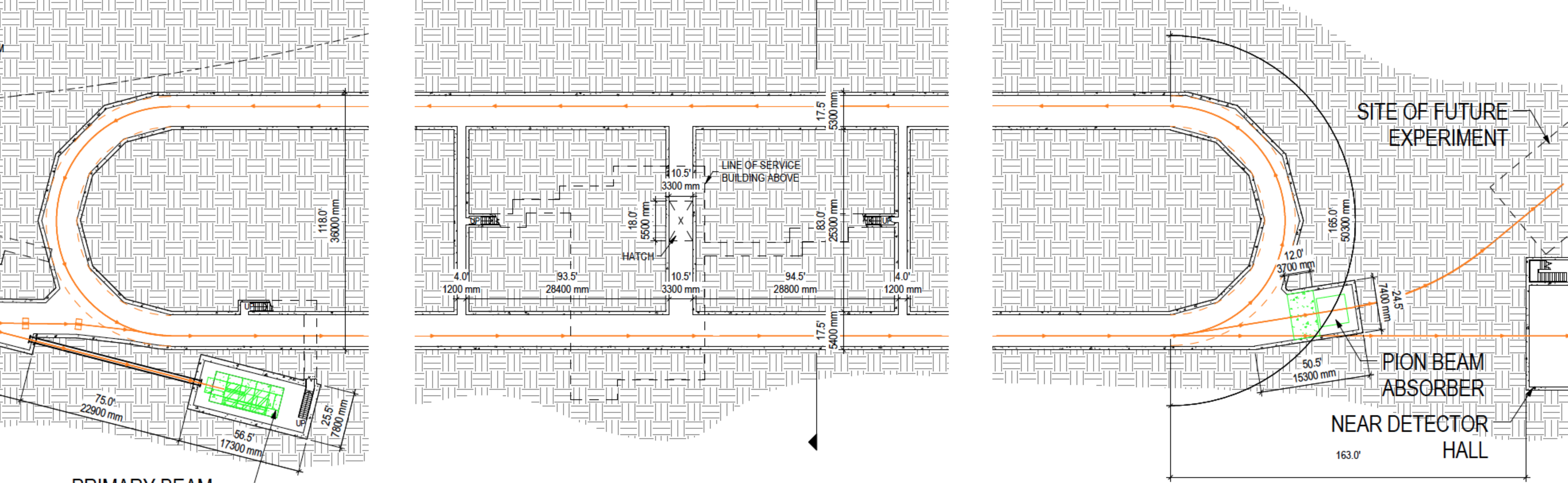}
  \end{center}
  \caption{Schematic of the nuSTORM decay ring.}
  \label{fig:Decay_ring}
\end{figure}

Fig.~\ref{fig:Decay_ring} shows a schematic of the decay ring.
As is described in section \ref{subsec:43}, 5\,GeV/c
pions are injected at the beginning of the straight section of the ring.
With the 185\,m length for the straight, $\sim 48$\% of the
pions decay in the injection straight.
Since the arcs are set for the central muon momentum of 3.8\,GeV/c,
the pions remaining at the end of the straight will not be transported
by the arc.
The power contained within the pion beam that reaches the end of the
injection straight is 2\,kW--3\,kW making it necessary to guide the
undecayed pion beam into an appropriate absorber.

As discussed earlier, another BCS, which is just a mirror reflection of the injection BCS, is placed at the end of the decay straight.  It extracts the residual pions and muons which are in the 5$\pm$0.5 GeV/c momentum range. These extracted muons will enter the absorber along with pions in this same momentum band.

\begin{figure}
  \begin{center}
    \includegraphics[width=0.52\textwidth]%
      {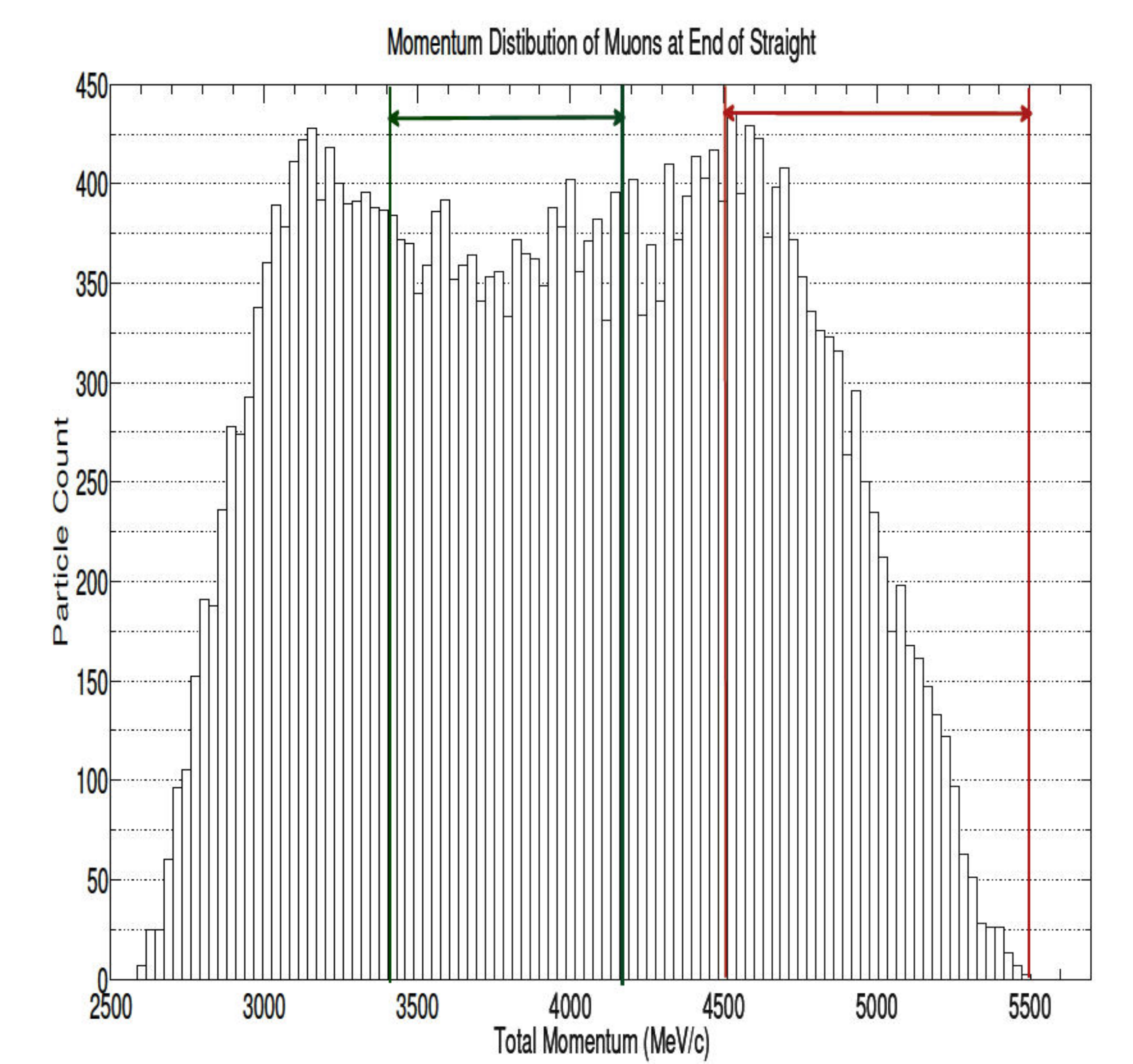} \quad\quad
    \includegraphics[width=0.36\textwidth]%
      {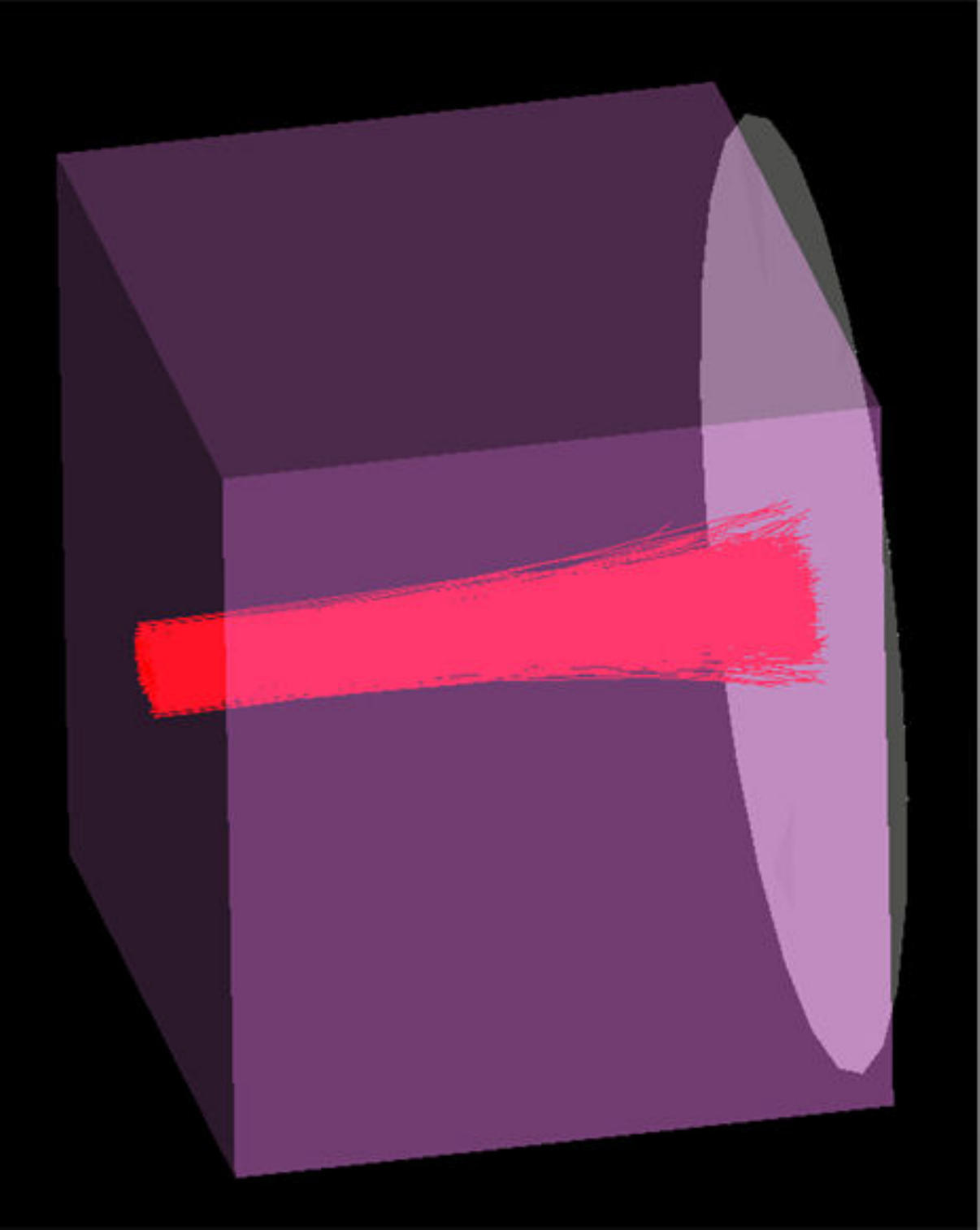}
  \end{center}
  \caption{
    Left panel: Momentum distribution of muons after the first
    straight.
    Right panel: Visualization of muons in the degrader.
  } 
  \label{fig:degrader}
\end{figure}

\begin{figure}
  \begin{center}
    \includegraphics[width=0.48\textwidth]%
      {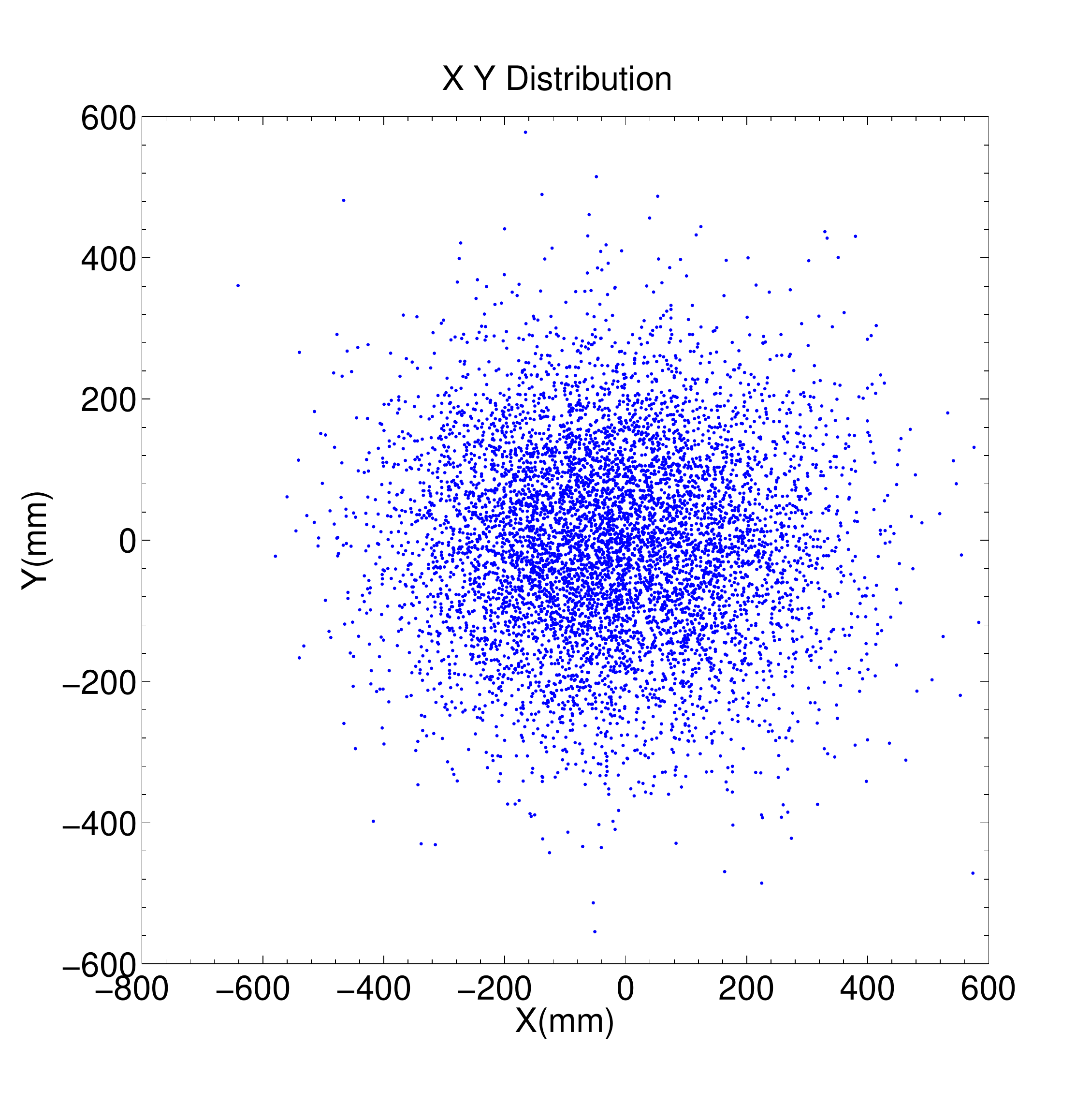} \quad\quad
    \includegraphics[width=0.46\textwidth]%
      {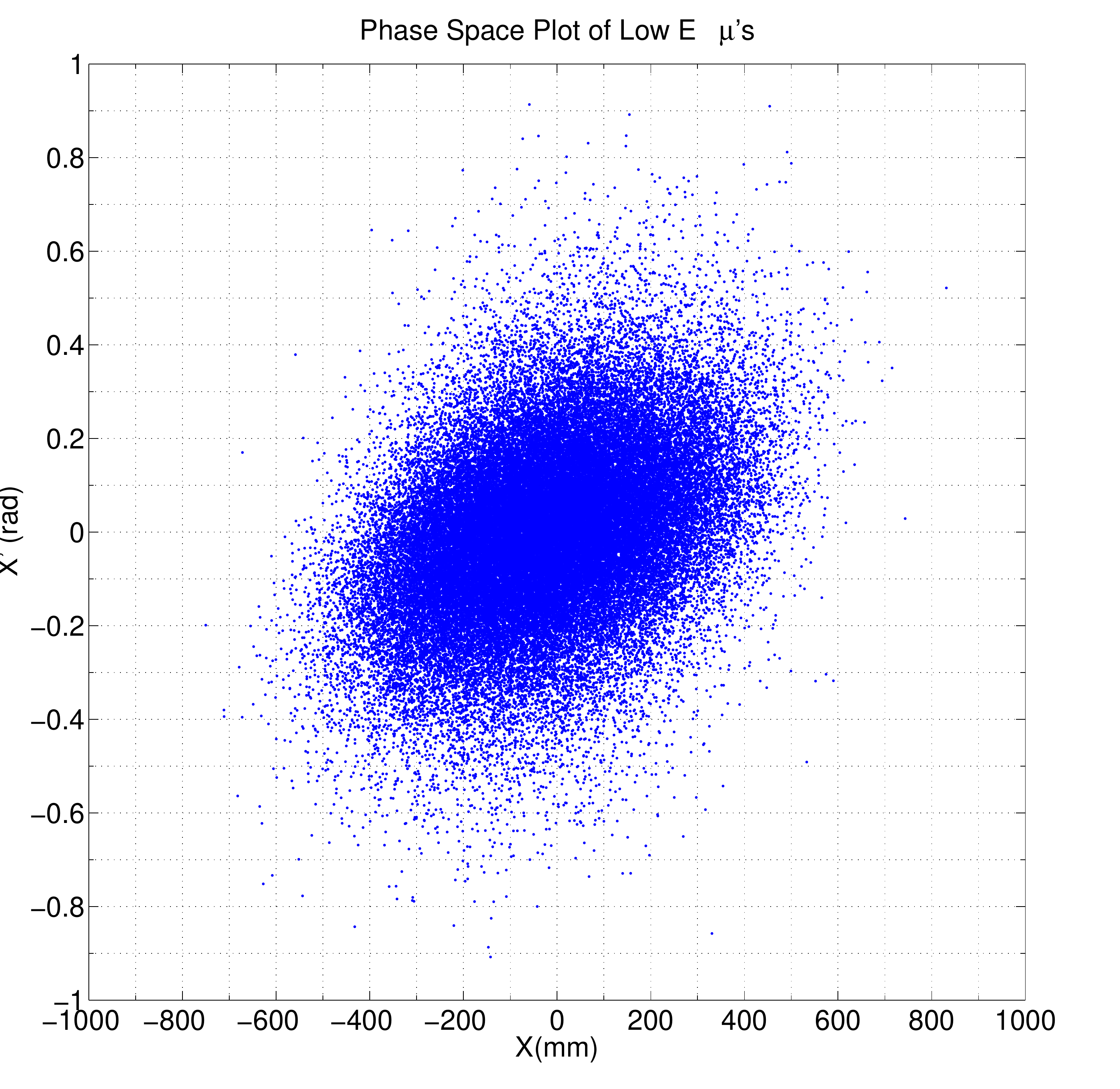}
  \end{center}
  \caption{
    Phase-space of the muon beam as it leaves the degrader.
    Left panel: $x-y$ distribution; Right panel: $x-x^\prime$ 
    distribution.
  }
  \label{fig:XY-dist}
\end{figure}

\begin{figure}
  \begin{center}
    \includegraphics[width=0.98\textwidth]%
      {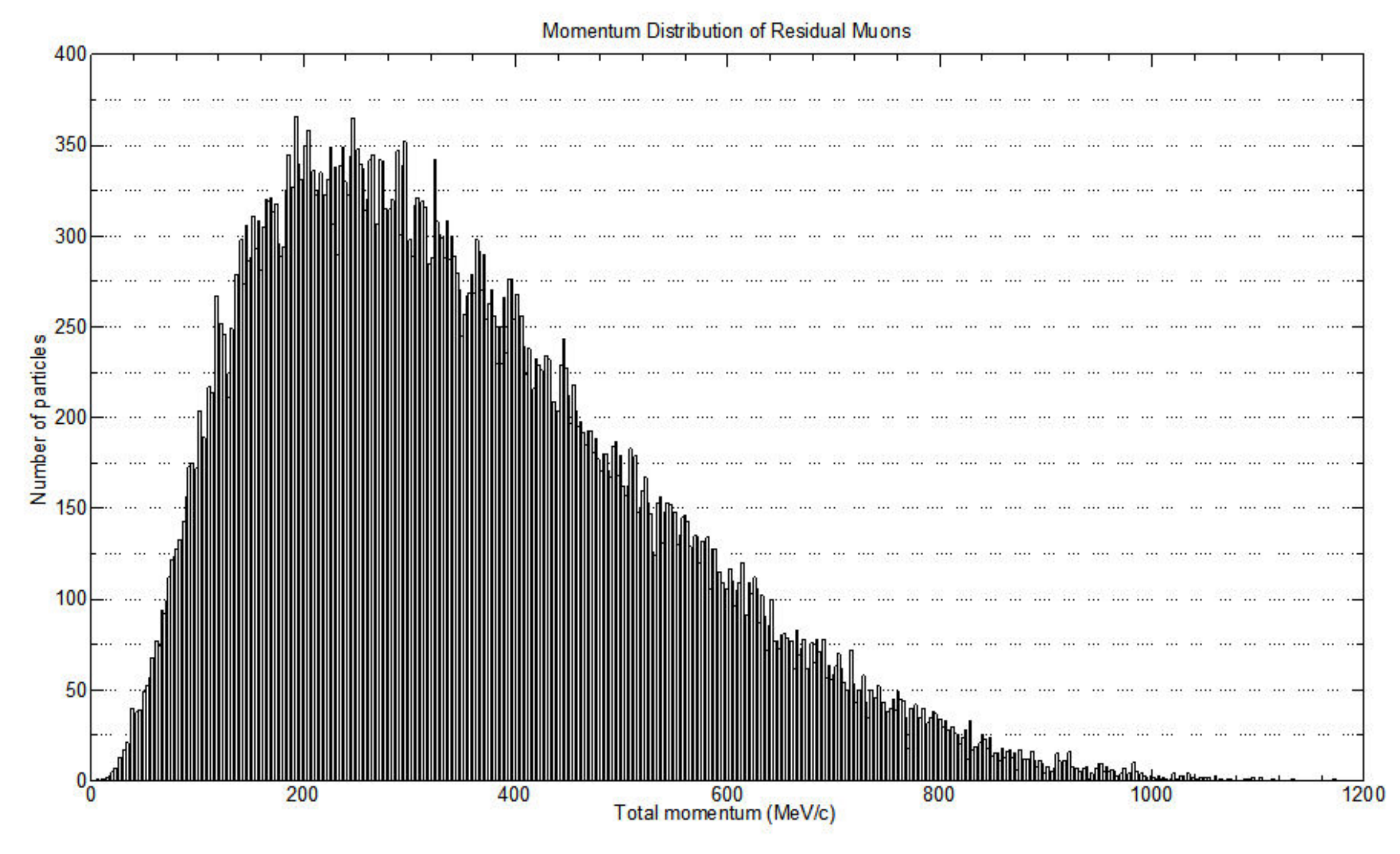}
  \end{center}
  \caption{Muon momentum distribution after degrader.}
  \label{fig:lowE}
\end{figure}

However, if the absorber is ``redefined'' to be a ``degrader" capable
of stopping the pions but allowing muons above a certain energy to
pass, then a low-energy muon beam appropriate for a 6D muon cooling
experiment can be produced.  
The left panel of Fig.~\ref{fig:degrader} shows the momentum
distribution for the first pass of muons at the end of the decay-ring
straight.
The green band indicates the momentum acceptance of the decay ring.  
The red band covers the same momentum band as the input pions and these
muons will be extracted along with the remaining pions.
If the degrader is sized appropriately, a muon beam of the desired
momentum for a 6D cooling experiment will emerge downstream of the
degrader.
The right panel of Fig.~\ref{fig:degrader} shows a visualization of a
G4Beamline \cite{G4bl} simulation of the
muons in the pion momentum band ($5\pm10\%$\,GeV/c) propagating
through a 3.48\,m thick iron degrader. 
The left panel of Fig.~\ref{fig:XY-dist} shows the $x-y$ distribution
of the muon beam exiting the degrader while the right panel shows the
$x-x^\prime$ distribution.
Figure \ref{fig:lowE} shows the muon momentum distribution of the
muons that exit the degrader.
Our initial estimate is that, in the momentum band of interest for a six-dimensional (6D)
cooling experiment (100--300\,MeV/c), we will have approximately
$10^{10}$ muons in the 1.6$\mu$sec spill.

\begin{figure}
	\centering
		\includegraphics[width=0.4\textwidth]{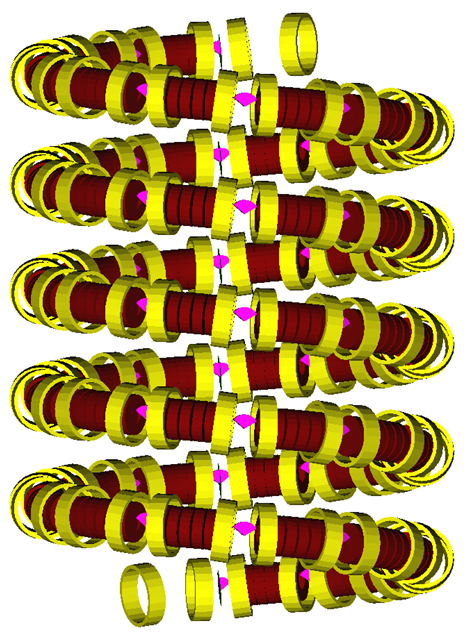}
	\caption{Layout of the initial stages of the Guggenheim cooling channel. Yellow: magnetic coils to generate focusing and bending field required for emittance exchange; magenta: liquid Hydrogen wedge absorbers to reduce momentum; red: RF cavities to restore momentum lost in the absorbers.}
	\label{fig:guggenheim}
\end{figure}

\begin{figure}
	\centering
		\includegraphics[width=0.6\textwidth]{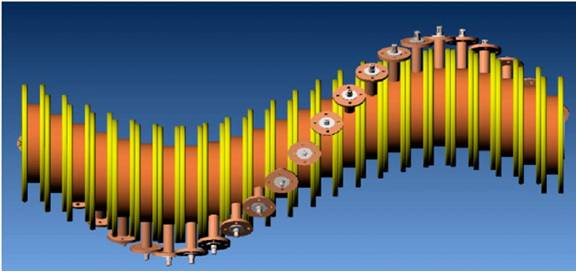}\\
		\includegraphics[width=0.6\textwidth]{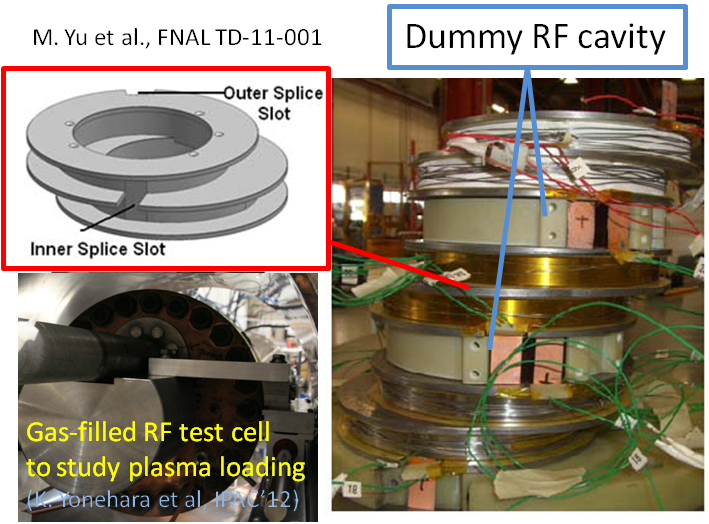}
	\caption{Top: conceptual drawing of the Helical Cooling Channel; bottom: test coil assembly producing helical solenoid field.}
	\label{fig:hcc}
\end{figure}

\begin{figure}
	\centering
		\includegraphics[width=0.49\textwidth]{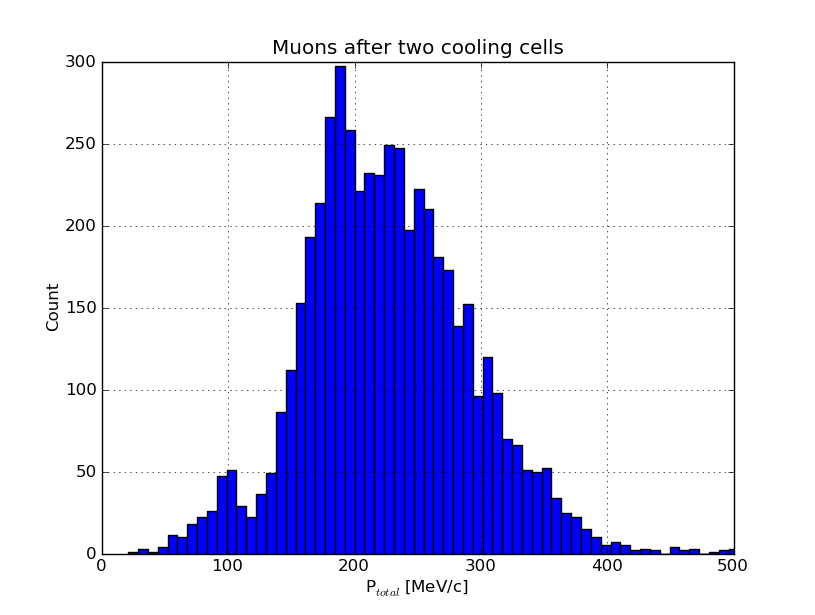}
		\includegraphics[width=0.49\textwidth]{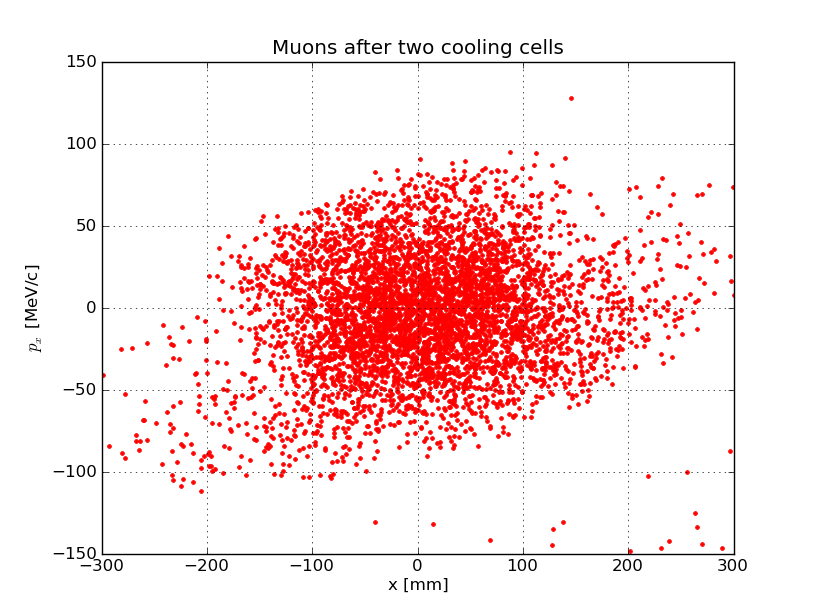}\\
		\includegraphics[width=0.49\textwidth]{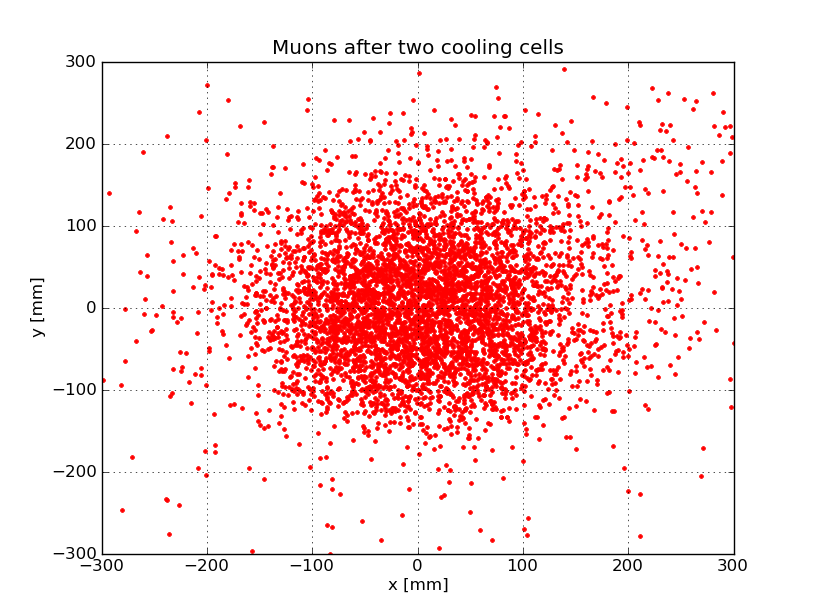}
	\caption{Top-left: muon momentum distribution after passing through two cells of the initial 201 MHz Guggenheim cooling channel; Top-right: $(x-p_x)$ phase portrait of the distribution; bottom: $(x-y)$ phase portrait of the distribution.}
	\label{fig:after_cooler}
\end{figure}

Advanced R\&D on the high intensity 6D ionization cooling channel required for a Muon Collider could be pursued using the nuSTORM facility and this muon beam.  The two key 6D cooling channels currently under detailed study can be tested at the nuSTORM facility without affecting the main neutrino activities: the Guggenheim and the Helical Cooling Channel (layouts in Fig.~\ref{fig:guggenheim}--\ref{fig:hcc}). After selection of one of these cooling schemes and a successful bench test, the hardware for the section of the cooling channel long enough to demonstrate 6D cooling could be set up at the nuSTORM facility in order to run a test demonstration experiment with the intense muon  beam.  Preliminary studies with one and two cells of both the initial (with 201 MHz RF) and the final (with 805 MHz RF) stages of the Guggenheim cooling channel suggest  promising muon transmission rates. These are summarized in Table~\ref{tab:transmission}. The corresponding momentum distribution and phase portraits are shown in Fig.~\ref{fig:after_cooler}.

\begin{table}
	\centering
	    \caption{Percentage of muons surviving one and two cells of the 201 MHz and 805 MHz Guggenheim channel, and its dependence on the degrader length.}
		\begin{tabular}{|c|c|c|c|c|}
		  \hline
			Degrader & 201 MHz, & 201 MHz, & 805 MHz, & 805 MHz, \\
			\mbox{[mm]} & one cell & two cells & one cell & two cells \\
			\hline
			3500 & 24\% & 8.6\% & 4.7\% & 0.6\% \\
			3480 & 24\% & 8.5\% & 4.9\% & 0.6\% \\
			3460 & 24\% & 8.4\% & 5.1\% & 0.6\% \\
			\hline
		\end{tabular}
	\label{tab:transmission}
\end{table}

\clearpage
\section{nuSTORM Conventional Facilities}
\label{sec:site}
The nuSTORM Conventional Facilities are anticipated to consist of six (6) functional areas consisting of the Primary Beamline, Target Station, Transport Line/Muon Decay Ring, Near Detector Hall, Far Detector Hall and the Site Work.

\begin{figure}[h]
  \centering{
    \includegraphics[width=0.9\textwidth]{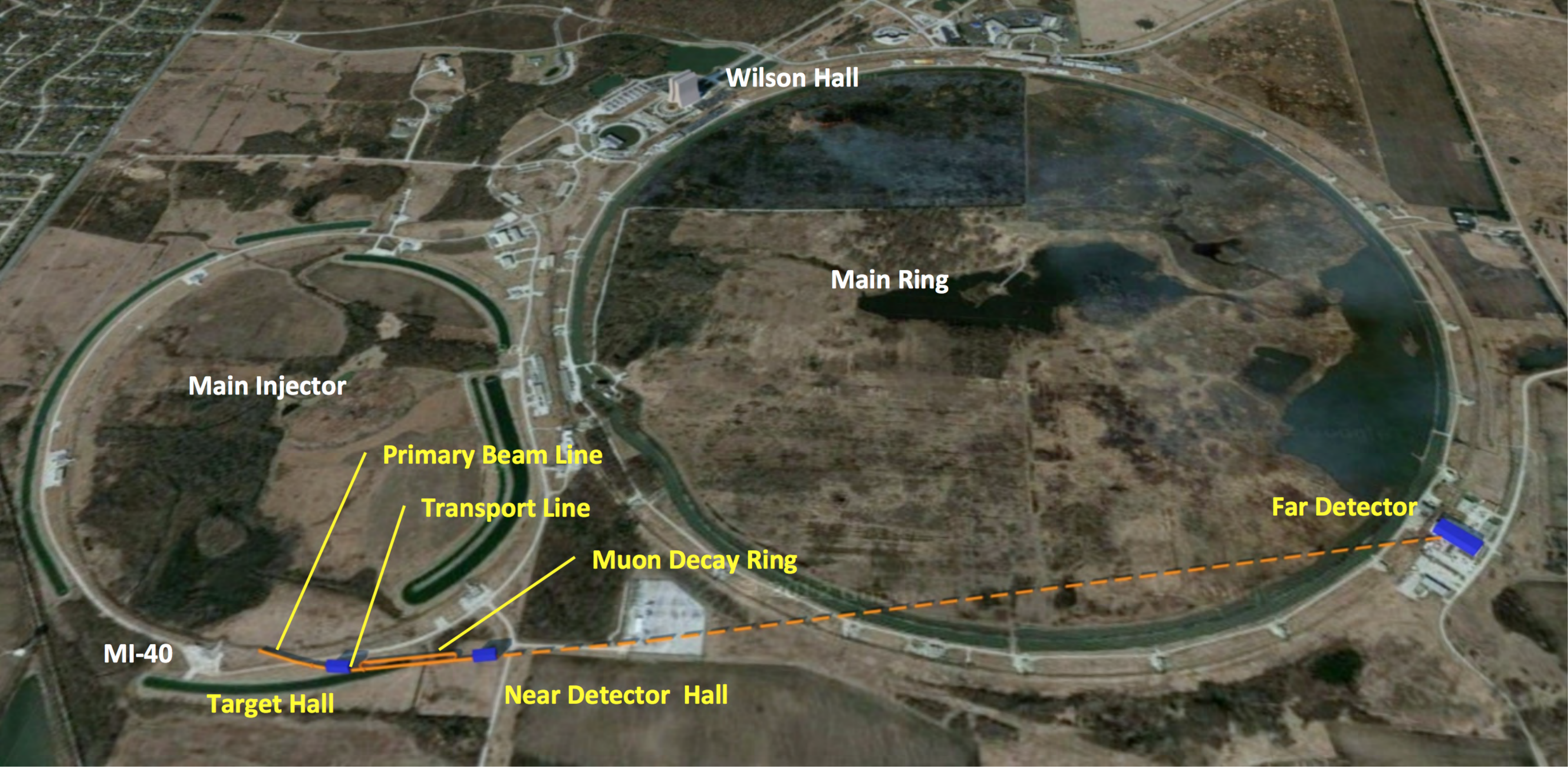}}
  \caption{Fermilab site view looking North and showing the nuSTORM facilities.}
  \label{fig:Site}
\end{figure}

\begin{figure}[h]
  \centering{
    \includegraphics[width=0.9\textwidth]{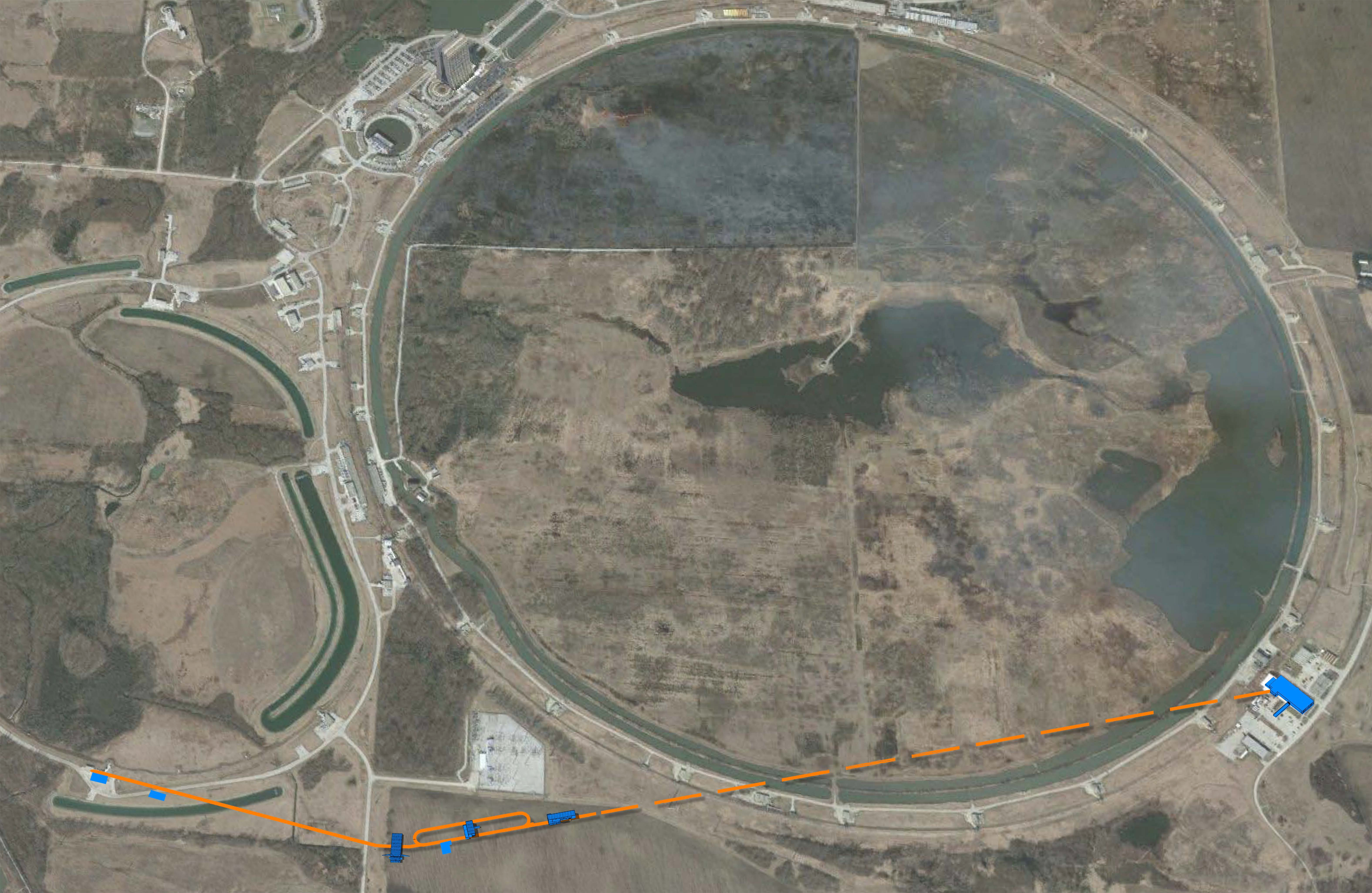}}
  \caption{Fermilab site view looking North and showing the nuSTORM facilities at the East site.}
  \label{fig:SiteEast}
\end{figure}

\begin{figure}[h]
  \centering{
    \includegraphics[width=0.9\textwidth]{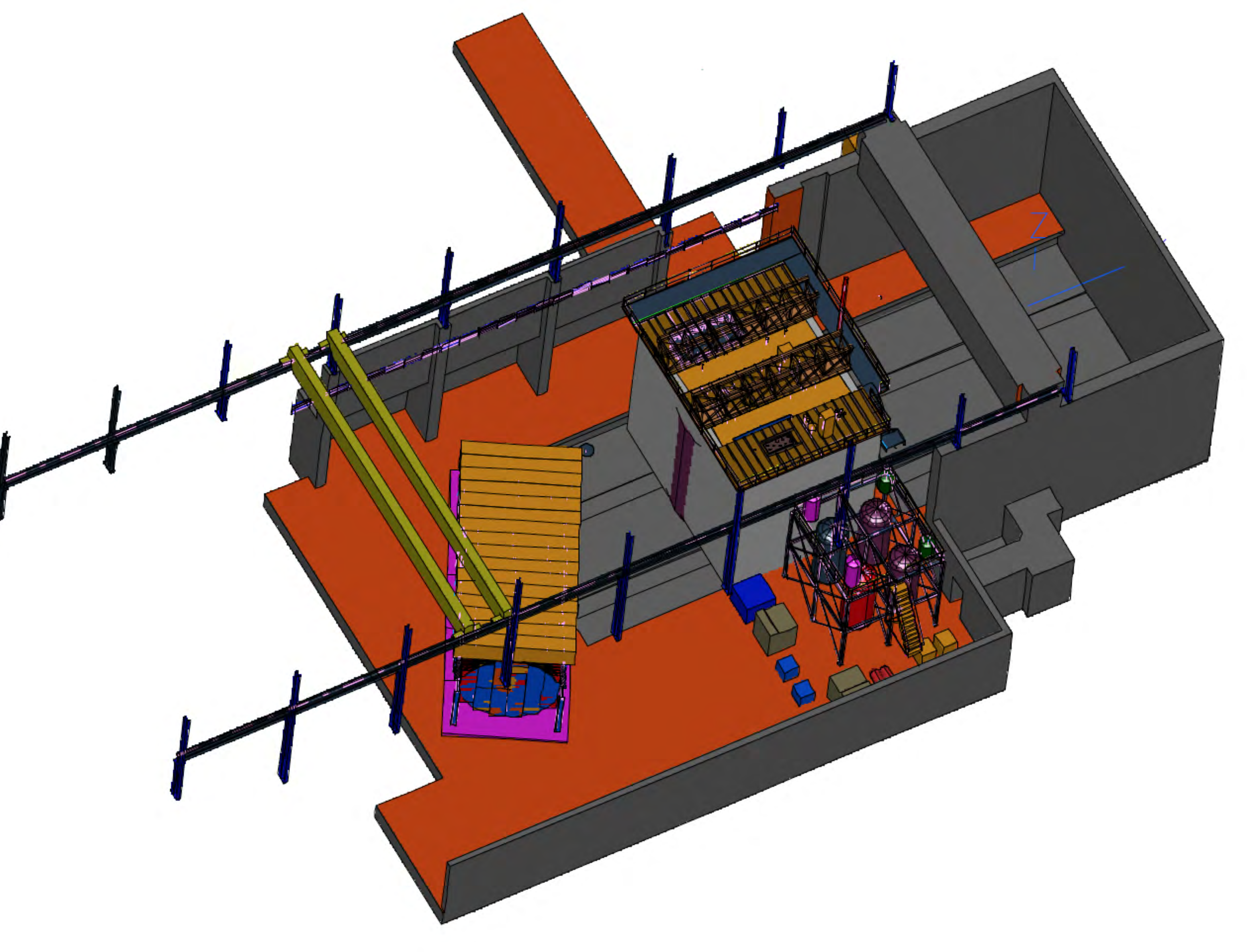}}
  \caption{Schematic of far detector hall (DAB) showing SuperBIND and demonstrating that there is room to add another detector (a conceptualized kT-scale magnetized LAr detector is shown) at some future date.}
  \label{fig:DAB}
\end{figure}

The facilities will be located in an area south of the existing Main Injector accelerator and west of Kautz Road on the Fermilab site.  In general terms, a proton beam will be extracted from the existing Main Injector at the MI-40 absorber, directed east towards a new below grade target station, pion transport line and muon decay ring.  The neutrino beam will be directed towards a Near Detector Hall located 20 m East of the muon decay ring and towards the Far Detector located approximately 1900 m away in the existing D0 Assembly Building (DAB).  Fig.~\ref{fig:Site}, below, shows a site photo with the nuSTORM Conventional Facilities superimposed.
We have also considered an East site for nuSTORM which would put all of the facility East of Kautz Road.   This eliminated any potential interference with existing infrastructure near the main site, in the event there would be significant future expansions to the nuSTORM facility (see  Fig.~\ref{fig:SiteEast}), but
requires a long ( $\simeq$ 510 m) primary proton beam line from the Main Injector.  The current West site does not impact existing infrastructure, while still allowing for future expansion.  Full details of the nuSTORM Conventional Facilities are given in the nuSTORM project definition report \cite{nuPDR:2013}.
DAB provides an ideal space for location of the far detector(s) for the short-baseline oscillation program.  With relatively minor retro-fitting, the DAB pit area can accommodate a 1.3 kT Fe-scintillator detector (SuperBIND, see Section~\ref{sec:far}) while providing enough space for a future kT-scale magnetized LAr detector.  Fig.~\ref{fig:DAB} shows a 3D view of the setup in DAB indicating the positions of SuperBIND (left) and a conceptualized LAr detector set next to the D0 experimental hall.   The height of the area allows  approximately 6' of heavy concrete overburden below the crane (indicated in the figure above the SuperBIND detector).

\clearpage
\section{Far Detector - SuperBIND}
\label{sec:far}
The Super B Iron Neutrino Detector (SuperBIND) is an iron and scintillator sampling 
calorimeter which is similar in concept to the MINOS detectors \cite{Michael:2008bc}.
We have chosen a cross section of  6 m in order to maximize the ratio of 
the fiducial mass to total mass.  The magnetic field will be toroidal as in MINOS and 
SuperBIND will also use extruded scintillator for the readout planes.  Details on the 
iron plates, magnetization, scintillator, photodetector and electronics are given below.
Fig.~\ref{fig:SuperBIND} gives an overall schematic of the detector.
\begin{figure}[hbtp]
  \centering{
    \includegraphics[width=0.9\textwidth]{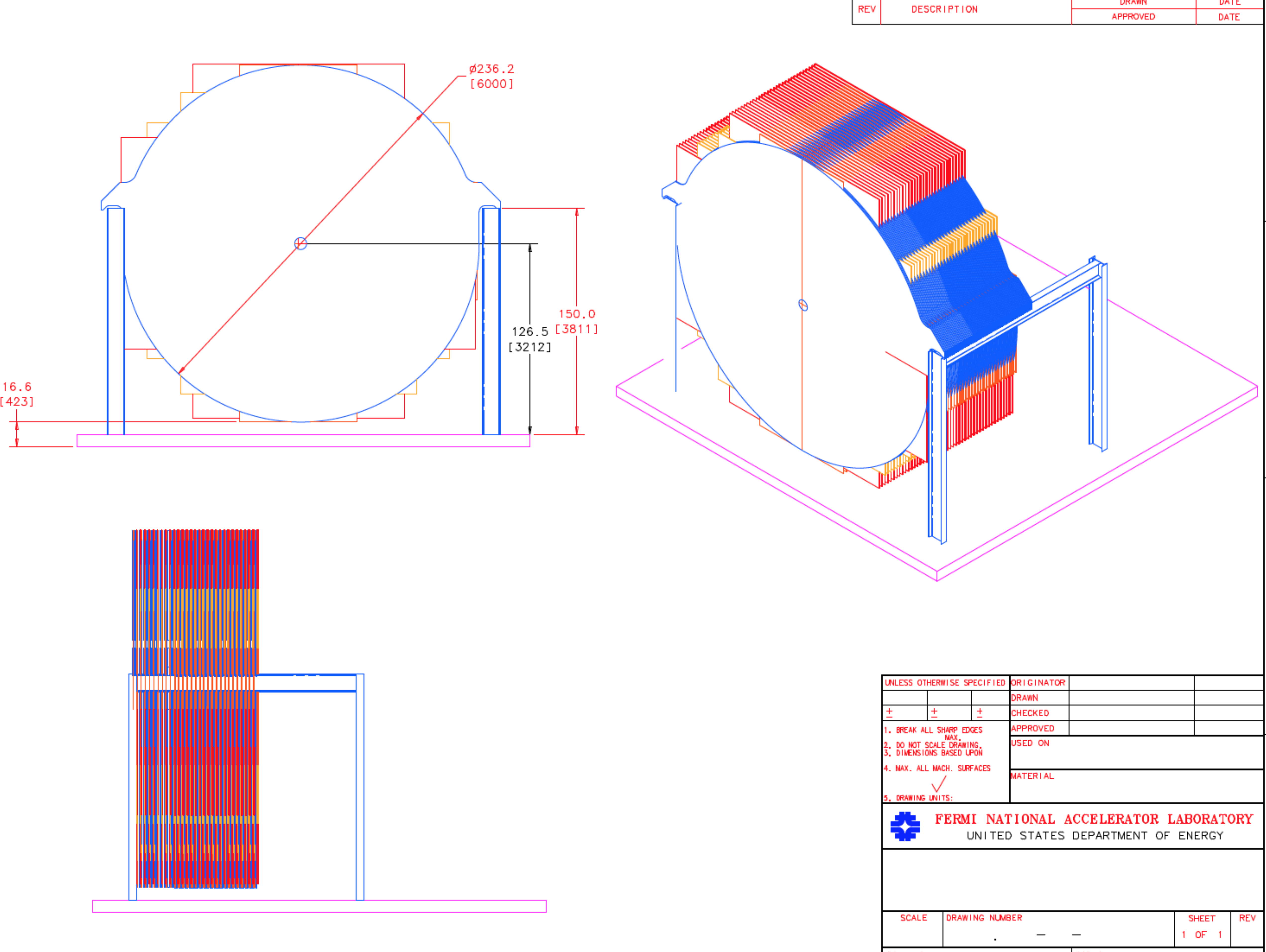}}
  \caption{Far Detector concept}
  \label{fig:SuperBIND}
\end{figure}
We note that within the Advanced European Infrastructures for Detectors at Accelerators (AIDA)
project \cite{AIDA}, whose time line runs from 2011 to 2015, detectors similar to those planned for
nuSTORM will be built and characterized at CERN. The motivation is to test the 
capabilities for charge identification of $\le$~5GeV/c electrons in a Totally Active 
Scintillator Detector and $\le$5~GeV/c muons in a Magnetized Iron Neutrino Detector (MIND). 
These detector prototypes will provide further experience in the use of STL technology, 
and SiPMs and associated electronics, to complement the already large body of knowledge 
gained through past and current operation of this type of detector.
\subsection{Iron Plates}
For the Iron plates in SuperBIND, we are pursuing the following design strategy.  
The plates are round with an overall diameter of 6 m and a thickness of 1.5 cm.
Our original engineering design used 1 cm plates, but we have simulated the detector
performance for  1 cm, 1.5 cm  and 2 cm thick plates.  The baseline is now 1.5 cm.  They are fabricated 
from two semicircles that are skip welded together.  We currently plan to hang the plates on ears as was done for MINOS, but
there is an option to stack the plates in a cradle using a strong-back when starting the 
stacking.  We envision that no R\&D on the iron plates will be needed.  Final specification 
of the plate structure would be determined once a plate fabricator is chosen.
\subsection{Magnetization}
\begin{figure}[hbtp]
  \centering{
    \includegraphics[width=0.9\textwidth]{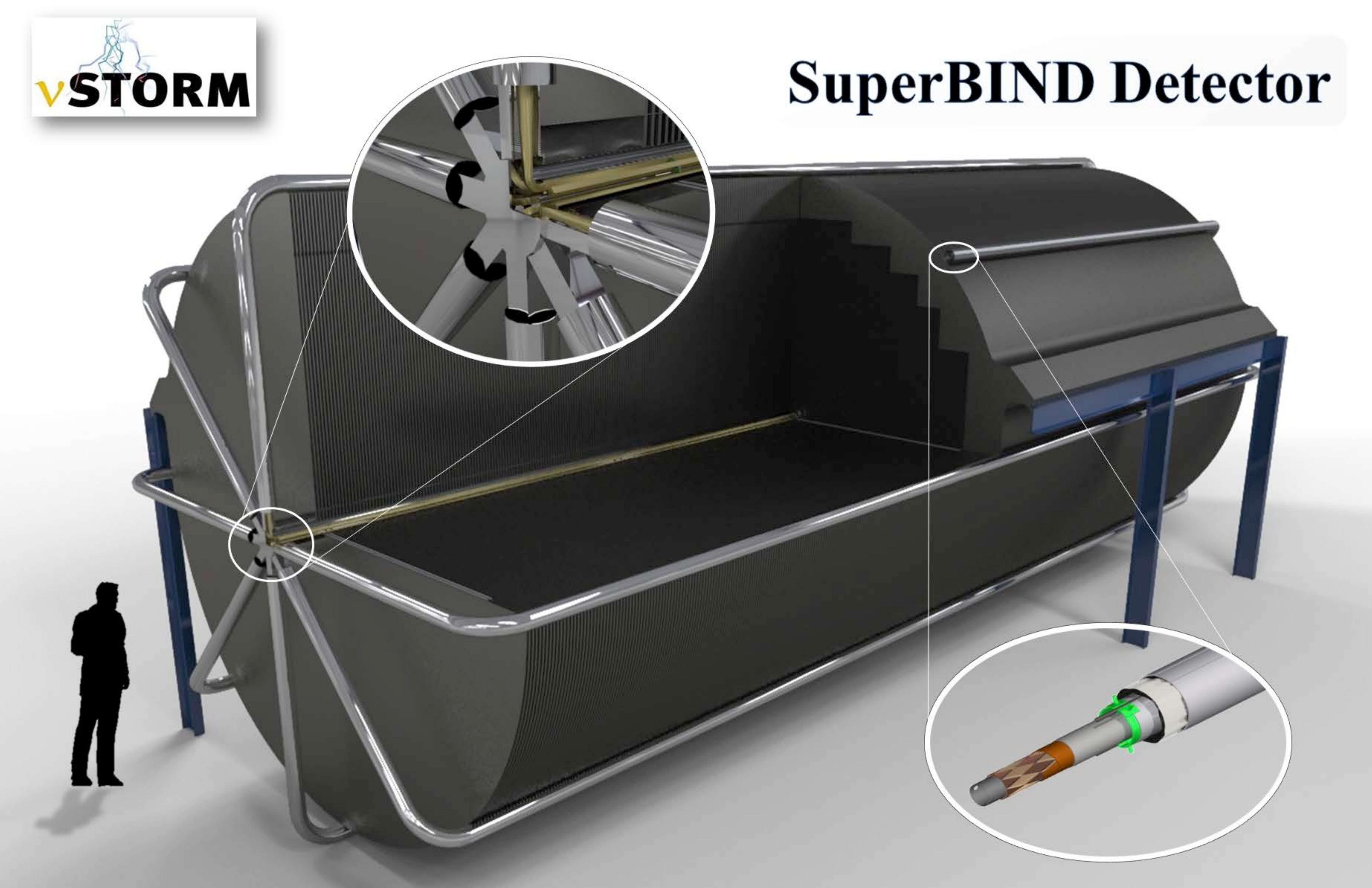}}
  \caption{Far Detector concept}
  \label{fig:SuperRend}
\end{figure}

As was mentioned above, MIND will have a toroidal magnetic field like that of MINOS.  
For excitation, however, we plan to use the concept of the Superconducting Transmission 
Line (STL) developed for the Design Study for a Staged Very Large Hadron 
Collider \cite{Ambrosio:2001ej}with possible extensions based on recent superconductor, cable-in-conduit development 
that has been carried out for ITER (see Section~\ref{sec:SCTLID}).  Minimization of the muon charge mis-identification rate requires the highest field
possible in the iron plates.  SuperBIND requires 
a much large excitation current per turn than that of the MINOS near detector 
(40 kA-turns).  We have simulated 240kA-turns (see Fig.~\ref{fig:SuperRend}).  The excitation circuit for SuperBIND will consist of 8 turns, each carrying 30kA.
The detailed magnetization implementation plan is described
in Section~\ref{sec:SCTLID}.
\subsubsection{Magnetic Field Map}
\begin{figure}[hbtp]
  \centering{
    \includegraphics[width=0.6\textwidth]{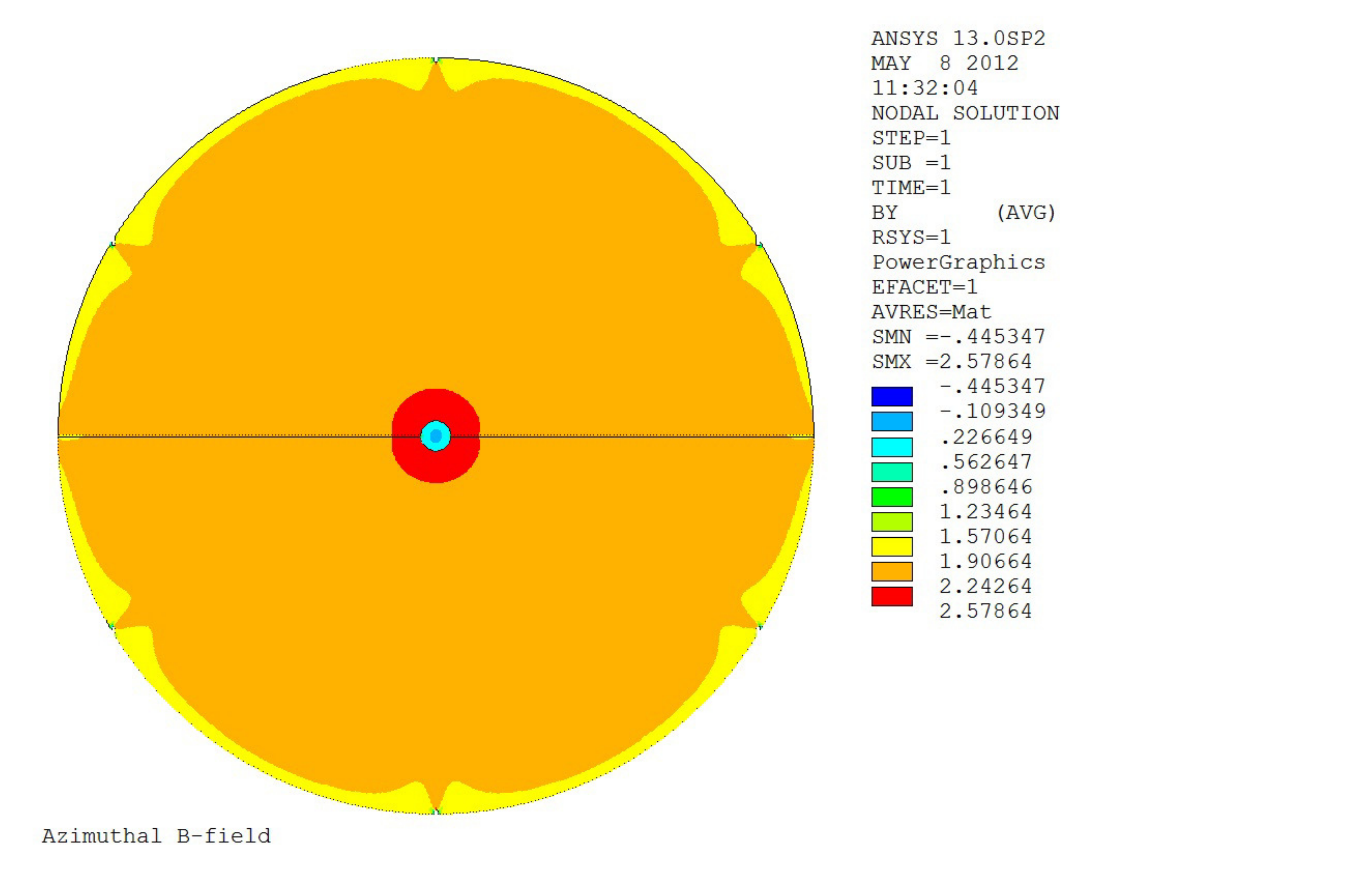}}
  \caption{Toroidal Field Map}
  \label{fig:b-Field}
\end{figure}

Utilizing the SuperBIND plate geometry shown in Fig.~\ref{fig:SuperBIND}, a 2-d finite 
element magnetic field analysis for the plate was performed.  Fig.~\ref{fig:b-Field} shows the 
results of those calculations.  For this analysis, a 20 cm diameter hole for the STL was assumed, the CMS steel 
\cite{Smith:2004uf} BH curve was used and an excitation current of 240 
kA-turn was assumed.
\subsubsection{SCTL implementation design}
\label{sec:SCTLID}
\begin{figure}[hbtp]
  \centering{
    \includegraphics[width=0.48\textwidth]{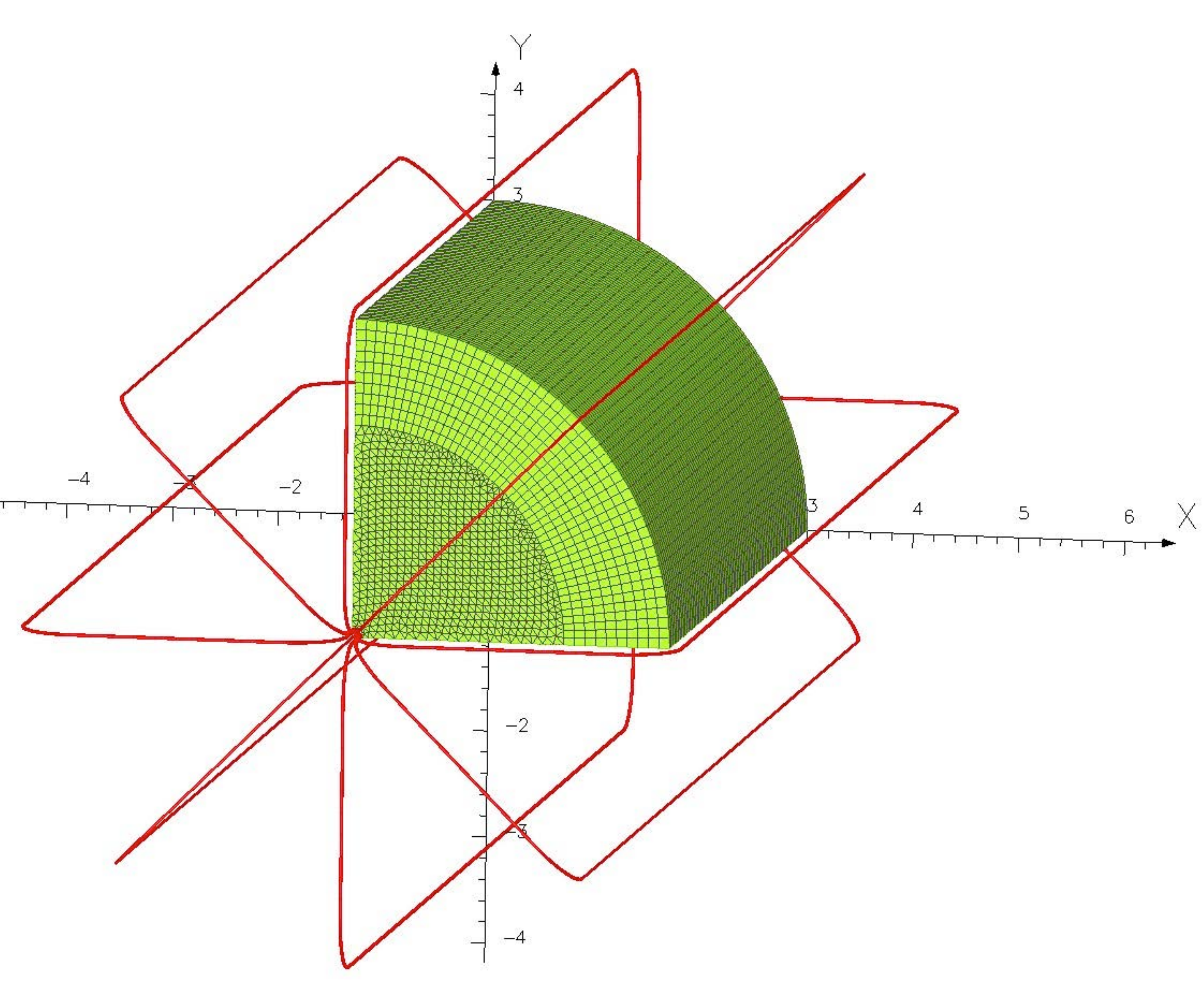}
    \includegraphics[width=0.48\textwidth]{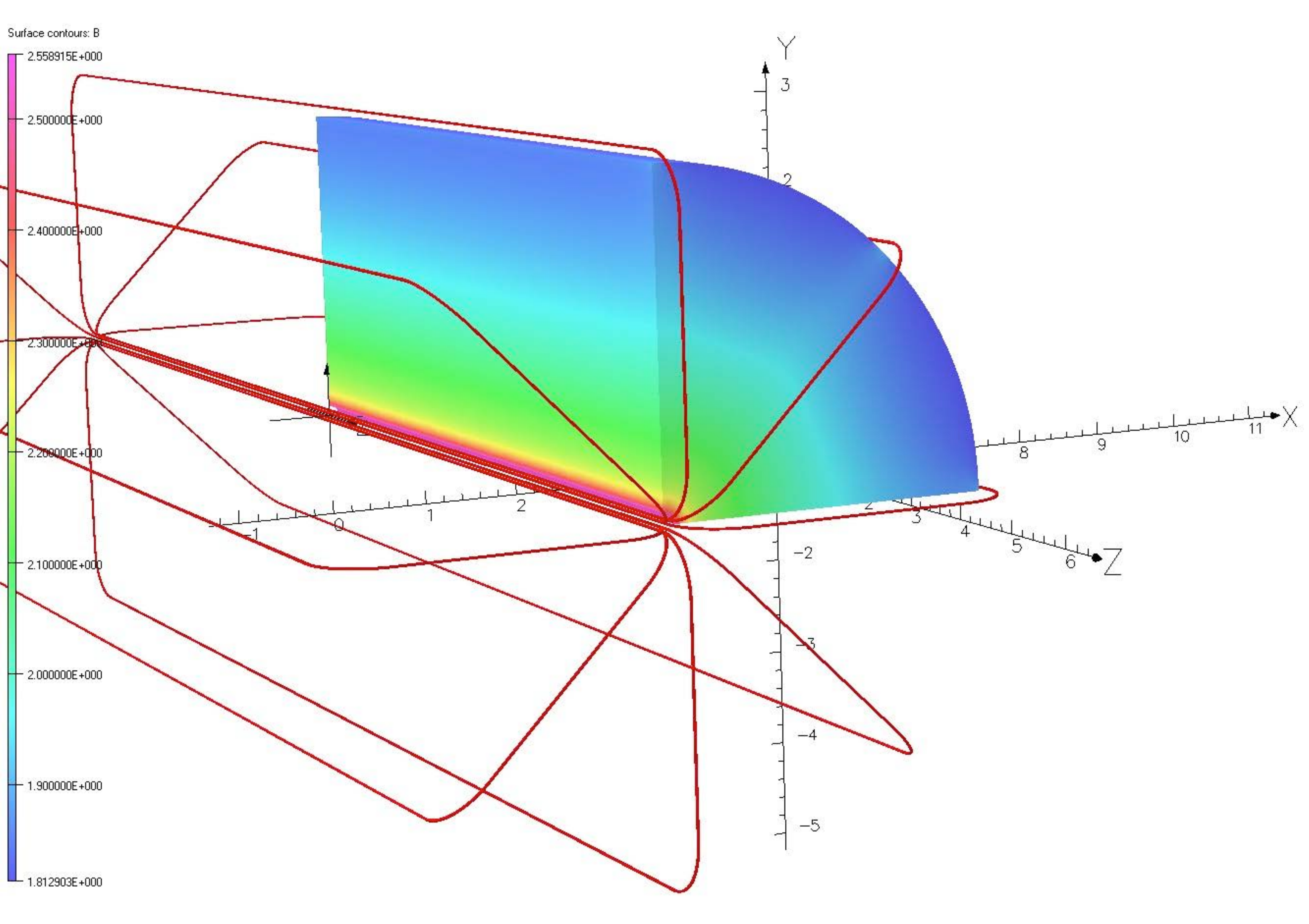}}
  \caption{Magnetic field calculation geometry (left), and flux density in the iron (right).}
  \label{fig:SBMag}
\end{figure}

\begin{figure}[hbtp]
  \centering{
    \includegraphics[width=0.9\textwidth]{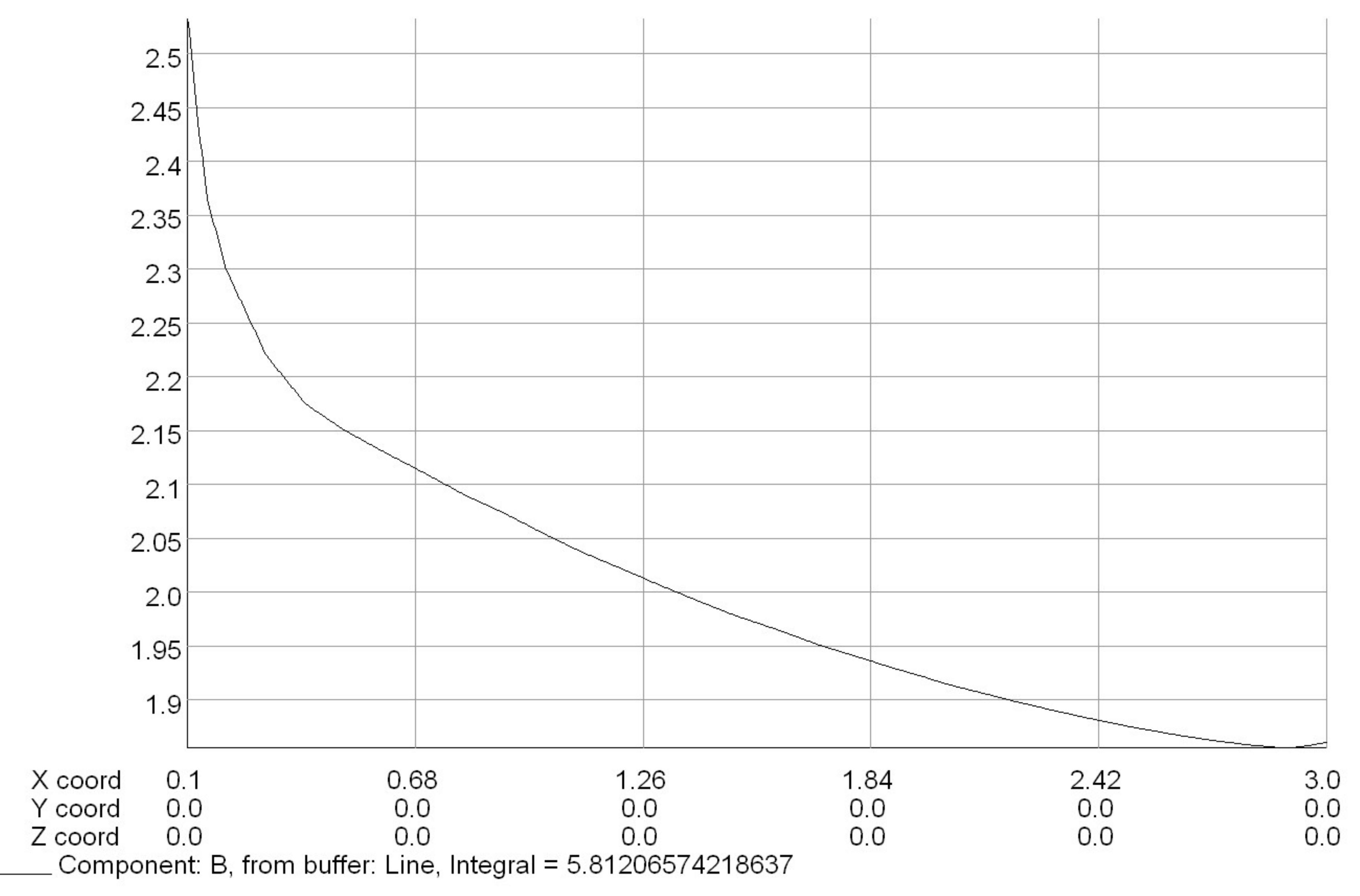}}
  \caption{Flux density distribution in the radial direction in the iron.}
  \label{fig:SBMag-flux}
\end{figure}

\begin{figure}[hbtp]
  \centering{
    \includegraphics[width=0.9\textwidth]{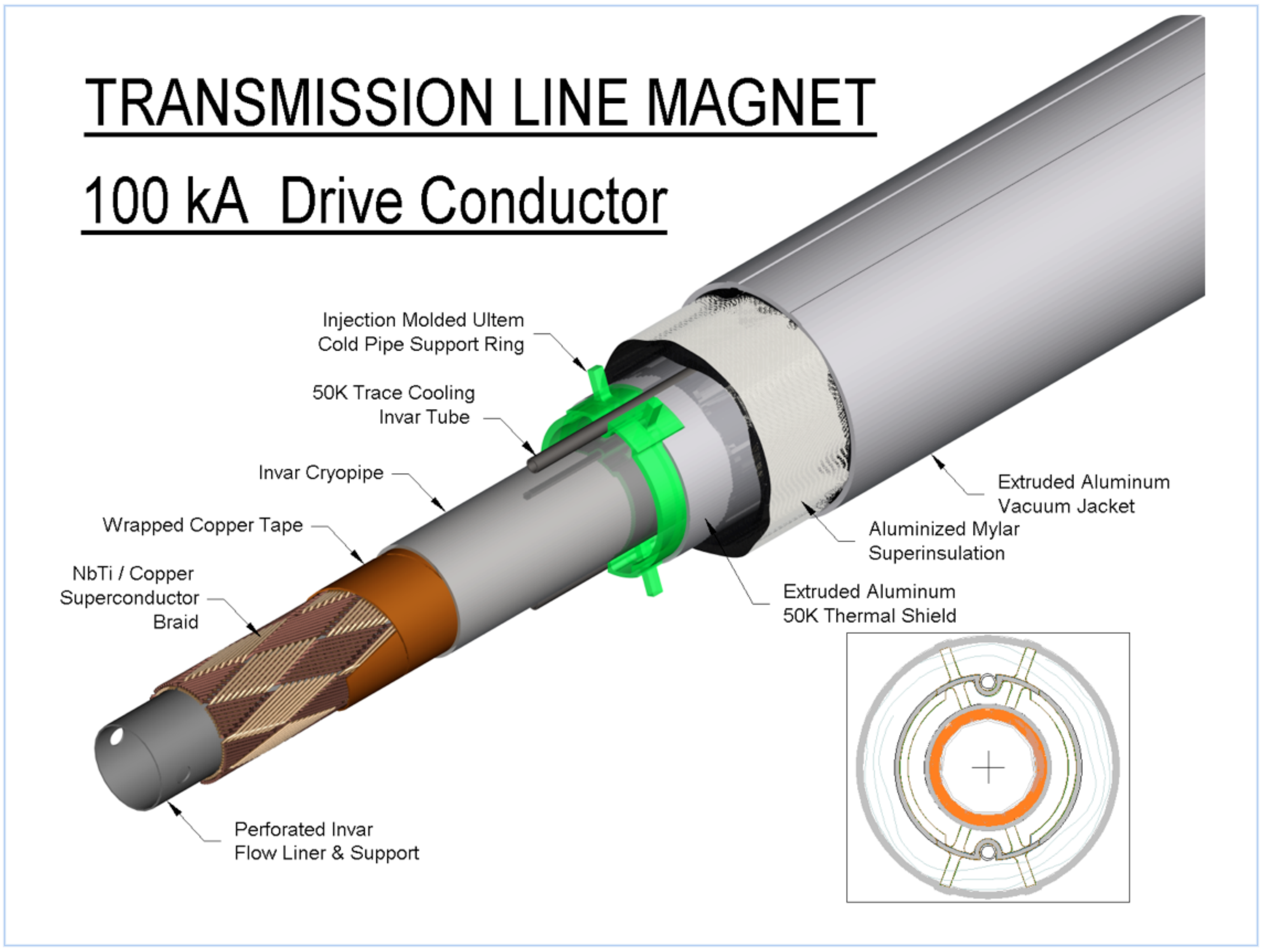}}
  \caption{Superconducting transmission line for the VLHC.}
  \label{fig:SCTL2}
\end{figure}

\begin{figure}[hbtp]
  \centering{
    \includegraphics[width=0.9\textwidth]{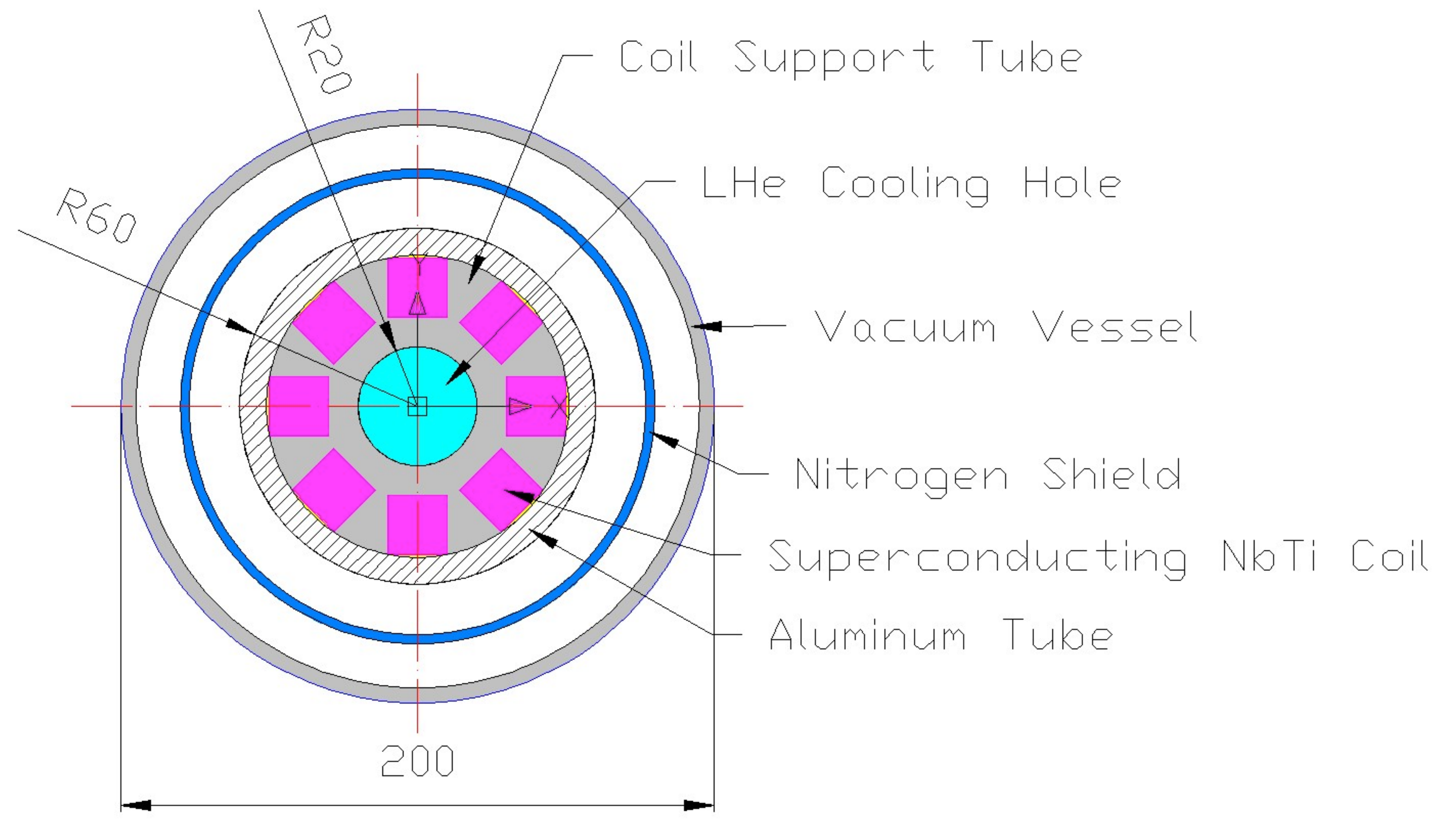}}
  \caption{The cross-section of the coil inner assembly.}
  \label{fig:InnerCryo}
\end{figure}

The superconducting magnet system for SuperBIND is based on NbTi superconductor cooled by the forced-flow LHe. Because of the dimensions 
of the detector (6 m diameter and 16 m length), it is reasonable to use the superconducting transmission line approach proposed by 
the Hon. G. W. Foster and successfully tested for the VLHC \cite{Ambrosio:2001ej,Piekarz:2005na}. This transmission line was capable of carrying up to 100 kA and was used to generate the 2 T field in the double aperture dipole magnet \cite{Foster:2001fx} of the VLHC. Because the line was very compact (80 mm OD), this approach is very cost effective in that it eliminates a large and expensive cryostat. The SuperBIND magnet is specified to be excited by 240 kA-turns. We have chosen a 30 kA current level for the transmission line superconductor in order to provide a more homogeneous field distribution in the iron core and to reduce the current and Lorentz forces on the superconductor. The total number of transmission line turns in this case is eight. The superconducting magnet system has parameters shown in Table~\ref{tab:SBmag}.
\begin{table}
\centering
\caption {Magnet parameters for SuperBIND.}
\label{tab:SBmag}
\begin{tabular}{|lrl|}
\hline
Name  & Unit & Value\\
\hline\noalign{\smallskip}
Iron core outer diameter &	6.0	& m\\
Iron core inner diameter	&	0.2	& m\\
Iron core length 		&	15.82	& m \\
Iron plate thickness		& 	15	& mm\\
Number of plates		&      440       & \\
Space between plates	&	21	& mm\\
Number of superconducting racetrack coils &	8	& \\
Superconducting cable length	& 320	& m\\
Racetrack coil current	& 30		& kA\\
Total current			& 240 &	 kA-turns\\
Peak field on the coil	& 0.83		& T\\
Inductance			& 40	& mH\\
Total stored energy		& 18		& MJ\\
\noalign{\smallskip}\hline
\end{tabular}
\end{table} 
The magnet was modeled by the TOSCA OPERA3d code. The model geometry is shown in Fig.~\ref{fig:SBMag}.
Because the iron plates are spaced by approximately 21 mm, the iron properties in the z direction were modeled as anisotropic, with a packing factor of 0.417. The effective flux density distribution is shown in Fig.~\ref{fig:SBMag-flux}.  The superconducting transmission line for the VLHC is shown in Fig.~\ref{fig:SCTL2}.
For the proposed toroidal magnet system configuration, the most critical parameters are the maximum field and the Lorentz forces on the superconducting coils. The peak effective magnetic field in the iron is 2.5 T which drops to 1.85 T at a radius of 3 m. The peak field on the superconductor is 0.83 T on the coil inner radius. The radial Lorentz force component on the inner conductor is -271 kN (-1700 kg/m), and on the outer, 40 kN (250 kg/m). The longitudinal force component for the radial conductors is 2.4 kN (80 kg/m). These force loads (except the force on the inner conductor) are in the range of the capability of the VLHC transmission line design where the most critical elements are the Ultem rings (spiders) supporting the cold cable mass inside the vacuum shell (See Fig.~\ref{fig:SCTL2}).
The inner part of coil will see large radial forces directed to the center. This is why a stainless steel slotted tube is needed in the center (See Fig.~\ref{fig:InnerCryo}).
The inner coil assembly cold mass has a 120 mm diameter and is cooled by forced-flow LHe through a central hole of  40 mm diameter. The cold mass is supported by Ultem rings in the same way as shown in Fig.\ref{fig:SCTL2}. Eight straight NbTi bars are mounted inside the slotted central tube with the outer aluminum tube. The whole cold-mass assembly is impregnated with an epoxy resin. Coil bars are electrically insulated from the supporting tube and an outer aluminum shell. During cool down, the Al shell provides a pre-stress for the coil, which increases the superconducting coil mechanical stability. The cold mass is mounted inside a stainless steel vacuum vessel of 200 mm diameter and includes a nitrogen shield.   Super-insulation is wrapped around the cold mass and the nitrogen shield.

The most complicated part of the design is the transition area between the inner part of the coil and eight transmission lines surrounding the iron core (detector), which are electrically connected in series. The final superconductor splices will have to be made after the assembly forming the eight turns is in place. The outer transmission line vacuum vessels should be rigidly attached to the detector iron with a high angular accuracy to eliminate transverse forces. An option that could be applied, in order to reduce the force on the outermost conductors, is to put them in slots in the iron.  

\begin{figure}[hbtp]
  \centering{
    \includegraphics[width=0.9\textwidth]{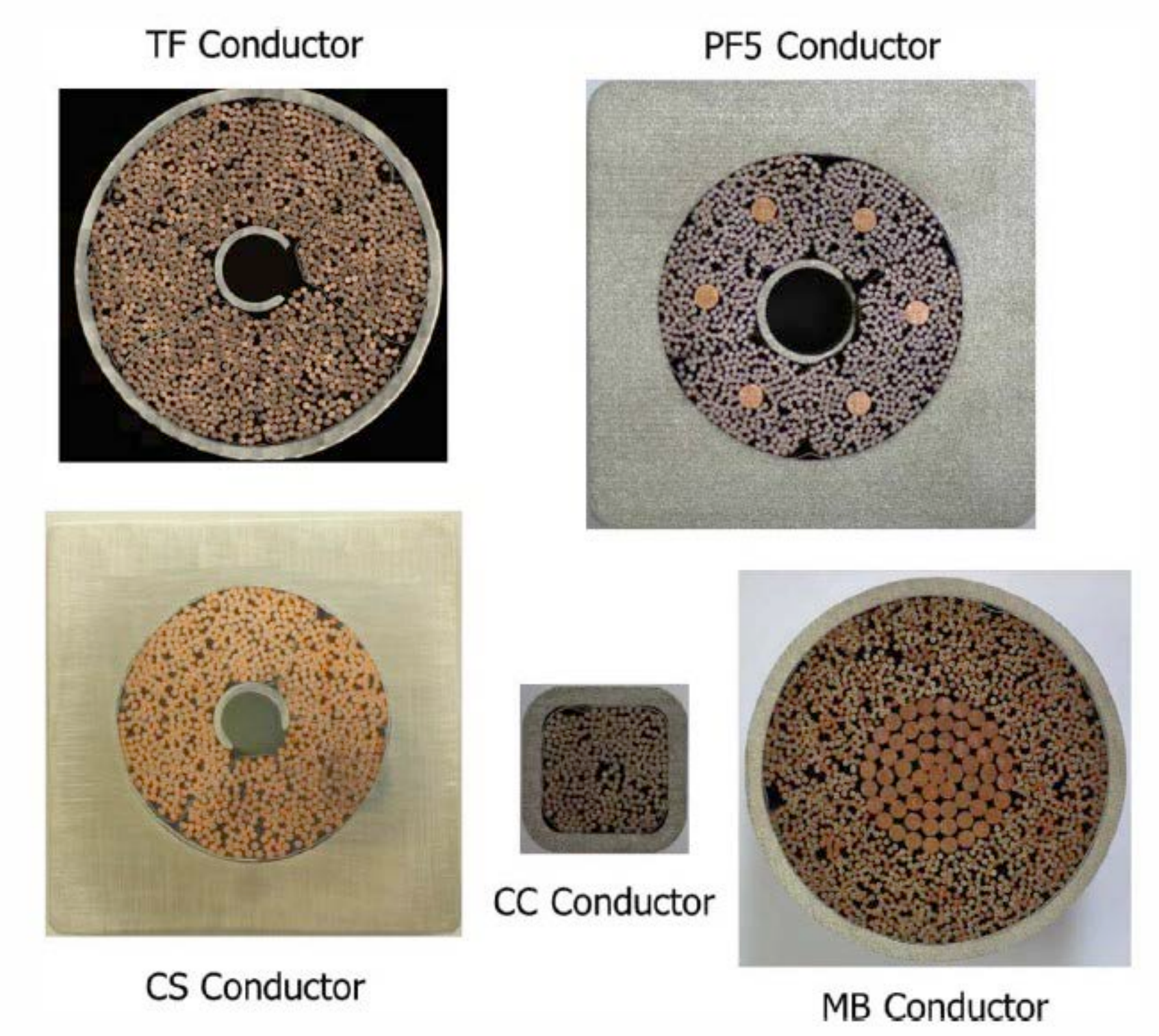}}
  \caption{ITER superconducting cables.}
  \label{fig:CICC}
\end{figure}

The NbTi superconducting coil cable could be assembled from SSC type Rutherford cables used in \cite{Foster:2001fx}, or an ITER type cable in a conduit conductor (CICC) \cite{Devred:2012x}. The NbTi CICC is widely used for ITER correction coils, poloidal coils, buses, manifold cables (See Fig.~\ref{fig:CICC}). The production technology and facilities are developed, and cables have been tested in Europe, Russia, and China \cite{Vysotsky:2012x,Foussart:2012x,Wesche:2009zz}. CICC has direct cooling through the cooling hole for large currents (Toroidal and Central Solenoid coils) or  cooling between cable strands (Correction coils, Busbars). 
The transmission line is one of the cost drivers of large superconducting magnet systems. For the VLHC it was estimated at \$500/m. For 320 m of total transmission line length in the case of SuperBIND, this would be \$160k.  The ITER project evaluated in-kind contributions for the correction coils cost of CICC at \$220/m \cite{Devred:2012x}. This is equivalent to the cost analysis that was done for the VLHC in that the estimation  also includes the cost of vacuum and shield tubes, Ultem spacers, and super-insulation. 
\subsection{Detector planes}

\subsubsection{Scintillator}

Particle detection using extruded scintillator and optical fibers is a mature technology.  
MINOS has shown that co-extruded solid scintillator with embedded wavelength shifting (WLS) 
fibers and PMT readout produces adequate light for MIP tracking and that it can be 
manufactured with excellent quality control and uniformity in an industrial setting \cite{Adamson:2002mj}.  Many 
experiments use this same technology for the active elements of their detectors, such as 
the K2K Scibar \cite{Maesaka:2003jt}, the T2K INGRID, P0D, and 
ECAL \cite{Kudenko:2008ia} and the Double-Chooz cosmic-ray veto detectors \cite{Greiner:2007zzd}.

Our  concept for the readout planes for SuperBIND is to have both a $x$ and a $y$ 
view between each Fe plate. Our simulations  are now using a scintillator extrusion 
profile that is 0.75 $\times$ 2.0 cm$^2$.  This gives both the required point resolution and 
light yield.  

\subsubsection{Scintillator extrusions}

The existing SuperBIND simulations have assumed that the readout planes will use an 
extrusion that is 0.75 cm $\times$ 2.0 cm.  A 2 mm hole down the centre 
of the extrusion is provided for insertion of the wavelength shifting fiber.  This is 
a relatively simple part to manufacture and has already been fabricated in a similar 
form for a number of small-scale applications.  The scintillator strips will consist 
of an extruded polystyrene core doped with blue-emitting fluorescent compounds, a 
co-extruded TiO$_2$ outer layer for reflectivity, and a hole in the middle for a WLS 
fiber.  Dow Styron 665 W polystyrene pellets are doped with PPO (1\% by weight) and 
POPOP (0.03\% by weight), which is the MINOS formulation. The strips have a white, co-extruded, 0.25 mm thick TiO$_2$ 
reflective coating.  This layer is introduced in a single step as part of a 
co-extrusion process.  The composition of this coating is 15\% TiO$_2$ 
in polystyrene.  In addition to its reflectivity properties, the layer facilitates 
the assembly of the scintillator strips into modules. The ruggedness of this coating 
enables the direct gluing of the strips to each other and to the module skins which 
results in labour and time savings.  This process has now been 
used in a number of experiments.

\begin{figure}[hbtp]
  \centering{
    \includegraphics[width=0.9\textwidth]{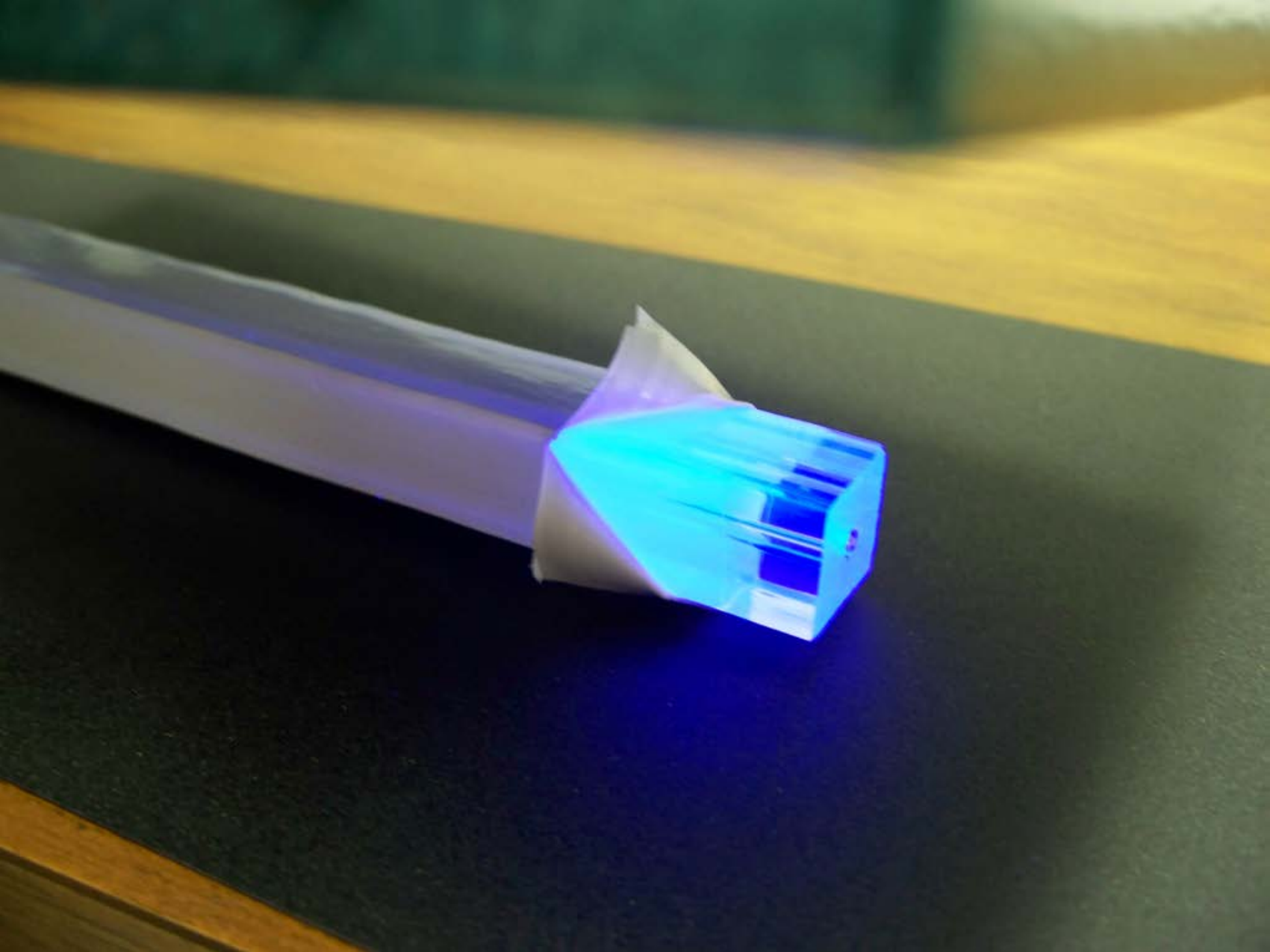}}
  \caption{TiO$_2$ doped LDPE coated scintillator extrusion.}
  \label{fig:PE_extru}
\end{figure}

Work is under way towards the development of a reflective coating using titanium dioxide (TiO$_2$) 
in low-density polyethylene (LDPE).  Because polyethylene (PE) and polystyrene (PS) are not miscible, 
the TiO$_2$-doped PE coating does not bind to the PS-based scintillator core during the co-extrusion process 
as it occurs in the current arrangement which uses a TiO$_2$-doped PS coating on the PS-based scintillator core.  
The TiO$_2$-doped PS coating provides a diffuse interface whereas the TiO$_2$-doped PE coating traps a layer of air; thus 
enabling a higher degree of total internal reflection of the light produced in the scintillator.  Preliminary 
light yield results using a radioactive source indicate that a PE coated scintillator sample has a light yield 
approximately 20\% higher than that of a scintillator sample wrapped in a layer of Tyvek which, until now, 
has been the best reflective material.  More effort is needed to determine the level of improvement in 
light yield, such as a direct comparison of the two TiO$_2$-doped  polymer coatings (PS and PE) and to study the conditions 
that a PE coating requires regarding gluing and assembly of the finished scintillator strips into detector modules.
Fig.~\ref{fig:PE_extru} shows a prototype scintillator extrusion with TiO$_2$-doped LDPE  cladding.

\subsection{Photo-detector}

Given the rapid development in recent years of solid-state photodetectors 
based on Geiger mode operation of silicon avalanche photodiodes, we have 
chosen this technology for SuperBIND.  Although various names are used for this 
technology, we will use silicon photomultiplier or SiPM.

\subsubsection{SiPM Overview}
SiPM, which stands for Silicon Photo Multiplier, is the common name for a relatively new type of photo detector formed by combining many small avalanche photodiodes operated in the Geiger mode to form a single detector  \cite{Sadygov:1996, Bacchetta:1996dc}.
The first generation of these detectors used a polysilicon resistor connected to each avalanche photodiode forming  a pixel. Pixel sizes vary, and can be anywhere from 10 $\times$ 10 microns to 100 $\times$ 100 microns. All the diodes are connected to a common electrical point on one side, typically through the substrate, and all the resistors are connected to a common grid with metal traces on the other side to form a two node device.  A typical SiPM will have from 100 to 10,000 of these pixels in a single device, with the total area of from 1 to 16 mm$^2$. Because all the diode and the individual quenching resistors are connected in parallel, the SiPM device as a whole appears as a single diode. In operation, the device appears to act somewhat like a conventional APD, but in detail it is radically different. Because the diodes are operated in the Geiger mode, and because every pixel of the SiPM device is nearly identical, the sum of the fired pixels gives the illusion of an analog signal that is proportional to the incident light, but it is an essentially digital device. Also, because the  individual pixels operate in Geiger mode, and are individually quenched, SiPMs are typically designed to operate at very large gains, of up to $10^6$, which significantly simplifies the electronics and reduces the costs and associated technical and schedule risks. 

\begin{figure}[hbtp]
	\begin{center}$
	  \begin{array}{cc}
		\includegraphics[width=0.4\textwidth]{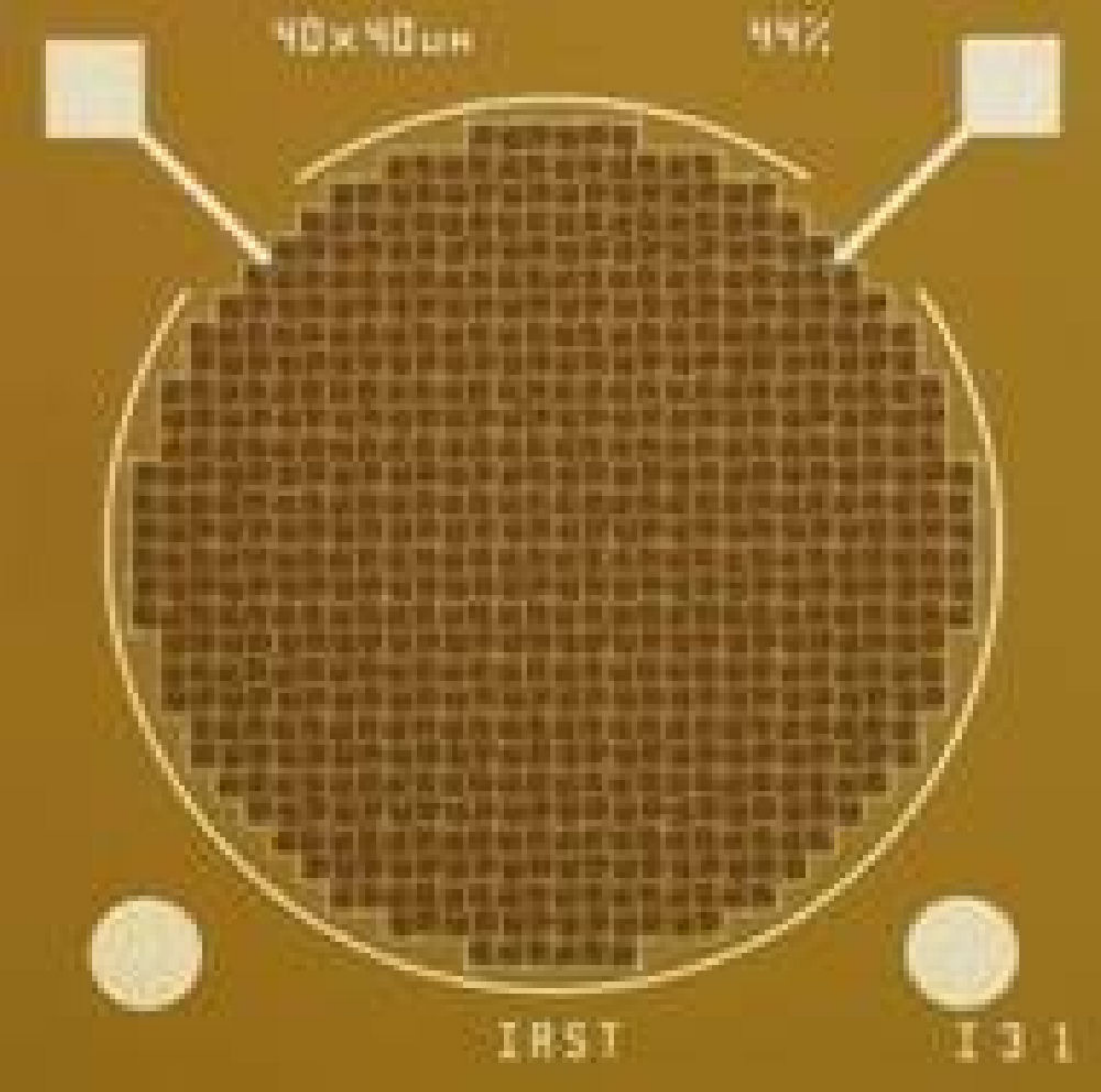} &
		\includegraphics[width=0.6\textwidth]{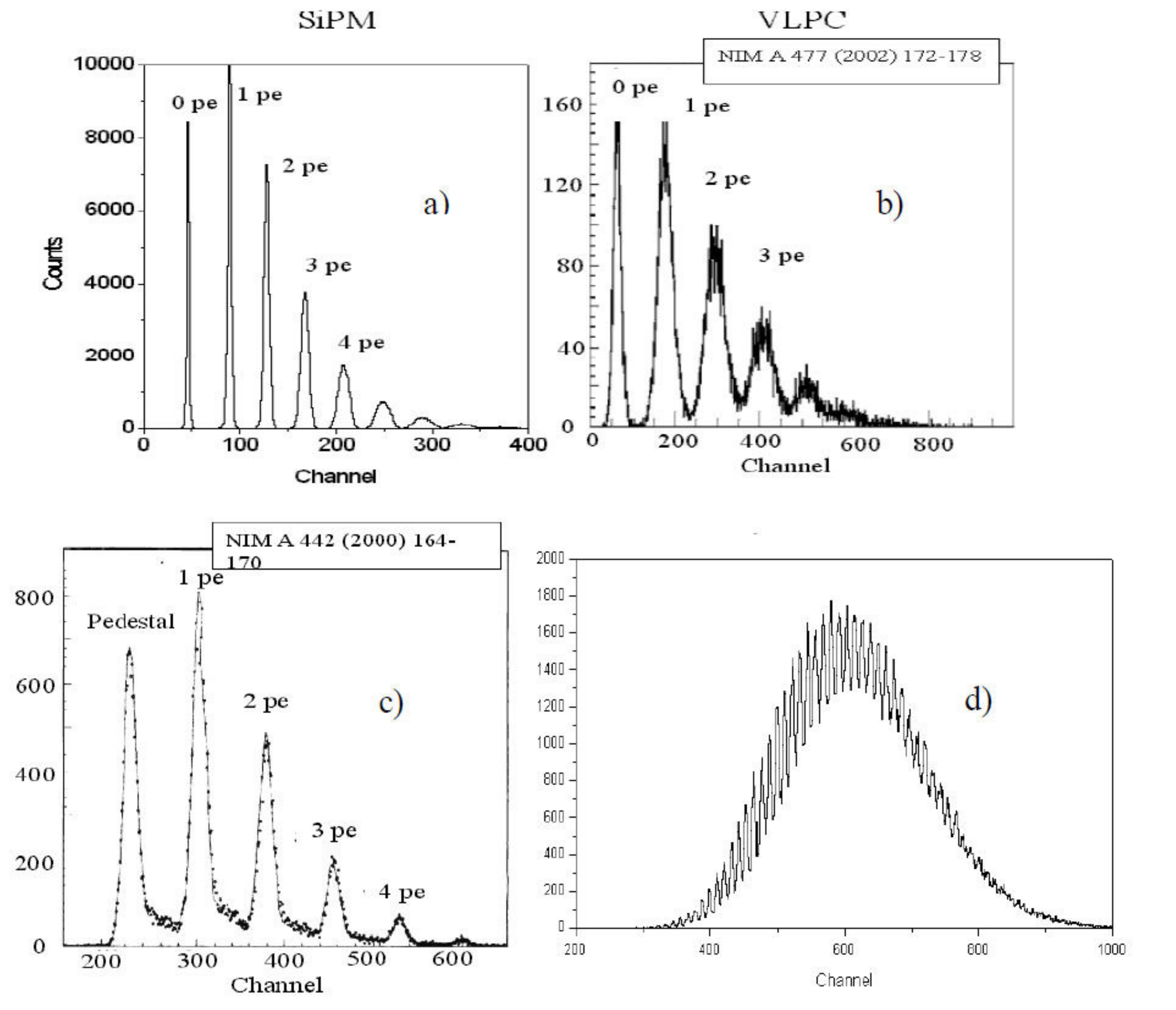}
	  \end{array}$
    \end{center}
  \caption{Photograph of SiPM (left) and SiPM photon counting capability (a) compared to 
    VLPC (b) and HPD (c) . The SiPM pulse height spectrum (d) for an intense 
	light burst with a mean photoelectron number of 46 is also shown.}
  \label{Fig:SiPM}
\end{figure}

SiPMs have a number of advantages over conventional photo multiplier tubes, including high photon detection efficiency, complete immunity to magnetic fields, excellent timing characteristics, compact size and physical robustness. They are immune to nuclear counter effects and do not age. They are particularly well suited to applications where optical fibers are used, as the natural size of the SiPM is comparable to that of fibers. But the most important single feature of the SiPM is that it can be manufactured in standard microelectronics facilities using well established processing. This means that huge numbers of devices can be produced without any manual labor, making the SiPMs very economical as the number of devices grows. Furthermore, it is possible to integrate the electronics into the SiPM itself, which reduces cost and improves performance. Initial steps have been taken in this direction, though most current SiPMs do not have integrated electronics. But it is widely recognized that this is the approach that makes sense in the long run for many applications. It improves performance and reduces cost, and can be tailored to a specific application. As the use of SiPMs spreads, so will the use of custom SiPMs with integrated electronics, just as ASICs have superseded standard logic in micro electronics. The photo counting capabilities of the SiPM are unmatched, as 
can be seen in Fig.~\ref{Fig:SiPM} (right) from \cite{Dolgoshein:2003nt}.
As is the case with other microelectronics devices, SiPMs can be customized for different applications, and currently there are a number of manufacturers providing different SiPMs of many different sizes and flavors. Some of the notable manufacturers that offer a variety of devices for purchase are (in alphabetical order): Advansid, Hamamatsu, Ketek and SensL.  At a recent AIDA meeting \cite{AIDA} focussing on SiPM development, over 100 vendors were identified world-wide that are working on SiPM technology or are considering entering the market.
\subsubsection{Readout Electronics}
The appropriate approach for instrumenting a large system revolves around the tradeoff between R\&D and production costs. For smaller systems, building a system around commercial, off-the-shelf (COTS) parts is an excellent choice that minimizes engineering labor. This is the approach taken by the mu2e experiment (approximately 16 thousand SiPMs). Another common approach for systems that are sufficiently large, is to develop an application specific integrated circuit (ASIC) for the particular experiment. This can provide levels of functionality and integration that are well beyond anything that can be achieved with COTS. This is the approach taken by NOVA. This will often result in exceptional performance, far beyond what can be achieved with COTS parts, but requires a great deal of engineering and significant technical and schedule risks that must be carefully managed. A variation of this approach, to adopt existing ASICs from other experiments, is an attractive option for medium sized experiments such as MINER$|nu$A and T2K.  At this point, the field has a great deal of institutionalized experience on which to draw, and this allows costs for the electronics to be projected with good accuracy, even at the early stages of a proposal. The key is that the SiPM is very well matched to the requirements of the detector- with very large gain, excellent photon detection efficiency (PDE) and immunity to magnetic fields. This makes the electronics much less challenging, meaning the risks are low and the costs well understood. 
SiPMs and related technologies continue to develop at a rapid pace. One of the most interesting approaches, with the greatest potential impact on the readout electronics is the development of SiPMs using standard CMOS processing. This implies that the photodetector could be integrated with the readout chip as one device. Currently, a number of companies are working on integrating electronics and SiPM detectors on the same wafer. The first such device was announced by Philips in 2009 and a complete system for evaluation of this technology is commercially available. The system features a fully digital SiPM with active quenching and it is reasonable to expect that this technology will continue to advance and new devices with lower costs and better performance will appear. 

\begin{figure}[hbtp]
\centering\includegraphics[width=0.95\textwidth]{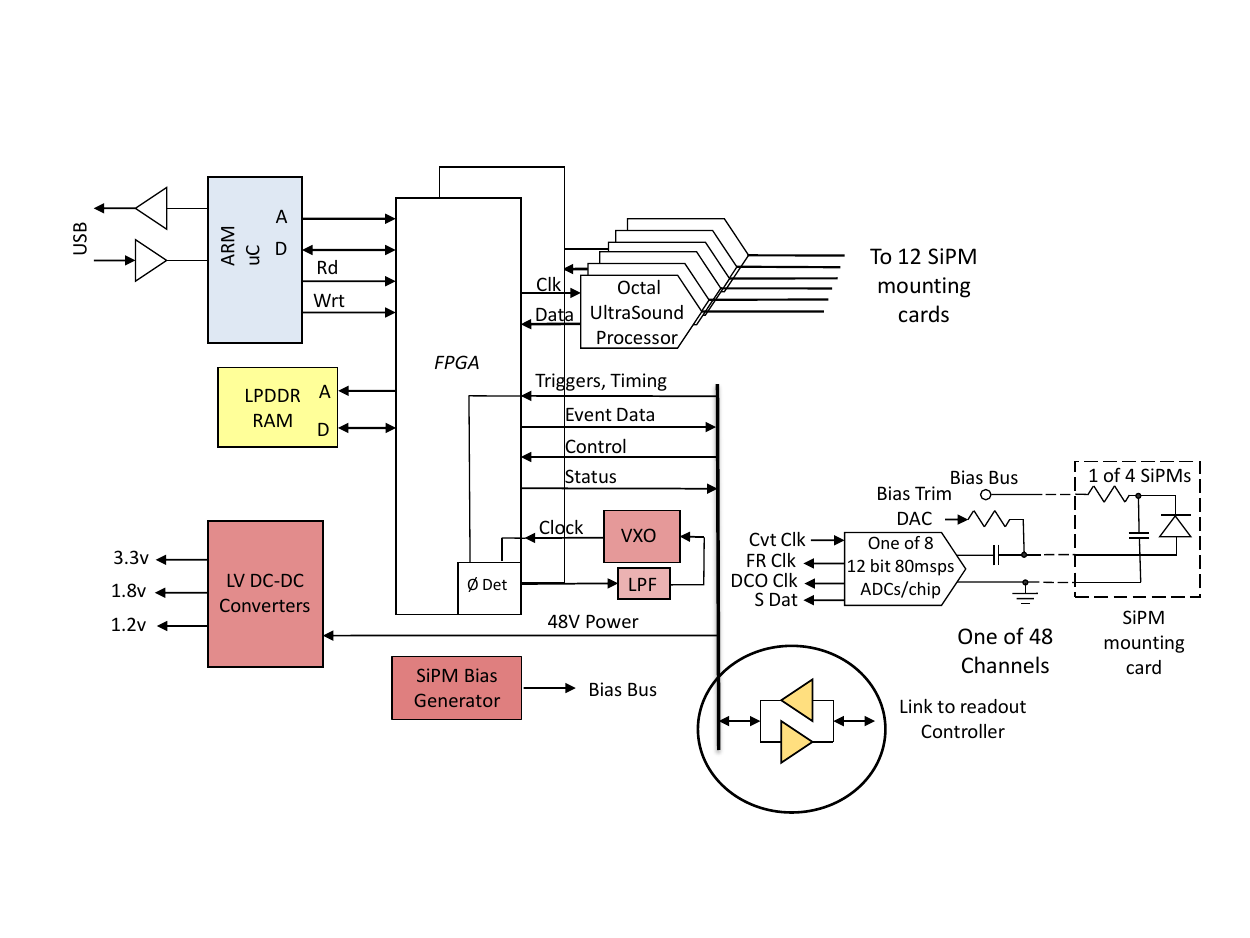}
\caption{ Example of a COTS readout card for SiPMs \cite{Sten:2565}.}
\label{fig:COTS}
\end{figure}

The first approach described above presents many options to use commercially available analog SiPMs coupled to COTS electronics.  Many excellent devices are available, and this is the approach taken so far by existing experiments and those planned for the near future. The mu2e experiment is considering this approach for the cosmic ray veto system (about 16k SiPMs in the baseline design.) This has the advantage of low technical risk and well understood cost. A typical implementation of the electronics might be based on commercial AFE (analog front end) chips and FPGAs, with Ethernet readout. An example of a preliminary design for mu2e is shown in Fig.~\ref{fig:COTS}.
The second approach, adopting commercially existing SiPMs to an ASIC designed specifically for SiPMs (or one that would work well with SiPMs). This approach requires a great deal less engineering than developing a custom chip and significantly reduces the technical risk, but requires a careful evaluation of the available devices and a clear understanding of the needs of the experiment. However, there are many similarities between different experiments in high energy physics and the popularity and interest in SiPMs is driving development for various applications. Some examples of ASICs that have been used (or are being developed for use) with SiPMs are the TriP-t (developed at Fermilab for D0, used by T2K), the DRS4 (developed by PSI), TARGET (developed for Cherenkov Telescope Array) as well as EASIROC and SPIROC and similar chips developed by the Omega group at IN2P3 in Orsay. The ASIC developed for NOVA might also be a viable option. Developing a new ASIC or optimizing an existing ASIC would also be an option. By evaluating the experience of recent projects such as Minerva, T2K, NOvA, we can estimate that the cost per channel for this approach is likely to be below \$10 per channel for the electronics (everything from photo detector to the DAQ, including bias, power, slow control, etc.). The total system costs, including R\&D, engineering, prototyping, firmware development, we estimate to be about \$30 per channel. A reasonable estimate for the cost of SiPMs, including yield, dicing and testing carried out by the manufacturer is about \$2 to \$5 per square mm.
\begin{figure}[hbtp]
\centering\includegraphics[width=0.6\textwidth]{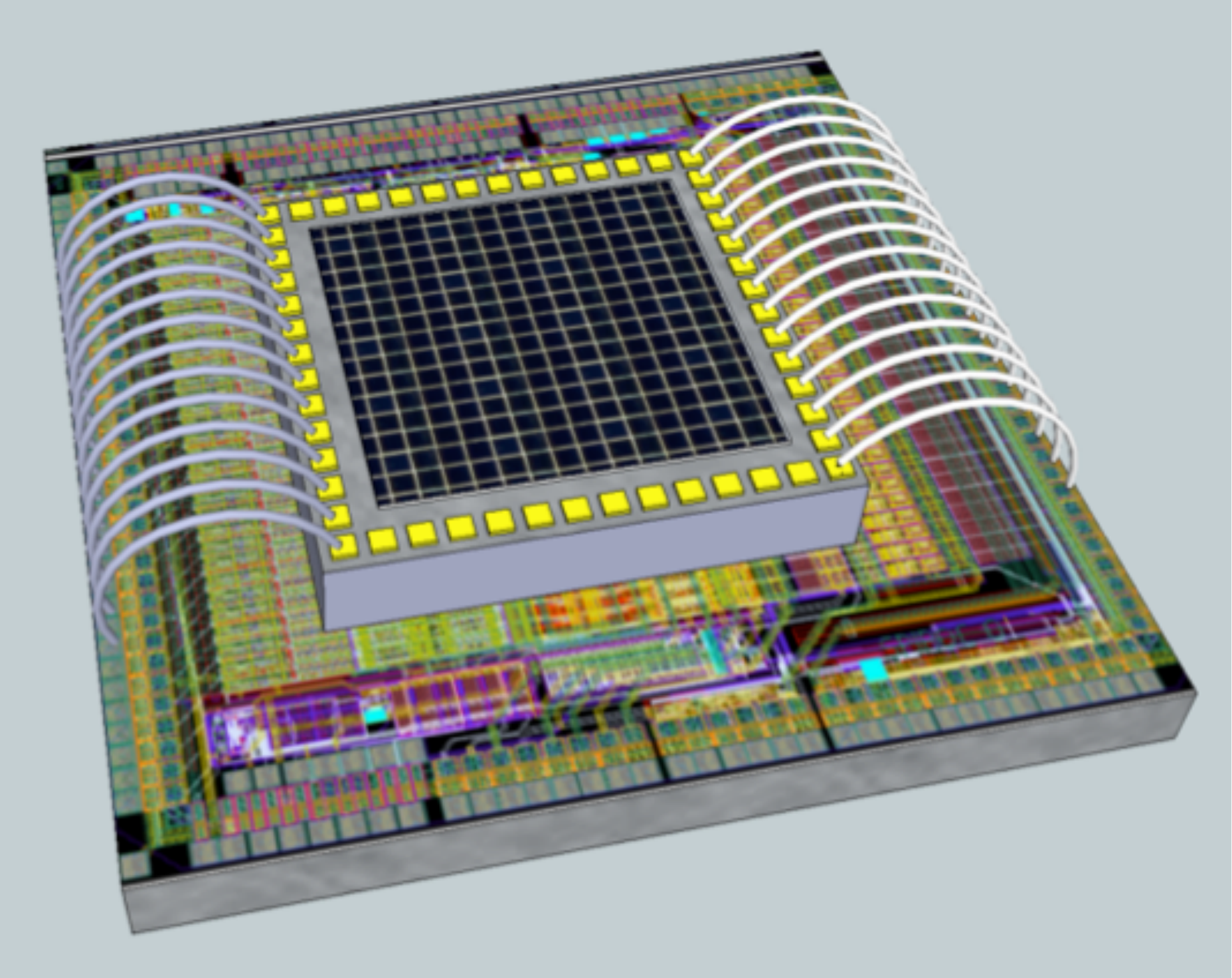}
\caption{A possible configuration for a hybrid approach is shown. The top chip is a SiPM, wire bonded to a readout chip on the bottom. A picture of the EASIROC is used for the bottom chip. The overall size of the device is 4mm by 4mm..}
\label{fig:SiPMHybrid}
\end{figure}

The third approach is to develop a fully custom solution, using either analog or digital SiPMs optimized specifically for the experiment. This approach could potentially significantly reduce the per channel cost of both the photodetector and electronics but involves higher technical risk and requires larger initial R\&D investment. This is clearly the best approach for a sufficiently large detector system.   An alternative approach would be to slightly modify an existing SiPM to allow many connections between the SiPM and a readout ASIC in order to develop a hybrid solution (a near-digital SiPM), where a few SiPM pixels are routed to an output pad so that they can be wire bonded to a readout chip.   This would provide most of the benefits of a full digital SiPM, but with a much shorter and simpler development effort.  A conceptual design for this approach is show in Fig.~\ref{fig:SiPMHybrid}.

\clearpage
\section{$\nu$ Flux at near detector hall}
\label{sec:near}
\providecommand{\nm}{\mbox{\boldmath $\nu_\mu$}}
\providecommand{\anm}{\mbox{\boldmath $\bar\nu_\mu$}} 
\providecommand{\nue}{\mbox{\boldmath $\nu_e$}} 
\providecommand{\ane}{\mbox{\boldmath $\bar\nu_e$}} 

\providecommand{\nmne}{\mbox{\boldmath $\nu_{\mu}\rightarrow\nu_e$}} 
\providecommand{\anmne}{\mbox{\boldmath $\bar\nu_{\mu} \rightarrow \bar\nu_e$}} 
\providecommand{\nmnt}{\mbox{\boldmath $\nu_{\mu}\rightarrow\nu_\tau$}} 
\providecommand{\nmnx}{\mbox{\boldmath $\nu_\mu\rightarrow\nu_x$}} 
\providecommand{\nent}{\mbox{\boldmath $\nu_e\rightarrow\nu_\tau$}} 
\providecommand{\nenx}{\mbox{\boldmath $\nu_e\rightarrow\nu_x$}} 

\providecommand{\thtwothree}{\mbox{\boldmath $\Theta_{23}$}}
\providecommand{\dmtwothree}{\mbox{\boldmath $\Delta m^2_{23}$}}
\providecommand{\sthtwothree}{\mbox{\boldmath $\sin^2 2\Theta_{23}$}} 
\providecommand{\wam}{\mbox{\boldmath $\sin^2 \theta_W$}} 

\providecommand{\enu}{\mbox{\boldmath $E_\nu$}}
\providecommand{\em}{\mbox{\boldmath $E_\mu$}}
\providecommand{\eh}{\mbox{\boldmath $E_{Had}$}}
\providecommand{\ne}{\mbox{\boldmath $\nu_e$}}

\begin{figure}[htbp]
  \begin{center}$
    \begin{array}{cc}
      \includegraphics[width=0.49\textwidth]{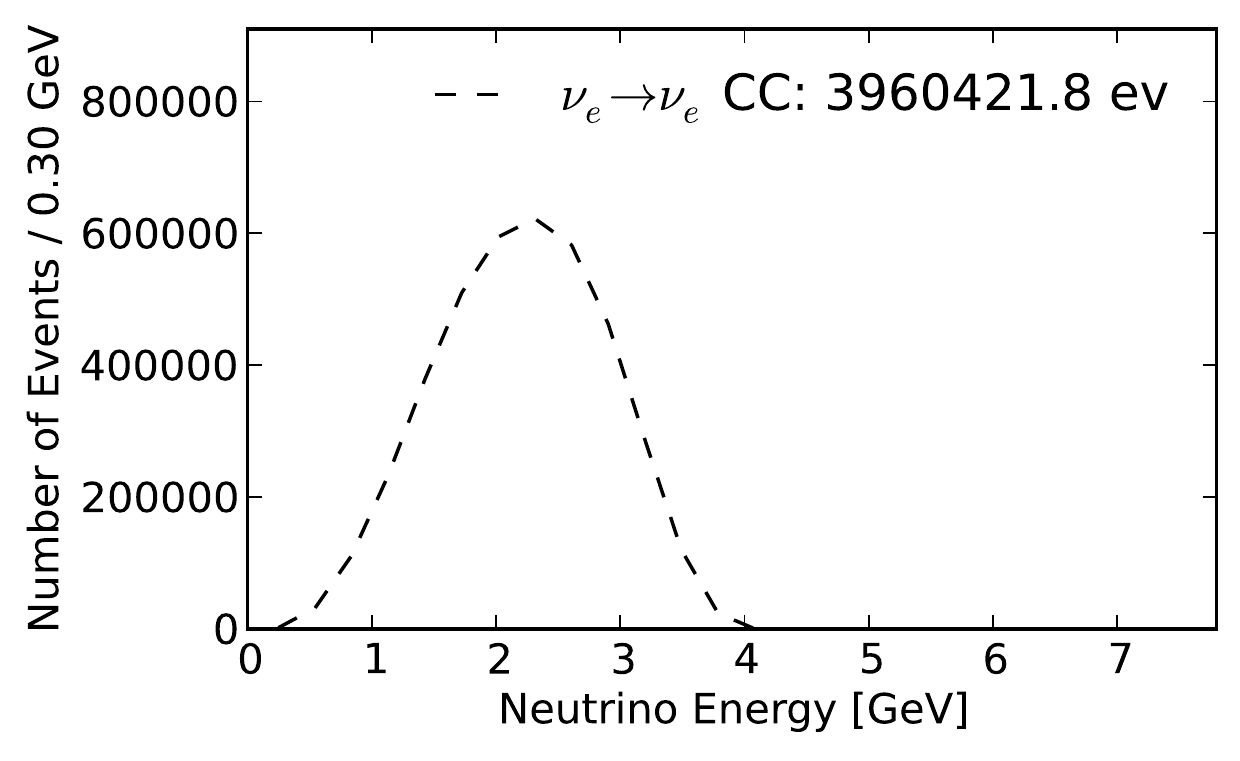}  &
      \includegraphics[width=0.49\textwidth]{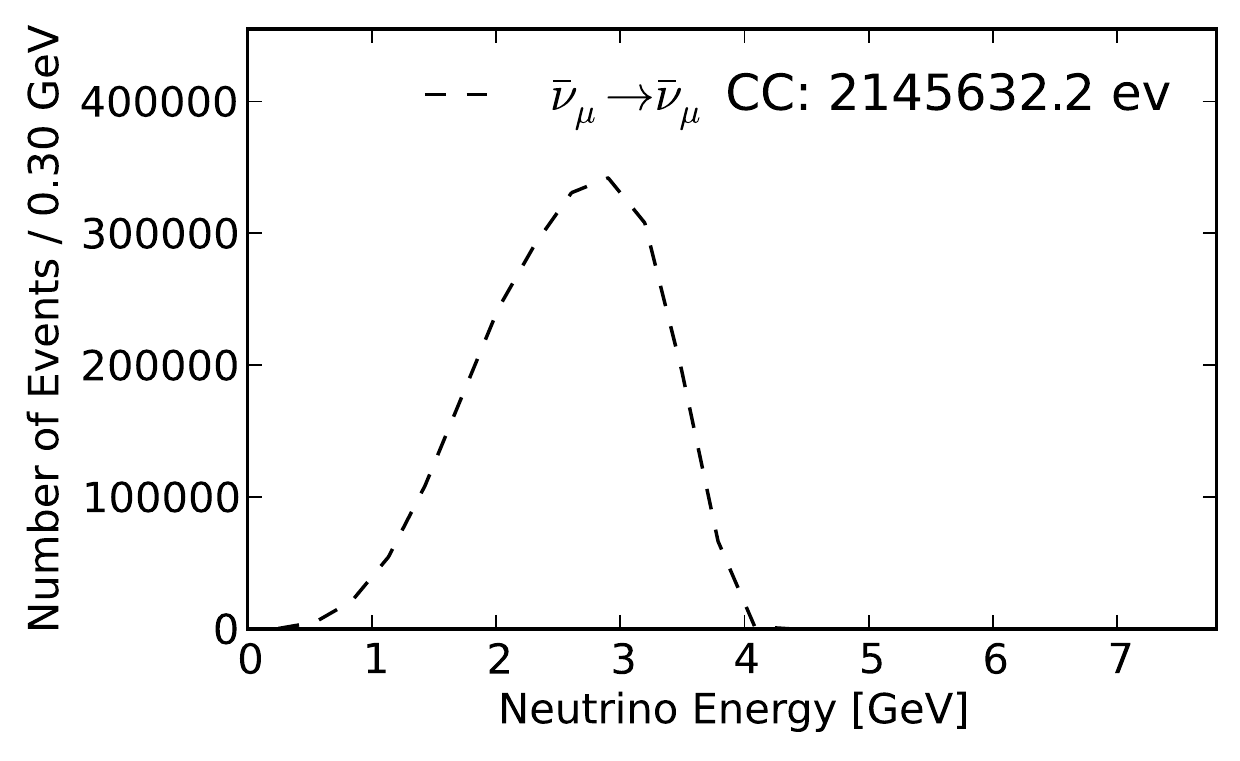} \\
    \end{array}$
  \end{center} 
\caption{$\nu_e$ spectrum at near detector (Left), $\bar{\nu}_\mu$ (Right).}
\label{fig:ND_rates}
\end{figure}

\providecommand{\pip}{\mbox{\boldmath $\pi^+$}} 
\providecommand{\pim}{\mbox{\boldmath $\pi^-$}} 
\providecommand{\piz}{\mbox{\boldmath $\pi^0 $}}
\providecommand{\gam}{\mbox{\boldmath $\gamma$}}
\providecommand{\vz}{\mbox{\boldmath ${\cal V}^0$}}
\providecommand{\mup}{\mbox{\boldmath $\mu^+$}}
\providecommand{\mum}{\mbox{\boldmath $\mu^-$}}
\providecommand{\kap}{\mbox{\boldmath $K^+$}}
\providecommand{\kam}{\mbox{\boldmath $K^-$}}
\providecommand{\kl}{\mbox{\boldmath $K^0_L$}}
\providecommand{\ks}{\mbox{\boldmath $K^0_S$}}
\providecommand{\nova}{\mbox{NO$\nu$A}}
\providecommand{\cohp}{\mbox{\boldmath ${\cal {Coh}} \pi^0$}}
The near detector hall at nuSTORM presents opportunities for both oscillation physics and neutrino cross section measurements.
For the calculations given below, 
we have assumed that the hall will be located at $\sim$ 50m from the end of the straight.  
However, the current siting plan (see \cite{nuPDR:2013}) has the near detector hall located 20 m from the end of the straight.
The neutrino flux at the 50 m position has been calculated and the representative number of events (per 100T fiducial mass) for a $10^{21}$ POT exposure is given in Fig.~\ref{fig:ND_rates}, left for $\nu_e$ and right for $\bar{\nu}_\mu$.  
\begin{table}[h]
\begin{minipage}[h]{.48\linewidth}
\begin{tabular}{l|r|}
Channel & $N_\textrm{evts}$ \hspace{5mm}\\
        \hline
$\bar{\nu}_\mu$ NC & 844,793 \\
$\nu_e$ NC & 1,387,698 \\
$\bar{\nu}_\mu$ CC & 2,145,632 \\
$\nu_e$ CC & 3,960,421 \\
\end{tabular}
\caption{Event rates at near detector (for 100T) with $\mu^+$ stored}
\label{tab:Near+}
	 \end{minipage} \hfill
\begin{minipage}[h]{.48\linewidth}
\begin{tabular}{l|r|}
Channel & $N_\textrm{evts}$ \\
        \hline
$\bar{\nu}_e$ NC & 709,576 \\
$\nu_\mu$ NC & 1,584,003 \\
$\bar{\nu}_e$ CC & 1,784,099 \\
$\nu_\mu$ CC & 4,626,480 \\
\end{tabular}
\caption{Event rates at near detector (for 100T) with $\mu^-$ stored}
\label{tab:Near-}
	 \end{minipage} \hfill
\end{table}	 
\subsection{For short-baseline oscillation physics}
A near detector is needed for the oscillation disappearance searches and our concept is to build a near detector that is identical to SuperBIND, but with approximately 200T of fiducial mass.  A muon ``catcher" will most likely be needed in order to maximize the usefulness of the ``as-built" detector mass.  Before a final specification for this near detector can be made, more detailed analyses of the disappearance channels will have to be done along with an evaluation of its use as a muon spectrometer for $\nu$ interaction 
experiments.
%
%



\section{Short-Baseline Experiment}

\subsection{ Neutrino Flux}
The number of muon decays ($N_\mu$) for nuSTORM can be defined in terms of the following:
\begin{equation}
N_\mu = (\text{POT}) \times (\text{$\pi$ per POT}) \times \epsilon_\text{col} \times \epsilon_\text{trans}  \times 
\epsilon_\text{inj}  \times (\text{$\mu$ per $\pi$}) \times A_\text{dyn} \times \Omega
\end{equation}
\noindent
where $(\text{POT})$ is the number of protons on target,  $\epsilon_\text{col}$ is the collection efficiency,
$\epsilon_\text{trans}$ is the transport efficiency, $\epsilon_\text{inj}$ is the injection efficiency,
$(\text{$\mu$ per $\pi$})$ is the chance that an injected pion results in a muon within the ring acceptance,
$A_\text{dyn}$ is the probability that a muon within the decay ring aperture is within the dynamic aperture, and
$\Omega$ is the fraction of the ring circumference that directs muons at the far detector.
The nuSTORM sensitivities described here
assume $10^{21}$ POT using 120 GeV protons.  From section ~\ref{subsec:43}, we obtain $\simeq$ 0.094 $\pi$/POT.  
From the G4Beamline simulation of pion capture, transport, injection
and propagation along the first straight, we obtain 8 $\times 10^{-3}$ muons at the end of the first straight (within the 3.8 GeV/c $\pm$ 10\% momentum band)
per proton on target based on the 0.094 $\pi$/POT.
From Section~\ref{subsubsec:441}, $A_\text{dyn}$ is 0.6 and  and $\Omega$ is 0.39.  This results in approximately 1.9 $\times10^{18}$ useful $\mu$ decays.
With a 1kT fiducial mass far detector located at approximately 2 km from the end of the decay ring straight, we have the 
following raw event rates:
%
\begin{center}
Neutrino mode with stored $\mu^+$.
\end{center}
\vspace{-7mm}
\begin{table}[h]
	\centering
\begin{tabular}{c|r|r|r|r}
Channel & $N_\textrm{osc.}$ & $N_\textrm{null}$ & Diff. & $(N_\textrm{osc.} - N_\textrm{null})/\sqrt{N_\textrm{null}}$ \\
        \hline
$\nu_e \to \nu_\mu$ CC & 332 & 0 & $\infty$ & $\infty$ \\ 
$\bar{\nu}_\mu \to \bar{\nu}_\mu$ NC & 47679 & 50073 & -4.8\% & -10.7 \\ 
$\nu_e \to \nu_e$ NC & 73941 & 78805 & -6.2\% & -17.3 \\ 
$\bar{\nu}_\mu \to \bar{\nu}_\mu$ CC & 122322 & 128433 & -4.8\% & -17.1 \\ 
$\nu_e \to \nu_e$ CC & 216657 & 230766 & -6.1\% & -29.4 \\ 
\end{tabular}
\begin{center}
Anti-neutrino mode with stored $\mu^-$.
\end{center}
\begin{tabular}{c|r|r|r|r}
Channel & $N_\textrm{osc.}$ & $N_\textrm{null}$ & Diff. & $(N_\textrm{osc.} - N_\textrm{null})/\sqrt{N_\textrm{null}}$ \\
        \hline
$\bar{\nu}_e \to \bar{\nu}_\mu$ CC & 117 & 0 & $\infty$ & $\infty$ \\ 
$\bar{\nu}_e \to \bar{\nu}_e$ NC & 30511 & 32481 & -6.1\% & -10.9 \\ 
$\nu_\mu \to \nu_\mu$ NC & 66037 & 69420 & -4.9\% & -12.8 \\ 
$\bar{\nu}_e \to \bar{\nu}_e$ CC & 77600 & 82589 & -6.0\% & -17.4 \\ 
$\nu_\mu \to \nu_\mu$ CC & 197284 & 207274 & -4.8\% & -21.9 \\
\end{tabular}
\caption{\label{tab:raw_evt}Raw event rates for $10^{21}$ POT (for stored $\mu^+$ and stored $\mu^-$) for best-fit values for the LSND anomaly figure-of-merit.}
\end{table}
In addition to the  $\mu$ decay beam, we also have a high-intensity
$\pi$ decay neutrino beam, $\parenbar{\nu}_\mu$, from the straight section (at injection into the ring) which can easily be time separated from the $\mu$ decay beam.  This 
$\parenbar{\nu}_\mu$ is roughly the same intensity as the integrated $\parenbar{\nu}_\mu$ beam from the stored $\mu$ decays.
\subsection{Monte Carlo}
A detailed detector simulation and reconstruction program has been
developed to determine the detector response for the far detector in the
short baseline experiment.  The simulation is based on software
developed for the Neutrino Factory Magnetized Iron Neutrino Detector
(MIND) \cite{Bayes:2012ex}. 

\subsubsection{Neutrino Event Generation and Detector Simulation}

An effort was made to ground the simulation in a software framework
common to other neutrino simulation efforts.  GENIE
\cite{Andreopoulos:2009rq} is used to generate neutrino events in
steel and scintillator.  Events are passed to a GEANT4-based
\cite{Apostolakis:2007zz} simulation for the propagation of the
final-state particles through successive steel and scintillator
layers. This simulation includes hadron interactions simulated by the
QGSP\_BERT physics list \cite{Apostolakis:2007zz}.

The simulation allows for the customization of the detector layers and
overall dimensions for the purpose of detector optimization studies.
For the purpose of the results described here, the detector is
composed of modules consisting of a 1.5~cm steel plate and a 1.5~cm
thick scintillator plane with a 0.5~cm air gap between modules. For a
detector 6~m in diameter, 343 modules are required to make up the
1.3~kTon far detector, for a total length of 12.4~m. Some optimization
of the detector configuration has been done including variations in
the steel plate thickness and the scintillator thickness and an equal
thickness of both provides the best outcome in reconstruction
efficiency and physics performance.

Hits in the scintillator are grouped into clusters, smearing the
detector hit position, and energy deposition of the accumulated hits
is attenuated in a simple digitization algorithm applied prior to
reconstruction. A double ended readout is assumed for this
digitization. It is assumed that each Silicon photo-multiplier readout
receives 1/4 of the energy deposited in a 2$\times$2~cm$^2$ unit cell,
attenuated over the distance between the hit position and the edge of
the detector. The detector performance is not strongly affected by the
cell size, which is determined by the transverse width of the
scintillator bars. Simulations have been run with scintillator bar
widths from 1~cm to 3~cm and there was very little change in the
performance. 

Some assumptions were made for the light yield and electronics
performance. The attenuation length in the wavelength shifting fibers
is assumed to be 5~m. The SiPMs are assumed to have an energy
resolution of 6\% with a conservative threshold of 4.7 photo-electrons
and a photodetector quantum efficiency of
$\sim$30\%\cite{Michael:2008bc}. A minimum ionizing particle produces
up to 80 photo-electrons in a scintillator bar so a high detection
efficiency is expected. Simulations ran with a lower collection
threshold of 2 photo-electrons find no significant change in the
detection efficiency.

\begin{figure}
  \begin{center}
    \includegraphics[width=0.85\textwidth]{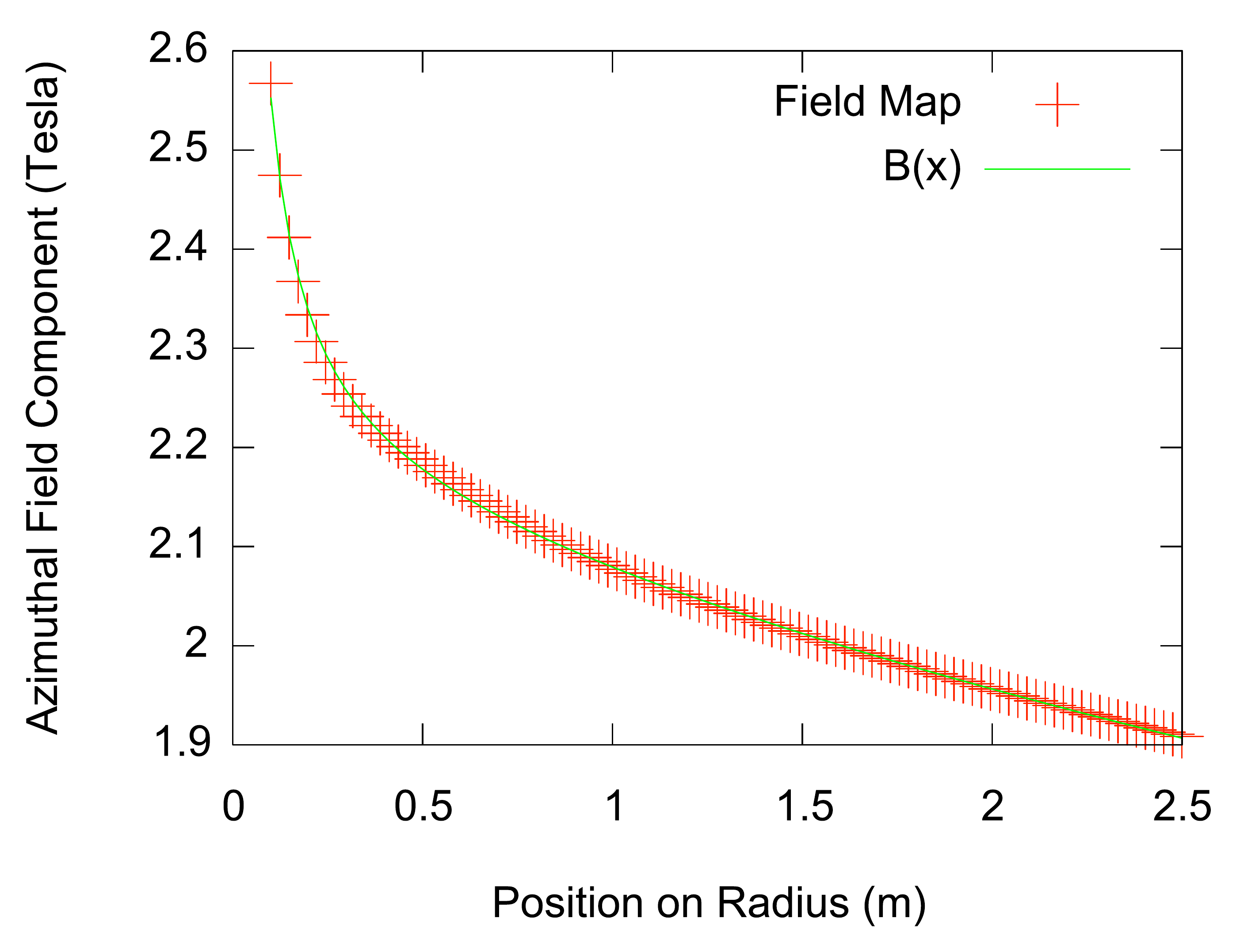}
  \end{center}
  \caption{
    The magnetic field magnitude as a function of radius along the
    45$^{\circ}$ azimuth with the parametrization used in the detector
    simulation.
  } 
  \label{fig:Bmodel}
\end{figure}

Magnetization within the steel is introduced with an empirically
derived formula. The magnetic field is parametrized as:
\begin{equation}
B_{\phi}(r) = B_{0} + \frac{B_{1}}{r} + B_{2}e^{-H r} ~ ;
\end{equation}
where $B_{0} = 1.36$\,T, $B_{1} = 0.04$\,T\,m, $B_{2} = 0.80$\,T, and
$H = 0.16$\,m$^{-1}$.  This parametrization and the field along the
45$^\circ$ azimuthal direction are shown in Fig.~\ref{fig:Bmodel}. The difference between the field map of the steel
plate and the parametrization is less than 1\%. The polarity of the
magnetic field was chosen to focus negative charges towards the center of
the detector for the study of $\nu_{\mu}$ interactions.

\subsubsection{Event Reconstruction}

The reconstruction uses multiple passes of a Kalman-filter algorithm
to identify muon trajectories within events and to determine the
momentum and charge of an identified track.  Multiple tracks are
identified within the simulation, and all identified tracks are fit to
determine the length of the tracks. The longest track is chosen as the
prospective muon track for the event and all other tracks are assumed
to be the result of pion production or electron showers. The secondary
tracks are important in the case of resonant pion production where
there may be a visible pion at a lower angle than the muon track.

The Kalman filtering and fitting algorithms are supplied by the
RecPack software package \cite{CerveraVillanueva:2004kt}.  Geometrical
information from the track including the length of the track, the
direction of bending in the magnetic field, and the pitch of the track
are used at various points in this procedure to provide initial state
information to the Kalman filter. The momentum and charge is
determined from the track curvature. 

The hadron reconstruction is not yet well developed, so the neutrino
energy is reconstructed by using the quasi-elastic approximation, if
no hadronization is visible. If there are secondary tracks or hits
that are not associated with a track, the true hadron energy is
smeared according to the results of the MINOS CalDet test beam
results\cite{Michael:2008bc}:
\begin{equation}
  \frac{\delta E_{had}}{E_{had}} = \frac{0.55}{\sqrt{E_{had}}} + 0.03.
\end{equation}
A future SuperBIND prototype, such as the AIDA MIND \cite{AIDA} prototype, is
anticipated to refine this measurement.

\subsection{Data Analysis}

Following the reconstruction, the events are analyzed to select events
with well reconstructed muon tracks, and remove tracks that are
mis-identified from pions or electron showers. This analysis for
appearance and disappearance experiments differs in the charge of
the muon that is selected for the signal. For the appearance analysis
the charge of the muon signal is opposite to that of the muon in the
storage ring, while the disappearance analysis signal is a like sign
muon. For the following discussion, a paired experiment with stored
$\mu^{+}$ is considered.

\subsubsection{Appearance}

To achieve the target of 10$\sigma$ significance for the LSND anomaly
the background efficiency must be reduced to less than a few parts in
$10^{5}$. This can be achieved with a cuts based analysis such as that
previously used for the neutrino factory \cite{Bayes:2012ex}, but a
multivariate analysis is preferred as it yields more robust results
overall. Both analyses are described here.

The cuts based approach relies on the sequence of cuts shown in Table
\ref{tab:cuts}.  The majority of these cuts are made to ensure that
the trajectory-fit is of good quality.  Cuts are made to remove events
that are not successfully reconstructed, with a starting position
closer than 1\,m from the end of the detector.  Events are rejected if
the reconstructed muon track has fewer then 60\% of all detector hits
assigned to the trajectory, the momentum is greater than 1.6 times the
maximum neutrino energy, or the charge of the fitted trajectory does
not match the charge derived from a geometric definition of the
curvature.

\begin{table}[htbp]
  \caption{The cuts used in an analysis of a nuSTORM $\nu_{\mu}$ appearance analysis with SuperBIND.}
  \centering
	\begin{tabular}{|rp{11cm}|}
	\hline\hline
	Variable & Description \\
	\hline
	Trajectory Identified & There must be at least one trajectory identified in the event.\\
	Successful Fit & The longest identified trajectory must be successfully fit.\\
	Maximum Momentum & The momentum of the longest trajectory is less than 6 GeV/c. \\
	Fiducial & The longest trajectory must start prior to the last 1~m of the detector. \\
	Minimum Nodes & the fit to the longest trajectory must include more than 60\% of the hits assigned to trajectory by the filter.\\
	Curvature Ratio & $(q_{init}/p_{range}) \times (p_{fit}/q_{fit}) > 0$\\
	Track Quality & $\log{P(\sigma_{q/p}/(q/p)|CC)}
        -\log{P(\sigma_{q/p}/(q/p)|NC)}  < -0.5$\\
        NC Rejection & $\log{P(N_{hits}|CC)}
        -\log{P(N_{hits}|NC)}  > 7$\\
	\hline\hline
	\end{tabular}
  \label{tab:cuts}
\end{table}

\begin{table}
  \caption{ The fraction of events left after cuts are applied to the
    simulations of the indicated species in the nominal SuperBIND
    detector when the appearance of a $\mu^-$ in an event is defined
    as the experimental signal. Determined from simulations of
    1.5$\times 10^{6}$ $\nu_{\mu}$~CC and $\bar{\nu}_{\mu}$~CC
    interactions and 5$\times 10^{6}$ $\bar{\nu}_{\mu}$~NC,
    $\nu_{e}$~CC, and $\nu_{e}$~NC interaction events.}
   \centering 
  \label{tab:cutsurvival}
  \begin{tabular}{|rccccc|}
    \hline\hline
    & \multicolumn{5}{c}{Interaction Type and Species} \\
    \cline{2-6}
    Event Cut & $\nu_{\mu}$ CC(\%) & $\bar{\nu}_{\mu}$ CC & $\nu_{e}$
    CC  $\bar{\nu}_{\mu}$ NC & $\bar{\nu}_{e}$ CC &$\bar{\nu}_{e}$ NC\\ 
    & & ($\times 10^3$)&($\times 10^3$)&($\times 10^3$)&($\times 10^3$)\\
    \hline
    Successful Reconstruction        & 71.1\% & 51.2 &    482 & 102 & 166\\
    Fiducial                                     & 70.8\% & 50.1 &    474 & 101 & 164\\
    Maximum Momentum               & 68.8\% & 32.1 &    397 & 82.6 & 135\\
    Fitted Proportion                       & 67.9\% & 30.4 &    388 & 80.5 & 132\\
    Curvature Ratio                         & 48.5\% & 12.7 &    266 & 52.0 & 85.5\\
    Track Quality                            & 41.7\% &  2.3  &    40.1& 13.8 & 47.5\\
    NC Rejection                             & 13.8\% & 0.004 & 0.008 & 0.003   & 0.008\\
    \hline\hline
  \end{tabular}
\end{table}

\begin{figure}
  \subfigure[Signal Efficiency]{
    \includegraphics[width=0.5\textwidth]{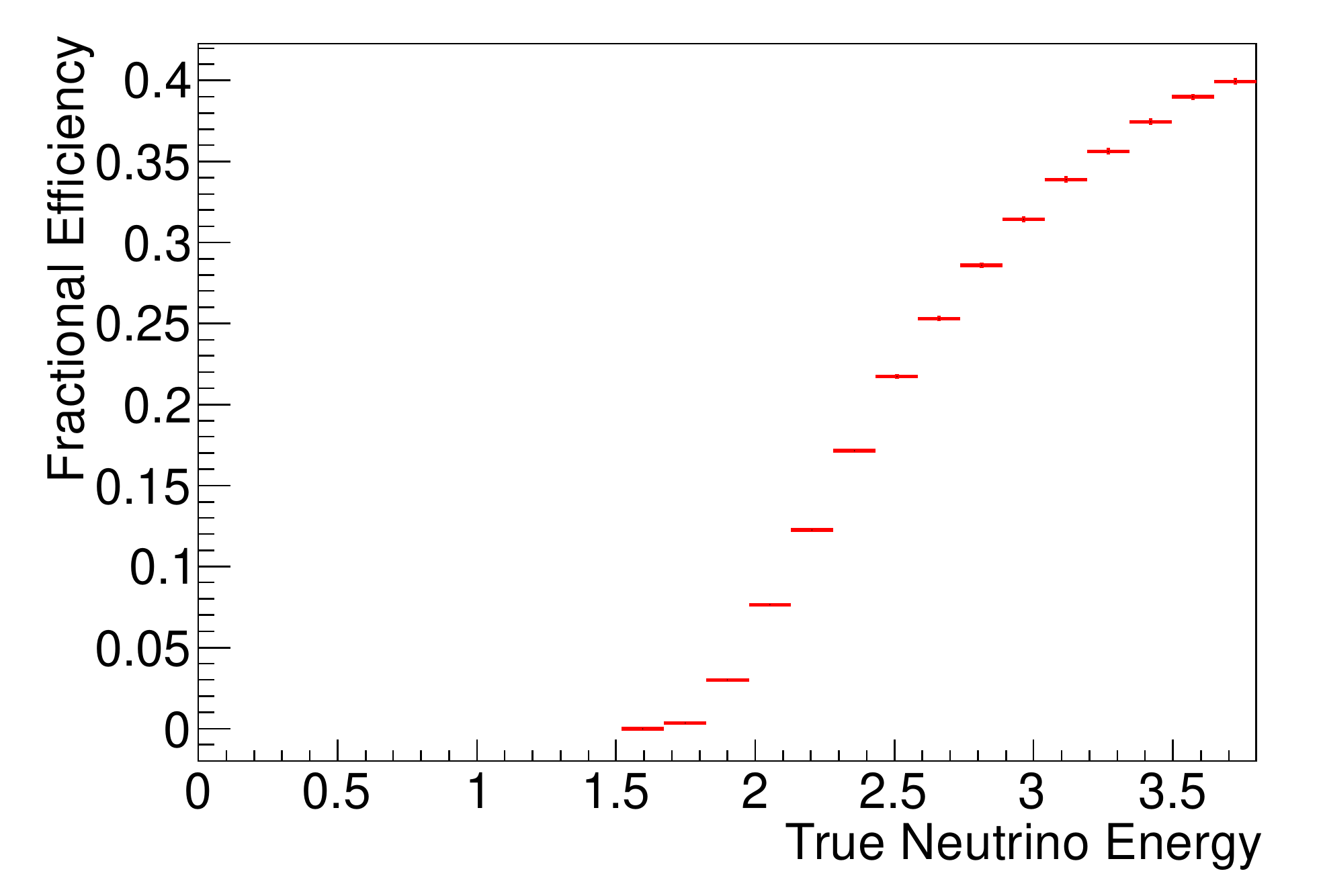}
  \label{fig:eff}
  }
  \subfigure[Background Efficiencies]{
    \includegraphics[width=0.5\textwidth]{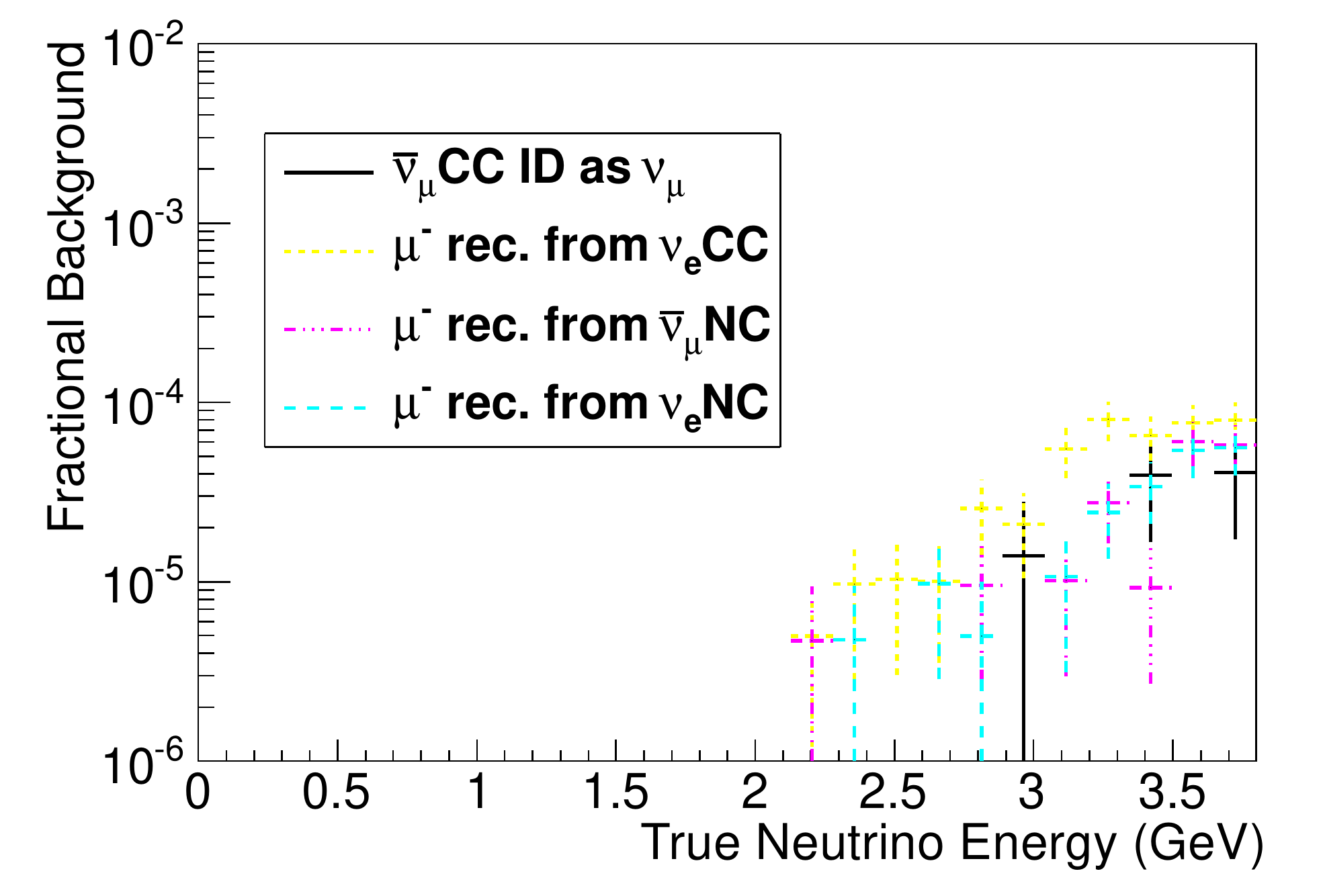}
  \label{fig:back}
}
\caption{ Signal and background efficiencies for the detection of a
  $\mu^-$ signal that will be present when $\mu^+$ are contained in
  the nuSTORM storage ring as determined by the cuts based
  analysis.}
\label{fig:cutsanal}
\end{figure}

Two further cuts---the track quality and NC rejection cuts---affect
the ratio of signal to background.  The track quality cut is based on
the relative error of the inverse momentum of the candidate muon
$|\sigma_{q/p}/(q/p)|$ where $q$ is the charge of the muon and $p$ is
its momentum.  Probability distribution functions,
$P(\sigma_{q/p}/(q/p))$, are generated from pure charged-current (CC)
and neutral-current (NC) samples.  A log-likelihood ratio,
$\mathcal{L}_{q/p}$, is created from the ratio of the CC and NC
probabilities for a given trajectory:
\begin{equation}
\mathcal{L}_{q/p} =
\log\left(\frac{P(\sigma_{q/p}/(q/p)|CC)}{P(\sigma_{q/p}/(q/p)|NC)}\right)~ .
\end{equation}
An event is accepted if $\mathcal{L}_{q/p} > -0.5$.  The NC rejection
cut is likewise defined using a log-likelihood ratio defined using the
number of hits used in a fit to a particular trajectory, $N_{hit}$,
for CC and NC samples.  It has found that the background rejection can
be reduced to below parts in 10$^{-5}$ if:
\begin{equation}
\mathcal{L}_{CC} =
\log\left(\frac{P(N_{hit}|CC)}{P(N_{hit}|NC)}\right) > 7 ~ .
\end{equation}
The impact of all of these cuts on the simulated event samples is
shown in Table \ref{tab:cutsurvival}.  The signal and background
efficiencies are shown in Fig.~\ref{fig:eff} and \ref{fig:back},
respectively.

An important limitation of this approach is the high energy threshold
and the lack of plateau in the signal efficiency. This is a result of
the restrictive nature of the cuts necessary to achieve the required
background rejection. The $\mathcal{L}_{CC}$ cut is particularly
responsible for the shape of this threshold as the number of hits in a
trajectory is a measure of the longitudinal momentum. The cut on
$\mathcal{L}_{CC}$ has the strongest impact on reducing the
background, so it can not be relaxed in this analysis. The alternative
is to introduce additional variables, but this has been demonstrated
to only reduce the signal efficiency in a cuts based framework. For
this reason a multi-variate analysis was tested for use in SuperBIND.

The multi-variate analysis is facilitated by the ROOT based
TMVA package\cite{Hocker:2007ht}. This analysis outperforms the cuts
based analysis by offering a lower energy signal
threshold and increasing the sensitivity of the experiment to
oscillations.


\begin{table}[htdp]
\caption{Variables used in the analysis of events in the SuperBIND
  simulation. Variables in \ref{sub:mva} are used in the definition
  of the classifier, while the cuts in \ref{sub:fixed} are fixed. }
\centering  
\subfigure[Variables used in the multivariate analysis.]{
	\begin{tabular}{|rp{9cm}|}
	\hline\hline
	Variable & Description \\
	\hline
	Track Quality & $\sigma_{q/p}/(q/p)$, the error in the trajectory curvature scaled by the curvature\\
	Hits in Trajectory & The number of hits in the trajectory \\
	Curvature Ratio & $(q_{init}/p_{range}) \times (p_{fit}/q_{fit})$: comparison of the initial guess of the curvature to the Kalman fit result. \\
	Mean Energy Deposition & Mean of energy deposition of hits in fit  of the trajectory \\
	Variation in Energy & $\sum^{N/2}_{i=0}\Delta E_{i} / \sum^{N}_{j=N/2}\Delta E_{j}$ where the energy deposited per hit $\Delta E_{i}~<~\Delta E_{i+1}$.\\
	\hline\hline 
	\end{tabular}
	\label{sub:mva}
          } \\
\subfigure[Preselection variables.]{
	\begin{tabular}{|rp{9cm}|}
	\hline\hline
	Variable & Description \\
	\hline
	Trajectory Identified & There must be at least one trajectory identified in the event.\\
	Successful Fit & The longest identified trajectory must be successfully fit.\\
	Maximum Momentum & The momentum of the longest trajectory is less than 6 GeV/c. \\
	Fiducial & The longest trajectory must start prior to the last 1~m of the detector. \\
	Minimum Nodes & The fit to the longest trajectory must include more than 60\% of the hits assigned to the trajectory by the filter.\\
	Track Quality & $\sigma_{q/p}/(q/p) < 10.0$\\
	Curvature Ratio & $(q_{init}/p_{range}) \times (p_{fit}/q_{fit}) > 0$\\
	\hline\hline
	\end{tabular}
	\label{sub:fixed}
         	}
\label{tab:var}
\end{table}%

\begin{figure}[htbp]
\begin{center}
\includegraphics[width=0.7\textwidth]{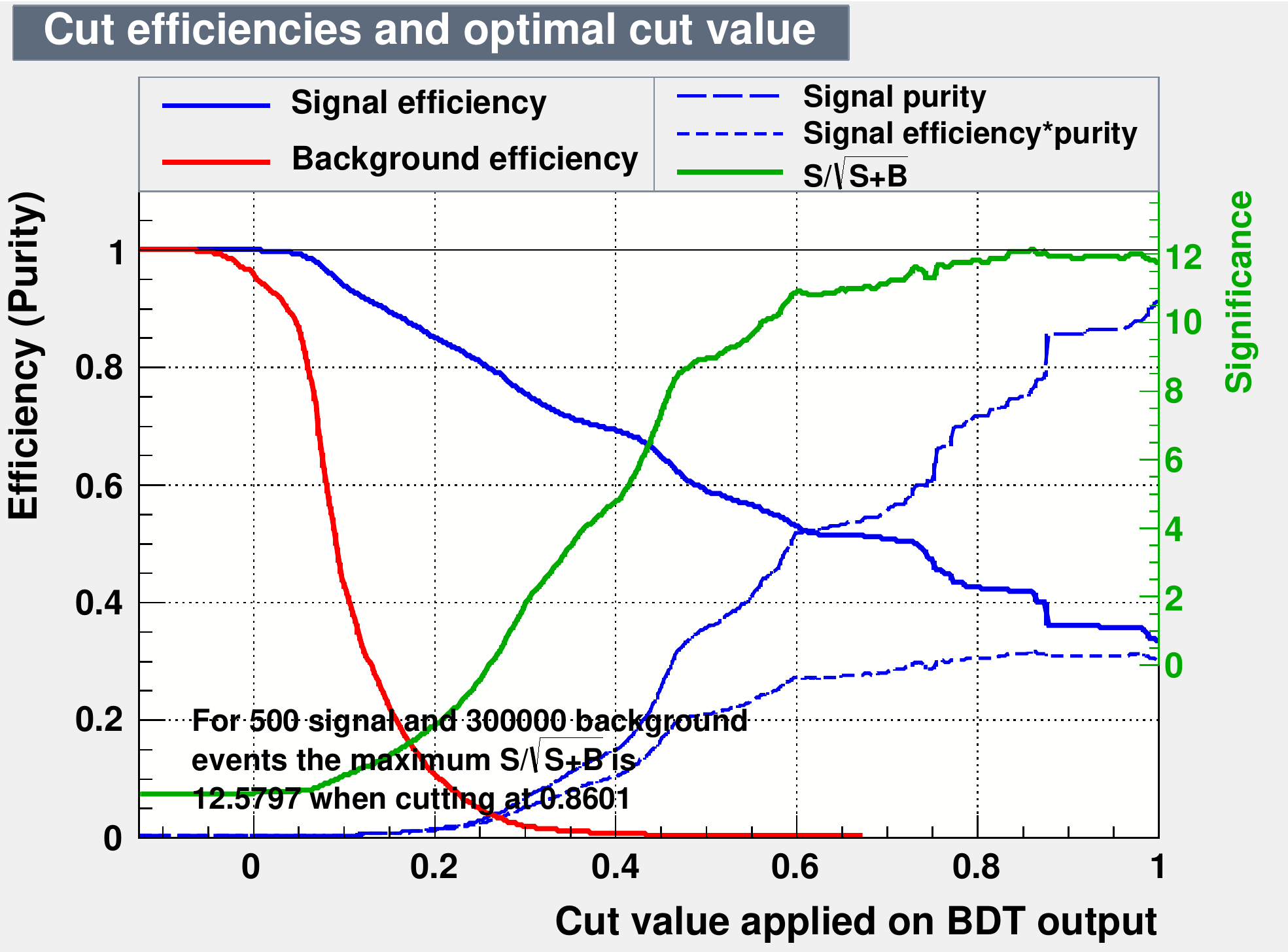}
\caption{Results from training the BDT method to simulations of
  $\nu_{\mu}$CC signal events and $\bar{\nu}_{\mu}$ background events,
  assuming a realistic number of events.}
\label{fig:mvaeff}
\end{center}
\end{figure}

\begin{figure}
  \subfigure[Signal Efficiency]{
    \includegraphics[width=0.5\textwidth]{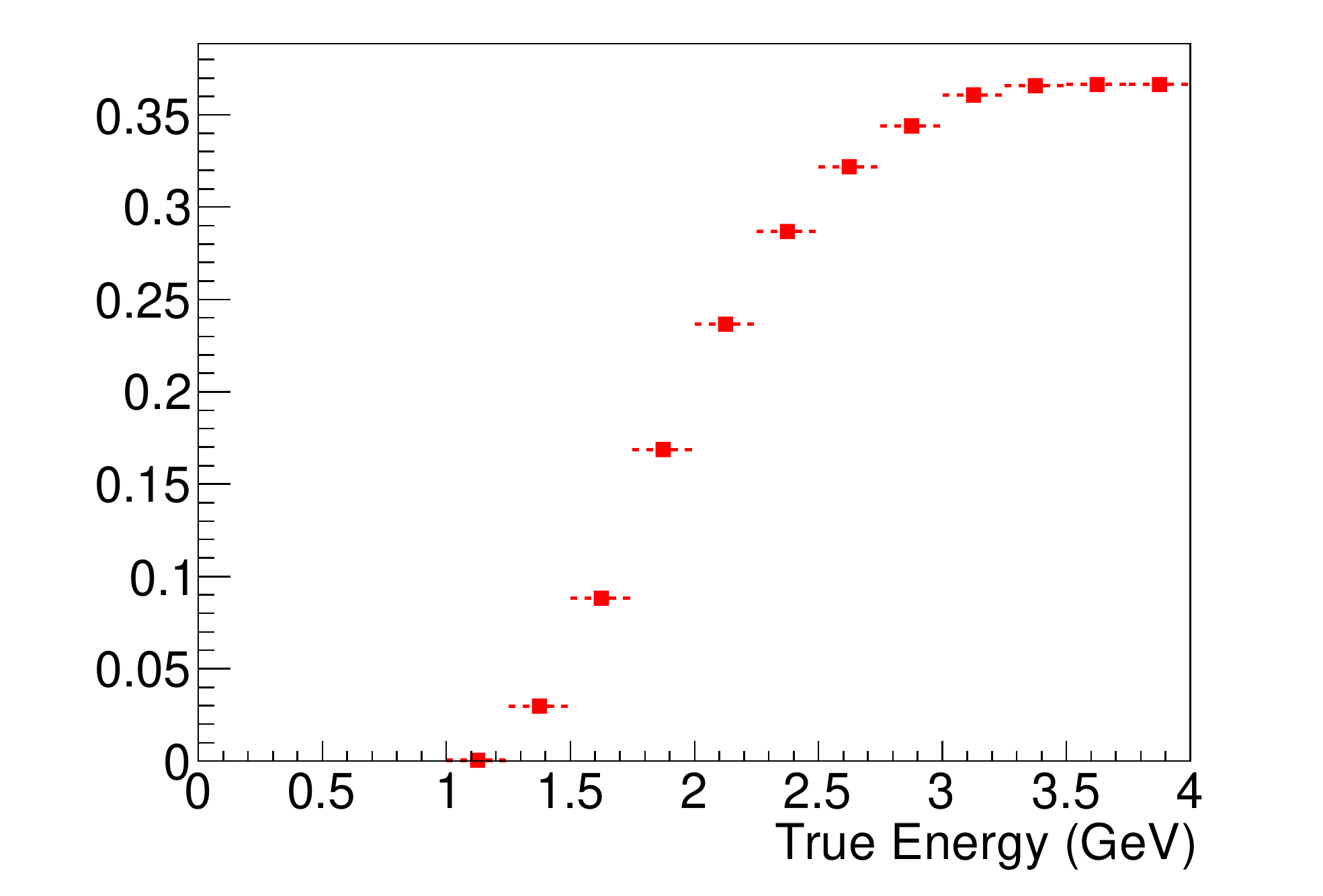}
  \label{fig:BDTeff}
  }
  \subfigure[Background Efficiencies]{
    \includegraphics[width=0.5\textwidth]{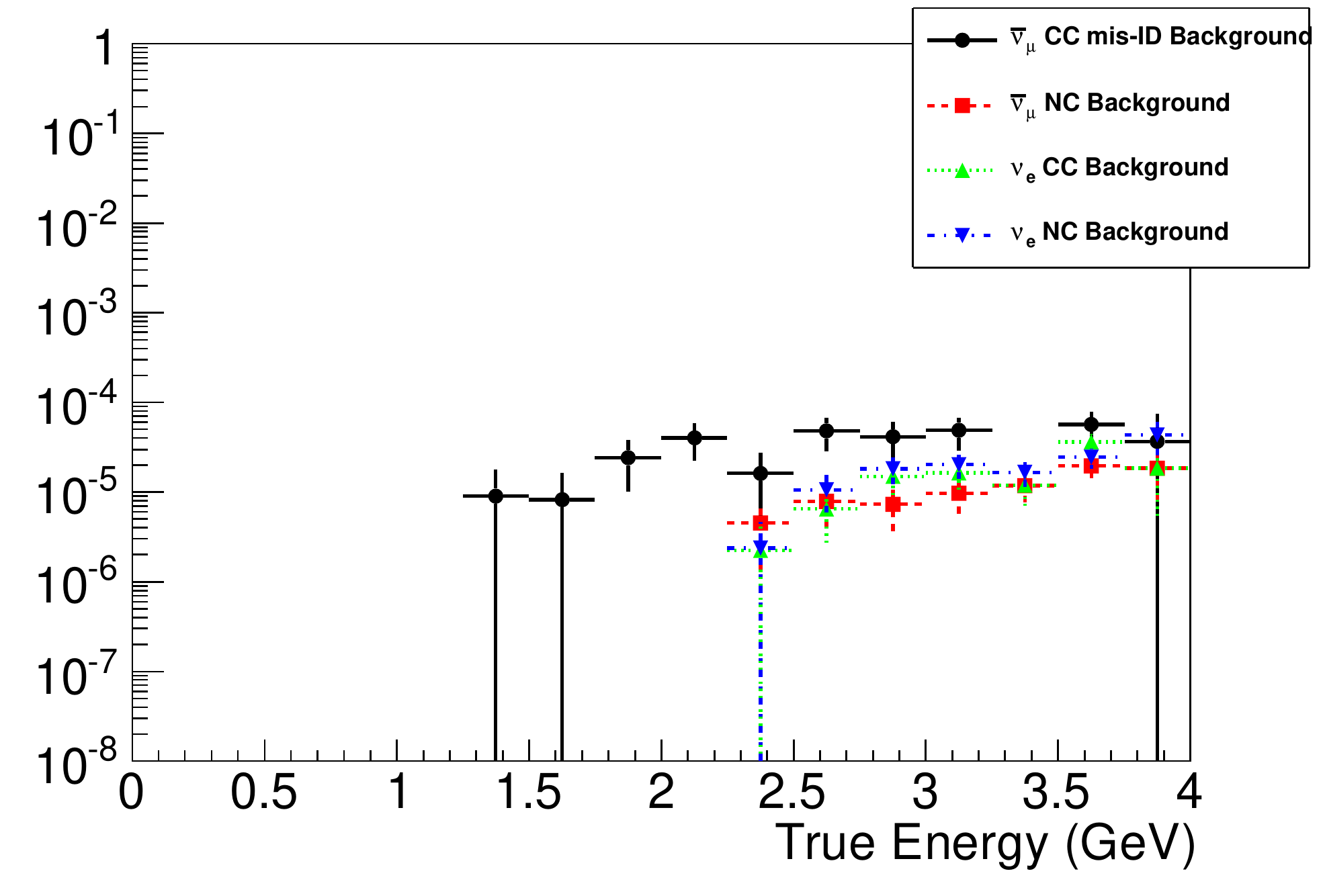}
  \label{fig:BDTback}
}
\caption{ Signal and background efficiencies for the detection of a
  $\mu^-$ signal that will be present when $\mu^+$ are contained in
  the nuSTORM storage ring as determined by a boosted decision tree
  multivariate analysis.}
\label{fig:BDTanal}
\end{figure}

The analysis is trained to discriminate between $\nu_{\mu}$ charge
current (CC) interaction signal events and $\bar{\nu}_{\mu}$ neutral
current (NC) interaction background events using a set of five
parameters to define a classifier variable. The majority of these
parameters were chosen from the experience of the MINOS experiment
\cite{Adamson:2010uj}. Table \ref{sub:mva} summarizes these
parameters.  A set of preselection cuts, detailed in
Table~\ref{sub:fixed} are applied to limit the analysis to the subset
of events containing useful data. These preselection cuts are common
to both analyses, as their purpose of removing poor tracks is
unchanged. The multi-variate analysis is trained using a variety of
methods, but the best performance is achieved using a Boosted Decision
Trees (BDT). Based on the performance of this method, shown in
Fig.~\ref{fig:mvaeff}, events are selected if the BDT classifier
variable is greater than 0.86. The optimized analysis produces the
efficiency contours in signal and backgrounds shown in
Fig.~\ref{fig:BDTanal}. Other methods including a neural network
(MLPBNN) and the k Nearest Neighbor (kNN) method were tested but they
do not produce sufficient background rejection to achieve the
statistical aims for the appearance analysis. The BDT method has a
much greater integrated efficiency than the previously developed cuts
based analysis with a lower energy threshold.

\subsubsection{$\bar{\nu}_{\mu}$ Disappearance}

\begin{figure}[htbp]
  \subfigure[Signal Efficiency]{
    \includegraphics[width=0.5\textwidth]{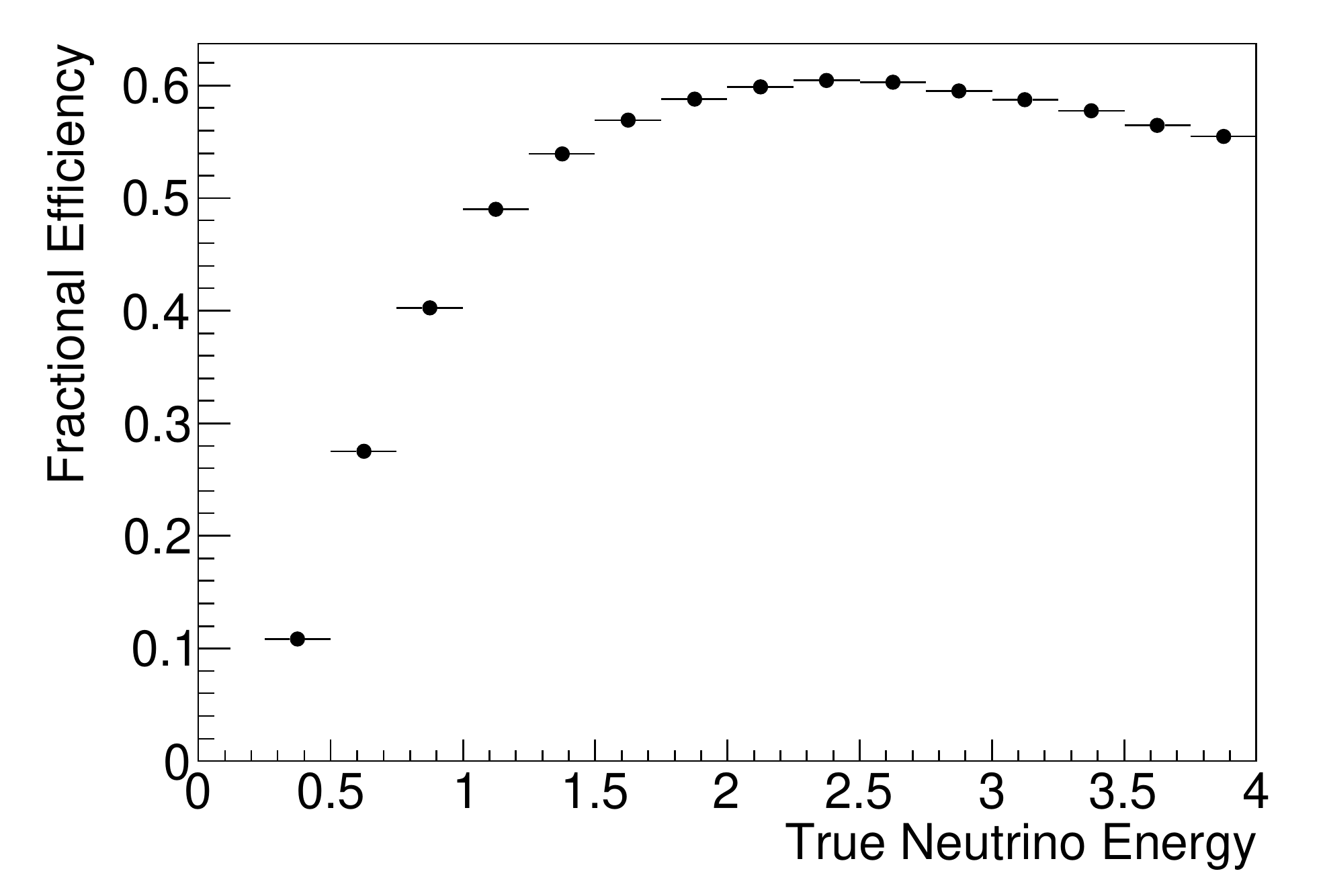}
    }
  \subfigure[Background Efficiency]{
    \includegraphics[width=0.5\textwidth]{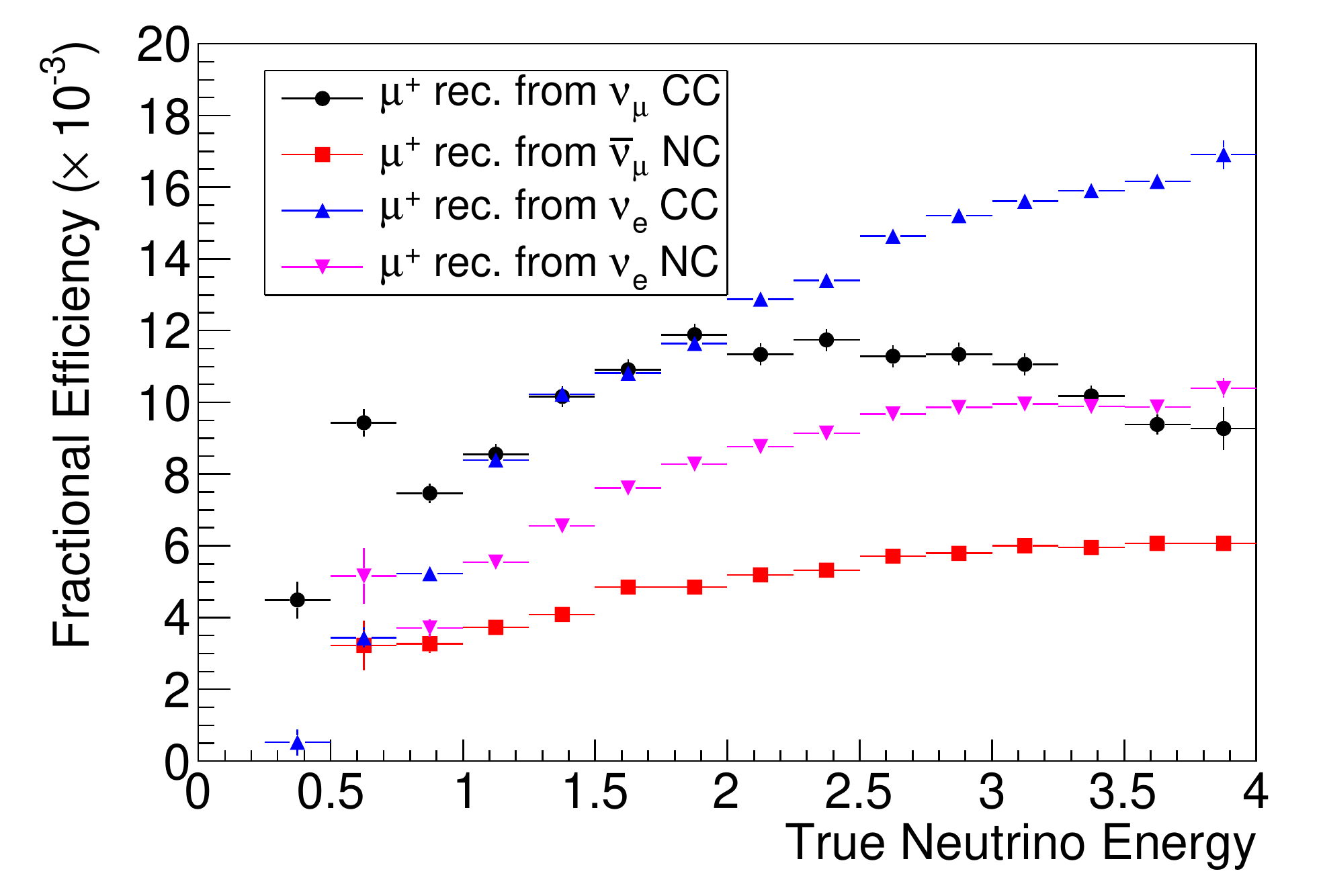}
  }
  \caption{Efficiency curves for the $\bar{\nu}_{\mu}$ disappearance
    analysis produced from the optimized analysis.}
  \label{fig:diseff}
\end{figure}

The $\bar{\nu}_{\mu}$ disappearance analysis has a different
optimization criteria from that of the appearance analysis.  A
disappearance experiment relies on the difference between the shape of
the oscillated spectrum versus that of the unoscillated spectrum. An
optimization based on a $\chi^{2}$ statistic comparing these two
spectra is used, assuming a 3+1 neutrino model with $\Delta m^{2}_{14}
= 0.89$~eV$^{2}$ and $\sin^{2}2\theta_{\mu\mu}=$0.09. The conclusion
of this optimization is that a neural network method (MLPBNN) shows
the best response when events with a classifier variable greater than 0.94
is selected. The background and efficiency curves for this
optimization are shown in Fig.~\ref{fig:diseff}. In this particular
case, the energy threshold is below 1~GeV because background
efficiencies on the order of 1\% are reasonable. A similar optimization
was not done using a cuts-based framework.

\subsection{Sensitivities}

The appearance of $\nu_\mu$, via the channel $\nu_{e}\to\nu_{\mu}$,
gives nuSTORM broad sensitivity to sterile neutrinos and directly
tests the LSND/MiniBooNE anomaly.  The oscillation probabilities for
both appearance and disappearance modes are:
\begin{eqnarray}
P_{\nu_e\to\nu_{\mu}} & = & 4|U_{e4}|^{2}|U_{\mu
  4}|^{2}\sin^{2}\left(\frac{\Delta m^2_{41}
  L}{4E}\right) \label{eq:app} ~ {\rm ;~and}\\
P_{\nu_{\alpha}\to\nu_{\alpha}} & = & 1 - [4|U_{\alpha 4}|^{2}(1 - |U_{\alpha
  4}|^{2})]\sin^{2}\left(\frac{\Delta m^2_{41} L}{4E}\right)
  ~. \label{eq:disapp} 
\end{eqnarray}
The detector is designed for the appearance signal
$\nu_{e}\to\nu_{\mu}$; the CPT conjugate of the channel with which
LSND observed an anomaly, $\bar{\nu}_{\mu}\to\bar{\nu}_{e}$. 
Although it is clear from equation \ref{eq:app} that the appearance
channel is doubly suppressed relative to the disappearance channel,
the experiment is much more sensitive to the appearance channel
because the backgrounds for wrong-sign muon searches can be suppressed
more readily. The systematic effects, while common to both channels,
will have different impacts on their respective sensitivities to a
sterile neutrino signal.

\subsubsection{Systematics}

Systematic uncertainties for short baseline $\nu_{\mu}$ oscillation
experiments at nuSTORM have been given some consideration
\cite{TunnellThesis}. The great benefit of using a muon beam as a
neutrino source is that many of the beam related systematics may be
reduced significantly. A list of systematics taken from the existing
literature is given in Table~\ref{tab:sys}, with the anticipated
systematic uncertainties for nuSTORM short baseline experiments. 

The leading systematic uncertainties for a short baseline experiment
at nuSTORM are in the prediction of the flux convoluted with the
cross section at the far detector. Based on the estimates made for the
25 GeV neutrino factory\cite{Abrams:2011fk}, the neutrino beam from
the muon decay ring should be known to 0.1\% based on the
characteristics of the muon beam. The characterization would be
accomplished with a combination of beam current transformers, beam
position monitors, polarimeters, and wire scanners. The low current in
the nuSTORM ring may reduce the precision of these measurements, and
the flux uncertainty, to 1-3\%, conservatively.

A leading cause of uncertainties particular to the cross sections
are differences in the quasi-elastic interactions between neutrinos
and anti-neutrinos for both $\nu_{e}$ and $\nu_{\mu}$. These
cross-sections are anticipated to be measured at the nuSTORM near
detector site using runs of stored $\mu^{+}$ and $\mu^{-}$. The
uncertainty of such measurements would then be dominated by the flux
uncertainty. The MINOS experiment sets an uncertainty on the convolved
flux with the cross-section of 4\% for the
signal \cite{Adamson:2010uj}. Background uncertainties are further
limited by the ability to reproduce the false signal backgrounds in
neutral current events. In this context the MINOS experiment assesses
a 40\% normalization error \cite{Adamson:2010uj} due to uncertainties
in the associated production mechanisms such as the resonant pion
production and a 15\% uncertainty in the hadronic model. Again it is
likely that refined measurements at nuSTORM will reduce these
systematic uncertainties.

Other systematics and backgrounds have also been considered. The
potential variation in the thickness of the steel plate has been
considered based on the MINOS construction specifications. These
variations have a direct impact on the magnetic field production, and
particle range in the detector. A simulation was run with random
variations in the steel plates. The simulation showed a negligible
impact on the ability of the apparatus to distinguish signal from
background. External backgrounds in the form of rock muons and cosmic
ray muons were considered through both analytical calculations and with  the CRY
cosmic ray generator \cite{Hagman:2012} with a GEANT4 simulation. It is believed that a
fiducial cut removing external events will reduce these backgrounds by
a factor of 10$^{8}$, yielding on the order of 1 muon over 5 years of
exposure.

It is anticipated that the contributing uncertainties can be reduced
to less than 1\% for signal events and 10\% for background events. The
combined existing systematic uncertainties can be as much as five
times these numbers. Thus an upper limit on the potential systematic
uncertainties of 5\% for signal and 50\% for background is assumed for
the sensitivity calculations.

\begin{table} 
  \centering
  \caption{Systematic uncertainties expected for a short baseline
    muon neutrino appearance experiment based at the nuSTORM
    facility.}
  \begin{tabular}{|rccccc|}
  \hline
    Uncertainty & \multicolumn{3}{c}{Known Measures} &
    \multicolumn{2}{c}{Expected Contribution} \\
    & Signal & Background & Reference& Signal & Background \\
    \hline
    Source luminosity & 1\% & 1\% &\cite{TunnellThesis} & 1\% & 1\% \\
    Cross section & 4\% & 40\% &\cite{Adamson:2009ju} & 0.5\% & 5\%  \\
    Hadronic Model & 0 & 15\% &\cite{Adamson:2006xv}& 0 & 8\% \\
    Electromagnetic Model & 2\% & 0 & \cite{Adamson:2006xv}& 0.5\% & 0 \\
    Magnetic Field & $<$1\% & $<$1\% &\cite{TunnellThesis} & $<$1\% & $<$1\%
    \\
    Steel & 0.2\% & 0.2\% & \cite{TunnellThesis} & 0.2\% & 0.2\% \\
    \hline
    Total & 5\% & 43\% & & 1\% & 10\%\\
    \hline
  \end{tabular}
  \label{tab:sys}
\end{table}

\subsubsection{$\nu_{\mu}$ Appearance}

The detector response derived from simulation is used to determine the
sensitivity of the appearance experiment to the presence of sterile
neutrinos.  The sensitivities and optimizations were computed using
GLoBES \cite{Huber:2004ka}. Modifications were made to simulate
accelerator effects such as the integration of the decay straight as
outlined in \cite{TunnellThesis,Tunnell:2012nu} and to include the
non-standard interactions for sterile neutrino
oscillations \cite{Kopp:2007ne}.  The detector response is summarized
as a ``migration'' matrix of the probability that a neutrino generated
in a particular energy bin $i$ is reconstructed in energy bin $j$.
Defined in this way, the migration matrix encapsulates both the
resolution of the detector and its efficiency.  Samples of all
neutrino interactions that could participate in the experiment are
generated to determine the response for each detection channel.  

\begin{figure}
  \begin{center}
    \includegraphics[height=0.6\textwidth,angle=270]{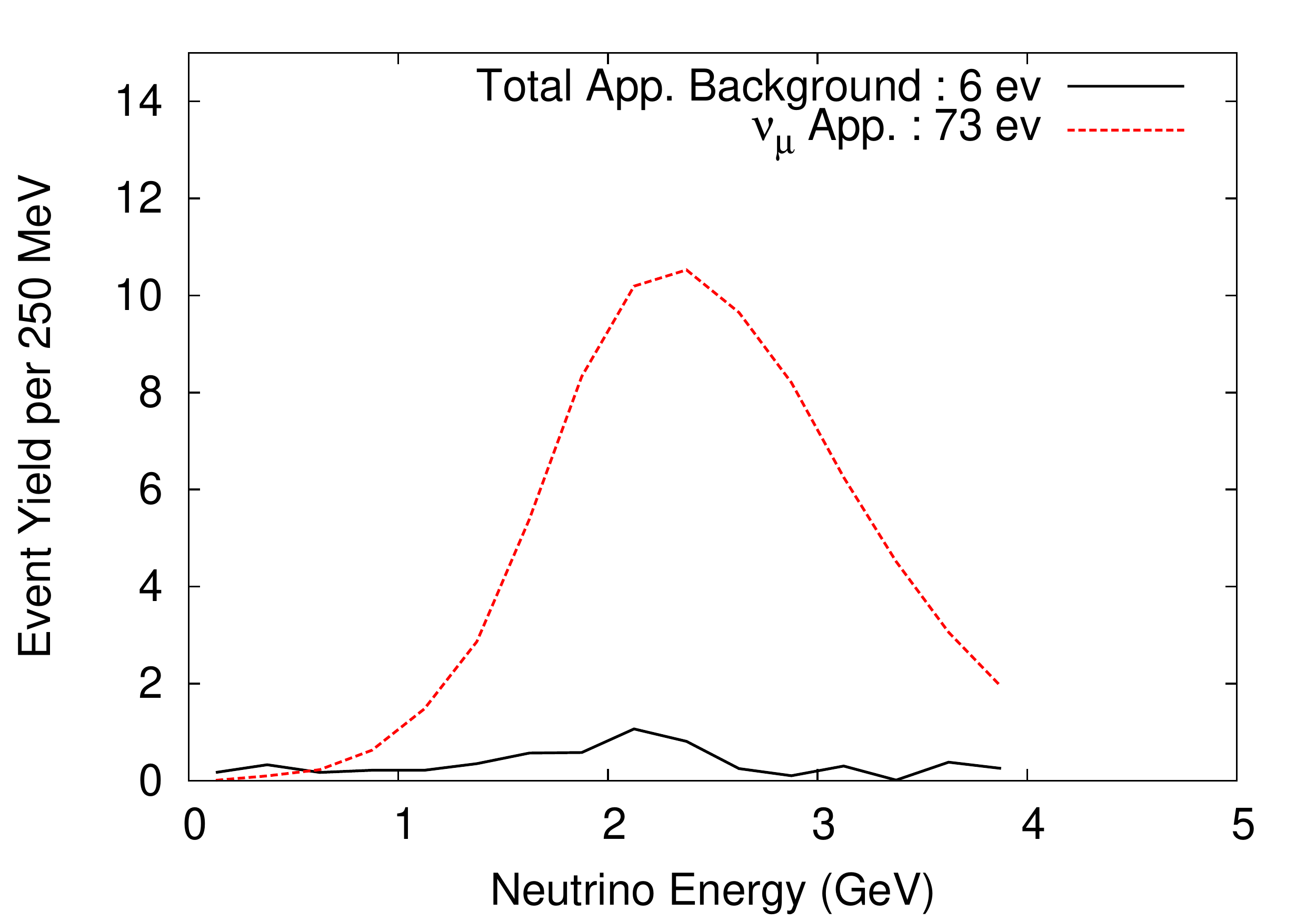}
  \end{center}
  \caption{ The neutrino spectrum of a $\nu_{\mu}$ appearance
    experiment measured at the SuperBIND far detector using the
    simulated detector response.  }
  \label{fig:spec}
\end{figure}

\begin{figure}
  \begin{center}
    \includegraphics[height=0.6\textwidth,angle=270]{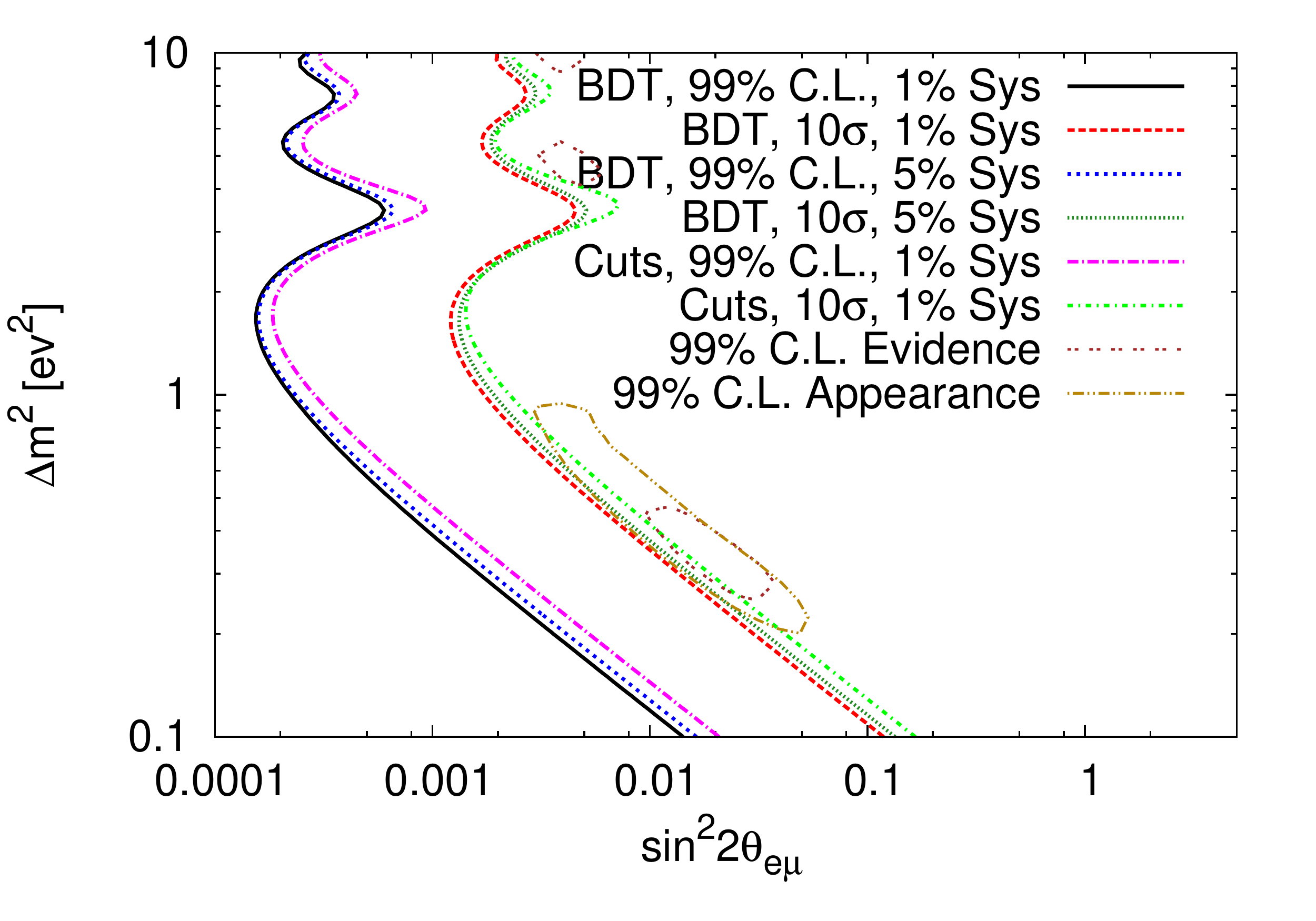}
  \end{center}
  \caption{ Contours of the $\chi^{2}$ deviation from the no-sterile
    neutrino hypothesis corresponding to a 99\% C.L. (2.575$\sigma$)
    and 10$\sigma$ variations. Two different sets of systematic
    uncertainties are represented; the default systematics with 1\%
    signal uncertainty and a 10\% background uncertainty and a
    conservative set that is five times larger. A cuts based analysis
    is also shown.  The 99\% confidence
    level contours from a global fit to all experiments showing evidence for unknown
    signals (appear + reactor + Gallium) and the contours derived from the accumulated data from all
    applicable neutrino appearance experiments \cite{Kopp:2013vaa} are also shown. }
  \label{fig:sens}
\end{figure}

The spectrum of expected signal and background for this simulation is
shown in Fig.~\ref{fig:spec} assuming 1.8$\times 10^{18}$ $\mu^{+}$
decays and a 2~km baseline.  A contour plot
showing the sensitivity of the $\nu_{\mu}$ appearance experiment to
sterile neutrinos  appears in Fig.~\ref{fig:sens} showing the
results of the multi-variate analysis with the ideal systematics and
inflated systematics as well as the results of the cuts based
analysis. These contours are shown with respect to the derived
variable $\sin^{2}2\theta_{e\mu} = 4|U_{e4}|^{2}|U_{\mu
  4}|^{2}$. Systematic uncertainties are included in the contour as
stated. The contours derived from the latest fits to existing
appearance data and to the data showing evidence of sterile neutrino
oscillations \cite{Kopp:2013vaa} are also shown. This study shows that
the multi-variate analysis is sensitive to the phase space covered by
the existing 99\% contours well above the 10$\sigma$ level, even with
the systematics inflated by a factor of five.


\subsubsection{$\bar{\nu}_{\mu}$ Disappearance}

As noted above, a disappearance experiment is potentially more
sensitive to the oscillation amplitudes than the appearance
analysis. However, it is also more sensitive to the signal
normalization than an appearance experiment. The neutrino flux is
extremely well understood for nuSTORM, but further understanding of
the measured spectrum resulting from the combination of efficiencies
and cross sections is required.  To control these effects, a near
detector identical to the far detector is required. This is the
motivation behind the 200 Ton version of SuperBIND at the near
detector site. A simulation of such a near detector is in progress, so
an approximation of the anticipated systematic must be made. Combined
fits of near and far detector rates are standard practice in
oscillation experiments \cite{Adamson:2007gu} and an optimization for
such an experiment at an early version of nuSTORM has already been
done \cite{Winter:2012sk} for an idealized $\nu_{e}$ disappearance
experiment assuming different beam properties. For the current
discussion, the systematic uncertainties are assumed to be identical
to the signal systematics described in the appearance context.

\begin{figure}[htbp]
  \centering
  \includegraphics[height=0.6\textwidth,angle=270]{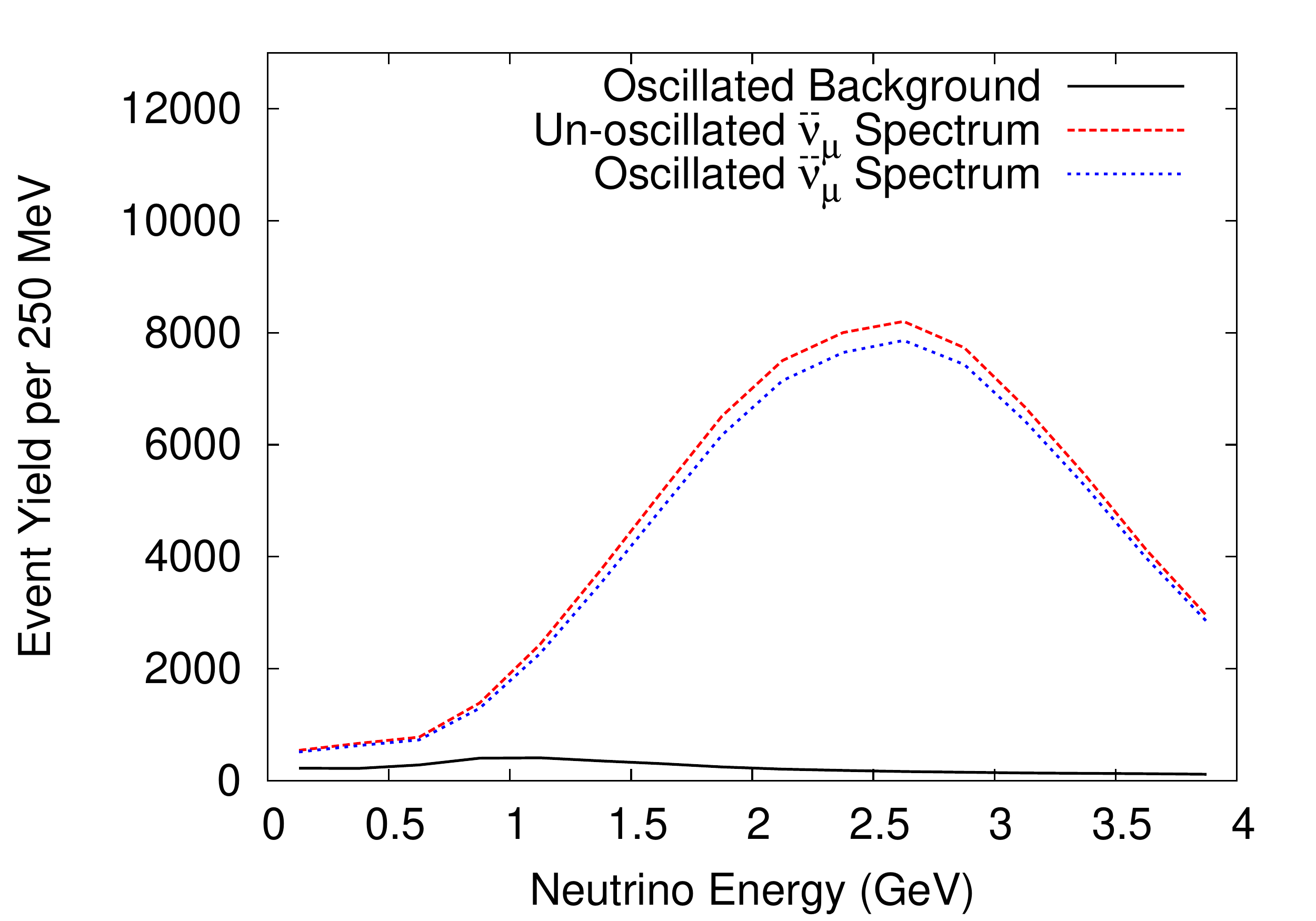}
  \caption{The expected number of events for a short baseline
    $\bar{\nu}_{\mu}$ disappearance experiment at nuSTORM assuming
    $\Delta m^{2}_{14} = 0.89$ and $\sin^{2}2\theta_{\mu\mu} =
    4U_{\mu4}^{2}(1 - U_{\mu4}^{2}) = 0.09$.}
  \label{fig:disspec}
\end{figure}

\begin{figure}[htbp]
  \centering
  \includegraphics[height=0.6\textwidth,angle=270]{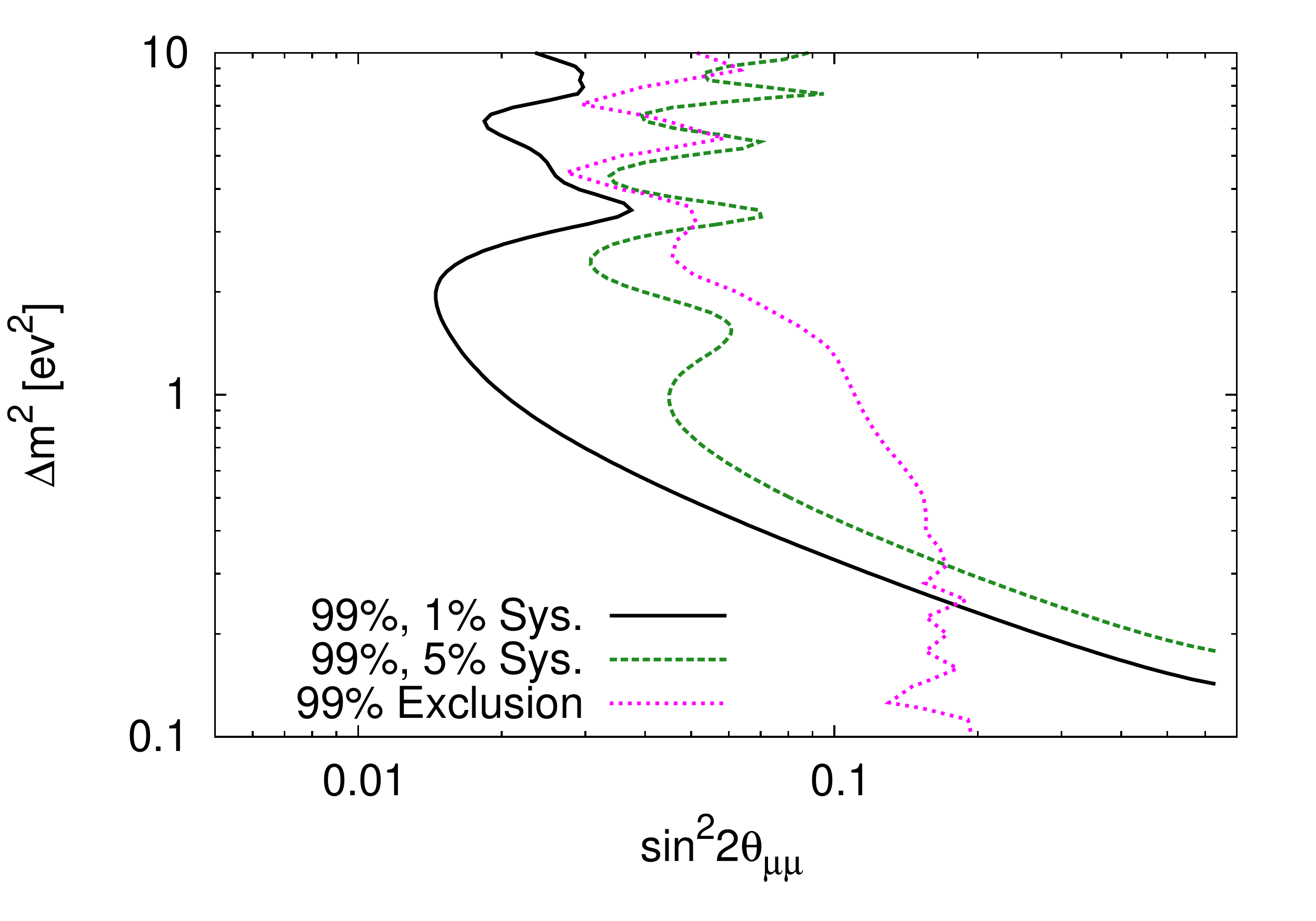}
  \caption{The sensitivity of a disappearance experiment at nuSTORM
    to the presence of a light sterile neutrino oscillation. Contour
    generated using GLoBES, assuming a 1.3~kTon far detector, 2~km
    away from the muon storage ring, exposed to neutrinos resulting
    from an exposure of 1.6$\times 10^{18}$ useful muon decays. A
    1\% signal and 10\% background systematic uncertainty is
    assumed by default.  The 99\% C.L. bound, assuming a systematic error multiplied by a
    factor of 5, is also shown with the existing experimental bounds \cite{Kopp:2013vaa}}
      \label{fig:dissens}
    \end{figure}

The sensitivity of the experiment is computed using GLoBES as in the
appearance experiment with the only difference in the selection of the
like sign muon.  The signal and background spectra, after the
application of detector response, with and without light sterile
neutrino oscillations are shown in Fig.~\ref{fig:disspec}. As with
the appearance analysis, sensitivity contours to short baseline
oscillations were generated assuming 3.2$\times 10^{17}$ $\mu^{+}$
decays per year collected over 5 years in a 1.3 kTon detector. The
contours with respect to the effective mixing angle
$\sin^{2}2\theta_{\mu\mu} = 4U_{\mu4}^{2}(1 - U_{\mu4}^{2})$ are shown
in Fig.~\ref{fig:dissens}. To improve on the phase space covered
with the exclusion limits from the fits to the existing
data\cite{Kopp:2013vaa}, the experiment needs better than the 5\%
signal systematic shown as an upper limit. This figure shows the great
sensitivity of the disappearance experiment to systematic effects that
is not as acute in the appearance analysis. The expected sensitivity
given the lower bound on the systematic uncertainties will improve on
the 99\% C.L. from the fits to the existing data \cite{Kopp:2013vaa}.

\subsubsection{Combined Appearance and Disappearance }
\label{subsubsec:CAnD}
Combining appearance and disappearance information from the muon
neutrino channels is important because the channels provide
complementary information and they will be observable in the
nuSTORM beam at the same time.  The migration matrices derived for
the two experiments are completely uncorrelated because of the
difference in the charge of the selected signal, i.e., like sign muon
signal rather than opposite sign muons. 

A simultaneous fit using both appearance and disappearance results
will make the best use of the available data. This is accomplished for
a given $\Delta m^{2}_{14}$ and effective mixing angle (either
$\sin^{2}2\theta_{e\mu}$ or $\sin^{2}2\theta_{\mu\mu}$) through the
use of a simple fit which finds the minimum $\chi^{2}$ between the
sterile hypothesis and the null hypothesis as a function of $|U_{\mu4}|$
(for $\sin^{2}2\theta_{e\mu}$) or $|U_{e4}|$ (in the case of
$\sin^{2}2\theta_{\mu\mu}$).  

\begin{figure}[htbp]
  \centering
  \subfigure[Improvement over appearance experiment]{ 
  \includegraphics[height=0.475\textwidth,angle=270]{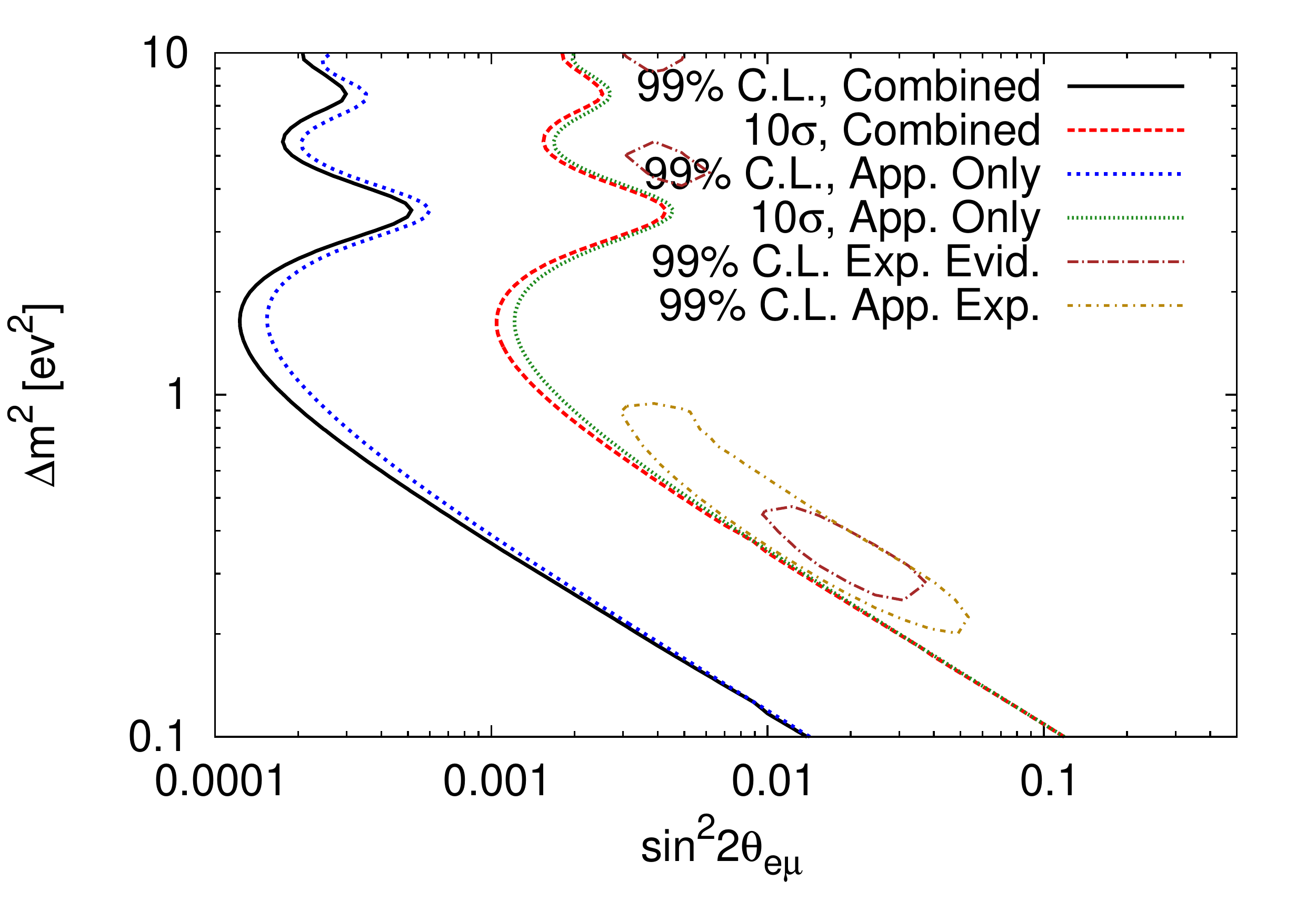}
  \label{fig:combsensapp}
  }
  \subfigure[Improvement over disappearance experiment]{
    \includegraphics[height=0.475\textwidth,angle=270]{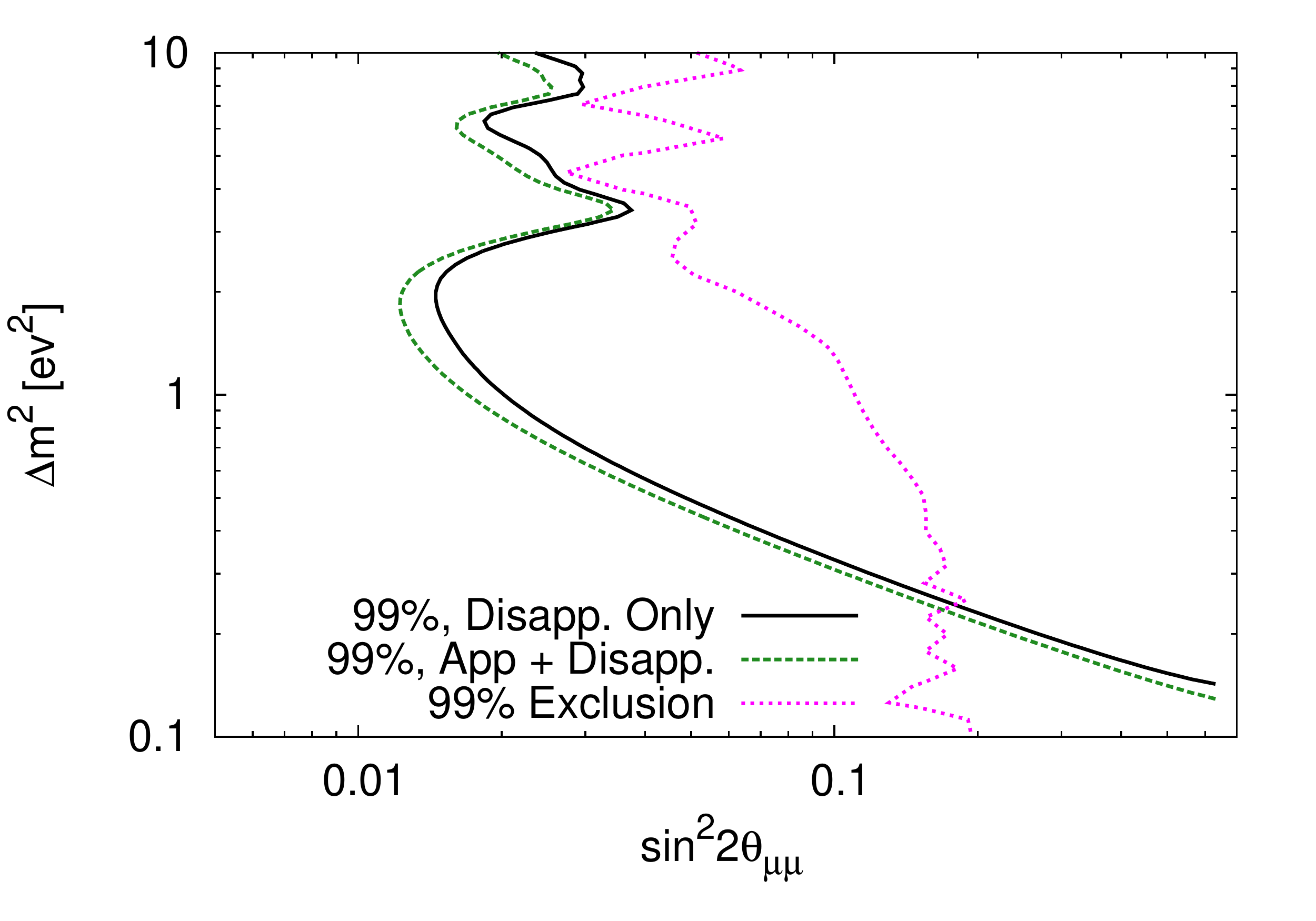}
  \label{fig:combsensdis}
    }
  \caption{The sensitivity of a combined fit between and $\nu_{\mu}$
    appearance experiment and a $\bar{\nu}_{\mu}$ disappearance
    experiment at nuSTORM. Contour generated using GLoBES, assuming
    a 1.3~kTon far detector, 2~km away from the muon storage ring,
    exposed to neutrinos resulting from an exposure of 1.6$\times 10^{18}$ useful muon decays. A 1\% signal and 10\% background
    systematic uncertainty is assumed.}
\end{figure}

The sensitivity contour derived from the combination of $\nu_{\mu}$
appearance and $\bar{\nu}_{\mu}$ disappearance is shown with respect
to $\sin^{2}2\theta_{e\mu}$ is shown in Fig.~\ref{fig:combsensapp}
and with respect to $\sin^{2}2\theta_{\mu\mu}$ in Fig.~\ref{fig:combsensdis}. The
combined sensitivity shows an improved performance over the appearance
experiment alone at $\Delta m^{2}_{14}>$0.8~eV$^{2}$. The improvement
is marginal but it does reduce the constraints on the systematic
uncertainties required to achieve the 10$\sigma$ benchmark for a short
baseline experiment. The improvement over the disappearance experiment
alone is more significant for all $\Delta m^{2}_{14}$.

\subsection{$\nu_e$ preliminaries}
\subsubsection{Shower reconstruction motivation}

In addition to the muon appearance-disappearance channels, the flavor content of the nuSTORM beam provides an option to investigate $\nu_{e}$ channels. $\nu_{e}$ appearance from the muon decay would prove challenging given the lack of charge identification available in SuperBIND for events likely to shower, but a $\nu_{e}$ disappearance channel is approachable. The CC-NC distinction required for these types of events could also provide an option to study NC disappearance, which would supply an especially powerful indicator for sterile neutrinos.

\paragraph{Electron shower reconstruction}

\begin{figure}[h]
\centering
\subfigure[$nu_{e}$ Charged-current quasi-elastic][Quasi-elastic]{
\includegraphics[width=0.45\textwidth]{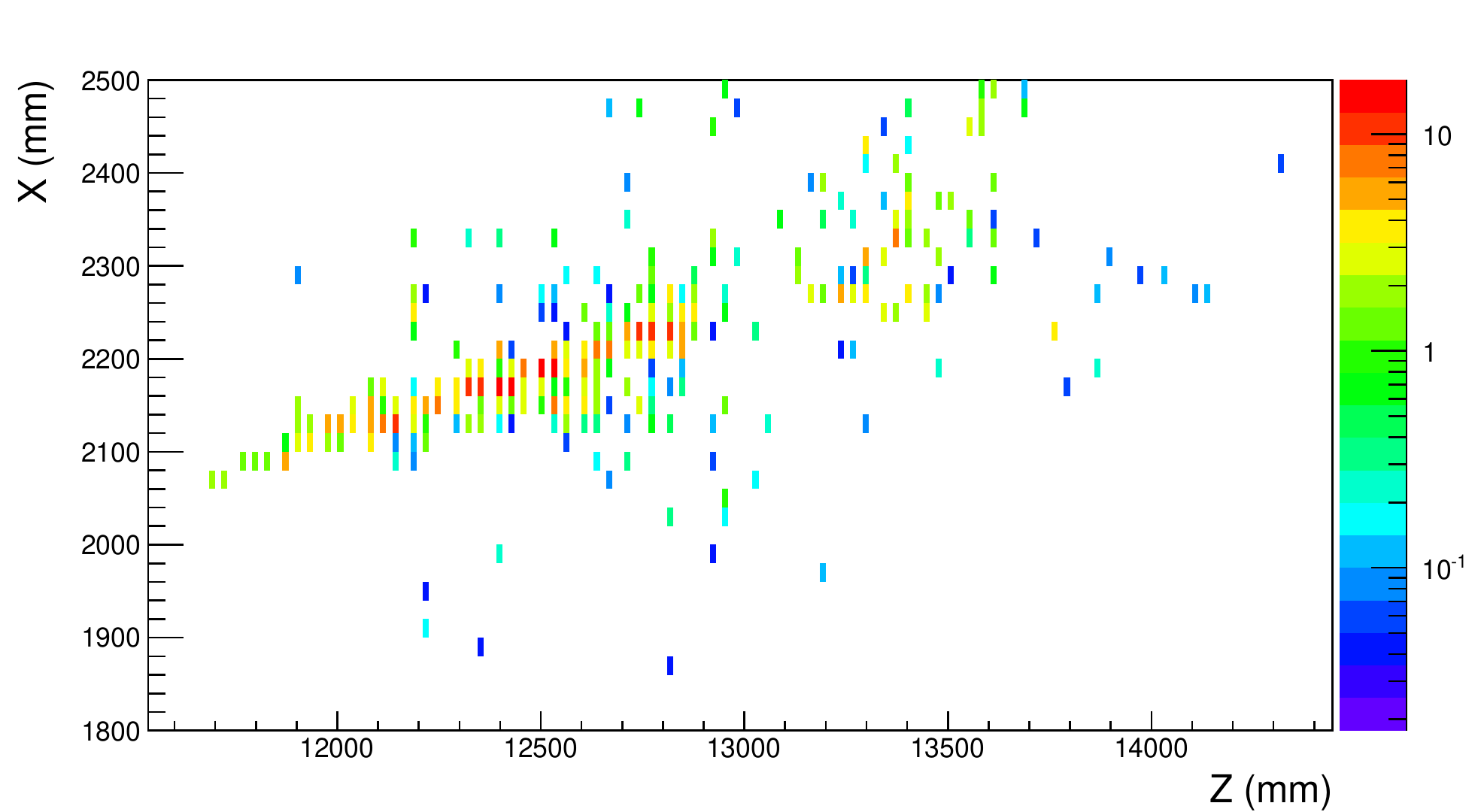}
\label{fig:subfig1}}
\subfigure[$nu_{e}$ Charged-current deep-inelastic][Deep-inelastic]{
\includegraphics[width=0.45\textwidth]{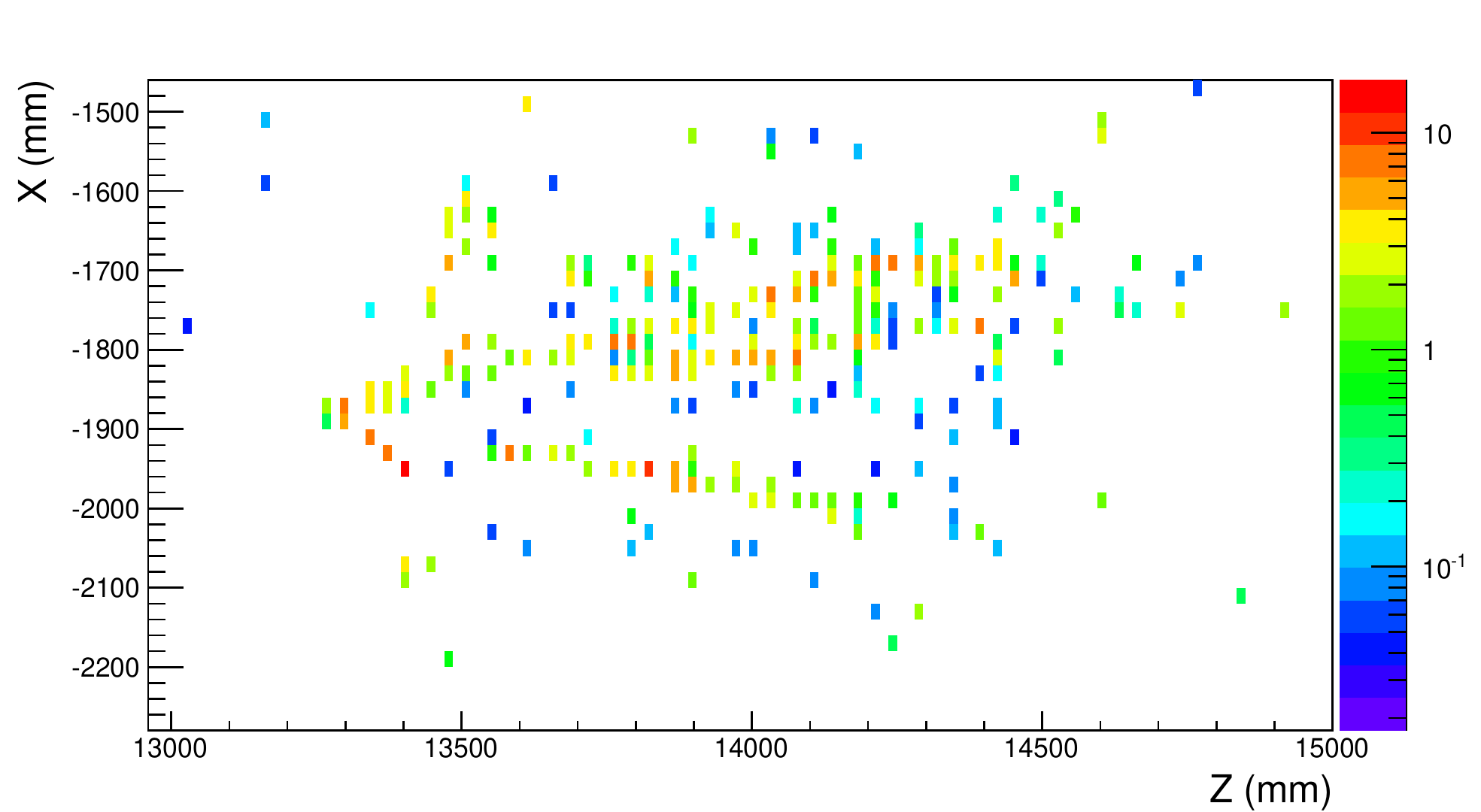}
\label{fig:subfig2}}
\qquad
\subfigure[$nu_{e}$ Charged-current resonance][Resonance]{
\includegraphics[width=0.45\textwidth]{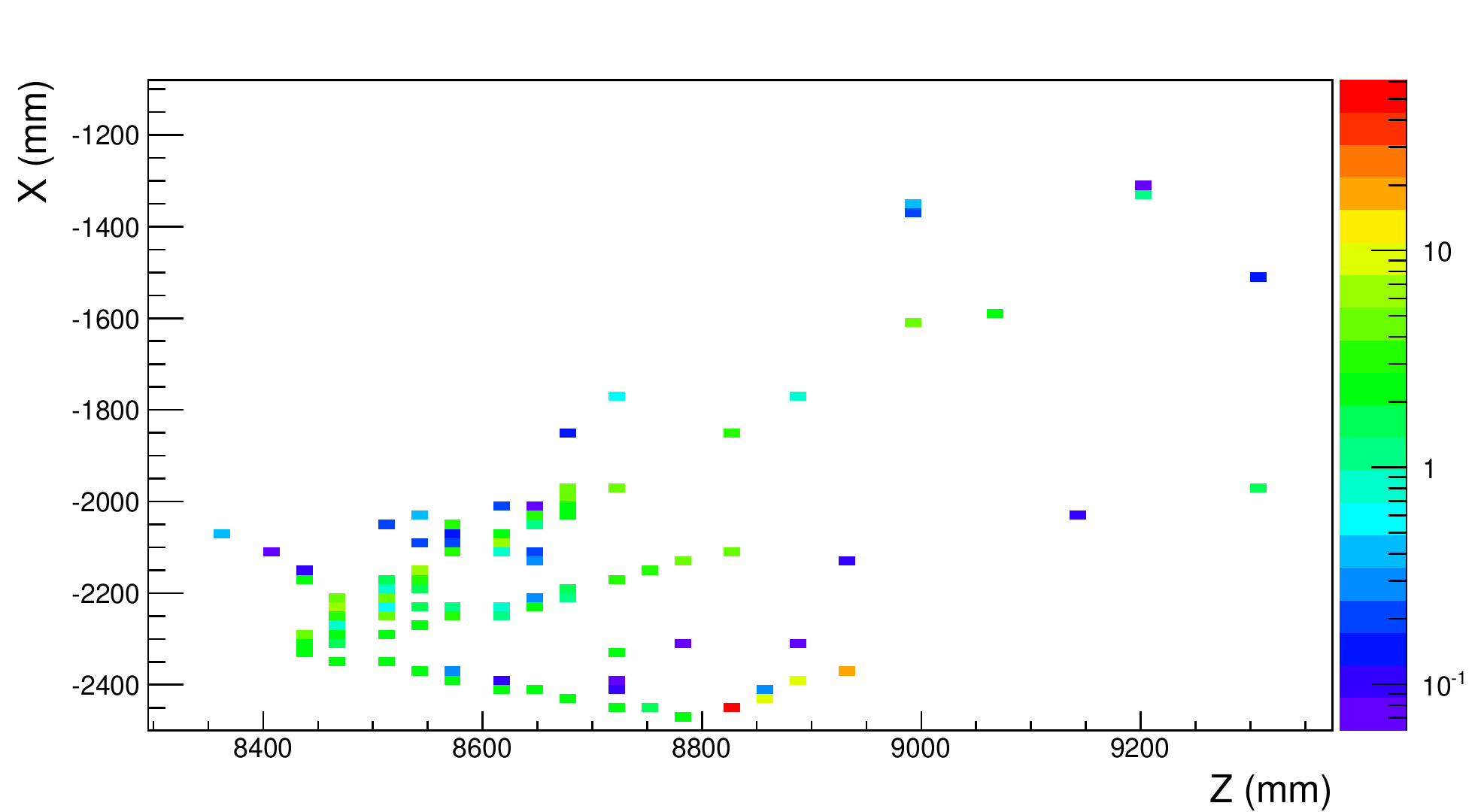}
\label{fig:subfig3}}
\subfigure[$nu_{e}$ Neutral-current quasi-elastic][NC Quasi-elastic]{
\includegraphics[width=0.45\textwidth]{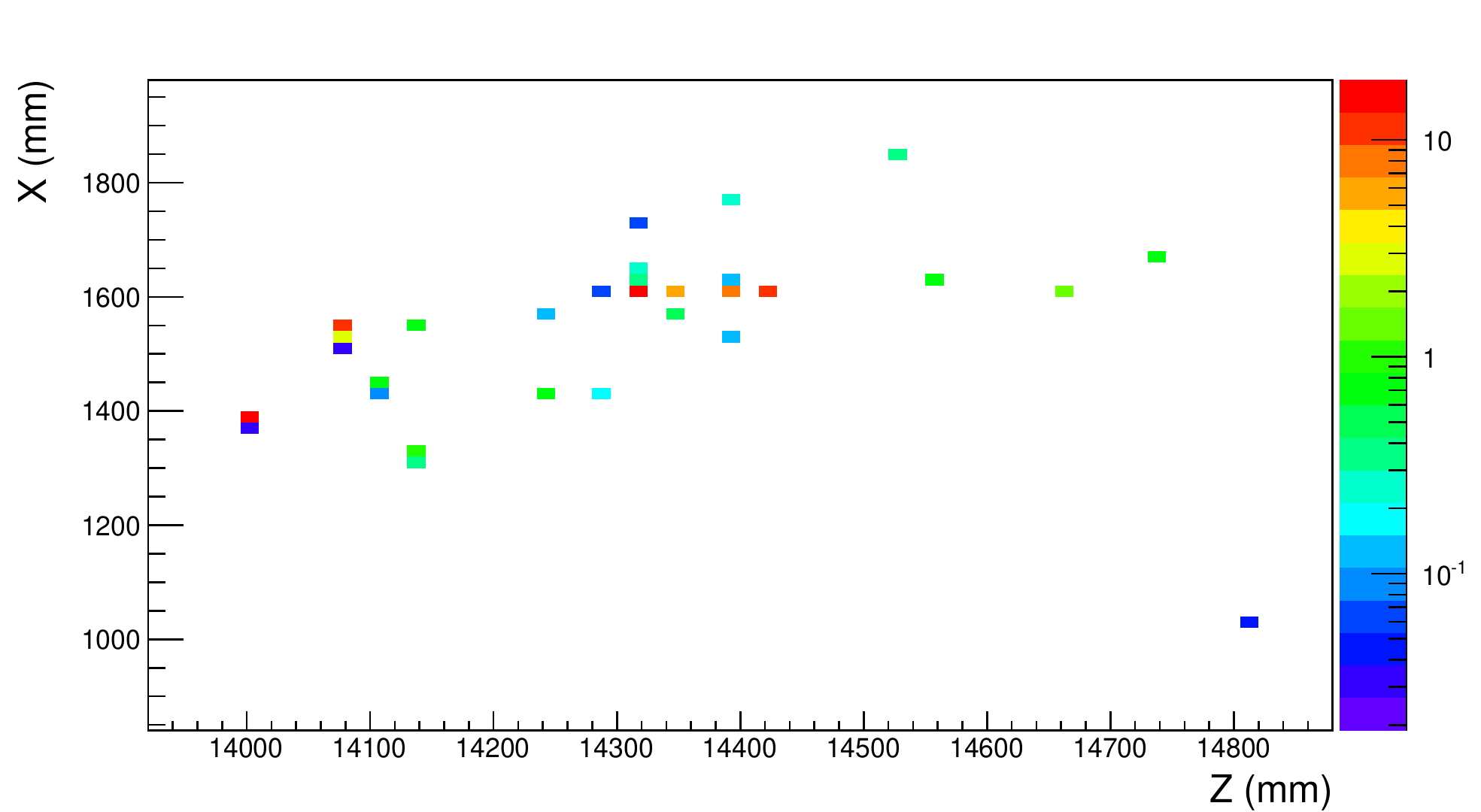}
\label{fig:subfig4}}
\caption{$\nu_{e}$ CC interaction examples in SuperBIND. The deposited energy in MeV is shown on the z-axis.}
\label{nue_cc-nc}
\end{figure}

The primary requirement placed upon the reconstruction for a $\nu_{e}$ analysis is the ability to identify the electron from the charged current interaction. In SuperBIND, the electron will quickly shower generating an event topology similar to that of neutral current interactions. Distinguishing between the electromagnetic shower from the $\nu_{e}$ CC and all other hadronic showers (both residually from the CC interactions through resonance, deep inelastic scattering etc., and all NC interactions) becomes the foremost concern. Examples of CC and NC event types for $\nu_{e}$ events in SuperBIND can be seen in Fig.~\ref{nue_cc-nc}.

\paragraph{Direction estimation}

The direction of the shower is estimated by taking the charge-weighted mean position of each plane and fitting with a straight line. In cases with little hadronic interaction this method is a reasonable approximation of the electron direction. However, any hadronic showers can interfere with the estimation of the electromagnetic shower direction, making this estimation poor. Where it is assumed the majority of the energy will be deposited within one Moliere radius, a deviation in the true shower direction of more than one radius prohibits current identification using the following method.

\paragraph{Current prediction}

\begin{figure}[h]
\centering
\includegraphics[width=0.8\textwidth]{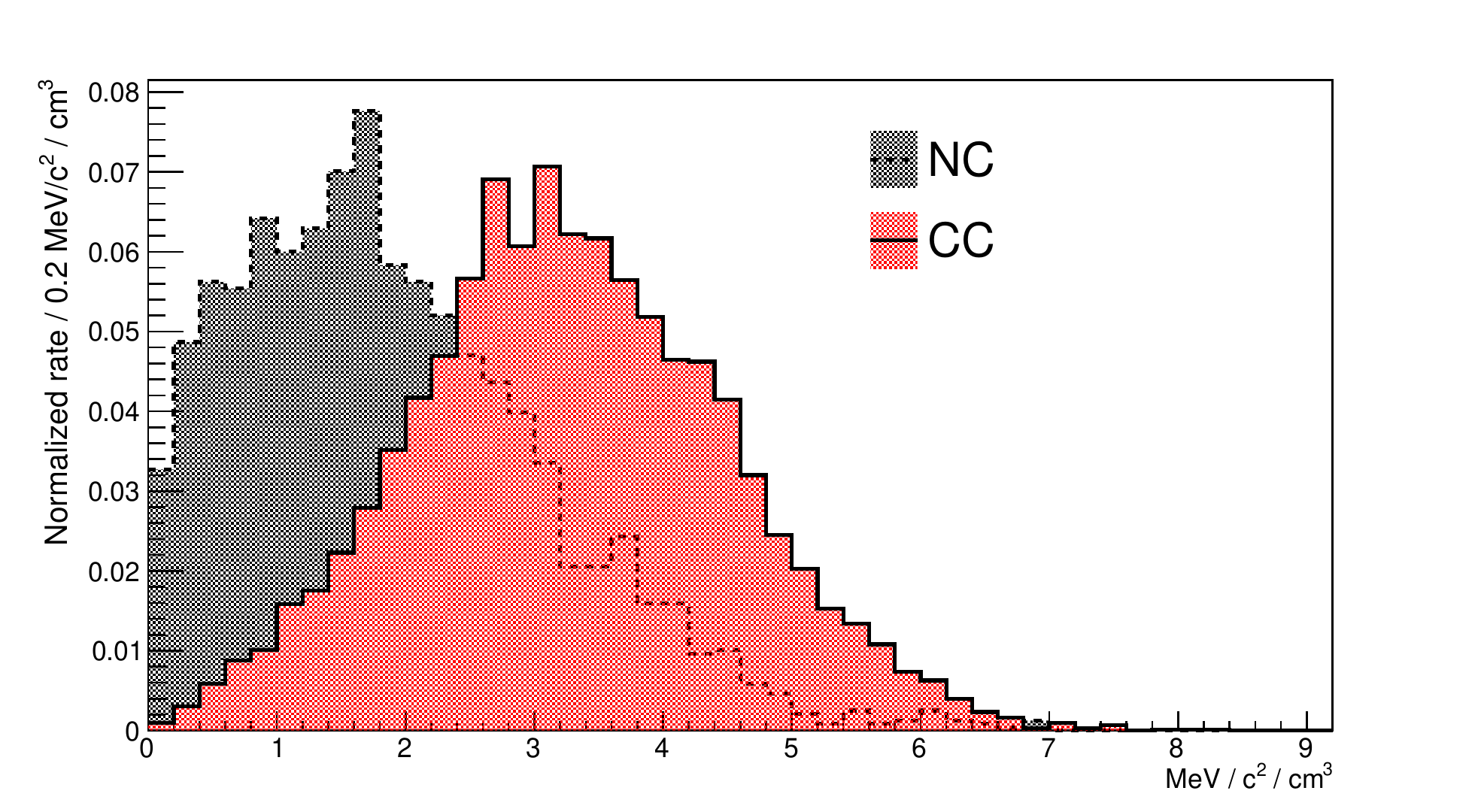}
\caption{Density of energy deposited in the scintillator for the total volume in which 50\% of the energy was found for $\nu_{e}$ CC and NC events.}
\label{cc-nc_density}
\end{figure}

By comparing the fraction of energy deposited in the central region of the shower, it is likely some distinction between the NC and CC events can be achieved. As it is expected that 90\% of the energy from an electro-magnetic shower should be contained within one Moliere radius, the volume which contains a majority of the energy deposited in the scintillator (here we have used 50\%) can be utilized as a figure of merit. The density of this region, which adds the further discrimination power of total energy deposited in the scintillator, can be seen in Fig.~\ref{cc-nc_density}. There are issues associated with successfully constraining this region, as the direction of the shower (which might need to be distinguished from multiple showers) can be difficult to reconstruct, and so is not included here. Therefore, the efficiency of this method is not currently optimal and is undergoing additional work. In any case, this shows some of the discrimination power available from a purely sampling approach in SuperBIND. 

It is expected that the shower density and other variables will be used in a multi-variate approach. Additional methods are under consideration, including the Library Event Matching employed by MINOS.

\clearpage
\subsection{Neutrino Interaction Study}
The requirements of the neutrino-nucleus-scattering program at the
nuSTORM facility were described in Section~\ref{sec:motivation:physics}.
Two studies have been carried out to demonstrate the potential of
nuSTORM.
A review of the systematic uncertainties that could be achieved at
nuSTORM is compared with the precision of a selection of present
and proposed experiments at conventional neutrino beams (section
\ref{SubSect:ScattSyst}).
A straw-man detector based on the MicroBooNE liquid-argon
time-projection-chamber is introduced in section
\ref{SubSect:nuScattPrec} and used to evaluate the precision with
which the charged-current quasi-elastic cross section could be
determined at the nuSTORM facility.
The results of the studies indicate that nuSTORM will be able to
provide definitive measurements of neutrino-nucleus scattering.

\subsubsection{Review of systematic uncertainties in neutrino-nucleus
            scattering studies}
\label{SubSect:ScattSyst}

The status of the data on neutrino-nucleus ($\nu N$) scattering was
reviewed in Section~\ref{sec:motivation:physics} together with a discussion of the
factors limiting the accuracy of the present generation of
experiments.
$\nu N$ scattering cross sections have been determined for incident
neutrino energies in the few-GeV range
\cite{miniboone_NCE_syst,miniboone_nu_e_nu_mu,miniboone_pi0,miniboone_cc_ddiff,argoneut_cc_numu,t2k,minerva}.
A summary of the systematic uncertainties affecting these measurements
is presented in Table~\ref{uncertainties}.
The flux uncertainties range from $\sim 7\%$ to $20\%$; in each case,
the flux uncertainty makes a substantial contribution to 
the total systematic error while in a number of cases the flux
uncertainty is the dominant source of systematic uncertainty.
\begin{table}
  \caption{
    Sources of systematic uncertainties for different experiments 
    \cite{miniboone_NCE_syst,miniboone_nu_e_nu_mu,miniboone_pi0,miniboone_cc_ddiff,t2k,minerva}. 
    The column headed ``Experiment'' reports the experiment and the
    channel (neutral-current elastic, NCE; charged-current
    quasi-elastic, CCQE; charged-current single $\pi^0$ production
    CC$\pi^0$; quasi-elastic, QE; and charged-current, CC).
    The incident neutrino flavor is also indicated.
    The systematic uncertainties are classified as uncertainties
    related to: the performance of the detector (column headed
    ``Detector''); the Monte Carlo simulation of the experiment
    (``Monte Carlo'') and ``Other''.
    The column headed ``Sub-total'' reports the combination of these
    uncertainties combined in quadrature.
    The flux uncertainty is reported in the column headed ``Flux''
    and the total systematic error in the column headed ``Total''.
  }
  \label{uncertainties}
  \begin{center}
    \begin{tabular}{|c|c|c|c|c||c|c|}
      \hline
                 & \multicolumn{6}{|c|}{Systematic uncertainty (\%)}         \\ \hline
      Experiment & Detector & Monte Carlo & Other & Sub-total & Flux & Total \\
      \hline
      MiniBooNE  & & & & & & \\
      NCE        & 15.6 & 6.4 & & 16.9 & 6.7 & 18.1 \\
      ($E_{\nu}\sim{1}$ GeV) & & & & & & \\
      \hline
      MiniBooNE            & & & & & & \\
      CCQE $\nu_{\mu}$ & 3.2 & 15.7 &      & 16.1 & 6.9 & 17.5 \\
      ($E_{\nu}\in 0.2-3.0$ GeV) & & & & & & \\
      \hline
      MiniBooNE              & & & & & & \\
      CCQE $\nu_{e}$ & 14.6 & 8.5 & & 16.1 & 9.8 & 19.5 \\
      ($E_{\nu}\in 0.2-3.0$ GeV) & & & & & & \\
      \hline
      MiniBooNE       & & & & & & \\
      CC$\pi^{0}$ $\nu_{\mu}$   & 5.8 & 14.4 & & 15.6 & 10.5 & 18.7 \\
      ($E_{\nu}\in 0.5-2.0$ GeV) & & & & & & \\
      \hline 
      MiniBooNE & & & & & & \\
      QE$\frac{d^{2}\sigma}{dT_{\mu}d\cos\theta_{\mu}}$  & 4.6 & 4.4 & &  6.4 & 8.7 & 10.7 \\
      ($E_{\nu}\in 0.5-2.0$ GeV)                       & & & & & & \\
      \hline
      T2K   & & & & & & \\
      Inclusive $\nu_{\mu}$ CC      & 0.7--12 & 0.4--9 & & 1.3--15 & 10.9 & 10.9--18.6 \\
      ($E_{\nu}\sim{1}$ GeV)   & & & & & & \\
      \hline
      Minerva & & & & & & \\
      $\bar{\nu}_{\mu}$ CCQE & 8.9--15.6 & 2.8 & 2--6 & 9.6--17 &  12 & 15.3--20.8 \\
      ($Q^{2} < $ 1.2 GeV$^{2}$) & & & & & & \\
      \hline
      LSND   & & & & & & \\
      $\bar{\nu}_{\mu} p \rightarrow \mu^{+} n$ & 5 & 12 &  &  13 &  15 & 20 \\
      $0.1 \text{GeV}$ & & & & & & \\
      \hline
    \end{tabular}
  \end{center}
\end{table}

\begin{figure}
 \begin{center}
  \includegraphics[width=0.9\textwidth]{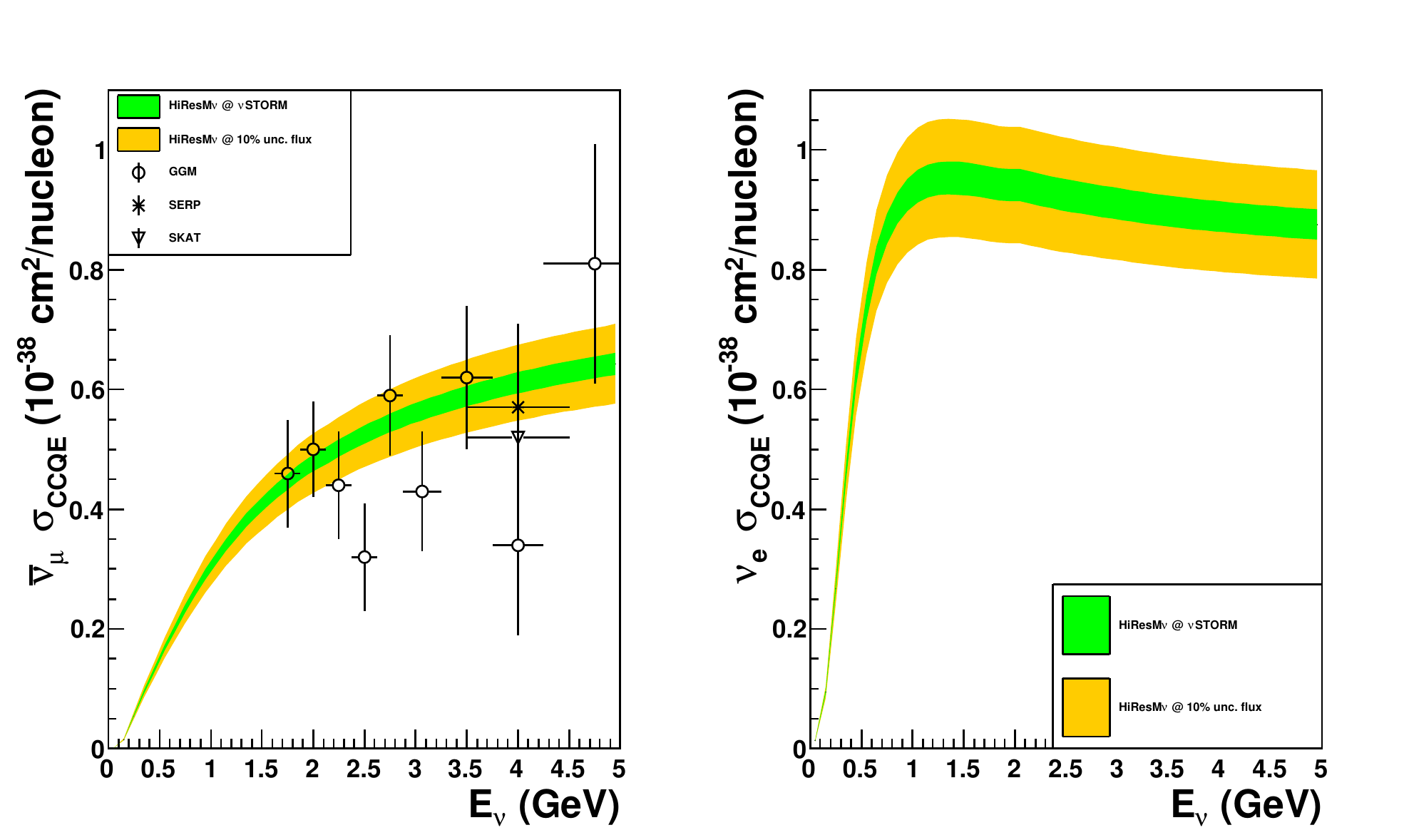} 
  \includegraphics[width=0.9\textwidth]{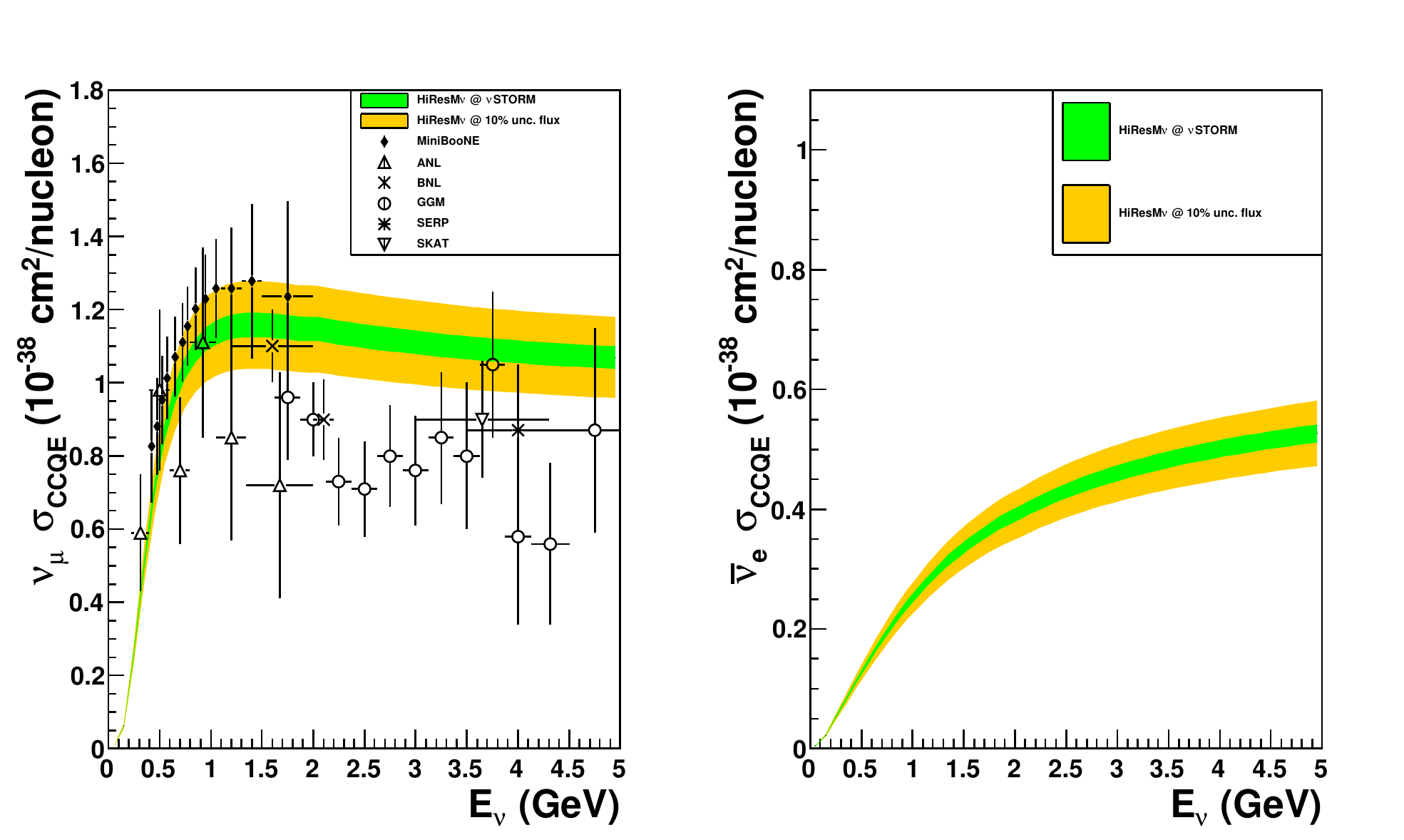}
 \end{center}
\caption{
 The CCQE cross section ($\sigma_{\rm CCQE}$) plotted as a function
  of incident neutrino energy ($E_\nu$).
  The cross sections that would be obtained with stored $\mu^+$ beams
  are shown in the top row; the $\bar{\nu}_\mu$ CCQE cross section is
  shown in the top left panel while the $\nu_e$ CCQE cross section is
  shown in the top right panel.
  The cross sections that would be obtained with stored $\mu^-$ beams
  are shown in the bottom row; $\nu_\mu$ CCQE cross section in the
  bottom left panel, $\nu_e$ CCQE cross section in the bottom right
  panel.
  The width of the colored bands represent the systematic
  uncertainty on the cross sections determined using the HiResM$\nu$
  detector at the nuSTORM facility (see text for details).
  The green band shows the detector uncertainties combined with the
  1\% uncertainty on the neutrino flux at nuSTORM.
  The yellow band shows the detector uncertainties combined with a
  flux uncertainty of 10\%.
  Measurements made by the MiniBoNE ($\blacklozenge$), ANL ($\bigtriangleup$), BNL ($\times$), 
  Gargamelle ($\bigcirc$), SERP ($\ast$) and SKAT ($\bigtriangledown$)
  collaborations are also shown
  \cite{miniboone_cc_ddiff,bnl_numubar,ggm_numu,lsnd,serp,skat,anl_data,bnl_numu}.
  The data can be found at \cite{neutrino_database}.
}
\label{genie_ccqe}
\end{figure}

At the nuSTORM facility, the flavor-composition of the neutrino
beam will be known and the flux will be determined with a precision of
1\% using the storage-ring instrumentation (see section
\ref{sec:Facility}).
Inspection of Table~\ref{uncertainties} indicates that, to make best
use of the excellent knowledge of the neutrino flux, a detector
capable of delivering measurements at the \% level will be required.
The HiResM$\nu$ detector \cite{hiresmnu} would be a suitable choice.
Table \ref{hires} lists the design parameters of the HiResM$\nu$
detector.
The performance of such a detector exposed to the nuSTORM flux is
illustrated in Fig.~\ref{genie_ccqe} where the charged-current
quasi-elastic (CCQE) cross section is plotted as a function of
neutrino energy $E_\nu$.
The figure shows the precision with which the cross section would be
measured if the systematic uncertainties estimated for the HiResM$\nu$
detector are combined with the 1\% flux uncertainty that nuSTORM
will provide.
For comparison, the performance of HiResM$\nu$ combined with a flux
uncertainty of 10\% is shown.
Fig.~\ref{genie_ccqe} also shows the present measurements of the
CCQE cross section; these measurements are only available for
muon-neutrino (and muon-anti-neutrino) beams.
The figure shows that nuSTORM has the potential to improve the
systematic uncertainty on muon-neutrino (muon-anti-neutrino) CCQE
cross section measurements by a factor of $\sim 5-6$.
The electron-neutrino- (electron-anti-neutrino-) nucleus cross section
measurements that can be made with nuSTORM will be a unique
contribution.
\begin{table}
  \caption{In order to take maximal advantage of the nuSTORM accurate 
  beam the detector errors need to be kept small. The HiresM$\nu$ small 
  uncertainties \cite{mishra_email} make it a suitable detector for this effect. The
  "Reconstruction" error refers to the track reconstruction error. It is 
  dominated by the proton-reconstruction in the QE event. The "Background" estimate
  corresponds to the contamination of resonant and DIS events. Finally, the "FSI error"
  estimation corresponds to the impact of final state interactions on the topology of the measured tracks.}
  \label{hires}
  \begin{center}
    \begin{tabular}{| c | c | c |}
      \hline
      Detector    & Types of Errors & Contribution (\%) \\
      \hline
                  & Reconstruction  & 0.8               \\
      HiResM$\nu$ & Background      & 2.1               \\
                  & FSI error       & 1.5               \\
      \hline
                  &  Total          & 2.9               \\
      \hline
    \end{tabular}
  \end{center}
\end{table}
\subsubsection{The potential of nuSTORM: a straw-man detector}
\label{SubSect:nuScattPrec}
\begin{figure}
  \begin{center}
    \includegraphics[width=0.8\textwidth]{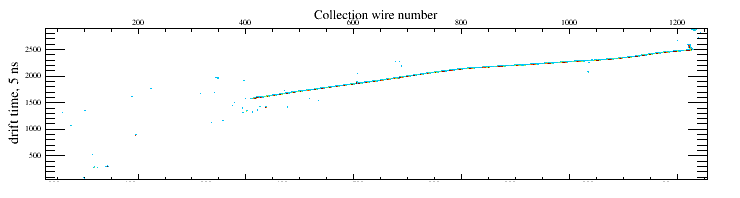}
  \end{center}
  \caption{
    Example of a $\bar{\nu}_{\mu}$ CCQE interaction. The track in this event display belongs to a 0.9 GeV $\mu^{+}$.
  } 
  \label{Fig:muCCQE}
\end{figure}

\begin{figure}
  \begin{center}
    \includegraphics[width=0.8\textwidth]{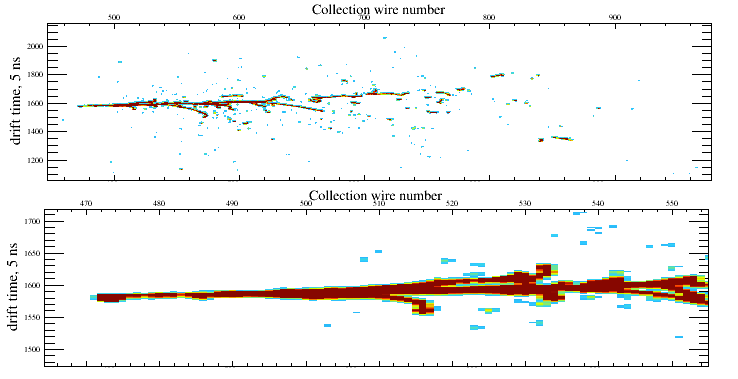}
  \end{center}
  \caption{
    Selection criteria for the analysis. The CCQE-like events are defined as
    containing one lepton and strictly no pions. Combining this with data in
    Table~\ref{efficiencies}, one can expect an efficiency loss for leptons with
    small momentum/short range. The resulting CCQE-like event count will
    include topologies where any given number of neutrons, and protons with
    less than 40\,MeV are present.}
  \label{eleCCQE}
\end{figure}

The MicroBooNE detector, an unmagnetized liquid-argon (LAr)
time-projection-chamber (TPC) with a mass of 170\,T
\cite{MicroBooNE} was used as the basis for the straw-man detector.
The LArSoft package \cite{LArSoft} was used to simulate 
$\bar{\nu}_\mu$ and $\nu_e$
events and the event samples
were combined to simulate the flux that will be delivered by
nuSTORM with a stored $\mu^+$ 
beam. 
Studies of the performance of LAr TPCs have demonstrated high spatial
resolution \cite{icarus_3d_tracking} and excellent energy resolution
\cite{argoneut}. The combination of particle range measurements with specific energy
loss ($dE/dx$) allows efficient particle discrimination.
The efficiencies and resolutions used in the LArSoft simulation are
summarized in Table~\ref{efficiencies}.
\begin{table}
  \caption{
    Parameters assumed in the LArSoft simulation implemented for this analysis.
  }
  \label{efficiencies}
  \begin{center}
    \begin{tabular}{|c|c|}
      \hline
      Effect & Value \\
      \hline              
       Momentum resolution of contained tracks & 3\% \\
       Angular resolution              & 3\% \\
       Minimum range for track finding       & 2\,cm \\
      \hline
    \end{tabular}
  \end{center}
\end{table}
Neutrino interactions were simulated within a LAr TPC enclosing a
volume of $256.35 \times 233 \times 1036.8$\,cm$^3$.   
The software used to simulate the neutrino interactions was LArSoft. 
LArSoft uses the GENIE neutrino interaction generator \cite{GENIE} to
produce interactions in the liquid argon volume, and stores the
final-state particles for further simulation.   
Particles were passed to Geant4 to be tracked through the detector.
Events were generated for incident neutrino energies uniformly
distributed between 0.5\,GeV and 4\,GeV, reweighted to the nuSTORM
flux and scaled to the number of interactions expected for an
exposure of $10^{21}$ POT on a 100\,T fiducial mass at a distance of 50\,m
from the end of the straight section (see section
\ref{sec:motivation:physics}). 
Fig.~\ref{Fig:muCCQE} and Fig.~\ref{eleCCQE} show typical CCQE in the
muon and electron channels in the MicroBooNE detector.
For this analysis, the parameters listed in Table~\ref{efficiencies}
were used in the LArSoft package.
Muon-neutrino induced CCQE events were selected by requiring a muon
candidate unaccompanied by proton or pion candidates.
Electron-neutrino induced CCQE events were selected by requiring a
single-electron candidate accompanied by at most one proton candidate
and unaccompanied by a pion candidate.
The events-selection criteria are summarized in table \ref{criteria}.
\begin{table}
  \caption{
    Selection criteria for the analysis. The CCQE-like events are defined as
    containing one lepton and no pions. Single electron candidates
    where the companion proton is not found are also accepted.}
  \label{criteria}
  \begin{center}
    \begin{tabular}{| c | c | c |}
      \hline
      Interaction Category & Required Topology \\
      \hline
      $\bar{\nu}_{\mu} \text{ CCQE}_{LIKE}$ & 1 $\mu^{+}$ + 0proton + 0pion   \\
      $\nu_{e} \text{ CCQE}_{LIKE}$         & 1 $e^{-}$ + 0/1 proton + 0pion    \\
      \hline
    \end{tabular}
  \end{center} 
\end{table}

\begin{figure}
  \begin{center}
    \includegraphics[width=0.9\textwidth]{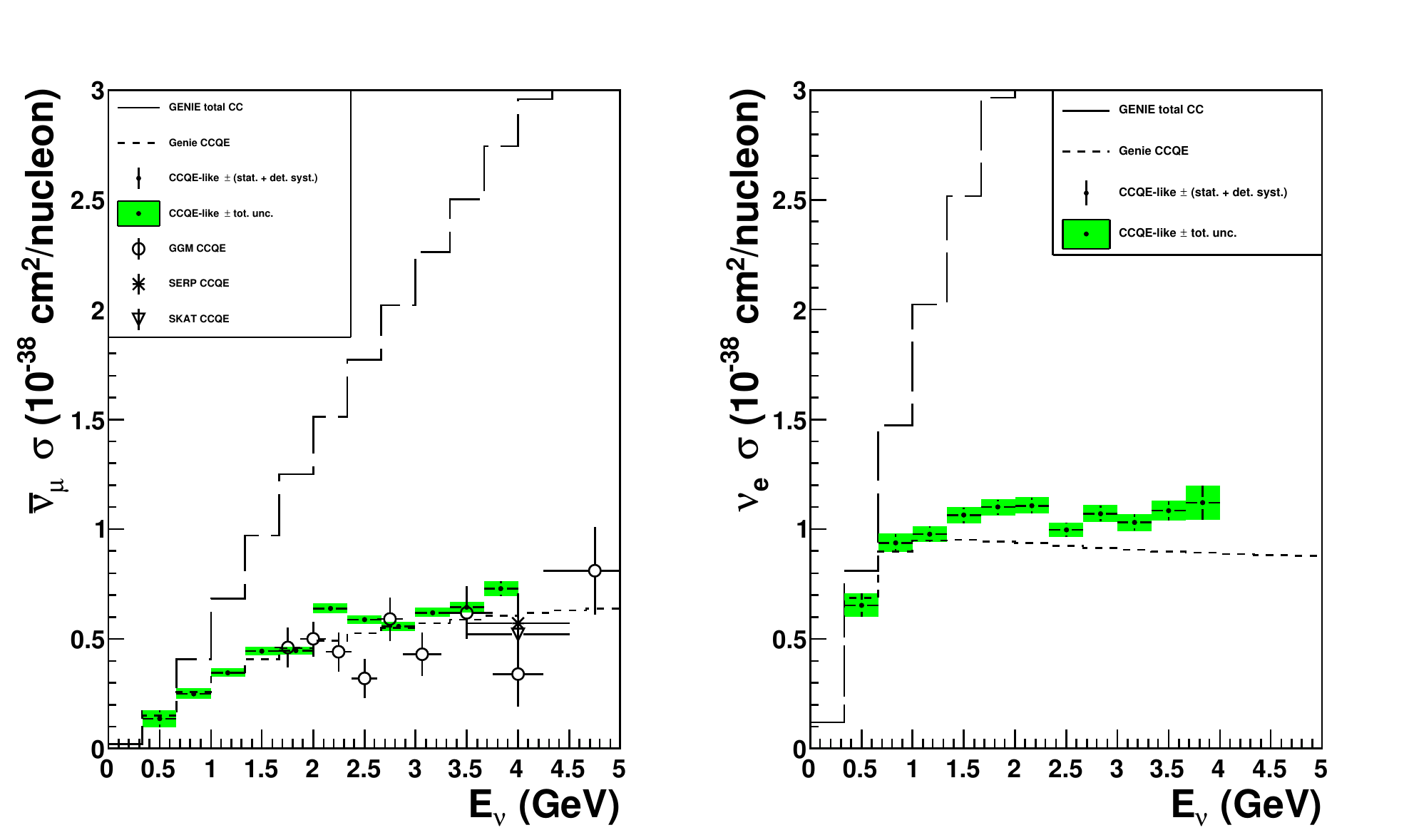}
  \end{center}
  \caption{
    $\bar{\nu}_{\mu}$ and $\nu_{e}$ CCQE-like cross sections
    obtained using the straw-man detector analysis described in the
    text.
    The uncertainties include statistical, detector systematic (3\%)
    uncertainties and the 1\% nuSTORM flux uncertainty. 
    The dashed lines show GENIE's total CC and Quasi-Elastic cross
    sections for comparison. 
  Measurements made by the
  Gargamelle ($\bigcirc$), SERP ($\ast$) and SKAT ($\times$)
  collaborations are also shown.
  }
  \label{final_result}
\end{figure}

In practice, the simple CCQE-like selection based on track finding will
also select events in which any number of hadrons are produced which
deposit less than $\sim 40$\,MeV in the detector.
Fig.~\ref{final_result} shows the cross sections for the production
of CCQE-like events using the criteria described above.
The event sample that can be accumulated using a detector with a
fiducial mass of 100\,T at nuSTORM is large enough that the
statistical uncertainty is significant only in the lowest $E_\nu$ bin
($\sim$ 0.5 GeV) and becomes negligible at higher energies.
The flux uncertainty of 1\% also makes a relatively small
contribution, leaving the detector systematic as the dominant source 
of uncertainty.
In the muon channel, nuSTORM offers a six-fold improvement in the
precision with which the CCQE cross sections can be measured.
In the electron channel the measurements that nuSTORM will provide
will be unique.
The extension of this analysis to other channels will be the subject
of future work.

\clearpage
\section{Outlook and conclusions}
\label{sec:OnC}
In this proposal we have presented a compelling case for the nuSTORM facility.  As mentioned in the introduction,
nuSTORM's motivation rests on three central themes: 1. A search for sterile neutrinos for unprecedented precision, 
2. Unique opportunities in $\nu$ interaction physics and 3. Presents a powerful technology test bed for muon
accelerator physics. 
With respect to search for sterile neutrinos, nuSTORM present the only facility that can do all of the following:
\begin{itemize}
\item Make a direct test of the LSND and MiniBooNE anomalies.
\item Provide stringent constraints for both $\nu_e$ and $\nu_\mu$ disappearance to
over constrain $3+N$ oscillation models and to test the Gallium and reactor anomalies
directly.
\item Test the CP- and T-conjugated channels as well, in order to obtain the relevant clues 
for the underlying physics model, such as CP violation in $3+2$ models.
\end{itemize}

With respect to $\nu$ interaction physics, nuSTORM presents the first opportunity to measure both  $\nu_e$ and $\nu_\mu$ cross sections at the the 1\% level
with a $\nu$ beam that can be characterized 5 to 10 times more precisely than conventional $\nu$ beams.

Lastly, with respect to accelerator R\&D, nuSTORM can provide muon beams suitable for the next generation of studies into muon ionization cooling 
which are so crucial to the viability of any $\mu^+\mu^-$ collider.  This can be done simultaneously while carrying out its neutrino physics program.
\subsection{Moving forward}
We are requesting Stage-1 approval for nuSTORM.  Stage-1 approval will allow us to strengthen our collaboration and will, in particular, provide a strong foundation for
our non-US collaborators to obtain support from their funding authorities to help prepare the next step for nuSTORM -- the preparation of a Conceptual Design Report.
Proton beams capable of serving the $\nu$STORM facility can also be
provided at CERN and with that realization, an Expression of Interest on nuSTORM has been submitted to the CERN SPS Committee.  It will come up for discussion at the next open meeting of the SPSC on June 25th of this year.  The EOI to CERN has requested resources to:
\begin{itemize}
  \item Investigate in detail how $\nu$STORM could be implemented at
    CERN; and 
  \item Develop options for decisive European contributions to the
    $\nu$STORM facility and experimental program wherever the 
    facility is sited.
\end{itemize}
The EoI defines a roughly two-year program which culminates in the delivery of a Technical Design Report which corresponds to the level of detail found in
Conceptual Design Reports as produced during the DOE CD process.

With support potentially from both the Fermilab PAC and the CERN SPSC, nuSTORM can proceed within a decidedly ``international-framework" with renewed vigor.   This is a rare, but not unique situation, and would allow the collaboration to make significant progress refining the scientific and technical cases for nuSTORM while at the same time performing an overall optimization of the facility and its cost.  This is truly a case where ``The whole is greater than the sum of its parts \cite{Aristotle}".
\subsection{Support request}
Table~\ref{tab:Support} lists the support we require over the next 12-18 months in order to make further progress on the work needed to be accomplished for the CDR.
\begin{table}[h]
\centering
\caption{Support request}
\label{tab:Support}
\begin{tabular}{|l|ccl|}
\hline
Task						&  Division		&  Effort type		&  FTE \\
\hline
$\pi$ production simulations	&  APC			&  S				&  0.15 \\
Inconel target studies			&  AD			&  E				&  1.0 \\
Proton beamline optimization	&  AD			&  S/E			&  0.3 \\
Decay ring lattice studies		&  AD			&  S				&  0.3 \\
Kicker design				&  AD			&  E				&  0.2 \\
Magnet design				&  TD			&  S/E/D			&  1.0 \\
Decay ring instrumentation design &  AD			&  E				&  0.5\\
\hline
\noalign{\smallskip}
\end{tabular}
\end{table}
S: Scientist, E: Engineer, D: Designer/drafter
\cleardoublepage
\begin{appendix}
\section{Long-Baseline Considerations}
A multitude of experiments have been proposed in order to observe CP violation
in the leptonic sector and determine the neutrino mass hierarchy (MH), see for instance ~\cite{Agarwalla:2012mz}.
In the U.S. context, the long baseline neutrino experiment (LBNE) is
being pursued. The present design consists of a
700\,kW conventional neutrino beam aimed at a 10\,kt liquid argon
(LAr) detector placed at $L=1300$\,km from the source~\cite{CDR}. Due
to its long baseline and relatively large matter effects, LBNE would
be able to determine the MH at $3\sigma$ for $75\%$ of the parameter
space. Its ability to measure the CP phase precisely and/or to 
discover CP violation is, however, limited due to a lack
of statistics. 

Measurement of the CP phase to a similar precision to that
achieved in the quark sector is only offered by a Neutrino Factory
(NF)~\cite{Coloma:2012wq,Coloma:2012ji}. In a NF, a highly collimated
beam of muon neutrinos and electron antineutrinos is produced from
muon decays in a storage ring with long straight
sections~\cite{Geer:1997iz}. 
The present NF design parameters~\cite{NF:2011aa} are $10^{21}$ useful
muon decays per $10^7$ seconds, aimed at a 100\,kton magnetized iron
detector (MIND) placed at 2000 km from the source, with a parent muon
muon energy of 10\,GeV. In order to form an intense muon beam for
acceleration and storage, muon phase space cooling is required for
this default configuration.  In addition, the fact that the neutrinos
in a NF are a tertiary beam implies significant proton driver
intensities; in this case, a 4\,MW proton beam plus its associated
target station.  

These technical challenges are to be contrasted with the advantages of
a NF -- there are no intrinsic backgrounds and the absolute neutrino
flux can be determined to better than 1\%.  Furthermore, the
presence of both muon and electron neutrinos in the beam  allows
for a measurement of all final flavor cross sections at the near
detector. A detailed analysis of the impact of systematic
uncertainties in neutrino oscillation experiments has been recently
performed in~\cite{Coloma:2012ji}, where the key systematics
affecting the different types of facilities were identified. In
particular, it was shown that the main sources of systematics
affecting a NF are matter density uncertainties (see also~\cite{Huber:2002mx}). One possibility to reduce the impact of the matter uncertainty would be to combine the results from the golden ($\nu_e\rightarrow\nu_\mu$) and the so-called platinum ($\nu_\mu\rightarrow\nu_e $) channels~\cite{Huber:2006wb}. 
It is well-known that this combination of channels is very effective in solving degeneracies, particularly in the case of large $\theta_{13}$ where background levels are not so relevant. However, the platinum channel is inaccessible in a MIND, because it requires the identification of the lepton charge in electron neutrino 
(and anti-neutrino) charged current events. 

In order to be able to access the platinum oscillation channels, a totally active-scintillator detector (TASD) or a Liquid Argon (LAr) detector would be needed. In this case, the big challenge would be the magnetization of the detector. At a NF it is mandatory to distinguish the sign of the charged lepton produced at the detector, in order to disentangle the appearance and disappearance signals. 
In~\cite{Bross:2007ts} the feasibility of electron charge identification was studied in the context of a low energy NF for a TASD 
in a 0.5\,T magnetic filed using a so-called magnetic cavern. It soon was
speculated that a magnetized LAr detector should be suitable as
well~\cite{Bross:2009zzb,Tang:2009wp,FernandezMartinez:2010zza,Ballett:2012rz}.
Note, that the detectors most likely will have to be deep underground
due to the large duty factor of stored particle beams in a NF.

In~\cite{Christensen:2013va} the physics performance of a low
luminosity ($10^{20}$ useful muon decays per year), low energy neutrino
factory, using a magnetized 10 kton LAr detector placed at 1300 km
from the source, was studied.  This choice of detector size and
baseline is obviously inspired by LBNE: it allows the reuse of the LBNE
facilities to the largest extent possible and thus makes this option
considerably more efficient in terms of resource usage. These
parameters correspond to an overall reduction in exposure by a factor of
100, and would in principle allow the NF to be based on existing
technology and already planned infrastructure. To give a specific
example -- consider phase~II of Project X, where initially a 3\,GeV proton beam with 1\,MW of power
will become available, which, according to studies
performed within the Muon Accelerator Staging Study (MASS), without muon
cooling could result in $10^{20}$ useful muon decays per year and
polarity.  This constitutes a reduction of luminosity by a factor 5
with respect to the default setup~\cite{NF:2011aa}. Then following ~\cite{Christensen:2013va} this permits the
detector mass to be reduced from 100\,kt to 10\,kt. This is but one
example of how to achieve this entry level luminosity.  Obviously, as
Project~X matures, future proton beam facilities may have much more
favorable parameters resulting in corresponding increases in
luminosity.  In an extended phase~II of Project~X, the beam power may
increase to 3\,MW and with the addition of cooling, $5.6\times10^{20}$
useful muon decays can be achieved.  Eventually, in phase~IV,  4\,MW at
8\,GeV will become available, which will yield $10^{21}$ useful muon
decays per year and polarity.

\begin{figure*}[t!]
  \includegraphics[width=1.0\columnwidth]{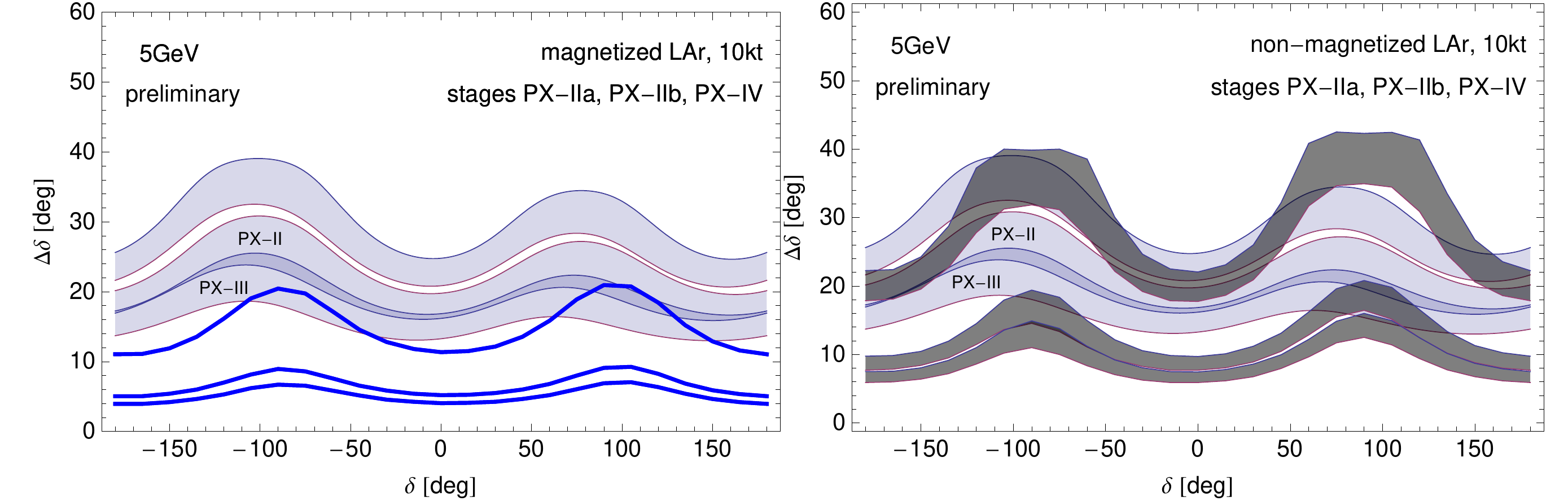}
  \caption{ Expected precision for a measurement of $\delta$ at $1\sigma$ (1 d.o.f.) as a function of the true value of $\delta$. Results are shown for the low luminosity, low energy neutrino factory (lines in the left panel, dark gray bands in the right panel) as well as for different exposures for LBNE (light bands). For LBNE the different bands correspond, from top to bottom, to 70, 110 and 230 MW$\times$kt$\times$yr. For the NF the different lines (bands) correspond, from top to bottom, to 1, 5.6 and 10 $\times 10^{20}$ useful muon decays per polarity and year, using a 10 kton LAr detector in all cases. The left panel shows the results when the detector is assumed to be magnetized; while the right panel shows the variation in the results when the charge identification efficiency is varied from 0.5 (upper edge of each band) to 0.9 (lower edge). See text for details.}
\label{fig:mass}
\end{figure*}

The results obtained in ~\cite{Christensen:2013va} for the
precision attainable for a measurement of the CP violating phase
$\delta$ are shown in the left panel of Fig.~\ref{fig:mass}, at
$1\sigma$ (1 d.o.f.). From top to bottom, the lines correspond to the
results for a NF using 1.0, 5.6 and 10.0$\times 10^{20}$ useful muon
decays per polarity and year. The results are compared with the
expected precision attainable at LBNE phase I (0.7 MW$\times$10
kt$\times$10 yr.), at LBNE using a beam produced in the second stage of
project X (1.1 MW$\times$10 kt$\times$10 yr.) and at LBNE using the
beam produced at the third stage of project X (2.3 MW$\times$10
kt$\times$10 yr.). Since the performance of a magnetized LAr detector is indeed
uncertain, in ~\cite{Christensen:2013va} the sensitivities using
the performance parameters of a TASD~\cite{Bross:2007ts}, were also
presented. Even though more conservative signal and background
rejection efficiencies were considered in this case, these only
resulted in a slight reduction in performance with respect to the
results using a LAr detector. The detector parameters are summarized
in Table~1 in~\cite{Christensen:2013va}.  Systematic uncertainties
have been implemented as in~\cite{Coloma:2012ji}, using the
default values for the systematic uncertainties listed in Table~2
therein. All correlations are taken into account, between different
channels as well as between the near and far detectors. All
simulations have been done using a modification of
GLoBES~\cite{Huber:2004ka,Huber:2007ji}, and marginalization over
not-shown oscillation parameters was performed as explained in~\cite{Coloma:2012ji}.
 
As already mentioned, one of the major technical challenges for this
setup is the magnetization of a large volume LAr detector. However,
even if magnetization would be the preferred option to achieve a very
clean charge ID for muons and electrons, this may not be mandatory. It
was already shown in ~\cite{Huber:2008yx} that, especially for
large values of $\theta_{13}$, a modest charge identification
efficiency would yield good performance for a full luminosity low
energy neutrino factory. Such efficiency may be attainable at a LAr
non-magnetized detector combining different signatures that are
different for neutrino and antineutrino events~\cite{Huber:2008yx}: 1)
in argon, the capture probability of a $\mu^-$ by an atom is about $
76\%$, whereas this process cannot take place for $\mu^+$; 2) the
angular distribution of the outgoing lepton is different for $\nu$ and
$\bar\nu$ events; 3) protons produced in QE $\nu$ events could in
principle be tagged in argon. In~\cite{Huber:2008yx} it was
estimated that, combining these signatures, a charge identification
efficiency between 50\% and 90\% could be achieved for a LAr detector.
Under the same assumptions, the performance of a low luminosity low
energy neutrino factory is presented in the right panel of
Fig.~\ref{fig:mass}, where the results are also compared to different
phases of LBNE. From top to bottom, the dark gray bands show the
performance for the setup described above, using the same useful muon
decays as in the left panel: 1.0, 5.6 and 10$\times10^{20}$ decays per
polarity and year, from top to bottom. The width of the bands shows
the variation in performance when the charge separation efficiency is
increased from 0.5 (upper edges) to 0.9 (lower edges). The same charge
separation efficiency is assumed for muons and electrons.

The results from~\cite{Christensen:2013va} (summarized in
Fig.~\ref{fig:mass}) show that a NF with 10 times fewer useful muon
decays and a 10 times smaller detector mass with respect to the
baseline NF scenario, would still outperform realistic super-beam setups like
LBNE. This setup stands as a viable upgrade path for the nuSTORM
facility, since this low luminosity can be achieved using existing
proton drivers at Fermilab and without muon cooling.  The reduced muon
energy with respect to the NF baseline design makes it possible to use a
baseline of around 1300\,km, and the use of a magnetized LAr
detector allows for full exploitation of the physics potential of the platinum
channel, which is crucial for the overall performance of the facility.
It has also been shown that, if magnetization is not viable, a modest
charge separation would still allow for reasonable performance,
comparable to that of LBNE. Finally, it should also be noted that,
once a 4\,MW lower energy proton beam becomes available from Project
X, and if muon cooling is added and the detector mass is increased by a
factor of 1-3, the performance of this facility would match or even
exceed that of the baseline NF.  Therefore, neither the initial
energy of 5\,GeV nor the baseline need to be changed at later
stages.

\section{Magnetized Totally Active Detector}
\label{sec:Appen}

We have shown in Sec.~\ref{sec:far} that a magnetized detector is required for nuSTORM if
we wish to study the $\parenbar{\nu}_\mu$ oscillation appearance channels and that this naturally lead to the choice of magnetized iron technology.  If one wanted to also look for $\parenbar{\nu}_e$ appearance,
then magnetized totally active detector technology would be an appropriate alternative.
Magnetic solutions for totally active detectors were studied within the International Scoping Study (ISS)~\cite{Abe:2007bi} in the context of investigating how very large magnetic volumes could be produced at
an acceptable cost.  A liquid Argon (LAr) or a totally-active  scintillator detector (TASD) could
be placed inside such a volume giving a magnetized totally active detector.
The following technologies were considered:

\begin{itemize}
\item Room Temperature Coils (Al or Cu)
\item Conventional Superconducting Coils
\item High Tc Superconducting Coils
\item Low Temperature Non-Conventional Superconducting Coils
\end{itemize}

Within the ISS, much larger detector masses were considered than the 1 kT needed for nuSTORM. 
However, we can consider using one of the 10 large solenoids (each 15 m diameter $\times$ 15 m long)
studied in the ISS for use with a 1 kT LAr detector.  The ISS concept of a ``magnetic cavern" is shown in Fig.~120. 

\begin{figure}[htbp]
  \centering{
    \includegraphics[width=0.6\textwidth]{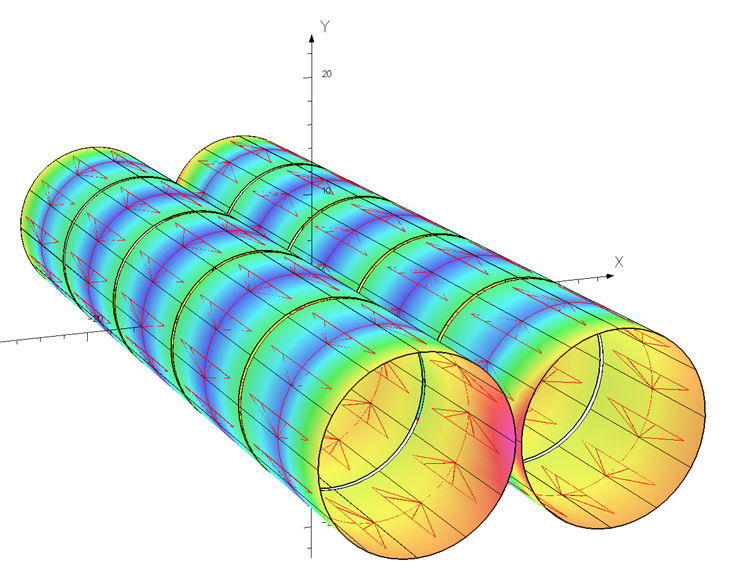}}
\caption{Magnetic Cavern Configuration}
\label{fig:Mag_cavern}
\end{figure}

\subsection{Conventional Room Temperature Magnets}

In order to get adequate field strength with tolerable power dissipation, conventional
room-temperature coils would have to be relatively thick.  We first considered coils of Al conductor
operating at 150 K.  We then determined the amount of conductor necessary to produce a
reference field of only 0.1 T.  In order to keep the current density at approximately 100A/cm$^2$,
10 layers of 1 cm$^2$ Al conductor would be required for our 15 m diameter $\times$ 15 m long
reference solenoid.  Using a \$20/kg cost for convention magnets~\cite{Green1}, the estimated cost 
for 1 solenoid is \$5M.  The power dissipation (assuming R=1 $\times$ 10$^{-8}$ Ohm-m) is
approximately 1 MW.  The operating costs for 1 MW of power would
be \$1.5M/year (based on typical US power costs).  The cost of the magnet system including 10
years of operation is then \$20M.  If one includes the cost of cooling the coils to 150 K, the costs
increase substantially.  Studies have shown~\cite{Green1} that there is little cost benefit to operating
non-superconducting (Al or Cu) coils at low temperature vs. room temperature.  If we consider that the
power dissipation at room temperature for Al coils triples (vs.~150 K operation), then the total magnet
cost increases to \$50M.

\subsection{Conventional Superconducting Coils}
Conventional superconducting solenoids are certainly an option for providing the large magnetic volumes
that are needed.  Indeed coils of the size we are considering were engineered (but never built) for the proposed
GEM experiment at the SSC.  A cylindrical geometry (solenoid) does imply that a fraction of the magnetic 
volume will be outside the volume of the active detector, which will likely be rectangular in cross section.  This is
certainly a disadvantage in terms of efficient use of the magnetic volume, but would provide 
personnel access paths to detector components inside the magnetic cavern.  It is certainly possible to consider
solenoids of rectangular cross section, and thus make more efficient use of the magnetic volume, but the engineering 
and manufacturing implications of this type of design have not been evaluated.

Technically, superconducting magnets of this size could be built, but the cost is not well known.
There have been a number of approaches to estimating the cost of a superconducting magnet and we will mention
two of those here.  The first comes from Green and St. Lorant \cite{Green:1993dg}.  They considered all the magnets that had 
been built at the time of their study (1993) and developed two formulas for extrapolating the cost of a superconducting
magnet: one scaling by stored energy and one scaling by magnetic volume times field.  They are given below:

\begin{equation}
C = 0.5(E_s)^{0.662}
\end{equation}
and
\begin{equation}
C = 0.4(BV)^{0.635}
\end{equation}
where $E_s$ is the stored energy in MJ, B is the field in Tesla, V is the volume in m$^3$ and C is the cost in M\$.  
The formulas given above give a cost for each 15 m diameter $\times$ 15 m long, 0.5T magnet of approximately
\$20M (based on E$_s$) and \$38M (based on magnetic volume). As another
reference point, we used the CMS coil \cite{Herve:2001kv} (B=4T, V=340 m$^3$, Stored energy = 2.7 GJ, ``As built cost"  = \$55M).
The Green and St. Lorant formulas give costs for the CMS magnet of  \$93M and \$41M based on stored energy and magnetic volume, respectively.
From these data we can make  ``Most Optimistic" and ``Most Pessimistic" cost extrapolations for our baseline NF solenoid.  The most optimistic
cost comes from using the formula, based on stored energy and assume that it over-estimates by a factor of 1.7 (93/55), based on the CMS
as built cost.  This gives a cost of  \$14M for each of our NF detector solenoids.  The most pessimistic cost extrapolation comes from using the
formula based on magnetic volume and conclude that it under-estimates the cost by a factor of 1.3 (55/41), based on the CMS as built cost.
This then gives a cost of  \$60M for each of our NF detector solenoids.  There is obviously a large uncertainty represented here.

Another extrapolation model was used by Balbekov {\it et al.}~\cite{Balbekov1} based on a model developed by A. Herve.  The extrapolation
formulae are given below:
\begin{equation}
P_0 = 0.33S^{0.8}
\end{equation}
\begin{equation}
P_E = 0.17E^{0.7}
\end{equation}
and
\begin{equation}
P = P_0 + P_E
\end{equation}
where P$_0$ is the price of the equivalent zero-energy magnet in MCHF, P$_E$ is the price of magnetization, and P is the total price.
S is the surface area (m$^2$) of the cryostat and E (MJ) is the stored energy.  This model includes the cost of power supplies, cryogenics
and vacuum plant.  From the above equations you can see that the model does take into account the difficulties
in dealing with size separately from magnetic field issues.  Balbekov et. al. used three ``as-builts" to derive the coefficients in the above
equations:

\begin{itemize}
\item ALEPH (R=2.65m, L=7m, B=1.5T, E=138MJ, P=\$14M)
\item CMS (R-3.2m, L=14.5m, B=4T, E=3GJ, P=\$55M)
\item GEM (R=9m, L=27m, B=0.8T, E=2GJ, P=\$98M)
\end{itemize}

The GEM magnet cost was an estimate based on a detailed design and engineering analysis.  Using this estimating model, we have for
one of the NF detector solenoids: P$_0$ = 0.33(707)$^{0.8}$ = 63MCHF,  P$_E$ = 0.17(265)$^{0.7}$ = 8.5MCHF.  The magnet
cost is thus approximately \$57M (which is close to our most pessimistic extrapolation given above).  One thing that stands out is that 
the magnetization costs are small compared to the total cost.  The mechanical costs involved with dealing with the large vacuum loading
forces on the vacuum cryostat assumed to be used for this magnet are by far the dominant cost.

\subsection{Low Temperature Non-Conventional Superconducting Coils}

In this concept we solve the vacuum loading problem of the cryostat by using the superconducting transmission line (STL)
that was developed for the Very Large Hadron Collider superferric magnets 
\cite{Ambrosio:2001ej}. The solenoid windings now consist
of this superconducting cable which is confined in its own cryostat.  Each solenoid consists of 150 turns 
and requires ~7500 m of cable. There is no large vacuum vessel and access to the detectors can be made through
the winding support cylinder, since the STL does not need to be close-packed in order to reach an acceptable
field. We have performed a simulation of the Magnetic Cavern concept using STL solenoids and the results are shown in
Fig.~121.  With the iron end-walls ( 1 m thick), the average field in the xz plane is approximately
\begin{figure}[htbp]
  \centering{
    \includegraphics[width=0.6\textwidth]{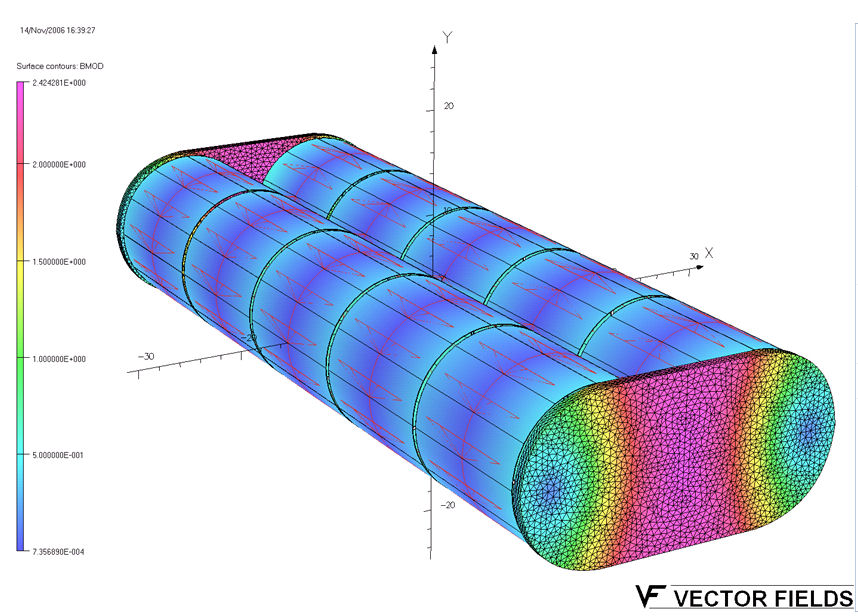}}
  \label{fig:Mag_cav_sim}
\caption{STL Solenoid Magnetic Cavern Simulation}
\end{figure}

0.58 T at an excitation current of 50 kA.   The maximum radial force is approximately 16 kN/m and the maximum axial force
approximately 40 kN/m.  The field uniformity is quite good with the iron end-walls and is shown in Fig.~122.

\begin{figure}[htbp]
  \centering{
    \includegraphics[width=0.6\textwidth]{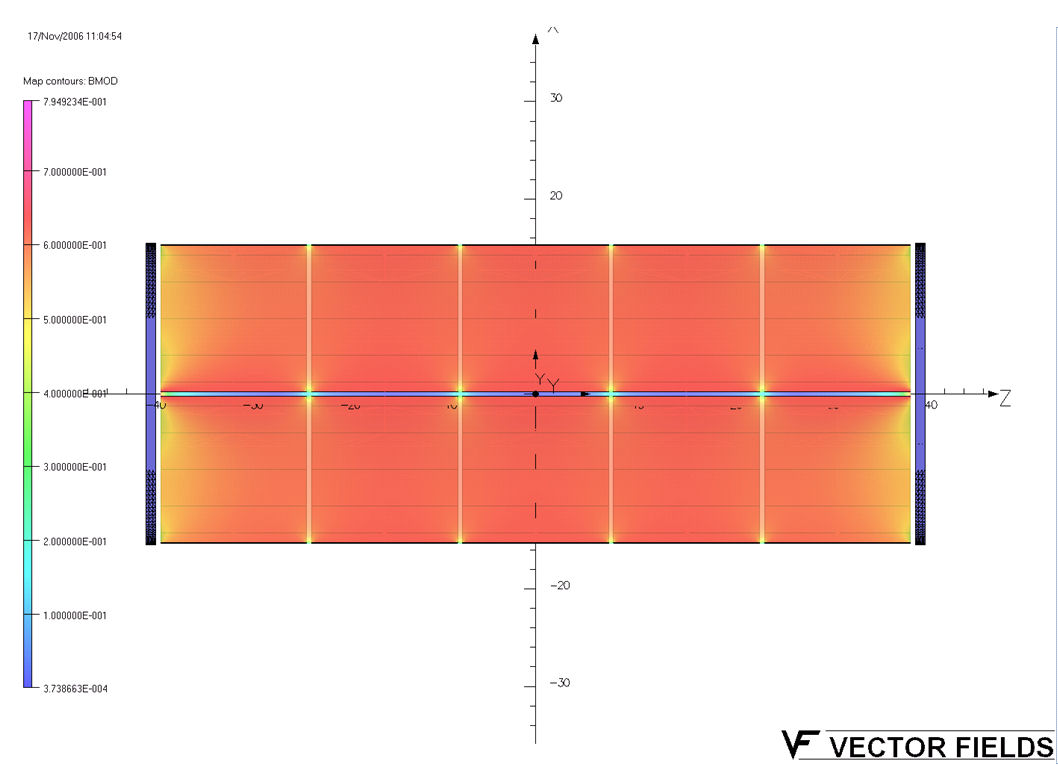}}
  \label{fig:Mag_cav_field_uni}
\caption{STL Solenoid Magnetic Cavern Field Uniformity in XZ plane}
\end{figure}

\subsection{Superconducting Transmission Line}

The superconducting transmission line (STL) consists of a superconducting cable inside a cryopipe cooled 
by supercritical liquid helium at 4.5-6.0 K placed inside a co-axial cryostat. It consists of a perforated Invar 
tube, a copper stabilized superconducting cable, an Invar helium pipe, the cold pipe support system, a thermal 
shield covered by multilayer super-insulation, and the vacuum shell. One of the possible STL designs developed 
for the VLHC is shown in Fig.~123.

\begin{figure}[htbp]
  \centering{
    \includegraphics[width=0.6\textwidth]{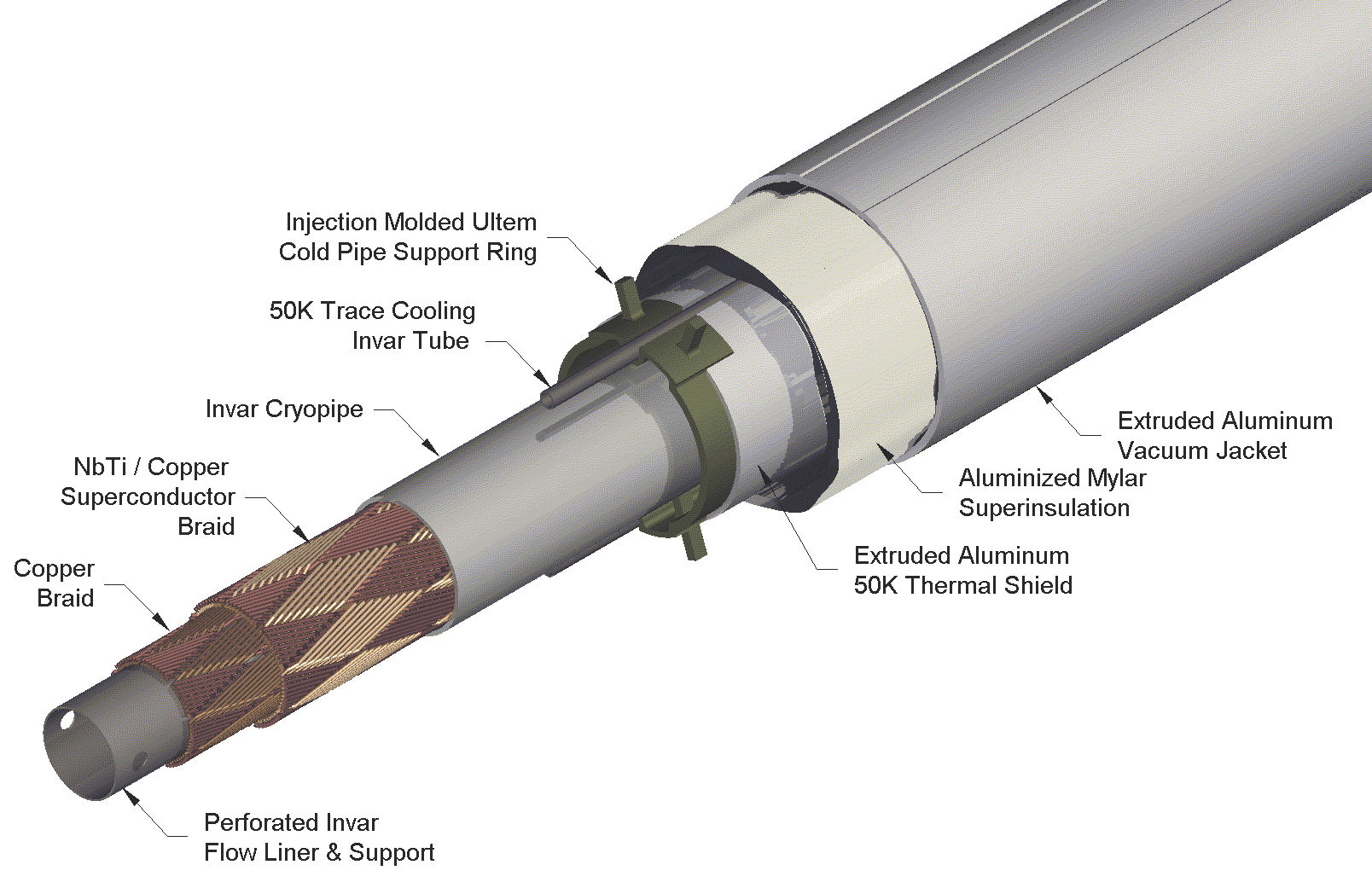}}
  \label{fig:STL}
\caption{Superconducting transmission line}
\end{figure}

The STL is designed to carry a current of 100 kA at 6.5 K in a magnetic field up to 1 T. This provides a 50\% 
current margin with respect to the required current in order to reach a field of 0.5T.  This operating margin
can  compensate for  temperature variations, mechanical or other perturbations in the system. 
The superconductor for the STL could be made in the form of braid or in the form of a two-layer spiral winding using 
Rutherford cable. The braid consists of 288 NbTi SSC-type strands 0.648 mm in diameter and arranged in 
a pattern of two sets of  24 crossing bundles with opposite pitch angle about the tube. A conductor made 
of Rutherford cables consists of 9 NbTi cables that were used in the SSC dipole inner layer. A copper braid is placed inside 
the superconductor to provide additional  current carrying capability during a quench. The conductor is
sandwiched between an inner perforated Invar pipe, which serves as a liquid helium channel, and an outer 
Invar pressure pipe that closes the helium space. Both braided and spiral-wrapped conductors and the 10 
cm long splice between them have been successfully tested with 100 kA transport current within the R\&D program
for the VLHC.  The STL has a 2.5-cm clear bore which is sufficient for the liquid helium flow in a loop up to 10 km in length. 
This configuration allows for cooling each  solenoid with continuous helium flow coming from a helium distribution box. 

The thermal shield is made of extruded aluminum pipe segments, which slide over opposite ends of each 
support spider. The 6.4-mm diameter Invar pipe is used for 50 K pressurized helium. It is placed in 
the cavities at the top and the bottom of both the shield and the supports. The shield is wrapped with 
40 layers of a dimpled super insulation. The vacuum shell is made of extruded aluminum or stainless steel.
Heat load estimates for the described STL are:

\begin{itemize}
\item Support system: 53 mW/m at 4.5 K and 670 mW/m at 40 K
\item Super insulation: 15 mW/m at 4.5 K and 864 mW/m at 40K
\end{itemize}

The estimated cost of the described STL is  approximately \$500/m. Further STL design optimization will be required to adjust the structure to the fabrication and operating conditions of the desired detector solenoids and to optimize its fabrication and operational cost.

\subsection{Conclusions}

Magnetizing volumes large enough to contain upwards of 1kT of LAr or totally active scintillator at fields up to 0.5T with the use of the STL concept would appear to be possible, but would require dedicated R\&D to extend the STL developed for the VLHC to this application.  It eliminates the cost driver of large conventional superconducting coils, the vacuum-insulated cryostat, and has already been prototyped, tested, and costed during the R\&D for the VLHC.  A full engineering design would still need to be done, but this technique has the potential to deliver the large magnetic volume required with a field as high as 1T, with very uniform field quality and at an acceptable cost.
\end{appendix}
\clearpage
\bibliographystyle{apsrev}
\bibliography{%
./00_Executive_summary/references_Executive_summary,%
./01_Overview/references_overview,%
./02_Motivation/02-01-Steriles/sterile-motivation,%
./02_Motivation/02-02-Neutrino-scattering/nscat_Motivation,%
./02_Motivation/02-03-RnD/RnD_references,%
./03_Facility/references_Facility,%
./03.5_Siting/references_site,%
./04_FarDetector/references_far,%
./05_NearDetector/references_near,%
./06_Performance/performance,%
./07_OnC/references_OnC,%
./08_Appendix/references_Appendix%
}
%
\end{document}